\documentclass[11pt,a4,oneside]{report}

\usepackage[ansinew]{inputenc}
\usepackage[T1]{fontenc}

\advance\topmargin by -2cm
\usepackage{float}
\usepackage[dvipdf]{graphicx}
\usepackage{enumerate}
\usepackage{amssymb}
\usepackage{comment}
\usepackage{makeidx}
\usepackage{graphicx}
\usepackage[plainpages=false, pdfpagelabels]{hyperref}
\usepackage[all]{hypcap}
\usepackage{fancyvrb}
\usepackage{subfig}
\usepackage{booktabs}

\includecomment{comm}

\usepackage{listings}

\usepackage{color}

\definecolor{bluekeywords}{rgb}{0.13,0.13,1}
\definecolor{greencomments}{rgb}{0,0.5,0}
\definecolor{redstrings}{rgb}{0.5,0,0}

\lstdefinelanguage{cyan}
{morekeywords={abstract, Any, Array, Boolean, boolean, break, Byte, byte, case, cast, Char, char,
const, default, delegate, Double, double, Dyn, each, else, end, enum,
extends, false, final, Float, float, for, func, heapalloc, if, implements,
import, in, Int, int, interface, it, let, local, Long, long, macro,
match, mixin,  mutable, Nil, object, of, overload, override,
package, private, protected, public, repeat, return, self, shared,
Short, short, slot, stackalloc, String, switch, true, type, until,
val, var, virtual, volatile, when, where,
while, with},
    keywordstyle=\color{bluekeywords},
  sensitive=false,
  morecomment=[l][\color{greencomments}]{//},
  morecomment=[s][\color{greencomments}]{{/*}{*/}},
  morestring=[b]",
  stringstyle=\color{redstrings}
}

\lstnewenvironment{cyan}[1][]{
    \lstset{
        basicstyle=\ttfamily,
        language=cyan,
        breaklines=true,
        numberstyle=\tiny\textbf,
        columns=fullflexible,
        numbers=none,
        showstringspaces=false,
        #1
    }
}{}

\lstnewenvironment{cyannum}[1][]{
    \lstset{
        basicstyle=\ttfamily,
        language=cyan,
        breaklines=true,
        numberstyle=\tiny\textbf,
        columns=fullflexible,
        numbers=left,
        showstringspaces=false,
        #1
    }
}{}

\makeindex

\title{The Cyan Language}
\author{José de Oliveira Guimarães\\
        Campus de Sorocaba da UFSCar\\
        Sorocaba, SP\\Brasil\\
        \href{mailto:jose@ufscar.br}{jose@ufscar.br}\\
        \href{josedeoliveiraguimaraes@gmail.com}{josedeoliveiraguimaraes@gmail.com} }

\begin{document}
\bibliographystyle{alpha}

\maketitle
\tableofcontents

\newcommand{\q} { \makebox[6ex]{} }
\newcommand{\p}[1] { \makebox[22ex][l]{#1} }
\newcommand{\rr} { \makebox[30ex]{} }
\newcommand{\nd}{\noindent}

\newpage

\chapter*{Foreword}

This is the manual of Cyan, a prototype-based statically-typed object-oriented language. The language introduces several novelties that make it easy to implement domain specific languages, blend dynamic and static code, reuse exception treatment, and do several other common tasks.
Although Cyan is an academic language, its design was made for programmers. Every aspect of it was designed to make programming fun.

The main novelties of Cyan are, in order of importance:
\begin{enumerate}[(a)]
\item the metaobject protocol, see the Thesis ``The Cyan Language Metaobject Protocol'' in the Cyan site, \url{http://www.cyan-lang.org};

\item generic objects with variable number of parameters, Chapter~\ref{generics};

\item an object-oriented exception handling system, Chapter~\ref{ehs};

\item context objects, Chapter~\ref{contextobjects};

\item static typing with optional dynamic typing, Chapter~\ref{dynamictyping};

\item clearly designed overloaded methods which are a restricted form of multi-methods, Section~\ref{multimethods};
\end{enumerate}


The address of the Cyan home page is  \url{http://www.cyan-lang.org}. Go to it for research articles and up-to-date changes in the language.

\newpage

\chapter{An Overview of Cyan}

Cyan is a statically-typed prototyped-based object-oriented language. As such, there is no class declaration. Objects play the role of classes and the cloning and {\tt new} operations are used to create new objects. Although there is no class, objects do have types which are used to check the correctness of message sending.
Cyan supports single inheritance, interfaces, generic prototypes, a completely object-oriented exception system, statically-typed anonymous functions, context objects (which are a generalization of anonymous functions), non-nullable types, optional dynamic typing, and a powerful metaobject protocol (MOP).

The Cyan MOP allows the control of the compilation process in many different ways making it relatively easy to produce and change code during the compilation. It is the biggest innotation of the language. The MOP is implemented through metaobject classes or prototypes. Only the main language package, cyan.lang, has around 80 of such classes (and growing). These metaobject classes are used, for example, to produce code for generic prototypes and to check arguments to methods. In particular, through a metaobject called {\tt grammarMethod} one can easily define Domain Specific Languages inside the Cyan code. And \textit{Concepts} \cite{Gregor:2006:CLS:1167515.1167499} for generic programming can be defined without any language support.

Although the language is based in prototypes, it is closer in many aspects to class-based languages like C++~\cite{Stroustrup:2013:CPL:2543987} and Java~\cite{Gosling:2014:JLS:2636997} than some prototype-based languages such as Self \cite{Ungar:1987:SPS:38807.38828} or Omega \cite{blaschek1994object}. For example, there is no workspace which survives program execution, objects have a {\tt new} method that creates a new object similar to another one (but without cloning it), and a Cyan program is given textually. In Omega, for example, a method is added to a class through the IDE. Cyan is close to Java and C++ in another undesirable way: its size is closer to these languages than to other prototype-based languages (which are usually small). However, several concepts were unified in Cyan therefore reducing the amount of constructs in relation to the amount of features. In particular, many constructs of other languages are implemented by methods in Cyan. For example, to throw an exception, to catch an exception, and to test if an object is of a certain prototype are all made through methods. We consider that Cyan, without the MOP, is not a big language. Since the MOP is not to be used extensively by regular programmers, we believe that learning the language will not be a problem.


In this Chapter we give an overview of the language highlighting some of its features.
An example of a program in Cyan is shown in Figures~\ref{a1} and \ref{a2}. The corresponding Java program is shown in Figure~\ref{je01}. It is assumed that there are classes {\tt In} and {\tt Out} in Java that do input and output (assume they are in package {\tt inOut}). The Cyan  program declares objects {\tt Person} and {\tt Program} (they should start with an upper-case letter). These objects are called {\it prototypes} to differentiate them from objects created during runtime. Object {\tt Person} declares a variable {\tt name} and methods {\tt getName} and {\tt setName}. Keywords {\tt var} and {\tt func} are used before a field (instance variable) and a method declaration.
Method {\tt getName} of {\tt Person} takes no argument and returns a {\tt String}. The return type appears after ``\verb|->|"\/. Inside a method, {\tt self} refers to the object that received the message that caused the execution of the method (same as {\tt self} of Smalltalk and {\tt this} of Java). The value returned by a method should appear after the keyword {\tt return}.
A package is a collection of prototypes (objects) and interfaces --- it is the same concept of Java packages and modules of other languages. All the public identifiers of a package become available after a ``{\tt import}"\/ declaration.

\begin{figure}
\begin{cyan}
package program

object Person
    private var String name = ""
    public  func getName -> String {
        return self.name
    }
    public  func setName: String name {
        self.name = name
    }
end
\end{cyan}
\caption{A Cyan program: {\tt Person}}
\label{a1}
\end{figure}

\begin{figure}
\begin{cyan}
package program

object Program
  public func run {
     var p = Person clone;
     var String name;
     name = In readLine;
     p setName: name;
     Out println: (p getName);
  }
end
\end{cyan}
\caption{A Cyan program: {\tt Program}}
\label{a2}
\end{figure}

\begin{figure}
\begin{verbatim}
package program;
import inOut;

private class Person {
  private String name;
  public String getName() {
      return this.name;
  }
  public void setName( String name ) {
      this.name = name;
  }
}

public class Program {
  public void run() {
     Person p;
     String name;
     p = new Person();
     name = In.readLine();
     p.setName(name);
     Out.println( p.getName() );
  }
}
\end{verbatim}
\caption{A Java program}
\label{je01}
\end{figure}

Prototype {\tt Program} declares a {\tt run} method. Assume that this will be the first method of the program to be executed. The first statement of {\tt run} is\\
\verb@     var p = Person clone;@\\
``{\tt Person clone}"\/ is the sending of message {\tt clone} to prototype {\tt Person}. ``{\tt clone}"\/ is called the ``{\it selector}"\/ \index{selector} \index{clone} of the message.
All objects have a {\tt clone} method.
This statement declares a variable called {\tt p} and assigns to it a copy of object {\tt Person}. The code\\
\verb@     var variableName = expr@\\
\nd declares a variable with the same compile-time type as {\tt expr} and assigns the result of {\tt expr} to the variable. The type of {\tt expr} should have been determined by the compiler using information of previous lines of code.

The next line,\\
\verb@     var String name;@\\
declares {\tt name} as a variable of \index{String} type {\tt String}. This is also considered a statement. In\\
\verb@     name = In readLine;@\\
there is the sending of message {\tt readLine} to prototype {\tt In}, which is an object used for input. Statement\\
\verb@     p setName: name;@\\
is the sending of message ``\verb@setName: name@"\/ to the object referenced to by {\tt p}. The message selector is ``{\tt setName:}"\/ and ``{\tt name}"\/ is the argument.
Finally\\
\verb@     Out println: (p getName);@\\
is the sending of message ``\verb@println: (p getName)@"\/ to prototype {\tt Out}. The message selector is ``{\tt println:}"\/ and the argument is the object returned by ``\verb@p getName@"\/.

The parameters\footnote{In this manual, we will use parameter and argument as synonymous.} that follow a selector in a method declaration may be surrounded by {\tt (} and {\tt )}. So method {\tt setName:} could have been declared as\\
\verb@    public func setName: (String name) { self.name = name }@\\
\nd This is allowed to increase the legibility of the code.

\section*{Definition and Declaration of Variables}

Statement\\
\verb@     var p = Person clone;@\\
could have been defined as\\
\verb@     var Person p;@\\
\verb@     p = Person clone;@\\   \index{variable}
Variable {\tt p} is declared in the first line and its type is  the prototype {\tt Person}. When an object is used where a type is expected, as in a variable or parameter declaration, it means ``the type of the object"\/.  By the type rules of Cyan, explained latter, {\tt p} can receive in assignments objects whose types are {\tt Person} or subprototypes of {\tt Person} (objects that inherit from {\tt Person}, a concept equivalent to inheritance of classes).

\section*{Inheritance}   \index{inheritance}

The type system of Cyan is close to that of Java although the first language does not support classes.
There are interfaces, single inheritance, and implementation of interfaces. The inheritance of prototype {\tt Person} from {\tt Worker} is made with the following syntax:
\begin{cyan}
object Worker extends Person
   private String company
   // other fields (instance variables) and methods
end
\end{cyan} \index{override}

If a method is redefined in a subprototype (be it public or protected), keyword ``{\tt override}"\/ should appear just after ``{\tt public}"\/ or ``{\tt protected}"\/. Methods of the subprototype may call methods of the superprototype using keyword {\tt super} as the message receiver:\\
\nd \verb@    super name: "anonymous"@\\

In order to a prototype to be inherited, its declaration must be preceded by identifier ``{\tt open}'' as in
\begin{cyan}
open
object Person
    // elided
end
\end{cyan}

Cyan has runtime objects, \index{new} \index{clone} created with {\tt new} and {\tt clone} methods, and objects such as {\tt Person}, {\tt Program}, and {\tt Worker}, which are created before the program execution. To differentiate them, most of the time the last objects will be called {\it prototypes}. However, when no confusion can arise, we may call them {\it objects} too.

It is important to bear in mind the dual role of a prototype in Cyan: it is a regular object when it appears in an expression and it is a type when it appears as the type of a variable, parameter, or return value of methods.

\section*{Interfaces}

Interfaces are similar to those of Java. One can write  \index{interface}
\begin{cyan}
interface Savable
    func save
end

open
object Person
    func init: String name, Int age {
        self.name = name;
        self.age = age
    }
    func getName -> String = name;
    func setName: String name { self.name = name }
    func getAge -> Int = age;
    func setAge: Int age { self.age = age }
    var String name
    var Int age
end

object Worker extends Person implements Savable
    private String company
    func save {
        // save to a file
    }
    ... // elided
end
\end{cyan}
Here prototype {\tt Worker} should implement method {\tt save}. Otherwise the compiler would sign an error. Unlike Java, interfaces in Cyan are objects too. They can be passed as parameters, assigned to objects, and receive messages.

\section*{Values}

The term ``variable"\/ in Cyan is used for local variable, field (instance variable or attribute), and parameter. A variable in Cyan is a reference to an object. The declaration of the variable does not guarantee that an object was created. To initialize the variable one has to use a literal object such as {\tt 1}, {\tt 3.1415}, \verb@"Hello"@, or to created an object with {\tt clone} or {\tt new}.

Object {\tt String} \index{String} is a pre-defined object for storing sequences of characters. A literal string can be given enclosed by \verb@"@ as usual: \verb@"Hi, this is a string"@, \verb@"um"@, \verb@"ended by newline\n"@.



\section*{Any}
\index{Any}
All prototypes in Cyan but {\tt Nil} inherit from prototype {\tt Any} which has some basic methods such as {\tt eq:} \index{eq:} (reference comparison), {\tt ==} (is the content equal?), {\tt asString}, and methods for computational reflection (get object information, get metadata). Method\\
\verb@     func eq: Any other -> Boolean@\\
\nd tests whether {\tt self} and {\tt other} reference the same object. Method {\tt ==} is equal to {\tt eq:}  by default but it should be redefined by the user.\footnote{For union types, {\tt ==} has a different behavior than {\tt eq:}.} Method {\tt eq:} cannot be redefined in subprototypes. Method {\tt neq:} retorns the opposite truth value of {\tt eq:}

\section*{Basic Types}

\index{Byte} \index{Short} \index{Int} \index{Long} \index{Float} \index{Double} \index{Char} \index{Boolean} \index{Nil} \index{String}
Cyan has the following  basic types:
{\tt Byte}, {\tt Short}, {\tt Int}, {\tt Long}, {\tt Float}, {\tt Double}, {\tt Char}, {\tt Boolean}, {\tt Nil}, and {\tt String} (no {\tt Void} prototype). Since Cyan will be targetted to the Java Virtual Machine, the language has one type for each of the basic types of Java except for {\tt void}. Unlike Java, all basic types in Cyan but {\tt Nil} inherit from prototype {\tt Any}. Therefore there are not two separate hierarchies for basic types, that obey value semantics,\footnote{The declaration of a variable allocates memory for the ``object"\/. Variables really contains the object, it is stack allocated. In reference semantics, variables are pointers. Objects are dynamically allocated.} and all other types, which obey reference semantics.

Methods {\tt eq:} and {\tt ==} of all basic types have the same semantics: they return true if the contents of the objects are equal:
\begin{cyan}
var Int I = 1;
var Int J = 1;
if I == J && I eq: J {
    Out println: "This will be printed"
}
\end{cyan}
Since the basic prototypes cannot be inherited, the compiler is free to implement basic types as if they obey value semantics. That is, a basic type {\tt Int} could be translated to {\tt int} of Java.\footnote{The compiler being built translates Cyan to Java.} There are cases in which this should not be done:
\begin{enumerate}[(a)]
\item when a basic type variable is passed as parameter to a method that accepts type {\tt Any}:
\begin{cyan}
object IntHashTable
    func key:  String aKey  value: Any aValue { ... }
    ...
end
...
IntHashTable key: "one" value: 1;
\end{cyan}
In this case the compiler will create a dynamic object of prototype {\tt Int} for the {\tt 1} literal;

\item when a basic type variable receives a message that correspond to a method of {\tt Any} such as {\tt prototypeName}:
\begin{cyan}
    // prints "Int"
Out println: (1 prototypeName);
\end{cyan}

But even in this case the compiler will be able to optimize the code since it knows which method should be called.

\end{enumerate}

In practice, the compiler could implement basic types as the basic types of Java almost all of the time. The overhead should be minimal.


\section*{{\tt Nil} and Union types}

There is a special type in the language, the union type. The type \verb@A|B@ is considered, in assignments and parameter passing, as a supertype of both {\tt A} and {\tt B}.
\begin{cyan}
var Int|String x;
x = 0; // ok
x = "Cyan"; // ok
\end{cyan}
The compiler automatically casts objects of {\tt Int} and {\tt String} to {\tt x}. To retrive the object stored in {\tt x} it is necessary to use the {\tt type} command:
\begin{cyan}
type x
    case Int n {
        Out println: "twice is " ++ 2*n
    }
    case String s {
        Out println: "first char is" ++ s[0]
    }
\end{cyan}
Inside first {\tt case} clause, ``\verb|case Int n|"\/, {\tt n} has type {\tt Int} and the value of {\tt x}. The same applies to the second {\tt case} and {\tt s}.

There is a special object called {\tt Nil} which is not subtype or supertype of anything. It somehow plays the role of
{\tt nil} of Smalltalk, {\tt NULL} of C++, and {\tt null} of Java/C\#. As in Smalltalk, \index{Nil} {\tt Nil} knows how to answer some messages --- it is an object. However, {\tt Nil} can only be assigned to a variable of type {\tt Nil}.

{\tt Nil} cannot be assigned to a variable whose type is a prototype that is not {\tt Nil} or an union.
\begin{cyan}
var String s;
var Person p;
s = Nil; // compile-time error
p = Nil; // compile-time error
\end{cyan}
To allow {\tt Nil} values in variables it is necessary to declare an union of {\tt Nil} with at least a regular prototype.
\begin{cyan}
var Nil|String s;
s = Nil; // ok
s = In readLine;  // readLine returns a String
type s
    case Nil {
        // in case s is Nil
    }
    case String s2 {
        // s2 is a String here
    }
\end{cyan}
Then the runtime error ``message send to {\tt Nil}"\/ does not happens in Cyan. Caveat: the compiler currently does not check if a variable is initialized or not before used. Then the Java compiler may signal a compile-time error ``variable was not initialized"\/ or there may be a run-time error ``Null pointer exception"\/. But this is a flaw of the compiler, not of the language.

A method that does not return anything can be declared as returning {\tt Nil}. It is equivalent to declare {\tt Nil} as the return value and do not declare a return value.

\section*{Constructors and Inheritance}

\index{constructor} \index{init}
Constructors have the name {\tt init} or {\tt init:} and may have any number of parameters. The return value should not be supplied (not even {\tt Nil}). For each method named {\tt init} or {\tt init:} the compiler adds to the prototype a method named {\tt new} or {\tt new:} with the same parameter types. Each {\tt new} or {\tt new:} method creates an object and calls the corresponding {\tt init} or {\tt init:} method. If the prototype does not define any {\tt init} or {\tt init:} methods, the compiler supplies an empty {\tt init} method that does not take parameters and calls the superprototype {\tt init} method (if any. If this method does not exist, a compiler error occurs). Consequently, a {\tt new} method is created too.

A subprototype should call one of the {\tt init} methods of the superprototype (if one was defined by the user) using keyword {\tt super}. This call should be the first statement of the method:
\begin{cyan}
open
object Person
    func init: String name { self.name = name }
    private String name
    ...
end

object Worker extends Person
    func init: String name, String job {
            // this line is demanded
        super init: name;
        self.job = job;
    }
    private String job
    ...
end
\end{cyan}
All {\tt new} methods return an object of the prototype. Therefore, {\tt Person} has a method\\
\nd  \verb@    Person new: String name@\/\\
\nd and {\tt Worker} has a method\\
\nd  \verb@    Worker new: String name, String job@\/\\
\nd To make it easy to create objects, there is an alternative way of calling methods {\tt new} and {\tt new:}.\\
\verb@    @ {\tt P(p1, p2, ..., pn)} is a short form for\\
\nd \verb@    (P new: p1, p2, ..., pn)@\\
\nd Therefore we can write either\\
\nd \verb@    var prof = Worker("John", "Professor")@\\
\nd or\\
\nd \verb@    var prof = Worker new: "John", "Professor"@\\
\nd Of course, if a prototype {\tt P} has a {\tt new} method that does not take parameters we can write just ``{\tt P()}"\/ to create an object.

\section*{Keyword Messages and Methods}

\index{method} \index{message send} \index{keyword message} \index{message!keyword} \index{selector}
Cyan was initially based on Smalltalk. As a result, it supports keyword messages, a message with multiple keywords as in\\
\verb@    var p = Point dist: 100.0 angle: 20.0;@\\
Like Smalltalk, Cyan calls {\tt dist:} and {\tt angle:} keywords. Following Smalltalk terminology, {\tt dist:angle:} is called a ``selector''.

Method calls become documented without further effort. Prototype {\tt Point} should have been declared as
\begin{cyan}
object Point
    func dist: (Float newDist) angle: (Float newAngle) -> Point {
        var p = self clone;
        p dist: newDist;
        p angle: newAngle;
        return p
    }
    public Float dist
    public Float angle
end
\end{cyan}
Unlike Smalltalk, after a single keyword there may be multiple parameters:
\begin{cyan}
object Quadrilateral
    func p1: (Int x1, Int y1)
               p2: (Int x2, Int y2)
               p3:  Int x3, Int y3
               p4:  Int x4, Int y4    {
        self.x = x1;
        ...
        self.y4 = y4
    }
    ...
    private Int x1, y1, x2, y2, x3, y3, x4, y4
end
...
var r = Quadrilateral p1: 0, 0    p2: 100, 10
                   p3: 20, 50  p4: 120, 70;
\end{cyan}
This example declares the parameters after the keywords in all possible ways. By ``keyword'' we mean method keywords, not Cyan reserved keywords.

We call the ``name of a method"\/ the concatenation of all of its keywords, each one followed by its number of parameters and a white space. The trailing white space should be removed. For example, methods
\begin{cyan}
    func key: (String aKey) value: (Int aValue) -> String
    func name: (String first, String last)
         age: (Int aAge)
         salary: (aSalary Float) -> Worker
\end{cyan}
\nd have names \verb|"key:1 value:1"| and \verb|"name:2 age:1 salary:1"|.

\section*{Abstract Prototypes}

\index{abstract}
An abstract prototype should be declared with keyword {\tt abstract} and it may have zero or more public abstract methods:
\begin{cyan}
public abstract object Shape
    public abstract func draw
end
\end{cyan}
An abstract prototype does not have any {\tt new} methods even if it declares {\tt init} methods. Abstract methods can only be declared in abstract objects. A subprototype of an abstract object may be declared abstract or not. However, if it does not define the inherited abstract methods, it must be declared as abstract too.

To call an object ``abstract"\/ seem to be a contradiction in terms since ``objects"\/ in prototype-based languages are concrete entities. However, this is no more strange than to have ``abstract"\/ classes in class-based languages: classes are already an abstraction. To say ``abstract class"\/ is to refer to an abstraction of an abstraction.

\section*{Final Prototypes and Methods}

A prototype whose declaration is not preceded by ``{\tt open}'' cannot be inherited. It is a {\tt final} prototype.
\begin{cyan}
object Int
    ...
end
...
object MyInt extends Int
    ...
end
\end{cyan}
There would be a compile-time error in the inheritance of the final prototype {\tt Int} by {\tt MyInt}.

A final method cannot be redefined. This allows the compiler to optimize code generation for these methods.
\begin{cyan}
open
object Person
    final func name -> String { return _name }
    final func name: String newName { _name = newName }
    ...
end
...
var Person p;
...
p name: "Peter"; // static call
\end{cyan}

\section*{Decision and Loop Methods and Statements}

The {\tt if} statement takes a boolean expression and is always followed by a sequence of statements between {\tt \{} and {\tt \}}. It is not necessary to put parentheses around the boolean expression. The {\tt else} part is optional. The {\tt while} statement also takes a boolean expression and a sequence of statements between {\tt \{} and {\tt \}}.
\begin{cyan}
if n%2 == 0 {
    s = "even"
}
else {    // the else part is optional
    s = "odd"
};
var i = 0;
while i < 5 {
    Out println: i;
    ++i
}
\end{cyan}

The {\tt repeat-until} command executes its statements until the expression is {\tt true}:
\begin{cyan}
var Int sum = 0;
var Int n = 1;
repeat
    sum = sum + n;
    ++n;
until n >= 4;
assert sum == 6;
\end{cyan}

Cascaded if´s are possible:
\begin{cyan}
if age < 3 {
    s = "baby"
}
else if age <= 12 {
    s = "child"
}
else if age <= 19 {
    s = "teenager"
}
else {
    s = "adult"
};
\end{cyan}

\section*{Cyan Symbols}

There is a special form of literal strings that start with {\tt \#} called simply ``symbol''.
\index{Symbol} \index{CySymbol}
A symbol one starts by a  {\tt \#} followed, without spaces, by letters, dot, digits, underscores, and ``{\tt :}''\/, starting with a letter or digit. These are valid symbols in Cyan:
\begin{cyan}
#name
#name:
#at:put:
#1
#711
\end{cyan}
The type of a symbol is {\tt String}.
\begin{cyan}
var String s;
s = #at:put:;
   // prints at:put:
Out println: s;
s = #7;
assert s == "7" && #at:put: == "at:put:";
\end{cyan}

\section*{Overloading}
There may be methods with the same method keywords but with different number of parameters and parameter types. For example, one can declare  \index{overloaded methods} \index{method!overloading}
\begin{cyan}
object Printer
    func print: Document d { ... }
    func print: String s, String form -> Boolean { ... }
    func print: Float s, Int beforeDot, Int afterDot { ... }
end
\end{cyan}
All of these methods are considered different. They can have different return value types, they have in fact different names. This is not true overloading.

True method overloading happens when the method keywords and the number of parameters are equal but the parameter types are different:
\begin{cyan}
object MyBlackBoard

    overload  // keyword that prefixes an overloaded method
    func draw: Square f   { ... }
    func draw: Triangle f { ... }
    func draw: Circle f   { ... }
    func draw: Shape f    { ... }
    private String name
end
\end{cyan}
There are four {\tt draw} methods that are considered different by the compiler. In a message send\\
\verb@     MyBlackBoard draw: fig@\\
\nd the runtime system searchs for a {\tt draw} method in prototype {\tt MyBlackBoard} in the textual order in which the methods were declared. It first checks whether {\tt fig} references a prototype which  is a subprototype from {\tt Square} (that is, whether the prototype extends {\tt Square}, assuming {\tt Square} is a prototype and not an interface). If it is not, the searches continues in the second method,\\
\verb@     draw: Triangle f@\\
\nd and so on. If an adequate method were not found in this prototype, the search would continue in the superprototype. Since all {\tt draw:} methods have the same number of parameters, it is necessary to prefix the first method with the keyword ``\verb|overload|"\/. This will be explained later.

\section*{Subtyping and Method Search}

\index{subtype}
The definition of subtyping in Cyan considers that prototype {\tt S} is a subtype of {\tt T} if {\tt S} inherits from {\tt T} (in this case {\tt T} is a prototype) or if {\tt S} implements interface {\tt T}. An interface {\tt S} is a subtype of interface {\tt T} if {\tt S} extends {\tt T}. This is a pretty usual definition of subtyping.

In the general case, in a message send\\
\nd \verb@    p draw: fig@\\
\nd the algorithm searches for an adequate method in the object the variable {\tt p} refer to and then, if this search fails, proceeds up in the inheritance hierarchy.
Suppose {\tt C} inherits from {\tt B} that inherits from {\tt A}. Variable {\tt x} is declared with type {\tt B} and refers to a {\tt C} object at runtime. Consider the message send\\
\verb@    x first: expr1 second: expr2@\\
At runtime a search is made for a  method of object {\tt C} such that:
\begin{enumerate}[(a)]
\item the method has keywords {\tt first:} and {\tt second:} and;
\item keyword {\tt first:} of the method takes a single parameter of type {\tt T} and the runtime type of {\tt expr1} is subtype of {\tt T}. The same applies to keyword {\tt second:} and {\tt expr2};
\end{enumerate}
The methods are searched for in object {\tt C} in the {\it textually} declared order. The return value type is not considered in this search. If no adequate method is found in object {\tt C}, the search continues at object {\tt B}. If again no method is found, the search continues at object {\tt A}. \index{runtime search}

The compiler makes almost exactly this search with just one difference: the search for the method starts at the declared type of {\tt x}, {\tt B}.

This unusual runtime search for a method is used for two reasons:
\begin{enumerate}[(a)]
\item it can be employed in dynamically-typed languages. Cyan was designed to allow a smooth transition between dynamic and static typing.  Cyan will not demand the declaration of types for variables (including parameters and excluding fields). After the program is working, types can be added. The algorithm that searches for a method described above can be used in dynamically and statically-typed languages;
\item it is used in the Cyan exception system. When looking for an exception treatment, the textual order is the correct order to follow. Just like in Java/C++/etc in which the catch clauses after a try block are checked in the order in which they were declared after an exception is thrown in the try block.
\end{enumerate}
The programmer should be aware that to declare two methods such that:
\begin{enumerate}[(a)]
\item they have the same keywords and;
\item for each keyword, the number of parameters is the same.
\end{enumerate}
will make message send much slower than the normal.

Methods that differ only in the return value type cannot belong to the same prototype. Then it is illegal to declare methods \verb@id -> Int@ and \verb@id -> String@ in the same prototype (even if one of them is inherited).

In the redefinition of a method in a subprototype, one can change the return value type of the subprototype method. This type can be a subprototype of the type of the return value of the method of the superprototype:
\begin{cyan}
open
object Animal
    func matchWhat -> Animal { ... }
end

object Cow extends Animal
       // ok, Cow is subprototype of Animal
    func matchWhat -> Cow { ... }
end
\end{cyan}

The search for a method in Cyan makes the language supports a kind of multi-methods. The linking ``message"\/-``method"\/ considers not only the message receiver but also other parameters of the message (if they exist). Unlike many  object-oriented languages, the parameter types are inspected at runtime in order to discover which method should be called.

\section*{Arrays and Maps}

Array \index{array} prototypes are declared using the syntax: \verb@Array<A>@ in which {\tt A} is the array element type. Only one-dimensional arrays are supported. A literal array object is created using \verb@[ element list ]@, as in the example:
\begin{cyan}
var n = 5;
var anIntArray = [ 1,  2, (Math sqr: n) ];
var Array<String> aStringArray;
aStringArray = [ "one", "t" ++ "wo" ];
\end{cyan}
This code creates two literal arrays. {\tt anIntArray} will have elements {\tt 1}, {\tt 2}, and {\tt 25}, assuming the existence of a {\tt Math} prototype with a {\tt sqr} method (square the argument). And {\tt aStringArray} will have elements \verb@"one"@ and \verb@"two"@. The array objects are always created at runtime. So a loop
\begin{cyan}
1..10 foreach: { (: Int i :)
   Out println: [ i-1,  i,  i + 1 ]
}
\end{cyan}
Creates ten different literal arrays at runtime.
The type of a literal array is {\tt Array<A>} in which {\tt A} is the type of the {\it first} element of the literal array. Therefore\\
\verb@     var fa = [ 1.0, 2, 3 ];@\\
declares {\tt fa} as a {\tt Array<Double>}. Since there is no automatic convertion of values, this results in a compile-time error: the compiler will not cast {\tt 2} to {\tt Double}.

Literal maps are defined in the following way:
\begin{cyan}
    let IMap<Int, String> map = [ 0 -> "zero", 1 -> "one", 2 -> "two" ];
    cast elem = map[0] {
        assert elem == "zero";
    }
\end{cyan}
{\tt map[0]} returns the element associated with {\tt 0}. Its type is a union of {\tt String} and {\tt Nil}. Statement {\tt cast} should be used to access the result. In the example, {\tt elem} has type {\tt String} and the {\tt assert} is only executed if {\tt map[0]} is a {\tt String}. It would be {\tt Nil} if {\tt map} did not associate a string to {\tt 0}.

\section*{Dynamic Typing}  \index{dynamic typing}

Although Cyan is statically-typed, it supports some features of dynamically-typed languages. A message send whose keywords are preceded by  {\tt ?} is not checked at compile-time. That is, the compiler does not check whether the static type of the expression receiving that message declares a method with those keywords. For example, in the code below, the compiler does not check whether prototype {\tt Person} defines a method with keywords {\tt name:} and {\tt age:} that accepts as parameters a {\tt String} and an {\tt Int}.
\begin{cyan}
var p Person;
...
p ?name: "Peter"  ?age: 31;
\end{cyan}  \index{message!non-checked}
This non-checked message send is useful when the exact type of the receiver is not known:
\begin{cyan}
    func printArray: Array<Any> anArray  {
        anArray foreach: { (: elem Any :)
            elem ?printObj
        }
    }
\end{cyan}
The array could have objects of any type. At runtime, a message {\tt printObj} is sent to all of them. If all objects of the array implemented a {\tt Printable} interface, then we could declare parameter {\tt anArray} with type \verb@Array<Printable>@. However, this may not be the case and the above code would be the only way of sending message {\tt printObj} to all array objects.

The compiler does not do any type checking using the returned value of a dynamic method. That is, the compiler considers that \\
\nd \verb@    if obj ?get { ... }@\\
\nd is type correct, even though it does not know at compile-time if {\tt obj ?get} returns a boolean value.

Dynamic checking with {\tt ?} plus the reflective facilities of Cyan can be used to create objects with dynamic fields. Object {\tt DTuple} of the language library allows one to add fields dynamically:

\begin{cyan}
var t = DTuple new;
  // add field "name" to t
t ?name: "Carolina";
   // prints "Carolina"
Out println: (t ?name);
   // if uncommented the line below would produce a runtime error
//Out println: (t ?age);
t ?age: 1;
   // prints 1
Out println: (t ?age);
   // if uncommented the line below would produce a
   // **compile-time** error because DTuple does not
   // have an "age" method
Out println: (t age);
\end{cyan}
Here fields {\tt name} and {\tt age} were dynamically added to object {\tt t}.

Type {\tt Dyn} is a virtual type used for dynamic typing. A variable of type {\tt Dyn} can receive in assignments an expression of any type. And an expression of type {\tt Dyn} can be assigned to a variable of any type. All message sends to an expression of type {\tt Dyn} is considered correct by the compiler.

\section*{Expressions in Strings}

In a string, a \verb@$@ not preceded by a \verb@\@ should be followed by a valid identifier. The identifier should be a parameter, local variable, or field of the current object. The result is that the identifier is converted at runtime to a string (through the {\tt asString} method) and concatenated to the string. Let us see an example:
\begin{cyan}
var name = "Johnson";
var n = 3;
var Float johnsonSalary = 7000.0F;
Out println: "Person name = $name, n = $n, salary = $johnsonSalary";
\end{cyan}
This code prints \\
\verb!     Person name = Johnson,  n = 3, salary = 7000.0!\\
\nd The last line is completely equivalent to
\begin{cyan}
Out println: "Person name = " ++ name ++ ", n = " ++ n ++
    ", salary = " ++ johnsonSalary;
\end{cyan}

\section*{Generic Prototypes}

Cyan also supports generic prototypes in a form similar to other languages but with some important differences. First, a family of generic prototypes may share a single name but different parameters. For example, there is a single name {\tt Tuple} that is used for tuples of any number of parameters (as many as there are in the library): \index{generic prototype}  \index{Tuple}
\begin{cyan}
var Tuple<String> aName;
var Tuple<String, Int>  p;
aName f1: "Lívia"
   // prints Lívia
Out println: (aName f1);
p f1: "Carol"
p f2: 1
   // prints "name: Carol  age: 1". Here + concatenates strings
Out println: "name: " ++ p f1 ++ "  age: " ++ p f2;
\end{cyan}
Second, it is possible to used field names as parameters:
\begin{cyan}
var Tuple<name, String> aName;
var Tuple<name, String, age, Int> p;
aName name: "Lívia"
   // prints Lívia
Out println: (aName name);
p name: "Carol"
p age: 1
   // prints "name: Carol  age: 1"
Out println: "name: " ++ p name ++ "  age: " ++ p age;
\end{cyan}
A generic prototype is considered different from the prototype without parameters too:
\begin{cyan}
object Box
    public Any value
end

object Box<T>
    public T value
end
...
var giftBox = Box new;
var intBox  = Box<Int> new;
\end{cyan}

\index{literal!tuple}
A unnamed literal tuple is defined between \verb"[." and \verb".]" as in
\begin{cyan}
var p = [. "Lívia", 4 .];
Out println: (p f1), " age ", (p f2);
   // or
var Tuple<String, Int> q;
q =  [. "Lívia", 4 .];
\end{cyan}
A named literal tuple demands the name of the fields:
\begin{cyan}
var p = [. name = "Lívia", age = 4 .];
Out println: p name ++ " age " ++ p age;
   // or
var Tuple<name, String, age, Int> q;
\end{cyan}

\section*{Anonymous Functions}

\index{block} \index{closure} \index{anonymous functions} \index{functions}
Cyan supports statically-typed anonymous functions, which are called blocks in Smalltalk. An anonymous function is a literal object that can access local variables and fields. It is delimited by {\tt \{} and {\tt \}} and can have parameters which should be put between \verb|(:| and \verb|:)| as in:
\begin{cyan}
var b = { (: Int x :) ^x*x };
   // prints 25
Out println: (b eval: 5);
\end{cyan}
Here \verb@{ (: Int x:) ^x*x }@ is a function with one {\tt Int} parameter, {\tt x}. The return value of the function is the expression following the symbol ``\verb@^@"\/. The return value type may be omitted in the declaration --- it will be deduced by the compiler.  This function takes a parameter and returns the square of it. A function is an literal object with a method {\tt eval} or {\tt eval:} (if it has parameters as the one above). The statements given in the function can be called by sending message {\tt eval} or {\tt eval:} to it, as in ``{\tt b eval: 5}"\/.

A function can also access a local variable:
\begin{cyan}
var y = 2;
var b = { (: Int x :) ^ x + y };
   // prints 7
Out println: (b eval: 5);
\end{cyan}

As full objects, functions can be passed as parameters:
\begin{cyan}
object Loop
    func until:  (Function<Boolean> test) do:  (Function<Nil> b)  {
        b eval;
        (test eval) ifTrue: { until: test do: b }
    }
end
...

// prints "i = 0", "i = 1", ... without the "s
var i = 0;
Loop until: { ^ i < 10 } do: {
    Out println: "i = $i";
    ++i
}
\end{cyan}
Here prototype {\tt Loop} defines a method {\tt until:do:} which takes as parameters
a function that returns a Boolean value (\verb@Function<Boolean>@) and a
function that returns nothing (\verb@Function<Nil>@). The second function
is evaluated until the first function evaluated to {\tt false} (and at least one time).
Notation \verb@"i = $i"@ is equivalent to \verb@("i = " ++ i)@.
Note that  both functions passed as parameters to method {\tt until:do:} use the
local variable {\tt i}, which is a local variable.

Functions are useful to iterate over collections. For example,
\begin{cyan}
var v = [ 1, 2, 3, 4, 5, 6 ];
    // sum all elements of vector v
var sum = 0;
v foreach:  { (: Int x :) sum = sum + x };
\end{cyan}
Method {\tt foreach:} of the array {\tt v} calls the function (as in ``{\tt b eval: 5}"\/) for each of the array elements. The sum of all elements is then put in variable {\tt sum}.

Sometimes we do not want to change the value of a local variable in a function. In these cases, we should use a constant instead of a variable or make a copy of the variable in the function.
\begin{cyan}
var y = 2;
var b = { (: Int x :)
        // make a copy of z
    var z = y;
    var sum = 0;
    while z > 0 {
        sum = sum + z;
        --z;
    };
    ^ x + sum;
};
(v eval: 5) println;

// make sure k is not changed
let k = 0;
var f = { (: Int n :)
    var sum = 0;
    var i = 0;
    while i < k {
        sum = sum + i;
        ++i;
    };
    ^ n + sum;
};

\end{cyan}

There are methods that can play the role of statements {\tt if} and {\tt while}.
\begin{cyan}
( n%2 == 0 ) ifTrue: { s = "even" } ifFalse: { s = "odd" };
var i = 0;
{^ i < 5 } whileTrue: {
    Out println: i;
    ++i
}
\end{cyan}
Anonymous functions in Cyan cannot have return statements. Then the functions that are parameters to methods {\tt ifTrue:ifFalse:} and {\tt whileTrue:} cannot have return statements.

\section*{Context Objects}

\index{context object} \index{object!context}
Context objects are a generalization of functions and internal (or inner) classes. Besides that, they allow a form of language-C-like safe pointers. The variables external to the function are made explicit in a context object, freeing it from the context in which it is used. For example, consider the function\\
\verb@     { (: Int x :) sum = sum + x }@\\
\nd It cannot be reused because it uses external (to the function) variable {\tt sum} and because it is a literal object. Using context objects, the dependence of the function to this variable is made explicit:
\begin{cyan}
   // Function<Int, Nil> is a function that takes an Int
   // as parameter and does not return anything
object Sum(Int &sum) extends Function<Int, Nil>
     func eval: Int x {
          sum = sum + x
     }
end

...
   // sum the elements of array v
var s = 0;
v foreach: Sum(s)
\end{cyan}
Context objects may have one or more parameters given between {\tt (} and {\tt )} after the object name. These correspond to the variables that are external to the function ({\tt sum} in this case). This context object implements interface
\verb@Function<Int, Nil>@ which represents functions that take an {\tt Int} as a parameter and returns nothing. Method {\tt eval:} contains the same code as the original function.  In line\\
\verb@     v foreach: Sum(s)@\\
\nd expression ``{\tt Sum(s)}"\/ creates at runtime an object of {\tt Sum} in which {\tt sum} represents the same variable as {\tt s}. When another object is assigned to  {\tt sum} in the context object, this same object is assigned to {\tt s}. It is as if {\tt sum} and {\tt s} were exactly the same variable.

Prototype {\tt Sum} can be used in other methods making the code of {\tt eval:} reusable. Reuse is not possible with functions because they are literals. Context objects can be generic, making them even more useful:
\begin{cyan}
object Sum<T>(T &sum) extends Function<T, Nil>
     func eval: T x {
          sum = sum + x
     }
end

...
   // sum the elements of array v
var v = [ 3.14, 2.71, 1.557 ];
var String s;
v foreach: Sum<String>(s);
\end{cyan}
Now context object {\tt Sum} is used to sum the {\tt Double} elements of vector {\tt v}.

A context-object parameter not preceded by {\tt \&} mean that it is a {\it copy parameter}. That means changes in the context-object parameter are not propagated to the real argument:

\begin{cyan}
object Sum(Int sum) extends Function<Int, Nil>
     func eval: Int x {
          sum = sum + x
     }
end

...
   // do not sum the elements of array v
var s = 0;
v foreach: Sum(s);
assert s == 0;
\end{cyan}
Macro {\tt assert} checks whether its argument returns {\tt true}. It issues a warning if not. In this example, the final value of {\tt s} will be {\tt 0}.

Parameters whose types are  preceded by {\tt \&} are called {\it reference parameters} (see first example).

Context objects are a generalization of both functions and nested objects, a concept similar to nested or inner classes. That is, a class declared inside other class that can access the fields and method of it. However, class {\tt B} declared inside class {\tt A} is not reusable with other classes. Class {\tt B} will always be attached to {\tt A}. In Cyan, {\tt B} may be implemented as a context object that may be attached to an object {\tt A} (that play the role of class {\tt A}) or to any other prototype that has fields of the types of the parameters of {\tt B}. Besides that, both referenced parameters and field parameters implement a kind of language-C like pointers. In fact, it is as if the context-object parameter were a pointer to the real argument:
\begin{verbatim}
   // C
int *sum;
int s = 0;
sum = &s;
*sum = *sum + 1;
   // value of s was changed
printf("%d\n", s);
\end{verbatim}

\section*{Grammar Methods}

\index{message!grammar} \index{method!grammar} \index{grammar method}
A method declared with selector {\tt s1:s2:} can only be called through a message send {\tt s1: e1 s2: e2} in which {\tt e1} and {\tt e2} are expressions. Grammar methods do not fix the selector of the message send. Using operators of {\it regular expressions} a grammar method may specify that some keywords can be repeated, some are optional, there can be one or more parameters to a given keyword, there are alternative keywords and just one of them can be used.

Two methods that take a variable number of {\tt Int} arguments are declared in prototype {\tt IntSet}. Each method is preceded by a metaobject annotation {\tt grammarMethod} (to be seen later). This annotation has a code of a Domain Specific Language between \verb|{*| and \verb|*}|.
\begin{cyan}
package grammar

object IntSet
    @grammarMethod{*
        (add: (Int)+)
    *}
    func addMany: Array<Int> array {
        for elem in array {
            set add: elem
        }
    }


    @grammarMethod{*
        (addEach: Int)+
    *}
    func addManyKeywords: Array<Int> array {
        for elem in array {
            set add: elem
        }
    }

    override
    func asString -> String = set asArray asString;

    let Set<Int> set = Set<Int>();

end
\end{cyan}
In the code of the first annotation, the {\tt +} after {\tt (Int)} indicates that after {\tt add:} there may be {\it one or more} {\tt Int} arguments:
\begin{cyan}
IntSet add: 0, 2, 4;
var odd = IntSet new;
odd add: 1, 3;
\end{cyan}

In the second method, {\tt addManyKeywords}, the code in the annotation {\tt grammarMethod} is different. The {\tt +} appears after the keyword {\tt addEach:} with the {\tt Int} parameter type. That means the keyword may be repeated:
\begin{cyan}
IntSet addEach: 0 addEach: 2 addEach: 4;
var odd = IntSet new;
odd addEach: 1 addEach: 3;
\end{cyan}

The annotated method, {\tt addMany} and {\tt addManyKeywords} in the example, should have a single parameter that matches the code of the DSL of the annotation {\tt grammarMethod}. The rules for calculating the type of this parameter are given in Chapter~\ref{grammarmethods}.

A grammar method may use all of the regular expression operators: {\tt A+} matches one or more {\tt A}´s, {\tt A*} matches zero or more {\tt A}´s, {\tt A?} matches {\tt A} or nothing ({\tt A} is optional), \verb"A | B" matches {\tt A} or {\tt B} (but not both), and \verb"A B" matches {\tt A} followed by {\tt B}. The \verb"|" operator may be used with types:
\begin{cyan}
    @grammarMethod{*
        (add: (Int|String)+)
    *}
    func addUnion: Array<Int|String> array {
        for elem in array {
            set add: elem
        }
    }

\end{cyan}
Method {\tt addUnion:} may receive as parameters a list of  {\tt Int}s and {\tt String}s.

Grammar methods are useful for implementing Domain Specific Languages (DSL). In fact, every grammar method can be considered as implementing a DSL. The advantages of using grammar methods for DSL are that the lexical and syntactical analysis and the building of the Abstract Syntax Tree are automatically made by the compiler. The parsing is based on the grammar method. The AST of the grammar message is referenced by the single parameter of the grammar method.

There is one problem left: grammar methods are defined using regular expression operators. Therefore they can only parse regular languages. Some languages that are not regular can be defined by using more than one grammar method. 

Another example is a domain specific language  for commanding a car.
\begin{cyan}
package grammar

object Car
    @grammarMethod{*
        (do:
           (on: | off: | left: | right: | move: Int)+
        )
    *}
    func carPlay: Tuple< Any,
                         Array< Union<f1, Any, f2, Any, f3, Any, f4, Any, f5, Int> >
                   >   t -> String {

        var s = "";
        for elem in t f2 {
            type elem
                case Any f1 { s = s ++ "car on " }
                case Any f2 { s = s ++ "car off " }
                case Any f3 { s = s ++ "car left " }
                case Any f4 { s = s ++ "car right " }
                case Int f5 { s = s ++ "car move($f5) " }
        }
        return s
   }

end
\end{cyan}
The car obeys commands related to movement such as
to turn left, turn right, move n centimeters, turn on, and off. The method does nothing but in a real setting it could, for example, send commands to a real remote controlled car:

\begin{cyan}
Car on:
    left:
    move: 100
    right:;

Car on:
    move: 200
    left:
    move: 50
    off;
\end{cyan}
These two message would cause the call of the same grammar method, {\tt carPlay}.

The uses of grammar methods are endless. They can define optional parameters, methods with variable number of parameters, and mainly DSL´s. One could define methods for SQL, XML (at least part of it!), parallel programming, graphical user interfaces, any small language. It takes minutes to implement a small DSL, not hours.

\section*{Methods as Objects}

\index{object!method} \index{method!object}
Method {\tt functionForMethod:} of prototype {\tt Any} allows one to consider methods as objects.
\begin{cyan}
object MySet
    func add: String elem { }
    ...
end
\end{cyan}
Method {\tt functionForMethod:} takes the name of a method and returns a anonymous function that represents the method. That is, when message {\tt eval} or {\tt eval:} is sent to the function, the method is called.
\begin{cyan}
    let Array<String> strArray = Array<String>();
    let Function<String, Nil> addMethod = strArray functionForMethod: "add:1";
    addMethod eval: "A";
    addMethod eval: "B";
    assert strArray size == 2;
\end{cyan}
Of course, {\tt addMethod} could be passed as a parameter. That simulates the passing of a method as parameter.

The ability of referring to a method is very useful in graphical user interfaces as the example below shows.
\begin{cyan}
object MenuItem
    func onMouseClick:  Function<Nil> b {
        ...
    }
end

object Help
    func show { ... }
    ...
end

object FileMenu
   func open { ... }
end

var helpItem = MenuItem new;
helpItem onMouseClick: (Help functionForMethod: "show" );
var openItem = MenuItem new;
openItem onMouseClick: (FileMenu functionForMethod: "open");
...
\end{cyan}

There may even exist a table containing methods as functions. Assume {\tt Test} declares methods {\tt add10:}, {\tt twice:}, and {\tt cube:}, each one taking an {\tt Int} and returning an {\tt Int}.
\begin{cyan}
        let Test t = Test();
        let IMap<String, Function<Int, Int>> methodMap = [
            "add10" -> (t functionForMethod: "add10:1"),
            "twice" -> (t functionForMethod: "twice:1"),
            "cube"  -> (t functionForMethod: "cube:1")
            ];
        cast f = methodMap["twice"] {
            assert f eval: 3 == 6;
        }
        cast f = methodMap["add10"] {
            assert f eval: 3 == 13;
        }
        cast f = methodMap["cube"] {
            assert f eval: 3 == 27;
        }
\end{cyan}

\section*{The Exception Handling System}

\index{exception handling system} \index{exception}
The exception handling system of Cyan was based on that of language Green~\cite{Jose:GreenSite:Online} \cite{Guimaraes:2006:GL:1646590.1646595}. However, it has important improvements when compared with the EHS of this last language. Both are completely object-oriented, contrary to all systems of languages we know of. An exception is thrown by using statement {\tt throw} that takes an exception object as parameter. The type of the exception object should be subprototype of {\tt CyException}.

An exception is caught using a {\tt try-catch-finally} statement.
\begin{cyan}
var age Int;
try
    age = In readInt;
    if age < 0 {
        throw ExceptionNegAge(age)
    }
catch { (: ExceptionNegAge e :)
   Out println: "Age ", e age, " is negative"
};
\end{cyan}
Here exception {\tt ExceptionNegAge} is thrown by \\
\nd \verb@    throw ExceptionNegAge(age)@\\
\nd in which ``\verb@ExceptionNegAge(age)@"\/ is a short form of ``\verb@(ExceptionNegAge new: age)@"\/.

After a {\tt catch} {\it clause} there should be an expression whose type declares at least one {\tt eval:} method that takes a parameter whose type is subprototype of {\tt CyException}. In this example, the anonymous function defines a method
\begin{verbatim}
  func eval: ExceptionNegAge
\end{verbatim}
Assume {\tt ExceptionNegAge} inherits from {\tt CyException}.
\begin{cyan}
   // '@init(age)' creates a constructor with
   // parameter age
@init(age)
open
object ExceptionNegAge extends CyException
    @property Int age
end
\end{cyan}

In this specific case, when the exception is thrown, \index{throw} \index{catch} \index{CyException} the control is transferred to the function given after {\tt catch}. The error message is then printed.

This example in Java would be
\begin{verbatim}
int age;
try {
    age = In.readInt();
    if ( age < 0 )
        throw new ExceptionNegAge(age);
} catch ( ExceptionNegAge e ) {
    System.out.println("Age " + e.getAge() + " is negative");
}
\end{verbatim}
There may  be as many {\tt catch} clauses as necessary, each one taking a single expression.
\begin{cyan}
var Int age;
try
    age = In readInt;
    if age < 0 {
        throw ExceptionNegAge(age);
    }
    else if age > 127 {
        throw ExceptionTooOldAge(age)
    }
catch { (: ExceptionNegAge e :)
   Out println: "Age ", e age, " is negative"
}
catch { (: ExceptionTooOldAge e :)
   Out println: "Age ", e age, " is out of limits"
};
\end{cyan}

The {\tt catch} expression may result in an object  with more than one {\tt eval:} method, each of them accepting one parameter whose type is subprototype of {\tt CyException}. So the following code is legal.
\begin{cyan}
var Int age;
try
    age = In readInt;
    if age < 0 {
        throw ExceptionNegAge(age)
    }
    else if age > 127 {
        throw ExceptionTooOldAge(age)
    }
catch ExceptionCatchAge;
\end{cyan}
Consider that {\tt ExceptionCatchAge} is
\begin{cyan}
object ExceptionCatchAge
    overload
    func eval: ExceptionNegAge e {
        Out println: "Age ", e age, " is negative"
    }
    func eval: ExceptionTooOldAge e {
        Out println: "Age ", e age, " is out of limits"
    }
end
\end{cyan}
This new implementation produces the same results as the previous one. When an exception {\tt E} is thrown in the function that reads the age, the runtime system starts a search in the object of the {\tt catch} clause, which is {\tt ExceptionCatchAge}. It searches for an {\tt eval:} method  that can accept {\tt E} as parameter in the textual order in which the methods are declared. This is exactly as the search made after a message send.
The result is exactly the same as the code with two functions passed as parameters to two {\tt catch} clauses.

The exception handling system of Cyan has several advantages over the traditional approach: exception treatment can be reused, {\tt ExceptionCatchAge} can be used in many places, exception treatment can be organized in a hierarchy ({\tt ExceptionCatchAge} can be inherited and some {\tt eval:} methods can be overridden. Other methods can be added), the EHS is integrated in the language (it is also object-oriented), one can use metaobjects with the EHS, and there can be libraries of treatment code. For short, all the power of object-oriented programming is brought to exception handling and treatment. Since the Cyan EHS has all of the advantages of the EHS of Green, the reader can know more about its features in an article by Guimarães \cite{journals/cj/Guimar04}.

\section*{Metaobjects}

\index{metaobject}
Compile-time metaobjects are objects that can change the behavior of the program, add information to it, or inspect the source code.  A compile-time metaobject is activated by a metaobject annotation that may appear before a prototype, a method, a field, a local variable, as a statement of a method, as an expression, etc.

A metaobject annotation starts with {\tt @} followed by the metaobject name:
\begin{cyan}
@checkStyle
object University
   @log
   func name -> String { return uName }
   ...
end
\end{cyan}
Metaobject {\tt checkStyle} is activated at compile-time in the first line of this example. It is attached to specific points of the compiler controlling the compilation of prototype {\tt University}. It could check whether the prototype name, the method names, the field names, and local variables follow some conventions for identifiers (prototype in lower case except the first letter, method keywords in lower case).
The compiler calls methods of the metaobject at some points of the compilation. It is as if the metaobject was added to the compiler. Which method is called at which point is defined by the Meta-Object Protocol (MOP). \index{Meta-Object Protocol} \index{MOP}

Metaobject {\tt log} would add code to the start of the method to log how many times it was called. This information would be available to other parts of the code. Again, a metaobject does not return anything. It is an object. What happens is that a method of the metaobject, not specified in the code, is called and it returns something.

\chapter{The Compiler, Packages, and File organization}  \label{compiler} \label{packages}   \index{package} \index{file}

This Chapter describes how the Cyan source files of a program are organized and how the compiler should be called. To explain that we need to define some terms.
We will call ``program unit"\/ a  prototype declaration or interface. Every source file is a ``compilation unit"\/ and may contain one prototype declaration. Then a compilation unit is a ``source file"\/ containing one prototype.

A Cyan program is divided in compilation units, program units, and packages that keep the following relationship:
\begin{enumerate}[(a)]
\item  Keyword {\tt public} may precede the program unit to indicate that it is public . A ``public"\/ program unit of a package {\tt pp} is visible in all compilation units that import {\tt pp}.
    Future versions of the language will allow visibility ``package"\/.
    A ``package"\/ program unit of a package {\tt pp} is only visible in the compilation units of {\tt pp}.
    Every file, with extension .cyan, declare exactly one program unit.  \index{program unit} Private program units are not currently supported by Cyan.
\begin{cyan}
object Person
   String name
   ... // methods
end
\end{cyan}
If no qualifier is used before ``{\tt object}"\/, then it is considered public.

The prototypes (which includes interfaces) defined in a source file hide any prototypes imported in the source file. So it is legal to define a prototype {\tt Test} in a source file and import another prototype {\tt Test} from a package.

\item every file should begin with a package declaration as ``{\tt package ast}"\/ in
\begin{cyan}
package ast

object Variable
   var String name
   var Type type
   ... // methods
end
\end{cyan}
The prototype declared in that file will belong to the package ``{\tt ast}"\/;

\item a package is composed by program units spread in one or more source files.
The name of a package can be composed by identifiers separated by ``."\/. All the source files of a
package should be in the same directory. The source files of a package  id1.id2. ... idn should be in a
directory idn which is a sub-directory of id(n-1), and so on. There may be packages {\tt id1.id2} and {\tt id1.id3} that share a directory {\tt id1}. Although a directory is shared, the packages are unrelated to each other. In a package id1.id2. ... idn, each idi should start with a lower-case letter.

\end{enumerate}

In the directory of a package the compiler and metaobjects may store files used by the compiler, the Cyan Metaobject Prototocol, and possibly some tools such as the IDE. These files may keep information on the source files and they link past and future compilations. This mechanism is called ``link past-future (LPF)''.  See Section ``Special Package Directories'' in the Thesis ``The Cyan Language Metaobject Protocol'' available in www.cyan-lang.org.

The information stored in the LPF files can be used to catch errors at compile time that would otherwise go undetected or to improve current error messages. Based on the information, the compiler could check:
\begin{enumerate}[(a)]
\item if the textual order of declaration of overloaded methods\footnote{Methods with the same name but different parameter types in each keyword.} was changed;
\item if methods were added to an overloaded method;

\item if a compilation unit (source code) changed between compilations in such a way that the changes were prohibit by a metaobject used in the previous version of the compilation unit. This is a research topic (await!);

\item the introduction of a new local variable may change the semantics of a method that accessed  a field with that same name.

\end{enumerate}

The ``info files"\/ could also store information collected at runtime. The compiler could insert code that checks, for example, if the numbers stored in {\tt Int} variables are dangerously near the limits allowed by this type. In the next compilation the programmer would receive a warning that {\tt Int} should be changed to a library prototype ``{\tt BigInt}"\/ or {\tt long}.


A package is a collection of prototype declarations and interfaces. Every Cyan prototype declared as {\tt object ObjectName ... end} must be in a file called ``{\tt ObjectName.cyan}"\/. Preceding the object declaration there must appear a package declaration of the form {\tt package packageName} as in the example given above.

Program units defined in a package {\tt packB} can be used in a source file of a package {\tt packA} using the import \index{import} declaration:\\
\begin{cyan}
package packA
import packB

object Program
   func run {
      ...
   }
end
\end{cyan}
The public program units of package {\tt packB} are visible in the whole source file. A program unit declared in this source file may have the same name as an imported program unit. The local one takes precedence. \verb|';'| is optional after the package name and the import list.

More than one package may be imported; that is, the word {\tt import} may be followed by a list of package names separated by commas. It is legal to import two packages that define two resources (currently, only prototypes) with the same name. However, to use one identifier (program unit) imported from two or more packages it is necessary to prefix it with the package name. See the example below.
\begin{cyan}
package pA

import pB, pC, pD

object Main
   func doSomething {
      var pB.Person p1;  // Person is an object in both packages
      var pD.Person p2;
      ...
   }
end
\end{cyan}
This same rule applies when package {\tt pA} and {\tt pB} define resources with the same name.

An object or interface can be used in a file without importing the package in which it was defined. But in this case the identifier should be prefixed by the package name:\\
\verb@    var v = ast.Variable;@\\
\verb@    var gui.Window window;@\\

There is a package called {\tt cyan.lang} which is imported automatically by every file. This package defines all the basic types, arrays, prototype {\tt System}, function objects, tuples, unions, etc. See Chapter~\ref{ilo}.

\label{pyan}

A program is described by a file with extension ``{\tt pyan}"\/. This file contains code of a Domain Specific Language called Pyan (Project cYAN) whose grammar follows.

\vspace{4ex}

\p{Program} ::=  \{ ImportList \} [ CTMOCallList ] ``program''\/ [ AtFolder ] \\
\rr [ ``main''\/ QualifId ]\\
\rr \{ \{ ImportList \} CTMOCallList Package  \}

\p{ImportList} ::= ``import''\/  QualifId AtFolder

\p{Package} ::= ``package''\/ QualifId [ AtFolder ]

\p{AtFolder} ::= ``at''\/ FileName

\p{CTMOCallList} ::= \{ CTMOCall \}

\p{CTMOCall} ::= ``@''\/ Id\\
\rr [ ``(''\/ ExprLiteral [ ``,''\/ ExprLiteral ] ``)''\/ ] \\
\rr [ LeftCharString TEXT RightCharString ]

\p{QualifId} ::= \{ Id ``.''\/ \} Id

\vspace{2ex}

Some itens are not described in the grammar: LeftCharString, QualifId, RightCharString, TEXT, and FileName.  LeftCharString is any sequence of the symbols \\
\begin{verbatim}
= ! ? $ % & * - + ^ ~ / : . \  | (  [  {  <
\end{verbatim}
Note that \verb">", \verb")", \verb"]", and \verb"}" are missing from this list.
RightCharString is any sequence of the same symbols of LeftCharString but with
\verb">", \verb")", \verb"]", and \verb"}" replacing
\verb"<", \verb"(", \verb"[", and \verb"{", respectively.
The compiler will check if the closing RightCharString of a LeftCharString is the inverse of it.

QualifId is a sequence of one or more Cyan identifiers separated by ``.''\/. An identifier is a sequence of letters, digits, underscore starting with a letter or underscore. The underscore alone is not considered an identifier. An identifier ending with two or more underscores is illegal (\verb|"name__"| is not valid).
TEXT is any text. It may include any character but end-of-file. FileName is a string with a file name. The character ``\verb|\|''\/ or ``\verb|/|''\/ is used to separate directories (folders). Any one of these characters may be used. ``\verb|\x|''\/ is not considered a escape character for any {\tt x}. Then a FileName can be \\
\verb|      "C:\Cyan Material\lib\cyan\lang"|

As an example, a program in Pyan could be\footnote{This could be a program in Pyan. But since metaobject {\tt option} is not yet implemented, it is not.}
\begin{verbatim}
import styles.default at "C:\Cyan\my"
@checkStyle
@option(addQualifier)
program at "C:\Cyan\example01"
    main main.Program
    package main at "C:\Dropbox\Cyan\cyanTests\general-0002\main"
    @option(no_dynamic)
    package bank at "C:\Cyan\tests\bank"
    package cyan.util at "C:\Cyan\cyan\util"
    package account
    package database at "database\util"
\end{verbatim}

Keyword {\tt program} starts the project. Optionally ``at''\/ specifies the path of the program. If not specified, the {\it default project directory} is that in which the project file is.
After it keyword {\tt main} may appear. It specifies the full path of the main prototype; that is, its package ``.''\/ its name. The execution starts in method {\tt run} of this prototype. If not specified, execution will start at prototype {\tt Program} of package {\tt main}.  After {\tt program} or {\tt main} (if present), there should appear one or more packages descriptions.

A package description  is keyword {\tt package}, the package name, and optionally ``{\tt at}''\/ followed by a string with the package directory. If this directory is not absolute, it is considered to be relative to the project directory. All source files of a package should be in the same directory. In the above example, package {\tt database} should be in directory\\
\verb|    C:\Cyan\example01\database|\\
\noindent
The compiler considers that the package is in a sub-directory whose name is the package name with ``.''\/ replaced by \verb|/| or \verb|\| (it depends on the separator the operating system uses). In Windows, a package {\tt cyan.util} should be in a directory (folder) ``\verb|cyan\util|''\/.
This directory is the one specified by ``{\tt at}''\/

As an example, package {\tt cyan.util} is in directory\\
\noindent \verb|    C:\Cyan\cyan\util|\\
\noindent In the above Pyan file, no directory is specified for package {\tt account}. Therefore it  is in the directory of the program in a sub-directory with the name of the package:
\noindent \verb|    C:\Cyan\example01\account|\\

The program and the packages may be preceded by zero or more metaobject annotations. These are of the for \verb|@meta| in which {\tt meta} is the metaobject name. These calls may have parameters and an attached text (as \verb|@meta(param){* text *}|). See the thesis on the Cyan MOP in the Cyan site for more details. In particular, the compiler options should be parameters of a metaobject {\tt options}.

{\tt ImportList} is a list of packages whose metaobjects are imported by this project file. The package should be in the directory that follows ``{\tt at}''\/. For example, package ``{\tt styles.default}''\/ should be in directory \\
\verb|    C:\Cyan\my\styles\default|\\
\noindent
Only the metaobjects are imported. These metaobjects may be used in this file. Then {\tt checkStyle} could be in package ``{\tt styles.default}''\/. Or it could be in the cyan.lang directory, which is always imported.

In a project file, the compiler considers that every directory of the program directory corresponds to a package, including sub-directories. The project file may use the {\tt package} keyword to include one package of the program directory, but that is optional. As an example, suppose file ``{\tt p.pyan}''\/ is in directory\\
\verb|    C:\Dropbox\Cyan\cyanTests\master|\\
\noindent which has the following directory tree:
\begin{verbatim}
    C:\Dropbox\Cyan\cyanTests\master
    C:\Dropbox\Cyan\cyanTests\master\generic
    C:\Dropbox\Cyan\cyanTests\master\generic\ga
    C:\Dropbox\Cyan\cyanTests\master\main
    C:\Dropbox\Cyan\cyanTests\master\shape
\end{verbatim}
File ``{\tt p.pyan}''\/ has the contents
\begin{verbatim}
program
    @checkStyle
    package shape
\end{verbatim}
Since there is no ``{\tt at}''\/ after ``{\tt program}''\/, the compiler considers that the program directory is\\
\verb|    C:\Dropbox\Cyan\cyanTests\master|\\
\noindent The compiler consider that the program has the packages {\tt generic}, {\tt generic.ga}, {\tt main}, and {\tt shape}. It was necessary to declare explicitly package {\tt shape} in ``{\tt p.pyan}''\/ because there is a metaobject annotation preceding it.

Appendix~\ref{thecompiler} shows how the compiler should be called, the compiler options, etc.

\section{Future Enhancements}

In the directory of a package there should be zero or more Cyan source files or source files of Domain Specific Languages. A file name ``Name.cyan''\/ should contain a public or package prototype {\tt Name}. There are special rules for names of source files with generic prototypes --- see Chapter~\ref{generics}.
A file with extension ``{\tt .syan}''\/ is a script file of a language called \index{ScriptCyan} ScriptCyan. This language has a slightly different grammar from Cyan. The source file does not start with the package declaration. It may start with import declarations. After that there are two options:
\begin{enumerate}[(a)]
\item statements that usually are inside a method;
\item methods, shared fields, and fields that usually are inside a prototype declaration. In this case there should be at least one method and the first statement after the import declarations (if any) should be a method. Since there is no ``{\tt object}''\/ keyword in the source, the user will have the impression that the methods are procedures and functions (not related to a prototype).
\end{enumerate}
In case (a) the compiler will add a prototype declaration, import package {\tt script}, and will insert the statements in a method \verb|run: Array<String> args|. Of course, parameter {\tt args} can be used inside the statements. As an example, suppose the code below is in a file called ``{\tt PrintArgs.syan}''\/ of a directory ``\verb|C:\Cyan\tests\myTest\argsTest|''\/. The project file informs  that the program directory is ``\verb|C:\Cyan\tests|''\/. Therefore the compiler deduces that the package name is ``{\tt myTest.argsTest}''\/. The compiler will transform code
\begin{cyan}
args foreach: { (: String s :)
    s println
};
\end{cyan}
into the code
\begin{cyan}
package myTest.argsTest
import script;

open
object PrintArgs extends ScriptCyan
    func run: Array<String> args {
        args foreach: { (: String s :)
            s println
        };
    }
end
\end{cyan}
Package {\tt script} will contain several prototypes for scripting. It has not been defined yet.

In case (b), the compiler will insert the methods, shared fields, and fields into a prototype that has the same name and package as in case (a).

It is expected that prototype {\tt ScriptCyan} declares several methods to make it easy to  build script files. These methods are not yet defined.

The file name of a ScriptCyan source file may contain a symbol ``{\tt -}''\/ followed by ``{\tt s}''\/ and a prototype name possibly preceded by a package:\\
\verb|     PrintArgs-s-mypack-MyScript.syan|\\
\nd In this case prototype {\tt PrintArgs} will have ``{\tt mypack.MyScript}''\/ as superprototype. {\tt mypack} should be a package of the project. Language {\tt ScriptCyan} is not supported by the current Cyan compiler yet.

\chapter{Basic Elements}

This chapter describes some  basic facts on Cyan such as identifiers, number literals, strings, operators, and statements (assignment, loops, etc). First of all, the program execution starts in a method called {\tt run} (without parameters or return value) or \\
\nd \verb@    run: Array<String>@\\
\nd of a prototype specified at compile-time through the ``{\tt .pyan}''\/ file (the project file). Method {\tt run} cannot be a true overloaded method (Section~\ref{overload}) or be inherited.
Type \verb@Array<String>@ is an array of strings. The arguments to {\tt run} are those passed to the program when it is called. In this text (all of it) we usually call {\tt Program} the prototype in which the program execution starts. But the name can be anyone. The program that follows prints all arguments passed to it when it is called.
\begin{cyan}
package main

object Program
    func run: Array<String> args {
        args foreach: { (: String elem :)
            Out println: elem
        }
    }
end
\end{cyan}

\section{Identifiers}

Identifiers \index{identifier} should be composed by letters, numbers, and underscore and they should start with a letter or underscore. However, an identifier without letters or numbers, only with underscores, is not valid. An identifier cannot end with two underscores too (like  ``\verb|one__|''\/ or ``\verb|one___|''\/). Upper and lower case letters are considered different.
\begin{cyan}
var Int _one;
var Long one000;
var Float ___0;
\end{cyan}
It is expected that the compiler issues a warning if two identifiers visible in the same scope differ only in the case of the letters as ``{\tt one}''\/ and ``{\tt One}''\/.

There is a restriction on identifiers in Cyan: prototype names should start with an upper-case letter and variables (local variables, parameters, and fields) names should start with a lower-case letter or underscore. For parameters, both the name and its type (but not both) are optional:
\begin{cyan}
    func with: Int do: action { ... }
\end{cyan}
The parameter of {\tt with:} does not have a name and therefore it cannot be referenced inside this method. Usually parameters of interfaces (Section~\ref{interfaces}) do not have names. The parameter of {\tt do:} has name {\tt action} but it does not have a type. It is assumed its type is {\tt Dyn}, the dynamic type. No message sends to expressions that have type {\tt Dyn} are checked by the compiler. And all assignments to and from expressions of this type are not checked too.

\section{Comments}

Comments are parts of the text ignored by the compiler. Cyan supports two kinds of comments:  \index{comment}
\begin{itemize}
 \item anything between \verb@/*@ and \verb@*/@. Nested comments
are allowed. That is, the comment below ends at line 3.

\nd \verb@1    /* this is a /* nested@\\
\nd \verb@2       comment */@\\
\nd \verb@3       that ends here */@\\

\item anything after \verb@//@ till the end of the line;

\end{itemize}
A comment may appear anywhere (maybe this will change).
A comment is replaced by the compiler by a single space.
\begin{cyan}
    var value = 1/* does value holds 10?  */0;
\end{cyan}
This code is the same as \verb"var value = 1 0" and therefore it causes a compile-time error instead of being an assignment of {\tt 10} to {\tt value}.

\section{Keywords}  \label{keywords}

Cyan uses the following keywords:

\begin{verbatim}
abstract   char       enum       implements     long       override                 void
Any        const      extends    import         macro      package    stackalloc    volatile
Array      default    false      in             match      private    String        when
Boolean    delegate   final      Int                       protected  switch        where
boolean    Double     Float      int            mixin      public     true          while
break      double     float      interface      mutable    return     type          with
Byte       Dyn        for        it             Nil        self       val
byte       each       func       let            null       shared     var
case       else       heapalloc  local          object     Short      virtual
Char       end        if         Long           of         short      Void
\end{verbatim}
Each of them should be preceded by space, the beginning of a line, or '{\tt (}' except {\tt Nil}, {\tt Boolean},    {\tt Char}, {\tt Byte}, {\tt Int},     {\tt Short}, {\tt Long}, {\tt Float}, {\tt Double}, {\tt String}, {\tt self},  {\tt true}, and {\tt false}. Each of them should be followed by space, end of line, end of file, or '{\tt )}'. A space is a character that makes method {\tt Character.isWhiteSpace(char ch)} of Java return true.

Note that a lot of reserved words are not currently used in the language.

\section{Assignments}

An assignment is made with ``{\tt =}''\/ as in\\  \index{assignment}
\verb"    x =  expr;"\\
\nd After this statement is executed, variable {\tt x} refer to the object that resulted in the evaluation of {\tt expr} at runtime. The compile-time type of {\tt expr} should be a {\it subtype} of the compile-time type of {\tt x}. See Section~\ref{typesandsubtypes} for a definition of subtype.

A variable may be declared and assigned a value:\\
\nd \verb"    var x =  expr;"\\
\nd The type of {\tt x} will be the compile-time type of {\tt expr}.  Both the type of the variable and the expression can be supplied:\\
\nd \verb"    var Int x = 100;"\\

\section{Basic Types}

\index{Byte} \index{Short} \index{Int} \index{Long} \index{Float} \index{Double} \index{Char} \index{Boolean}
Cyan has one basic type, starting with an upper case letter, for each of the basic types of Java: {\tt Byte}, {\tt Short}, {\tt Int}, {\tt Long}, {\tt Float}, {\tt Double},
{\tt Char}, and {\tt Boolean}.
Besides that, there are prototypes {\tt Nil} and {\tt String}, also considered basic types.

Unless said otherwise, Cyan literals of the basic types are defined as those of Java. In particular, the numeric types have the same ranges as the corresponding Java types.
{\tt Byte}, {\tt Short}, and {\tt Long} literals should end with {\tt B} or {\tt Byte}, {\tt S} or {\tt Short}, and {\tt L} or {\tt Long}, respectively as in \\  \index{literals}
\begin{cyan}
var aByte  = 7B;
var aShort = 29Short;
var aLong  = 1234567L;
var bLong  = 37Long;
var anInt  = 223Int;
\end{cyan}
{\tt Int} literals may optionally end with {\tt I} or {\tt Int}.
All basic types but {\tt Nil} inherit from  {\tt Any}. Therefore there are not two separate hierarchies for basic and regular types. All types obey ``reference''\/ semantics. Conceptually, every object is allocated in the heap. However, objects of basic types such as {\tt 1}, {\tt 3.1415}, and {\tt true} may be allocated in the stack.

Integral literal numbers without a postfixed letter are considered as having type {\tt Int}. Numbers with a dot such as {\tt 10.0} as considered as {\tt Double}´s. {\tt Float} literals can end with {\tt F} or {\tt Float}. {\tt Double} literals should end with {\tt D} or {\tt Double}.
There is no automatic conversion between types:
\begin{cyan}
var Int age;
var Byte byte0;
var Float height;
var Double width;
   // ok
age = 21;
   // compile-time error, 0 is Int
byte0 = 0;
   // ok
byte0 = 0B;
   // ok
height = 1.65F;
   // compile-time error
height = 1;
   // compile-time error, 1.65 is Double
height = 1.65;
   // ok
height = 1F;
width = 1.65;
width = 1.65Double;
\end{cyan}

 Underscores can be used to separate long numbers as in\\
\nd \verb@     1_000_000@\\
\nd Two or more underscores cannot appear together as in \\
\nd \verb@     1__0@\\
\nd The first symbol cannot be an underscore: \verb@_1_000@ would be considered an identifier by the compiler.

The {\tt Boolean} type has two enumerated constants, {\tt false} and {\tt true}, with \verb@false < true@. When {\tt false} is cast to an {\tt Int}, the value returned is {\tt 0}. {\tt true} is cast to {\tt 1}. {\tt Char} literals are given between \verb"'" as in \\
\nd \verb@    'A'  '#'  '\n'@\\

Prototype {\tt Nil} has a special status in the language. It is the only prototype that does not inherit from {\tt Any}, the superprototype of anyone (this will be explained latter). {\tt Nil} is not supertype or subtype of anything. Then to a variable of type {\tt Nil} can only be assigned prototype {\tt Nil} and it can only be assigned to a variable or parameter of type {\tt Nil} or {\tt Dyn}. Of course, {\tt Nil} cannot be inherited from a prototype.

Methods that do not declare a return type, as \\
\verb@     func set: Int newValue { ... }@\\
\nd in fact return a value of type {\tt Nil}. Therefore this declaration is equivalent to\\
\verb@     func set: Int newValue -> Nil { ... }@\\
\nd Any method that has {\tt Nil} as the return type always return {\tt Nil} at the end of its execution.
The {\tt return} statement (explained later) is required in methods that return anything other than {\tt Nil}.

Since {\tt Nil} does not have subtypes, a method returning {\tt Nil} can be implemented as not returning a value. After all, it always return the same value.

Prototype {\tt String} represents a immutable string. It has several methods such as {\tt at: []} (for indexing) and {\tt ==} (equality). String literals can be expressed in three forms: enclosed  by \verb@"@ as in C/C++/Java, enclosed by \verb|"""|, and starting with {\tt \#}.

A literal string delimited by \verb@"@ has the same syntax as in C/C++/Java: \verb@"Hi, this is a string"@, \verb@"um"@, \verb@"ended by newline\n"@. Cyan strings and literal characters support the same escape characters as Java.
There is a literal string that spans throught multiple lines. It starts with three characters \verb|"| as in the example:
\begin{cyan}
var s = """
     This is a text
     that uses
     more than one
     line""";
s println;
\end{cyan}

A special multi-line literal string starts with \verb@|"""@, the \verb@|@ character before the quotes. Both \verb@|@ and all characters before it are eliminated. Every \verb@|@ should be in a different line, although there may be empty lines between two lines.
All \verb@|@ characters should be in the same column. All white spaces before the closing \verb|"""| are removed.\footnote{{\tt ch} is a white space if the Java method {\tt Character.isWhitespace(ch)} returns true.}
The result is a multi-line string in which
\verb@|@ delimited the start of each line.
\begin{cyan}
        var String s;

        s = |"""
            |first5

            |second5
               """;

        s = |"""   |first3
                   |second3
                """;

        s = |"""
                   |first2
                   |second2
               """;


        s = |"""
         |first4
         |second4
               """;

        s = |"""
         |first12

         |second12

               """;

        s = |"""|first13

         |second13""";
\end{cyan}

A literal string can also be represented by ``symbols''. A symbol starts with
{\tt \#} followed, without spaces, by letters, dot, digits, underscore, and any number of {\tt :}´s, as in  these examples:  \label{symboldef}
\begin{cyan}
#f      #age    #age:
#123    #_0     #field001
#foreach:do:    #main.package
\end{cyan}

A single-quote literal string may start with \verb|n"| or \verb|N"| to disable any escape character inside the string:\\
\verb#   var fileName = n"D:\User\Carol\My Texts\text01"#\\
\nd In this case ``\verb#\t#''\/ do not mean the tab character. Of course, this kind of string cannot contain the character \verb@'"'@. Three quoted literal strings disable escape characters by default.
Then \verb|"""\n"""| has two characters.

The value of a variable can be inserted in a literal string at runtime by preceding its name, inside the literal string, by \verb|$|. This does not work with symbols because they cannot contain \verb|$|.
\begin{cyan}
var n = 5, k = n*n;
assert "n = $n, k = $k" == "n = 5, k = 25";
var s = """
     The values of n and k are
         $n
         $k
     """;
s println;
\end{cyan}


Method  {\tt eq:} of {\tt String} returns true if the argument and {\tt self} have the same contents. It always give the same result as method {\tt ==}.
\begin{cyan}
var s = "cyan";
var p = s;
assert p == s;
assert "cyan" == """cyan""";
assert #cyan == "cyan" && """cyan""" == #cyan;
assert s == "cyan" && #cyan == s && "cyan" == s;
assert p == s && p eq: s;

assert "\\n" == """\n""";

assert p eq: s;
assert "cyan" eq: """cyan""";
assert #cyan eq: "cyan" && """cyan""" eq: #cyan;
assert s eq: "cyan" && #cyan eq: s && "cyan" eq: s;
assert p eq: s && s eq: p;

\end{cyan}

Method {\tt neq:} returns the negation of the result of {\tt eq:}. For basic type objects, it always return the same value as {\tt !=}.
\begin{cyan}
assert "cyan" != #green;
assert """Cyan""" != "green";
assert ("cyan" != #green) == ("""Cyan""" != "green");
assert (#green != """cyan""") == (#green eq: """cyan""");
\end{cyan}

Types {\tt Byte}, {\tt Short}, {\tt Int}, {\tt Long}, {\tt Float}, and {\tt Double} support almost the same set of arithmetical and logical operators \index{arithmetical operator} \index{logical operator} as the corresponding types of Java. The {\it binary or} operator in Cyan is \verb@|||@ instead of \verb@|@ because the later is used exclusively for union types.

We show just the interface of {\tt Int}. Types {\tt Float} and {\tt Double} do not support methods \verb@&@, \verb@|@, \verb@~|@, and \verb@!@. All basic types are automatically included in every Cyan source code because they belong to package {\tt cyan.lang}. \label{btd}
\begin{cyan}
package cyan.lang

   // method bodies elided

final object Int
    func eq:  (Dyn other) -> Boolean
    func neq: (Dyn other) -> Boolean
    func +  (Int other) -> Int
    func -  (Int other) -> Int
    func *  (Int other) -> Int
    func /  (Int other) -> Int
    func %  (Int other) -> Int
    func <  (Int other) -> Boolean
    func <= (Int other) -> Boolean
    func >  (Int other) -> Boolean
    func >= (Int other) -> Boolean
    func pred -> Int
    func succ -> Int
    func odd -> Boolean
    func even -> Boolean
    func prime -> Boolean
    func isMultiple: Int n -> Boolean
    func maxValue -> Int
    func minValue -> Int
    func == (Dyn other)  -> Boolean
    func != (Dyn other)  -> Boolean
    func <=> (Int other) -> Int
    func .. (Int theEnd) -> Interval<Int>
    func ..< (Int theEnd) -> Interval<Int>
    func -  -> Int
    func +  -> Int
    func & (Int other) -> Int
    func ||| (Int other) -> Int
    func ~| (Int other) -> Int
    func ~ -> Int
    func <.< (Int other) -> Int
    func >.> (Int other) -> Int
    func >.>> (Int other) -> Int
    func |> Function<Int, Int> f -> Int
    func asByte    -> Byte
    func asShort   -> Short
    func asLong    -> Long
    func asFloat   -> Float
    func asDouble  -> Double
    func asChar    -> Char
    func asBoolean -> Boolean
    func asInt     -> Int
    func asString  -> String
    func to: (Int max) do: (Function<Int, Nil> b)
    func to: (Int max) into: (InjectObject<Int> injectTo)
    func times: Function<Nil> b
    func repeat: Function<Int, Nil> b
    func to: (Int max)
    func in: (Iterable<Int> container) -> Boolean
    func between: Interval<Int> inter -> Boolean
    func hashCode -> Int
    func defaultValue -> Int
end


...


abstract object InjectObject<T> extends Function<T, Nil>
    override
    abstract func eval: T
    abstract func result -> T
end

interface Iterator<T>
    func hasNext -> Boolean
    func next -> T
end

\end{cyan}

Some of the basic types have have methods to simulate pipes:
\begin{cyan}
(5 |> Sqr |> { (: Int elem :) ^2*elem } ) print;
\end{cyan}
Assuming that {\tt Sqr} extends \verb|Function<Int, Int>| and calculates the square of its parameter, this command will print {\tt 50}. It is equivalent to:
\begin{cyan}
({ (: Int elem :) ^2*elem } eval: (Sqr eval: 5) ) print;
\end{cyan}

Prototype {\tt Boolean} uses \verb|=>| as an {\tt if} command:
\begin{cyan}
i < 3  => { "baby" println };
i > 19 => { "adult" println };
\end{cyan}
This method takes a literal function as argument.

Variables of types {\tt Byte}, {\tt Char}, {\tt Short}, {\tt Int}, and {\tt Long} may be preceded by {\tt ++} \index{++} \index{--} or {\tt --}. When {\tt v} is a private field or a local variable, the compiler will replace {\tt ++v} by \\
\nd \verb@    (v = v + 1)@\\
\nd Idem for {\tt --}.

A prototype may declare an \index{[]} \index{indexing} operator {\tt []} and use it just like an array (see Section~\ref{indexing}).

Each basic prototype {\tt T} but {\tt Float}, {\tt Double}, and {\tt Nil} has an {\tt in:} method that accepts an object that implements \verb"Iterable<T>" as parameter. This call method {\tt foreach:} of this parameter comparing each element with {\tt self}. It returns {\tt true} if there is an element equal to {\tt self}. It can be used as in
\begin{cyan}
var Char ch;
ch = In readChar;
( ch in: [ 'a', 'e', 'i', 'o', 'u' ] ) ifTrue: {
    Out println: "$ch is a vowel"
};
var Array<Int> intArray = [ 0, 1, 2, 3 ];
var List<Int> intList = List<Int> new;
intList add: 0;
intList add: 1;
var Int n = In readInt;
if n in: intArray  ||  n in: intList {
    Out println: "$n is already in the lists"
}
\end{cyan}
The parameter to {\tt in:} can be any object that implements {\tt Iterable} of the correct type. In particular, all arrays whose elements implement this interface.

Intervals implement the \verb|Iterable<T>| interface. Then we can write
\begin{cyan}
var Char ch;
ch = In readChar;
( ch in: 'a'..'z' ) ifTrue: {
    Out println: "$ch is a lower case letter"
};
var age = In readInt;
if age in: 0..2 { Out println: "baby" }
else if age in: 3..12 {
    Out println: "child"
}
else if age in: 13..19 {
    Out println: "teenager"
}
else {
    Out println: "adult"
}
\end{cyan}

\verb|&&| and \verb@||@ are not methods of prototype {\tt Boolean}. They are instead operators of the language and use the regular short-circuit evaluation. That is, \verb|aa && bb| is {\tt false} if {\tt aa} is {\tt false}. In this case {\tt bb} is not even evaluated. So the {\tt if} statement below is safe.
\begin{cyan}
    if index < array size && array[index] == x {
        Out println: "found $x"
    }
\end{cyan}
Prototype {\tt Boolean} has a logical not, {\tt !}.
\begin{cyan}
    if ! ok { Out println: "fail" }
    if age < 0 || age > 127 { Out println: "out of limits" }
\end{cyan}

Method {\tt ++} defined in {\tt Any}, the superprototype of every one but {\tt Nil}, concatenate the string of the receiver plus the string of the argument:
\begin{cyan}
    assert 1 ++ 2 == "12" &&
           1 ++ 'A' == "1A" &&
           0 ++ "1" == "0" ++ 1;
\end{cyan}

Prototype {\tt String} support the {\tt in:} method:
\begin{cyan}
    func daysMonth: (String month, Int year) -> Int {
        if month in: [ "jan", "mar", "may", "jul", "aug", "oct", "dec" ] {
            return 31
        }
        else if month in: [ "apr", "jun", "sep", "nov" ] {
            return 30
        }
        else if month == "fev" {
            if leapYear: year { return 29 } else { return 28 }
        }
        else {
            return -1
        }
    }
\end{cyan}

\section{Operator and Keyword Precedence}

\index{precedence}
\begin{figure}
\begin{verbatim}
    ||
    ~||
    &&
    =>  ==>
    !
    ==   <=   <   >   >=   !=   ===   !==   <=>   ~=
    non-unary message send
    |>
    ++  -- (binary)
    ..  ..<
    +   -
    /   *   %
    |||  ~|   &
    <.<   >.>    >.>>
    .*  .+  .%
    unary message send
    +   -   !  ~   (unary)
\end{verbatim}
\caption{Precedence order from the lower (top) to the higher (bottom)}
\label{precedence}
\end{figure}

Cyan has special precedence rules for methods and operators whose names are the symbols given in Figure~\ref{precedence}.
The precedence is applied to every message send that uses some of these symbols. So a message send\\
\nd \verb@    x + 1 < y + 2@
\nd will be considered as if it was \\
\nd \verb@    (x + 1) < (y + 2)@

Then when we write
\begin{cyan}
    if age < 0 || age > 127  { Out println: "out of limits" }
    if index < array size && array[index] == x {
        Out println: "found $x"
    }
\end{cyan}
the compiler interprets this as
\begin{cyan}
    if  (age < 0) || (age > 127)  { Out println: "out of limits" }
    if  (index < array size) && (array[index] == x) {
        Out println: "found #x"
    }
\end{cyan}
In a message send, unary selectors have precedence over multiple keywords. Then\\
\nd \verb"    obj a: array size"\\
\nd is the same as\\
\nd \verb"    obj a: (array size)"\\

Every operator but \verb|+|, \verb|-|,  \verb|*|, \verb|/|, \verb|%|,
\verb|~|, \verb|!|, \verb|..|, and \verb|..<| should be preceded and followed by
a white space. That is, all binary operators but the arithmetical
ones (\verb|+, -, *, /, %|) should be surrounded by white spaces. Note that not
all operators are used by the Cyan basic types (\verb|.*|, for example).

Unary methods associate from left to right. Then\\
\nd \verb"    var String name = club members first name;"\\
\nd is the same as:\\
\nd \verb"    var String name = ((club members) first) name;"\\

The method names of the last line of the Figure~\ref{precedence} are unary. All other methods and operators are binary and left associative.
That means a code\\
\nd \verb"    ok = i >= 0 && i < size && v[i] == x;"\\
\nd is interpreted as \\
\nd \verb"    (ok = i >= 0 && i < size) && v[i] == x;"\\
\nd This is true even when {\tt Boolean} is not the type of the receiver.

The compiler does not check the type of the receiver in order to discover how many parameters each keyword should use.
When the compiler finds something like
\begin{cyan}
obj s1: 1 s2: 1, 2 s3: 1, 2, 3
\end{cyan}
it considers that the method name is {\tt s1:s2:s3} and that {\tt si} takes  {\tt i} parameters. This conclusion is taken without consulting the type of {\tt obj}. Therefore, code
\begin{cyan}
   // get: takes two parameters
var k = matrix get: (anArray at: 0), 1;
\end{cyan}
cannot be written
\begin{cyan}
var k = matrix get: anArray at: 0, 1;
\end{cyan}
This would mean that the method to be called is named {\tt get:at:} and that {\tt get:} receives one parameter, {\tt anArray}, and {\tt at:} receives two arguments, {\tt 0} and {\tt 1}. To know the reason of this rule, see Chapter~\ref{dynamictyping}.

\section{Loops, Ifs, and other Statements}  \label{whiletrue}

Currently each message send, assignment, and local variable declaration should end with a semicolon (``;''\/). However we expect to make the semicolon optional as soon as possible, at least in some cases. The semicolon is optional for all statements such as {\tt if}, {\tt while}, {\tt for}, and {\tt type-case}.

Decision and loop statements that not use the return statement can be implemented using message sends to {\tt Boolean} objects and to function objects. There are four methods of prototype {\tt Boolean} used as decision statements: {\tt ifTrue:}, {\tt ifFalse:}, {\tt ifTrue:ifFalse:}, and {\tt ifFalse:ifTrue:}. \index{decision statement} \index{ifTrue:} \index{ifFalse:}
\begin{cyan}
( n%2 == 0 ) ifTrue:   { s = "even" };
( n%2 != 0 ) ifFalse:  { s = "even" };
( n%2 == 0 ) ifTrue:   { s = "even" } ifFalse: { s = "odd" } ;
( n%2 != 0 ) ifFalse:  { s = "even" } ifTrue:  { s = "odd" } ;
\end{cyan}
They are self explanatory. Besides that, there are methods {\tt t:f:} and {\tt f:t:} in {\tt Boolean} that return an expression or another according to the receiver:
\begin{cyan}
var String s;
var Any any = ( n%2 == 0 ) t: "even"  f: "odd";
type  any
    case String s2 {
        s = s2
    }
any = ( n%2 != 0 ) f: "even"  t: "odd";
type any
    case String s2 {
        s = s2
    }
\end{cyan}
If the expression is true, the expression that is parameter to {\tt t:} is returned. Otherwise it is returned the parameter to {\tt f:}. Since the return type of methods {\tt t:f:} and {\tt f:t:} is {\tt Any}, the example uses command {\tt type-case} to cast
{\tt any}  to {\tt String}.

As a future improvement, a metaobject  {\tt checkTF} will check whether the arguments of both keywords have the same type\footnote{For the time being, one cannot be subtype of another.} (both are strings in this case).  This metaobject will also cast the value returned to the correct type, {\tt String} in the example. The return type is {\tt Any}. That is why it is necessary to use command {\tt type-case}  in the example.

Note that an {\tt if} statement that needs a {\tt return} cannot be implemented using message sends:
\begin{cyan}
(i == 0) ifTrue: {
    return false;  // compile-time error.
};
\end{cyan}
The {\tt return} statement cannot appear inside a function.

Function objects that return a {\tt Boolean} value have a {\tt whileTrue:} and a {\tt whileFalse:} methods. \index{whileTrue:} \index{whileFalse:}
\begin{cyan}
var i = 0;
{^ i < 5 } whileTrue: {
    Out println: i;
    ++i
}
var i = 0;
{^ i >= 5 } whileFalse: {
    Out println: i;
    ++i
}
\end{cyan}
Of course, {\tt whileTrue} calls the function passed as parameter while the function that receives the message is true. {\tt whileFalse} calls while the receiver is false.

The {\tt if} and the {\tt while} statements were added to the language to make programming easier. The syntax of these statements are shown in this example:  \index{if} \index{while}
\begin{cyan}
if  n%2 == 0  {
    s = "even"
}
else {    // the else part is optional
    s = "odd"
}
var i = 0;
while  i < 5  {
    Out println: i;
    ++i
}
\end{cyan}
The {\tt \}} that closes a {\tt while} statement should be either in the same line as keyword {\tt while} or in the same column as it.  There should be no semicolon after the closing {\tt \}}.

There is a {\tt repeat-until} statement that executes its statements until the {\tt until} expression evaluates to {\tt true}.
\begin{cyan}
var Int sum = 0;
var Int n = 1;
repeat
    sum = sum + n;
    ++n;
until n >= 4;
assert sum == 6;
\end{cyan}

Cascaded if´s are possible:
\begin{cyan}
if  age < 3  {
    s = "baby"
}
else if  age <= 12  {
    s = "child"
}
else if  age <= 19  {
    s = "teenager"
}
else {
    s = "adult"
}
\end{cyan}
Unlike the languages of the C family, the parentheses around the boolean expression are not necessary and the {\tt \{} and {\tt \}} are required. The \verb|';'| after {\tt if} and {\tt while} statements are not necessary. There are several restriction on the formatting of  {\tt if} statements:
\begin{enumerate}[(a)]
\item the column of the {\tt \{} that follows keyword {\tt if} should be in a column greater or equal than
the column of the {\tt if} keyword;
\item the column of the first symbol of the {\tt Boolean} {\tt if} expression should be in a column greater or equal than the column of the {\tt if} keyword;

\item if the {\tt \}} that closes the statement list that follows an {\tt if} is in line {\tt lineR} and
column {\tt columnR}, than {\tt lineR} should be equal to the line of the {\tt if} or the line of the previous {\tt else} or {\tt columnR} should be equal to the column of the {\tt if} or the column of the previous {\tt else};

\item {\tt else} should be in the same line or column of the previous {\tt else} or {\tt if} keywords;

\item the {\tt \}} that closes the {\tt else} statements should be in the same line as the {\tt else} or in the same column;
\item the {\tt \}} that closes an {\tt if} or {\tt else} may be in the same line as the next {\tt else}:
\begin{cyan}
    if 0 < 1 {
        "ok" println;
    } else if 1 < 2 {
        "ok" println;
    } else {
        "not ok" println;
    }
\end{cyan}
In this case, the {\tt \}} that closes the \textbf{next} {\tt else} or {\tt if} should be in the same column as the \textbf{previous} {\tt \}}.

\end{enumerate}

These rules assert that {\tt if} statements are almost always clearly formatted. See the examples with
formatting errors.
\begin{cyan}

    if 0 < 1 { }
    else if 0 < 4 {
    }
    else if 0 < 1 {
       }   // '}' is not in the same column or line as 'else'


    if 0 < 1 { }
    else if 0 < 4 {
    }
         else if 0 < 1 {
    }  // '}' should be in the same line or column as 'if'

    if 0 < 1 { } else
        if 0 < 5 { }
        else
            if 0 < 4 { } else if
                     // if expression in a column smaller than the if column
                0 < 1  { }
            else {
        };


    if 0 < 1 { } else
        if 0 < 5 { }
        else
            if 0 < 4 { } else if
                              0 < 1            {
                              }
               // 'else' must be in the same line or column
               // as the previous 'if'
            else {
        };


    if 0 < 1 { } else
        if 0 < 5 { }
        else
            if 0 < 4 { } else if
                              0 < 1            {
                              }
                              else {
           // this '{' closes an 'else' statement and should be
           // in the same line or column as it
        };


    if 0 < 1 { } else
        if 0 < 5 {
               // the '}' that closes an 'if' should be in the same
               // line or column as it
            }
            else { }

    if 0 < 5 { 0 println }
           // 'else' must be in the same line or column as the previous 'if'
        else { 1 println }


    if 0 < 5 { 0 println
    }
        // else should be in the same column as '{'
      else { 1 println }


\end{cyan}
These examples have no errors:
\begin{cyan}
        if 0 < 6 { } else { }

        if 0 < 1 { }
        else if 0 < 4 {
             }
             else if 0 < 1 {
             }

        if 0 < 1 { }
        else
             if 0 < 4
             {

             }
             else if 0 < 1 {
             }



        if 0 < 1 { } else
            if 0 < 5 { }
            else { }

        if 0 < 1 { } else
            if 0 < 5 { }
            else
                if 0 < 4 { } else if
                                  0 < 1            {
                                  }
                                  else {
                                  };

        if 0 < 1 { } else
            if 0 < 5 {  }
            else { }


        if 0 < 5 { 0 println }
        else { 1 println }

        if 0 < 1 { } else if 0 < 5 { }
                     else { }

        if 0 < 1 {
        } else if 1 < 2 {
        } else {
        }

\end{cyan}

Statement {\tt for} can be used to iterate over any object that implements a method
\begin{cyan}
    iterator -> Iterator<T>
\end{cyan}
Its syntax is
\begin{cyan}
    for [Type ] elem in list {
        // statements
    }
\end{cyan}
The {\tt Type} of {\tt elem} is optional.
{\tt elem} will assume the elements given by method {\tt next} of the iterator returned by method {\tt iterator} of {\tt list}. When method {\tt hasNext} of the iterator returns {\tt false} the loop ends. The {\tt Type} should be exactly the type of the elements returned by the iterator. It cannot be a supertype of that type.

This command can be used as in
\begin{cyan}
    for elem in [ 2, 3, 5, 7 ] {
        "$elem is prime" println
    }
    for ch in 'a'..'z' {
        ("letter " ++ ch) println;
    }
    var sum = 0;
    for i in 0..< 10 {
        sum = sum + i
    }
    sum println;
    sum = 0;
    for i in 1..10 {
        for j in 1..10 {
            sum = sum + i*j;
        }
    }
\end{cyan}

The {\tt \}} that closes a {\tt for} statement should be either in the same line as keyword {\tt for} or in the same column as it.  There should be no semicolon after the closing {\tt \}}.

{\tt elem} should not have been previously declared as a local variable or parameter. {\tt elem} is alive only inside the statements of this {\tt for}. Its type is
deduced by the compiler if it is not given.

Since variable {\tt elem} is only alive inside the {\tt for} command, it can be reused:
\begin{cyan}
    for elem in [ 2, 3, 5, 7 ] {
        "$elem is prime" println
    }
    for elem in 0..< 10 {
        elem println;
    }
\end{cyan}

There are other kinds of loop statements, which are supplied as message sends: \index{loop} \index{repeatUntil}
\begin{cyan}
i = 0;
   // the function is called forever, it "never" stops
{
  ++i;
  Out println: i
} loop;
\end{cyan}

Prototype {\tt Int} also defines some methods that act like loop statements: \index{repeat:} \index{to:do:}
\begin{cyan}
        // this code prints numbers 0 1 2
    var sum = 0;
    var i = 0;
    3 repeat: { (: Int n :)
        sum = sum + n
    };
    assert sum == 3;
        // this code prints numbers 0 1 2
    3 repeat: { (: Int j :)
        Out println: j
    };
    var aFunction = { (: Int j :) Out println: j };
        // this code prints numbers 0 1 2
    3 repeat: aFunction;

        // prints 0 1 2
    i = 0;
    1 to: 3 do: { (: Int n :)
        n println;
    };
        // prints 0 1 2
    0 to: 2 do: { (: Int j :)
        Out println: j
    };
\end{cyan}
Prototype {\tt Char} also has equivalent {\tt repeat:} and {\tt to:do:} methods:
\begin{cyan}
'a' to: 'z' do: { (: Char ch :)
     Out println: ch
};
\end{cyan}

\section{Arrays} \index{Array} \label{array}

{\tt Array} is a generic prototype that cannot be inherited for sake of efficiency. It has methods that mirror those of class {\tt ArrayList} of Java:

\begin{cyan}
package cyan.lang

final object Array<T> implements Iterable<T>
    func == (Dyn other) -> Boolean
    func != (Dyn other) -> Boolean
    func add: (T elem)
    func add: (Int i, T elem)
    func clear
    func isEmpty -> Boolean
    func remove: (Int i)
    func [] at: Int index -> T
    func [] at: Int index put: (T elem)
    func last -> T
    func asString: (Int ident) -> String
    func slice: (Interval<Int> interval) -> Array<T>
    func concat: Array<T> other -> Array<T>
    func size -> Int
    func foreach: Function<T, Nil> b
    func filter: Function<T, Boolean> f -> Array<T>
    func filter: Function<T, Boolean> f foreach: Function<T, Nil> b
    func map: Function<T, T> f -> Array<T>
    func iterator -> Iterator<T>
    func contains: T elem -> Boolean
    func indexOf: T elem -> Int
    func apply: (String message)
    func apply: (String message) select: (String slot) -> Dyn
    func .* (String message)
    func .+ (String message) -> Dyn
end
\end{cyan}

Arrays supports some interesting methods: {\tt apply:}, \verb|.*|, and \verb|.+|. The first two ones applies an operation given as string to all array elements. Method \verb|.+| sums all array elements or return the first element if the array has just one element. It is assumed that the type of the array element supports a binary operation \verb|"+"|.
\begin{cyan}
var Array<Int> v = [ 2, 3, 5, 7, 11 ];
v apply: #print;   // print all array elements
v .* #print;       // print all array elements
(v .+ "+") print;  // print the sum of all array elements
(v .+ "*") print;  // print the multiplication of all array elements
\end{cyan}

Intervals can be arguments to {\tt slice:} which allows the slicing of arrays:
\begin{cyan}
var letters = [ 'b', 'a', 'e', 'i', 'o', 'u', 'c', 'd' ];
var vowels  = letters slice: 1..5;
   // print a e i o u
Out println: vowels;
\end{cyan}

\section{Maps} \index{Map} \label{map}

A literal map is delimited by {\tt [} and {\tt ]} as an array and uses \verb|->| for mapping a {\it key} to \ {\it value}.
\begin{cyan}
    let IMap<String, Int> map = [ "Newton" -> 1642, "Gauss" -> 1777 ];
      // method asArray return an array with tuple elements
    for elem in map asArray {
        Out println: elem key, " was born in the year ", elem value;
    }

    cast year = map["Newton"] {
        "Newton was born in $year" println;
    }

\end{cyan}
The type of the literal map is \verb|IMap<K, V>| in which {\tt K} is the type of the {\it key} (appears before \verb|->|) and {\tt V} is the type of the values (after \verb|->|). The expression \verb|map[key]| returns a value of type \verb|Nil|V| and, therefore, a {\tt cast} statement (Section~\ref{cast}) is need for retrieving the value.
All the keys should have the same type and all the values should have the same type. {\tt IMap} is an interface. The literal object will have a prototype that implements {\tt IMap}.
\begin{cyan}
    let errorMap = [ Any -> "Any", Int -> "Int" ];
\end{cyan}
Although {\tt Int} is a subprototype of {\tt Any}, the compiler will sign an error in this code.

Method {\tt asArray} of \verb|IMap<K,V>| returns an array with all elements of the map. Each element is a tuple
\begin{cyan}
    Tuple<key, K, value, V>
\end{cyan}
The first example of this section iterates over the map elements using a {\tt for}. A single element can be got using indexing or the {\tt get} method of {\tt IMap}. The element returned has type \verb"V|Nil". If the {\it key} passed as parameter is in the map, the value returned is the {\it value} associated to it. If the {\it key} is not in the map, {\tt Nil} is returned.

\chapter{Main Cyan Constructs}

\index{object} \index{object!slot}
A prototype may declare zero or more {\it slots}, which can be variables (called fields) shared field variables (to be seen later), and methods (called instance methods). In Figure~\ref{obj01}, there is one field, {\tt name}, and two methods, {\tt getName} and {\tt setName}. Keywords {\tt public}, {\tt private}, {\tt package}, and {\tt protected} can precede a method declaration. Currently only keyword {\tt private} (or none) can precede a field declaration. A public method can be accessed anywhere the prototype can. A private method and field can only be used inside the prototype declaration. Protected methods can be accessed in the prototype, its subprototypes, sub-subprototypes, and so on. A subprototype inherits from a prototype ---  that will soon be explained.

In the declaration of a field, there are four optional parts:
\begin{enumerate}
\item the visibility (only {\tt private});
\item keywords ``{\tt var}''\/ or ``{\tt let}''\/  that may precede the type;
\item  ``{\tt ;}''\/, that may follow the declaration;
\item and ``\verb| = expr;|''\/, that may follow the variable name.
\end{enumerate}
The only non-optional parts are the type and the name. There are restriction on the expression {\tt expr}. See Section~\ref{init}.

A read-only field is declared with the word {\tt let} ou without any keyword:
\begin{cyan}
object Person
    let String name
    Int age
    ...
end
\end{cyan}
Both {\tt name} and {\tt age} are read-only. A read-only variable can receive a value in its declaration and it can receive a value in the constructor (to be seen). But a regular method cannot assign a value to it. This is the default, used if no keyword precedes the field declaration. If a variable should change its value after the object is created, declare it with {\tt var} as in
\begin{cyan}
object Person
    func init: String name, Int age {
        self.name = name;
        self.age = age
    }
    var String name
    var Int age
    func getName -> String = name;
    func setName: String other { self.name = other }
    func getAge -> Int = age;
    func setAge: Int other { self.age = other }
end
\end{cyan}

A single type can be used for more than one field:
\begin{cyan}
object Rectangle
    ...
    Int x1, y1, x2, y2
end
\end{cyan}
But we can only assign a value to a declaration with a single value:
\begin{cyan}
object Rectangle
    ...
    Int x1 = 0;  // ok
    Int y1 = 0, x2, y2  // compile-time error
end
\end{cyan}

\begin{figure}[H]
\begin{cyan}
package bank

object Client
  func getName -> String {
     return self.name
  }
  func setName:  String name {
     self.name = name
  }
  func print {
      Out println: name
  }
  private var String name = "";
end

\end{cyan}
\caption{An object in Cyan}
\label{obj01}
\end{figure}


\index{field}

Package {\tt cyan.lang} supplies a metaobject {\tt property} that creates methods for getting and setting the value of a field.
\begin{cyan}
package people

object Person
    func init: String p_name, Int p_age {
        _name = p_name;
        age = p_age;
    }
    @property var String _name
    @property var Int age
end
\end{cyan}

The name of the methods created depend on whether the variable name starts with underscore or not. If it does, as {\tt \_name}, the methods created do not start with {\tt get} and {\tt set}. Otherwise methods starting with {\tt get} and {\tt set} are created. In this example, the methods would be
\begin{cyan}
    func name -> String = _name;
    func name: String other { self._name = other }

    func getAge -> Int = age;
    func setAge: Int other { self.age = other }
\end{cyan}

{\tt property} used with a read-only variable causes the creation of a {\tt get} method only:
\begin{cyan}
package people

object Person
    func init: String p_name, Int p_age {
        _name = p_name;
        age = p_age;
    }
    @property String _name  // only get is created
    @property let Int age   // only get is created
end
\end{cyan}


\index{method!public} \index{method!private} \index{method!protected}
A method declared without a qualifier is considered {\tt public}.
A prototype declared without a qualifier is considered {\tt public} (currently prototypes are always public).
A field without a qualifier is considered {\tt private} (currently they are always private). Then, a declaration
\begin{cyan}
package Bank

object Account
      // constructor, to be seen later
    func init: Client client { self.client = client }

    func set: Client client {
        self.client = client
    }
    func print {
        Out println: (client getName)
    }

    var Client client
end
\end{cyan}
is equivalent to
\begin{cyan}
package Bank

public object Account
      // constructor, to be seen later
    public
    func init: Client client { self.client = client }

    public
    func set: Client client {
        self.client = client
    }
    public
    func print {
        Out println: (client getName)
    }
    private Client client
end
\end{cyan}


\index{local variable} \index{variable!local}
The declaration of local variables is made with the syntax:
\begin{cyan}
    var String name;
    var Int x1, y1, x2, y2;
\end{cyan}
The last line declares four variables of type {\tt Int}. Keyword \index{var} \index{keyword!var} {\tt var} or {\tt let} is demanded in the declaration of local variables. Variables declared with {\tt let} are read-only and should receive an expression in the declaration:
\begin{cyan}
    let Double pi = 3.1415;
    let Int maxSize = 100;
    let Char endLine; // compile-time error
\end{cyan}
Further assignments to the variable are forbidden.

A declaration can have a single type and several variables as that of {\tt x1}, {\tt y1}, {\tt x2}, and {\tt y2} above. But a variable that receives an expression should be in its own declaration:
\begin{cyan}
    var String name;
    var Int x1 = 0;
    var Int y1, x2 = 0, y2; // compile-time error
\end{cyan}

We will call {\tt block} a sequence of statements delimited by {\tt \{} and {\tt \}} that appear in a command like {\tt while}, {\tt if}, and {\tt type-case}. A anonymous function is not a block.
The scope of a local variable is from where it was declared to the end of the function or block in which it was declared:
\begin{cyannum}
    func p: Int x {
        var String iLiveHere;
        if  x > 0  {
            var Int iLiveInsideThenPart;
            doSomething: {
                var String iLiveOnlyInThisFunction;
                ...
            }
            ...
        }
    }
\end{cyannum}
Then {\tt iLiveHere} is accessible from line 2 to line 11 (before the \verb"}"). Variable {\tt iLiveInsideThenPart} is live from line 4 to 10 (before the \verb"}"). The scope of {\tt iLiveOnlyInThisFunction} is the function that in between lines 6 and 8 (after the declaration and before the \verb"}").

For short, you cannot declare a variable if there is another visible at that point:
\begin{cyannum}
    func nothing: Int p -> Int {
        var Int n = 0;
        if 0 < p {
            var Int k = n;
            var Int n = 2; // redeclaration
            return k + n
        }
        else {
            var Int k = n + 1; // ok
            return k
        }
    }
\end{cyannum}
The two declarations of {\tt k} are correct because the scope of the first {\tt k} ends in line 7. The scope of {\tt n} ends in the closing {\tt \}} of the method. Therefore the declaration of {\tt n} in line 5 is illegal.

The type of a variable should be a prototype or an interface (explained later). In the declaration\\
\verb@    var String name;@\\
\nd prototype ``{\tt String}''\/ plays the role of a type. Then a prototype name can play two roles: objects and types. If it appear in an expression, it is an object, as ``{\tt String}''\/ in:\\
\verb@    anObj = String;@\\
\nd If it appears as the type of a variable or return value type of a method, it is a type. Here ``variable''\/ means local variable, parameter, or field.

A local variable can be \index{assignment!declaration} declared and assigned a value:\\
\verb@      var Int n = 0;@\\
Both the type and the assigned value can be omitted, but not at the same time. If the type is omitted, it is deduced from the expression at compile-time. If the expression is omitted, to the variable should  be assigned a value before it is used.  When the type is omitted, the syntax\\
\verb!    var variableName = expr!\\
\nd should be used to define the variable as in:\\
\verb@    var n = 0;@\\
\nd Variable {\tt variableName} cannot be used inside {\tt expr}. It it could, the compiler would not be able to deduce the type of {\tt expr} in some situations such as\\
\nd \verb!     var n = n;!\\

In an assignment ``\verb"var n = expr"''\/, the type of the expression is deduced by the compiler using information collected in the previous lines of code. The Hindley-Milner \index{Hindley-Milner} inference algorithm is not used.

All prototypes, including the basic types, are objects in Cyan. Then {\tt Int} is an object which happens to be an ... integer! And which integer is {\tt Int}? It is the default value of type {\tt Int}. So the code below will print {\tt 0} at the output:\\
\verb@    Out println: Int;@\\

A method is declared with keyword {\tt func} \index{func} \index{keyword!func} followed by the method keywords and parameters, as shown in Figure~\ref{obj01}. Following Smalltalk, there are two kinds of methods in Cyan: unary and keyword methods. \index{method!unary} \index{method!keyword}
A unary method does not take any parameters and may return a value. Its name should be an identifier not followed by a ``{\tt :}''\/. For example, {\tt print} in Figure~\ref{obj01} is a unary method.

When a method takes parameters it should have one or more keywords, each one ending with ``{\tt :}''\/ as in
\begin{cyan}
    func at: Int n { ... }
    func at: Int n put: String s { ... }
    func with: Int n
         with: Int another
         concat: String s -> String { ... }
\end{cyan}

This kind of method is called a keyword method or a method with keywords.

There may be a method that is not unary but that does not take parameters:
\begin{cyan}
    func open: String
         read: { ... }
\end{cyan}
Smalltalk does not allow that. However, it is illegal to declare a method
without parameters that is a keyword method:
\begin{cyan}
        // compile error
    func read:  -> String { ... }
\end{cyan}

An optional return value \index{return value} \index{method!return value} type can be given after keyword {\tt func}. The return value should be given by the {\tt return} command. The return expression should be subtype (Section~\ref{typesandsubtypes}) of the return value type of the method.
Using {\tt Nil} as the return value type is the same as to omit the return type.

Methods without return type or declaring {\tt Nil} always return {\tt Nil}. Therefore one can
write
\begin{cyan}
   (0 println) println
\end{cyan}
``\verb@0 println@''\/ returns {\tt Nil}. Message {\tt println} is therefore sent to {\tt Nil}. It will be printed\\
\verb#    0Nil#\\

Objects are used through methods \index{method} \index{message send} and only through methods. A method is called when a message is sent to an object. A message has the same shape as a method declaration but with the parameters replaced by real arguments. Then method {\tt setName:} of the example of Figure~\ref{obj01} is called by\\
\verb@    Client setName: "John";@\\
\nd This statement causes method {\tt setName:} of {\tt Client} to be called at runtime.

\section{self}

Inside a method of a prototype, pseudo-variable {\tt self} \index{self} can be used to refer to the object that received the message that caused the execution of the method. This is the same concept as {\tt self} of Smalltalk and {\tt this} of C++/Java. A field {\tt age} can be accessed in a method of a prototype by its name or by the name preceded by ``{\tt self.}''\/ as in
\begin{cyan}
  func getAge -> Int {
     return self.age
  }
\end{cyan}
Then we could have used just ``{\tt age}''\/ in place of ``{\tt self.age}''\/.

\section{clone Methods}

A copy of an object  is made with the {\tt clone} \index{clone} method. Every prototype {\tt P} has a method\\
\verb@     func clone -> P@\\
\nd that returns a {\it shallow} copy of the current object. In the shallow copy of the original to the cloned object, every field of the original object is assigned to the corresponding variable of the cloned object.

In the message send\\
\verb@    Client setName: "John";@\\
\nd method {\tt setName} of {\tt Client} is called. Inside this method, any references to {\tt self} is a reference to the object that received the message, {\tt Client}. In the last statement of
\begin{cyan}
var Client c;
c = Client clone;
c setName: "Peter";
\end{cyan}
method {\tt setName} declared in {\tt Client} is called because {\tt c} refer to a {\tt Client} object (a copy of the original {\tt Client} object, the prototype). Now the reference to self inside {\tt setName} refers to the object referenced to by {\tt c}, which is different from {\tt Client}.

The {\tt clone} method of an object can be redefined to provide a more meaningful clone operation. For example, this method can be redefined to return {\tt self} in an {\tt Earth} prototype (since there is just one earth) or to make a deep copy of the {\tt self} object.

In language Omega \cite{blaschek1994object}, \index{Omega} \index{language!Omega} the pseudo-type {\tt Same} means the type of {\tt self}, which may vary at runtime. Method {\tt clone} declared in the {\tt Object} prototype returns a value of type {\tt Same}. That means that in object {\tt Object}, the value returned is of type {\tt Object} and that in a prototype {\tt P} the return value type of {\tt clone} is {\tt P}. In Cyan the compiler adds a new {\tt clone} method for every prototype {\tt P}. This is necessary because there is nothing similar to {\tt Same} in the language.

\section{init and new Methods}  \label{init}

\index{new} \index{init} \index{initShared} \index{method!new} \index{method!init} \index{method!initShared}

A prototype may declare one or more methods named {\tt init} or {\tt init:}. All of them have special meaning: they are used for initializing the object. For each method named {\tt init} the compiler adds to the prototype a method named {\tt new}. For each method named {\tt init:} the compiler adds to the prototype a method named {\tt new:} with the same parameter types. Each {\tt new} method creates an object without initializing any of its slots and calls the corresponding {\tt init} method (idem for {\tt init:} and {\tt new:}). If the prototype does not define any {\tt init} or {\tt init:} method, the compiler supplies an empty {\tt init} method that does not take parameters and calls the superprototype {\tt init} method (if there is one. If not, an error occurs).

Some rules apply to the {\tt init} and {\tt init:} methods. They:
\begin{enumerate}[(a)]
\item should be declared with no return type (it cannot be {\tt Nil});
\item cannot be called using reflection at runtime. These methods are used to create {\tt new} and {\tt new:} method and them discarded. They do not exist at runtime. However, {\tt new} and {\tt new:} can be called
    using reflection. The do exist at runtime;
\item should not be preceded by keyword {\tt override};
\item should not be abstract or final;
\item should not be indexing methods (See Section~\ref{indexing});
\item an {\tt init:} method should take at least one parameter;
\item cannot be declared in interfaces;
\item two {\tt init:} methods can have the same number of parameters. However, there should be at least a number {\tt n} such that the {\tt n}$^{th}$ parameter type in one method is not subtype or supertype of the {\tt n}$^{th}$ parameter type of the other method.
    If this were allowed, there would be an ambiguity when calling method new:
    \begin{cyan}
    object Pet
        func init: Animal animal { ... }
        func init: Dog dog { ... }
        ...
    end

    // in other prototype:
        let Dog meg = Dog("Meg");
        let Pet myPet = Pet new: meg;
    \end{cyan}
    In the creation of object {\tt Pet}, both the first and second {\tt init:} methods could be called (if the compiler did not signal an error in this code --- but it does).

    To solve this problem, one can use unions:
    \begin{cyan}
    object Pet
        func init: Dog|Animal animal { ... }
        ...
    end
    \end{cyan}
    The following declaration is illegal because {\tt Wrong(0, 0, 0)} would be ambiguous.
    \begin{cyan}
    object Wrong
        func init: Any a, Int b, Any c { ... }
        func init: Any a, Any b, Any c { ... }
        ...
    end
    \end{cyan}
    The following declaration is legal because the third parameter allows the compiler to differentiate between the two constructors.
    \begin{cyan}
    object Fine
        func init: Any a, Int b, Int c { ... }
        func init: Any a, Any b, String c { ... }
        ...
    end
    \end{cyan}

\item can only be called by immediate subprototypes using {\tt super} as the message receiver. That is, if {\tt C} inherits from {\tt B} that inherits from {\tt A}, then {\tt C} cannot call the {\tt init} or {\tt init:} methods of {\tt A}.
    To call these methods of the immedidate superprototype, use ``{\tt super init}''\/, and ``{\tt super init: args}''\/ as the first statement of the {\tt init} or {\tt init:} method of the subprototype. Currently {\tt init} methods of the same prototype cannot be called (this will change, of course).

\end{enumerate}

There are further restrictions related to methods {\tt init}, {\tt init:}, {\tt new}, and {\tt new:}, given below.
\begin{enumerate}[(a)]
\item A user-declared method cannot have name {\tt new} or {\tt new:}.
\item No keyword of any user-declared method can be {\tt init:} or {\tt new:}. A future addition to the language would be to allow a constructor to start with {\tt init:} and have other keywords:
\begin{cyan}
object Person
    func init: String name { self.name = name; age = 0; salary = 0Float; }
    func init: String name
         age: Int age
         salary: Float salary {
        self.name = name;
        self.age = age;
        self.salary = salary
    }
    String name
    Int age
    Float salary
end
\end{cyan}
The compiler would create, for this prototype, methods ``{\tt new: String}''\/ and \\
\verb|    |```{\tt new: String age: Int salary: Float}''\/.

\end{enumerate}

Methods {\tt new} and {\tt new:} are only accessible through prototype objects. That means one cannot send a {\tt new} or {\tt new: args} message to an expression that is not a prototype:
\begin{cyan}
object Test
    func init: String s { ... }
end

object Program
    func run {
         var t = Test clone;
         var Test u;
             // Ok !
         u = t clone;
             // compile-time error
         u = t new: "hi!";
             // ok
         u = Test new: "hi!";
    }
end
\end{cyan}

{\tt init} and {\tt init:} methods can have visibility {\tt public}, {\tt private}, {\tt protected}, and {\tt package}. The {\tt new:} or {\tt new} method created from the {\tt init:} or {\tt init} method has the same visibility as this one. A {\tt private} method can only be called inside the prototype in which it was declared. A {\tt protected} method can only be called in the prototype it was declared and subprotototypes.
A {\tt package} method can be called in all prototypes of the same package in which it was declared. A {\tt public} method is accessible anywhere.

A singleton prototype is created by declaring a single {\tt private} {\tt init} method, a redefined {\tt clone} method, and no {\tt init:} methods.
\begin{cyan}
object Earth
    private func init { }
    private func clone -> Earth = Earth;
    // other regular methods
    // that do not create Earth objects
end
\end{cyan}

Unlike most object-oriented languages, Cyan demands that all fields be initialized before used. To assure that, the language puts severe restrictions on constructors:
\begin{enumerate}[(a)]
\item every {\tt init} or {\tt init:} method of a prototype should initialize every field of the prototype that is not initialized in its declaration. Shared fields are not considered because they have already been initialized by {\tt initShared} methods or in their declarations. Then
\begin{cyan}
object Manager
    func init: String name { self.name = name }
    var String name
    var Float salary = 1000F;
end
\end{cyan}
is legal but the following prototype is illegal.
\begin{cyan}
object Manager
        // salary is not initialized
    func init: String name { self.name = name }
    var String name
    var Float salary
end
\end{cyan}

The initialization of a field should be made in the top-level statement list of the method. The current Cyan compiler is not smart enough to deduce {\tt category} is initialized in the following example. Initialize a local variable inside the {\tt if} statement and assign it to {\tt category}.
\begin{cyan}
object Person
    func init: String name age: Int age {
        self.name = name;
        if age >= 18 {
            category = "adult"
        }
        else {
            category = "minor"
        }
    }
    String name
    String category
end
\end{cyan}

A field can be initialized in its declaration with an expression. But this expression should be a ``safe expression''\/ (SE), defined recursively as:
\begin{enumerate}[(i)]
\item a basic type (such as {\tt Int}, {\tt Nil}, or {\tt String}) is a SE;
\item a value of a basic type (such as {\tt 0} or \verb|"Hello"|) is a SE;
\item an unary message to a literal value of a basic type (such as {\tt -0} or \verb|+3.14|) is a SE;
\item a literal array, a literal map, or a literal tuple whose elements are SE is a SE;
\item an object creation of a prototype if the arguments used are SE. For example, \\
\verb|    Array<Int>(5)| \\
\noindent or \\
\verb|    Array<String> new: Int new|\\
\noindent is a SE. Unlike the {\tt initShared} methods, there is no resctriction on the package of the prototype, it can be anyone.
\end{enumerate}

If this is not required, a variable could be initialized with a call to a method of the same prototype that accesses a non-initialized variable:
\begin{cyan}
object Test
    Int first = self next
    var Int nextValue = 9
    func next -> Int {
        ++nextValue;
        return nextValue
    }
end
\end{cyan}
In this example, the intention  was to set {\tt first} to {\tt 10} but at runtime, when {\tt next} is called, variable {\tt nextValue} has not been initialized and {\tt next} returns a non-initialized variable {\tt nextValue}.

\item suppose {\tt S} is the superprototype of a prototype {\tt P}. If {\tt S} defines a method {\tt init:} but not a method {\tt init}, then every {\tt init} or {\tt init:} method of {\tt P} should call
    a method  {\tt init:} of {\tt S}. A method {\tt init} or {\tt init:} of {\tt P} may not have a
    super call to a method {\tt init} or {\tt init:} of {\tt S} if {\tt S} has a method {\tt init} (the compiler will insert a call to {\tt init} of {\tt S}). This
    method {\tt init} of {\tt S} may have been added by the compiler;

\item inside an {\tt init} or {\tt init:} method, fields can only be used in expressions after they have been initialized;

\item {\tt init} and {\tt init:} methods cannot use {\tt self} anywhere except in two situations:
    \begin{enumerate}[(i)]
    \item when the prototype is {\it final} and after all prototype fields have been initialized;
    \item the method to be called is annotated with \\
    \verb|    @accessOnlySharedFields|\\
     \noindent
     The method that would be called need not to be {\tt final}.
  These annotated methods can only access shared fields and {\tt self} cannot be leaked.
  The overridden subprototype method of a superprototype method annotated with \\
    \verb|    @accessOnlySharedFields|\\
     \noindent should also be annotated with this same annotation.

    \end{enumerate}

\end{enumerate}

There is an order of initialization of field of a prototype. When an object is created with an {\tt init} or {\tt init:} method, first the fields initialized in their declarations are set (in textual order). Then the statements of method {\tt init} or {\tt init:} are run. The example below shows the order of initialization. {\tt one} is initialized before {\tt two} and so on in an expression ``\verb|Order new|''\/.
\begin{cyan}
object Order
    func init {
        three = 3
    }
    Int three
    Int one = 1;
    func print {
        (one + two + three) println
    }
    Int two = 2;
end
\end{cyan}

If a prototype does not declare an {\tt init} or {\tt init:} method, the compiler will supply one if every field is initialized in its declaration and the superprototype (if any) has an {\tt init} method. The method added by the compiler to the prototype is
\begin{cyan}
    func init {
        super init
    }
\end{cyan}
Of course, considering that there is a superprototype.

If there is any field of a prototype that is not initialized in its declaration, then the prototype should declare at least one {\tt init} or {\tt init:} method. If the superprototype of a prototype {\tt P} does not define an {\tt init} method but defines a method {\tt init:}, then {\tt P} should declare an {\tt init} or {\tt init:} method. The compiler could not create a method
\begin{cyan}
    func init {
        super init: args
    }
\end{cyan}
because it would not know which parameters {\tt args} to pass in the call  {\tt super init: args}.

Metaobject {\tt init} \label{initmeta} automatically create an {\tt init} method
that initializes fields. Consider a prototype {\tt Proto} that declares fields  {\tt p1}, {\tt p2}, ..., {\tt pn} of types {\tt T1}, {\tt T2}, ..., {\tt Tn}. Then a metaobject annotation\\
\nd \verb"     @init(p1, p2, ..., pn)"\\
\nd can be put before the declaration of {\tt Proto}. When  the compiler finds this metaobject annotation, it will add the following method to the prototype
\begin{cyan}
    func init: (T1 p1), (T2 p2), ... (Tn pn)  {
        self.p1 = p1;
        ...
        self.pn = pn;
    }
\end{cyan}
So, a prototype
\begin{cyan}
@init(name, location)
object University
    @property String name
    @property Int age

end
\end{cyan}
can be used as
\begin{cyan}
var p = Person new: "Carol", 1;
p name println;
\end{cyan}  \label{atinit}

There are abbreviations for calling methods called {\tt new} or {\tt new:} of a prototype. Expressions
\begin{cyan}
P new
P new: a
P new: a, b, c
\end{cyan}
can be replaced by
\begin{cyan}
P()
P(a)
P(a, b, c)
\end{cyan}
Using prototypes {\tt Test} and {\tt Person} we can write  \label{newwithpar}
\begin{cyan}
var t1 = Test(0);
var Test t2 = Test("Hello");
var Person p = Person("Mary", 1);
var q = Person("Francisco", 5);
\end{cyan}

Using the short form for object creation, we can easily create a net of objects. In this example, {\tt BinTree} inherits from {\tt Tree} (Section~\ref{inheritance}).   \label{easy}
\begin{cyan}
open
object Tree
end

@init(left, value, right)
open
object BinTree extends Tree
    @property Tree left, right
    @property Int value
end

@init(value)
open
object No extends Tree
    @property Int value
end
...

var tree = BinTree( No(-1), 0, BinTree(No(1), 2, No(3)) );
\end{cyan}

\section{Limitations on the Use of Prototypes as Objects}

Prototypes in Cyan are objects but with restrictions. Unlike every other prototype-based language, not every method of a prototype can be called and it is not always legal to assign it to a variable (this includes parameter passing). But why? Because a prototype is an object whose fields may not have been initialized. Let us see an example.

\begin{cyan}
object Person

    func init: String name, Int age {
        self.name = name;
        self.age = age
    }
    @property
    let String name

    @property
    let Int age

end
\end{cyan}
Annotation {\tt property} creates get methods for {\tt name} and {\tt age}. If methods {\tt getName} or {\tt getAge} of prototype {\tt Person} is called before any other method, there will be an error: a field will be accessed before it has been initialized.
\begin{cyan}
        // access non-initialized field 'name'
    Person getName println;
    var Any any = Person;
        // access non-initialized fields 'name' and 'age'
    var personCopy = any clone;
        // method 'call:' may access non-initialized fields
        // 'name' and 'age'
    otherObject call: Person;
        // ok!
    var protoName = Person prototypeName;
    if any isA: Person {
        "oh!!, a person!" println;
    }
\end{cyan}
To prevent such runtime errors, the language has some rules regarding the use of prototypes. If a prototype has an {\tt init} method, it can be used as a regular object. There are no restrictions in its use. The prototype itself is created using the {\tt new} method of itself. So all of its fields are properly initialized.

If a prototype does not have an {\tt init} method, it:
\begin{enumerate}[(a)]
\item cannot receive messages unless the corresponding method is
    \begin{enumerate}
    \item {\tt new} or {\tt new:};
    \item a {\tt final} method declared in prototype {\tt Any} with annotation {\tt canBeCalledOnPrototypes}:
    \begin{cyan}
    @annot(canBeCalledOnPrototypes)
    final
    func prototypeName -> String { ... }
    \end{cyan}

    \end{enumerate}

\item can be used in an expression if:
    \begin{enumerate}
    \item it is {\tt Nil};
    \item it is an argument to method {\tt isA:} or {\tt notIsA:}
    \begin{cyan}
        if any isA: Person { ... }
        if any notIsA: Cicle { ... }
    \end{cyan}
    \end{enumerate}
\end{enumerate}
These restrictions aim to prevent prototype fields from being accessed before they are initialized.

A prototype, as an object, is initialized by its method {\tt init} (if there is one). Therefore, during the execution of its {\tt init} method, references to the prototype will result in a runtime error.
\begin{cyan}
object FailInInit
    func init {
        message = "prototype refer to itself";
        FailInInit println;
    }
    var String message;
end
\end{cyan}
The use of {\tt FailInInit} inside {\tt init} will cause a {\tt NullPointerException} and then an exception\\
\verb|    java.lang.ExceptionInInitializerError|\\
\noindent This is because the prototype name is translated into a static field called {\tt prototype} of a class {\tt \_FailInInit} that represents the prototype. This field is in process of being initialized when it is used inside {\tt init}. In Java, what happens is:
\begin{verbatim}
    prototype = new FailInInit(); // call 'init'
\end{verbatim}
Before this assignment, {\tt prototype} is {\tt null} (in Java). Inside {\tt init}, \begin{cyan}
    FailInInit println;
\end{cyan}
is translated into the Java code
\begin{verbatim}
    _FailInInit.prototype._println();
\end{verbatim}
{\tt prototype} is {\tt null} because it is being initialized.

Cyan prohibits references to a prototype inside its {\tt init} method. However, this does not solve the problem. The prototype may be indirectly referenced by methods called inside method {\tt init}.

\section{Shared Variables and Method {\tt initShared}}

A prototype may declare a field as {\tt shared}, as {\tt today} in \index{variable!shared} \index{shared variable}
\begin{cyan}
object Date
    ... // methods

       // shared
    @property shared var Date today

    @property Int day
    @property Int month
    @property Int year
end
\end{cyan}
Variable {\tt today} is shared among all {\tt Date} objects. The {\tt clone} message does not duplicate shared variables. By that reason, we do not call shared variables  ``instance''\/ variables. They are similar to  ``class variables''\/ of some languages and ``static''\/ variables of C++/Java/C\#. Every shared variable of a prototype should be initialized in its declaration or in a special method called {\tt initShared}.

There are several restrictions on an {\tt initShared} method. It
\begin{enumerate}[(a)]
\item should be declared with no return type (it cannot be {\tt Nil});
\item should be private (this may change in the future);
\item should not be preceded by keyword {\tt override} (it is private);
\item should not be abstract or final (it is private);
\item should not be indexing methods (See Section~\ref{indexing});
\item cannot be declared in interfaces;
\item cannot initialize non-shared fields.
\end{enumerate}

A {\it Restricted Safe Expression}, RSE, is recursively defined as
\begin{enumerate}[(i)]
\item a basic type (such as {\tt Int}, {\tt Nil}, or {\tt String}) is a RSE;
\item a literal value of a basic type (such as {\tt 0} or \verb|"Hello"|) is a RSE;
\item an unary message to a literal value of a basic type (such as {\tt -0} or \verb|+3.14|) is a RSE;
\item a literal array, a literal map, or a literal tuple whose elements are RSE is a RSE;
\item an object creation of a prototype of package {\tt cyan.lang} is a RSE if the arguments used are RSE. For example, \\
\verb|    Array<Int>(5)| \\
\noindent or \\
\verb|    Array<String> new: Int new|\\
\noindent is a RSE.
\end{enumerate}

A RSE can be assigned to a shared field in its declaration. There should be an {\tt initShared} method that initializes all shared fields not initialized in their declarations. Method {\tt initShared}
should have only assignments to shared variables and these should be initialized with
{\it safe expressions} as defined for {\tt initShared} methods, which are called SE'.

This prevents that a shared variable or a field of another prototype be accessed before it has been initialized. Of course, these restrictions will be relaxed as soon as possible. They prevent very common patterns such as to create an object and assign it to a shared variable --- see example below.
\begin{cyan}
object SolarSystem
    ...
    private func initShared {
          // compile-time error in these two lines
        earth  = Planet new: "earth";
        saturn = Planet new: "saturn"
    }
    shared Planet earth, saturn
    ...
end
\end{cyan}

This prototype should use an union type and there should be a method, called before {\tt SolarSystem} is used, that initializes the shared variables {\tt earth} and {\tt saturn}.
\begin{cyan}
object SolarSystem
    ...
    private func initShared {
        earth = Nil;
        saturn = Nil
    }

    func completeinitShared {
        earth = Planet new: "earth";
        saturn = Planet new: "saturn"
    }

    shared Nil|Planet earth, saturn
    ...
end
\end{cyan}
Unfortunately, now every use of {\tt earth} or {\tt saturn} should test if the variable is {\tt Nil}. We expect that this will change in a near future. The compiler will build a graph of package and prototype dependences and produce initializing code, to be called in the beginning of the program execution, that respects the dependence order among the packages and prototypes.

Another example of a correct use of {\tt initShared} is given below.
\begin{cyan}
object NameServer
    ...
    private func initShared {
        varName = ""
    }
    shared Int nextVarNumber = 0;
    shared String varName
end
\end{cyan}

\section{Shared Methods}

A method can be declared as {\tt shared} meaning that it can only access shared fields. A shared method:
\begin{enumerate}[(a)]
\item should not be preceded by keyword {\tt override} or \tt{overload};
\item should not be abstract or final (it does not make sense);
\item cannot be declared in interfaces;
\item cannot use non-shared fields;
\item should have a name different from any other method or field of the same prototype;
\item should have a name different from any non-private method inherited from superprototypes.
\end{enumerate}
An example of declaration follows
\begin{cyan}
object MyMath
    shared
    func getMax -> Int = 12;
    shared
    func fatorial: Int n -> Int {
        if n == 0 { return 1 }
        else {
          return n*(fatorial: n-1)
        }
    }
    ...
end
\end{cyan}

The receiver of a message passing whose corresponding method is shared should be the prototype in which the method is declared.
\begin{cyan}
    n = 2* MyMath getMax;
    (MyMath fatorial: 10) println;
\end{cyan}

or none. In the last case, the receiver is not specified:
\begin{cyan}
    n = 2*getMax;
    (fatorial: 10) println;
\end{cyan}
There cannot be a shared and a non-shared method in a prototype with the same name. Shared methods should also have different names

\section{Keyword Methods and Selectors}

The example below shows the declaration of a method. The method body is given between {\tt \{} and {\tt \}}.
\begin{cyan}
    func withdraw: Int amount -> Boolean {  // start of method body
        Boolean ret = true;
        (total - amount >= 0) ifTrue: {
            total = total - amount
            }
            ifFalse: {
                ret = false
            };
   }  // end of method body
\end{cyan}

Command \verb@return@ returns \index{return} \index{method!return} the method value. The execution of the function is ended by the \verb@return@ command.

Method {\tt withdraw} takes an argument {\tt amount} of type {\tt Int} and returns a boolean value (of type {\tt Boolean}). It uses a field {\tt total} and sends message\\
\verb@     ifTrue: { .. } ifFalse: { ... } @\\
\nd to the boolean value {\tt total - amount >= 0}. The message has two function arguments, \\
\verb@     { total = total - amount }@\\
\nd and\\
\verb@     { ret = false }@\\
A message like this is called a {\it keyword message} and is similar to Smalltalk keyword messages.
As another example, an object Rectangle can be initialized by\\
\verb@     Rectangle width: 100 height: 50@\\
This object should have  been defined as
\begin{cyan}
@init(w, h, x, y)
object Rectangle
    func width: Int w height: Int h {
        self.w = w;
        self.h = h;
    }
    func set: (Int x, Int y) { self.x = x; self.y = y; }
    func getX -> Int = x;
    func getY -> Int = y;
    Int w, h    // width and height
    Int x, y    // position of the lower-left corner
    ...
end
\end{cyan}

Each identifier followed by a ``{\tt :}''\/ is called a \index{keyword} \index{method!keyword} {\it keyword}. So {\tt width:} and {\tt height:} are the keywords of the first method of {\tt Rectangle}.
Sometimes we will use ``method with multiple keywords''\/ instead of ``keyword method''\/. The concatenation of the method keywords is called the method selector. Then {\tt width:height:} is a method selector.

The signature \index{signature} \index{method!signature} of a method is composed by its keywords, parameter types, and return value type. Then the signature of method ``{\tt width:height:}''\/ is\\
\verb@     width: Int height: Int@\\
\nd The return type is {\tt Nil} and it may not appear. The signature of {\tt getX} is\\
\verb@     getX -> Int@\\

It is important to note that there should be no space before ``{\tt :}''\/ in a keyword. Then the following code is illegal:\\
\verb@    (i > 0) ifTrue : { r = 1 }  ifFalse : { r = 0 }@\\
And so is the declaration\\
\verb@   func width : Int w height : Int h {@\\

To make the declaration of a keyword method clearer, parenthesis can be used to delimit the parameters that appear after a keyword:  
\begin{cyan}
object Rectangle
   func width: (Int w) height: (Int h) {
      self.w = w;
      self.h = h;
   }
   func set: (Int x, y) {
       self.x = x; self.y = y;
   }
   ...
end
\end{cyan}
Parameters \index{variable!parameter} \index{parameter} are read-only. They cannot appear in the right-hand side of an assignment. 


\section{Operator Methods}

Operators can be method names in Cyan. However, there are limitations: binary operators should take one parameter, unary operators should not take parameters. Operators {\tt +} and {\tt -} can be both binary and unary.
\begin{cyan}
object Complex
    func init: Double re, Double im {
        self.re = re;
        self.im = im
    }
      // unary -
    func - -> Complex =
       Complex(-re, -im);

    func + (Complex other) -> Complex =
       Complex(re + other getRe, im + other getIm);
      // binary -
    func - (Complex other) -> Complex =
       Complex(re - other getRe, im - other getIm);

    func * (Complex other) -> Complex =
       Complex( re*(other getRe) - im*(other getIm),
                re*(other getIm) + im*(other getRe) );

    override
    func asString -> String = "($re, $im)";

    func getRe -> Double = re;
    func getIm -> Double = im;
    Double re;
    Double im;
end
\end{cyan}

Operators are frequently used for goals not linked to the expected meaning of them. For example, to use {\tt -} for removing a number from a list. The expected meaning would be to subtract something from the number. To prevent such kind of misuse, Cyan limits the use of the following operators:
\begin{cyan}
+    -    *    /    <<    >>    >.>>    %
\end{cyan}
Prototypes that declare methods with these operators should be read-only.  That is, all fields should be declared with {\tt let} or without the keyword {\tt var}. The goal of this is to limit the use of these operators to mathematical structures, which are usually read-only. This is controversial, we know that.

\section{On Names and Scope}

Unary methods and fields of an object should \index{scope} have different names. Fields and shared variables can have names equal to local variables (which includes parameters):
\begin{cyan}
func setName: String name {
    self.name = name
}
\end{cyan}


An object can declare methods ``{\tt value}''\/ and ``{\tt value:}''\/ as in the following example:
\begin{cyan}
object Store
    var Int _value = 0;
    func value -> Int = _value;
    func value: Int newValue {
        self._value = newValue
    }
end
\end{cyan}
And a method name can be a language keyword followed by ``{\tt :}''\/
\begin{cyan}
    func while: Boolean expr { ... }
\end{cyan}

Usually we will not use get and set methods. Instead, we will use the names of the attributes as the method names as in\\
\begin{cyan}
var Fish fish = Fish new;
fish name: "Cardinal tetra";
fish lifespan: 3;
Out println: "name: ", fish name, " lives up to: ", fish lifespan;
\end{cyan}
{\tt Fish} could have been declared as
\begin{cyan}
object Fish
    String _name;
    Int _lifespan;
    func name -> String = _name;
       // parameter with the same name as field
    func name: _name String { self._name = _name }
    func lifespan      -> String = _lifespan;
    func lifespan: Int _lifespan { self._lifespan = _lifespan }
end
\end{cyan}

\section{Operator []}  \label{indexing}

It is possible to define operator \verb@[]@ for indexing:
\begin{cyan}
object Table
     func [] at: Int index  -> String {
         return anArray[index]
     }
     func [] at: Int index  put: String value {
         anArray[index] = value
     }
     Array<String> anArray
end
...

var t = Table new;
t[0] = "One";
t[1] = "Two";
    // prints "One Two"
Out println: t[0], " ", t[1];
\end{cyan}
This operator can only be used with  methods {\tt at:} and {\tt at:put:}. Each keyword should have only one parameter. When \verb|t[expr]| appears inside an expression, it is considered the same as \verb|(t at: expr)|. When \verb|t[expr]| appears  in the left-hand side of an assignment, \\
\verb|    t[expr] = rightExpr|\\
\noindent this is considered as \\
\verb|    t at: (expr) put: (rightExpr)|\\
\noindent

One or both methods can be declared. But when both are declared, the type of keyword {\tt at:} should be the same. The allowed signatures of these methods are:
\begin{cyan}
    at: T -> U
    at: T put: W -> Nil
    at: T put: W -> U
\end{cyan}
Only one of the last two signatures may be used. Usually, {\tt U = W}. But these types can be different from each other.

Indexing methods cannot be abstract and they should be public.

\section{Inheritance}   \label{inheritance}

A prototype may extends another one using the syntax\\  \index{inheritance}
\verb@    object Student extends Person ... end@\\
This is called inheritance. {\tt Student} inherits all methods and fields defined in {\tt Person}. {\tt Student} is called a sub-object or subprototype. {\tt Person} is the superprototype or superprototype. The declaration of a superprototype should be preceded by identifier ``{\tt open}'':
\begin{cyan}
package main

open
object Person
    // elided
end
\end{cyan}
Note that ``{\tt open}'' is not a keyword. To restrict the subprototypes to the package of the prototype one can use ``{\tt open(package)}'' as in

\begin{cyan}
package main

open(package)
object Person
    // elided
end
\end{cyan}

Every field of the subprototype should have a name different from the names of the public and protected methods of the superprototype (including the inherited ones) and different from the names of the methods and other fields of the subprototype.
Since the name of a non-unary method includes the ``:''\/, there may be field {\tt iv} and method {\tt iv:}.

A public method of a prototype is visible anywhere the prototype is. A protected method is visible in the prototype and its subprototypes. A protected method is declared with the syntax
\begin{cyan}
    protected
    func getList -> List<Int> { ... }
\end{cyan}


A {\tt package} method is visible only in the package of the prototype in which the method is.
\begin{cyan}
package main
open
object Employee
    package
    func getSalary -> Double { ... }
    // elided
end
 // other source file:

package company

object Manager extends Employee

    package
    func getSalary -> Double { ... }
    // elided
end
\end{cyan}
In this case, prototype {\tt Manager}, that inherits from {\tt Employee}, is trying to override method
{\tt getSalary}. This results in a compile-time error: the first method is visible only in package {\tt main} and the {\tt getSalary} of {\tt Manager} is visible only in package {\tt company}. Therefore there is a compilation error.

A public, package, or protected method of a subprototype may have the same keywords, parameter types and number of parameters for each keyword than a method of the superprototype.
\begin{cyan}
open
object MovieList
    ...
    func add: String movieName
         director: String director
         year: Int year {
        ...
    }
    func search: String movieName
         year: Int -> Movie {
         ...
    }
end

object LoveMovieList extends MovieList
    ...
    override
    func add: String movieName
         director: String director
         year: Int year {
        ...
    }
    override
    func search: String movieName
         year: Int -> LoveMovie {
         ...
    }
end
\end{cyan}
Here method
\begin{cyan}
    func add: String movieName
         director: String director
         year: Int year
\end{cyan}
of {\tt MovieList} is redefined in {\tt LoveMovieList}, its subprototype. Each keyword takes the same parameters, which means the same number of parameters and the same types. The subprototype method is preceded by the Cyan keyword {\tt override}. The return value type of the subprototype method that was overridden may be a sub-type of the return value type of the superprototype method. For example, method {\tt search:year:} of {\tt LoveMovieList} overrides a superprototype method and returns a {\tt LovieMovie} that is a subprototype (then a sub-type) of {\tt Movie}, which is the return type of the method of the superprototype. Assume that {\tt LovieMovie} is a subprototype of {\tt Movie}. The sub-type relationship is defined in Section~\ref{typesandsubtypes}.

A method can only be overridden by a method with the same visibility. That is, a public method can only override a public method. And a protected method can only override a protected method. A public or protected method of a subprototype with the same name as a private method of the superprototype is not overridden  (of course!).

The Cyan keyword {\tt override} should follow the qualifier {\tt public} or {\tt protected} if any of these are present. Currently this order is enforced.

\begin{cyan}
open
object Person
    func init: String name, Int age {
        self.name = name;
        self.age  = age;
    }
    override
    func print {
         Out println: "name: $name (age $age)"
    }
    @property String name
    @property Int age
end

object Student extends Person
    func init: String name, Int age , String school {
        super init: name, age;
        self.school = school
    }
    override
    func print {
        super print;
        Out println: " School: ", school
    }
    func nonsense {
            // compile-time error in this line
            // new: cannot be called
        var aPerson = super new: "noname", 0;
            // compile-time error in this line
            // init: cannot be called
        var aPerson = super init: "noname", 0;
            // ok, clone is inherited
        var johnDoe = super clone;
    }
    @property String school
end
\end{cyan}

There is a keyword called {\tt super} \index{super} \index{keyword!super} used to call methods of the superprototype. In the above example, method {\tt print} of {\tt Student} calls method {\tt print} of prototype {\tt Person} and then proceeds to print its own data.

Methods {\tt init}, {\tt init:}, {\tt new}, {\tt new:}, and {\tt initShared} are never inherited. However, {\tt init} or {\tt init:} methods of a subprototype may call {\tt init} or {\tt init:} methods of the superprototype using {\tt super}:.

Keyword {\tt override} should not be used in the declaration of method {\tt init:} of {\tt Student} because {\tt init:} of {\tt Person} is not inherited. The compiler adds to prototype {\tt Person} a method\\
\verb@     Person new: String name, Int age@\\
\nd and to {\tt Student}\\
\verb@     Student new: String name, Int age, String school@\\
Since methods {\tt init:} and {\tt new:} are not inherited, there will be compile-time errors in method {\tt nonsense}. See Section~\ref{init} for the many restrictions on {\tt init}, {\tt init:}, and {\tt initShared} methods.

A prototype may be declared as ``final''\/, which means that it cannot be inherited: \index{final} \index{object!final}
\begin{cyan}
public final object Int
    ...
end
\end{cyan}
There would be a compile-time error if some prototype inherits {\tt Int}. The prototypes
{\tt Byte}, {\tt Short}, {\tt Int}, {\tt Long}, {\tt Float}, {\tt Double}, {\tt Char}, {\tt Boolean},  and {\tt String} are all final.

A method declared as ``{\tt final}''\/ cannot be redefined in subprototypes:
\begin{cyan}
public object Car
    final func name: String newName { _name = newName }
    final String func name = _name;
    String _name
    ...
end
\end{cyan}
Final methods should be declared in non-final prototypes (why?). Final methods allow some optimizations. The message send of the code below is in fact a call to method {\tt name} of {\tt Car} since this method cannot be overridden in subprototypes. Therefore this is a static call, much faster than a regular call.
\begin{cyan}
var Car myCar;
...
s = myCar name;
\end{cyan}

The table below summarizes the allowed combination among keywords in a method declaration. Keyword {\tt abstract} is explained in Section~\ref{abstract}.

\begin{table}[h]
\begin{tabular}{lllllll}
          & public & protected & private & override & abstract & final \\
public    &        &           &         & Y        & Y        & Y     \\
package   &        &           &         & Y        & Y        & Y     \\
protected &        &           &         & Y        & Y        & Y     \\
private   &        &           &         &          &          &       \\
override  & Y      & Y         &         &          & Y        & Y     \\
abstract  & Y      & Y         &         & Y        &          &       \\
final     & Y      & Y         &         & Y        &          &
\end{tabular}
\caption{Keyword combination in method declaration}
\end{table}

The order in which the method qualifiers can appear is rigid. It is
{\tt public}/{\tt private}/{\tt protected}/{\tt package}, {\tt final}, {\tt override}, {\tt abstract}, and {\tt overload}. Then all the declarations below cause compile-time errors.
\begin{cyan}
    abstract protected func abs: Int n
    override final func getList -> List { ... }
    abstract override func parse: String code -> AST { .. }
\end{cyan}

Future versions of the language may demand the these keywords appear in alphabetical order: {\tt abstract}, {\tt final}, {\tt overload}, {\tt override}, {\tt public}/{\tt private}/{\tt protected}. Or without any order.

In order to prevent some traps explained by Huang, Yang, and Chan \cite{DBLP:journals/jss/ChanYH04}, a prototype that defines at least one method with {\tt package} visibility can only be inherited by prototypes in the same package.

\section{Downcasting with type-case and cast statements}  \label{cast}

Downcasting is to change the type of an object of a supertype to the type of a subtype.
Usually this is made by a construction that does the casting and throws an exception if it is not possible. Cyan does not support constructions that does exactly that. But it does support command {\tt type-case} for safe downcasting without exception signalling. Its syntax is
\begin{cyan}
type expression
    case Type1 [ varName1 ] {
        statement list
    }
    // any other number of case clauses
    [ else {
        statement list
      }
    ]
\end{cyan}
The {\tt varName1} is optional and so is the {\tt else} part.
There may be one or more {\tt case} clauses. A \verb|';'| after the last {\tt case} clause or {\tt else} clause is optional.
At runtime, if {\tt expression} has type {\tt Type1}, it is cast to {\tt Type1} and assigned to {\tt varName1}. Then the statements of the {\tt case} are executed. If {\tt expression} has not type {\tt Type1}, the following {\tt case} clauses are tested, in textual order.

As an example, the following code will print \verb|1 2 true B|
\begin{cyan}
    var Dyn d = 0;
    for elem in [ d, "Hi", 5.0, 'A' ] {
        type elem
            case Char ch {
                (ch succ ++ " ") print
            }
            case String s {
                (s size ++ " ") print
            }
            case Int n {
                (n + 1 ++ " ") print
            }
            case Double n {
                ((n equal: 5.000001) ++ " ") print
            }
    }
\end{cyan}

The type of a {\tt case} clause cannot be subtype of the type of a previous {\tt case} clause. Then the two {\tt type-case} of the next example produce errors. Assume that {\tt Circle} inherits from {\tt Shape}.
\begin{cyan}
    var Any elem = 0;
    type elem
        case Any {
        }
        case Int n {   // error here
            ("An int equal to $n") print
        }
    var Shape shape = Circle(100.0, 10.0);
    type shape
        case Shape s {
        }
        case Circle  {  // error here
            "This will never be printed" println
        }
\end{cyan}
An {\tt else} clause may follow the last {\tt case} clause with the obvious meaning. A {\tt type-case} statement should have at least one {\tt case} clause.

Statement {\tt cast} is a special downcast statement whose syntax is:
\begin{cyan}
cast  [ Type1 ] Id1 = Expr1, ..., [ Typen ] Idn = Exprn {
    statementListTrue
}
[
else {
    statementListFalse
}
]
\end{cyan}
If the type of {\tt Expri} is \verb@T|Nil@ or \verb@Nil|T@, with {\tt T} different from {\tt Nil}, {\tt Typei} is optional and assumed {\tt T}. If the cast succeeds, {\tt statementListTrue} is executed. Otherwise, {\tt statementListFalse} is executed. Exemple:
\begin{cyan}
var Int|Nil intNil = 0;
var Any any = "ok";
cast elem = intNil, String s = any {
    assert elem == 0 && s == "ok";
}
else {
    assert false;
}
cast String s = any {
    "s = $s" println
}
\end{cyan}
{\tt elem} is assumed to have type {\tt Int}.

\section{Interfaces} \label{interfaces}

Cyan supports {\it interfaces}, a concept similar to Java interfaces. The declaration of an interface lists zero or more method signatures as in  \index{interface}
\begin{cyan}
interface Printable
    func print
end
\end{cyan}

The {\tt public} keyword is not necessary since all signatures are public. {\tt func} is not necessary but it is demanded for sake of clarity (should it be eliminated too?).

 An interface has two uses:
\begin{enumerate}[(a)]
\item it can be used as the type of variables, parameters, and return values;
\item a prototype can {\it implement} an interface. In this case, the prototype should implement the methods described by the signature of the interface. A prototype can implement any number of interfaces. Name collision in interface implementation is not a problem.
\end{enumerate}
Interfaces are similar to the concept of the same name of Java.

As an example, one can write
\begin{cyan}
interface Printable
    func printObj
end

@init
open
object Person
    @property String name
    @property Int age
end

object Worker extends Person implements Printable
    @property String company
    override
    func printObj {
        Out println: "name: " ++ name ++ " company: " ++ company
    }
    ... // elided
end
\end{cyan}
Here prototype {\tt Worker} should implement method {\tt printObj} because this prototype is {\it implementing} interface {\tt Printable} that defines a {\tt printObj} method. Otherwise the compiler would sign an error. Note that the {\tt printObj} method of {\tt Worker} is preceded by {\tt override}. This is demanded by the compiler.

Interface {\tt Printable} can be used as the type of a variable, parameter, and return value:
\begin{cyan}
var Printable p;
p = Worker clone;
p print;
\end{cyan}

An interface may extend any number of interfaces:
\begin{cyan}
interface ColorPrintable extends Printable, Savable
    func setColor: Int newColor
    func colorPrint
end
\end{cyan}
Therefore Cyan supports a limited form of multiple inheritance. An interface that does not explicitly inherits from any other in fact inherits from prototype {\tt Any} as any other prototype.

An interface is a regular prototype for many uses. It can receive messages for example. But there will be a runtime error if the message corresponds to a method declared in the interface. There will not be a runtime error if the method is inherited from {\tt Any}.

The method signatures declared in an interfaces are transformed into public methods by the compiler. These methods throw exception {\tt ExceptionCannotCallInterfaceMethod}:
\begin{cyan}
   // interface ColorPrintable as a prototype
object ColorPrintable extends Printable, Savable
    func setColor: Int newColor {
        throw: ExceptionCannotCallInterfaceMethod("ColorPrintable::setColor");
    }
    func colorPrint {
        throw: ExceptionCannotCallInterfaceMethod("ColorPrintable::colorPrint");
    }
end
\end{cyan}

Interfaces are then objects with full rights: they be assigned to variables, passed as parameters, and receive messages. However, an interface declaration cannot be preceded by ``{\tt open}'' or ``{\tt open(package)}''. Interfaces are already open. To restrict the use to a package one can declare it with visibility {\tt package}. In future versions of Cyan, this is not working yet.

Although interfaces are objects, the compiler puts some restrictions on their use and declaration.
\begin{enumerate}[(a)]


\item An interface can only extend another interface. It is illegal for an interface to extend a non-interface prototype.
\item Interfaces cannot declare any  {\tt init} or {\tt init:} methods. No object will ever be created from them. But the interface itself may receive messages and it may be cloned.
\item A regular prototype cannot inherit from an interface.
\item If the type of an expression is an interface {\tt I}, then the compiler checks whether the messages sent to it match those method signatures declared in the interface, super-interfaces, and {\tt Any} (See Section~\ref{any}).
\item If interface {\tt Inter} declares a method signature and prototype {\tt P} implements interface  {\tt Inter}, {\tt P} should declare a method with that same signature except that the return value type can be a subtype of the return value type of the method signature of the interface. It is possible that {\tt P} inherits the method. In this case {\tt P} may not declare the method. But whenever {\tt P} declares a method of a implemented interface, it should be preceded by keyword ``{\tt override}''\/.

\item An interface cannot declare a true overloaded method (Section~\ref{overload}). If prototype {\tt P} implements interface {\tt Inter} that defines a method signature {\tt ms}, {\tt P} should define a method {\tt m} with that signature. {\tt m} cannot belong to an overloaded method.

\end{enumerate}

Besides that, method {\tt isInterface} inherited from {\tt Any} returns {\tt true} when the receiver is an interface.
The examples that follow should clarify these observations.
\begin{cyan}
  // ok
var Printable inter = Printable;
  // ok, asString is inherited from Any
Out println: (inter asString);
  // ok, Printable is a regular object
Out println: (Printable asString);
var Any any = Printable;
  // ok
Out println: (any asString);
  // it is ok to pass an interface as parameter
assert: (any isA: Printable);
assert: (any isInterface && Printable isInterface &&
         inter isInterface);
\end{cyan}

\section{Method Overloading} \label{multimethods}

\label{overload}  \label{messagepassing}

There may be methods with the same keywords but with different number of parameters. For example, one can declare  \index{overloaded methods} \index{method!overloading}
\begin{cyan}
object MyPanel
    func print: Circle c, Int x, Int y -> Circle { ... }
    func print: String s, String format -> Boolean { ... }
    func print:  { ... }
    func print: Int n  -> Int { ... }
end
\end{cyan}
There are four {\tt print} methods that are considered different by the compiler. In a message send\\
\verb@     MyPanel print: anObj@\\
\nd there is no ambiguity on which method should be called. It can only be the last method, which is the only one that takes just one parameter. Since all methods are considered different, they may have different return value types. This is not true method overloading.

One may also use several keywords:
\begin{cyan}
object Test
    func at: Int i, Int j  put: Int k     -> String { }
    func at: Int i put: Int k -> Int  { ... }
    func at: String s put: String s, Int k -> Boolean { ... }
    ...
end
\end{cyan}
This object could be used as in
\begin{cyan}
var f = Test;
f at: 0, 1 put: 2;
f at: #1 put: "one", 0;
f at: 0 put: 1;
\end{cyan}

\label{nameofmethod}

We call the ``name of a method''\/ the concatenation of all of its keyword names, each one followed by its number of parameters and a white space. The trailing white space should be removed. For example, methods
\begin{cyan}
    func key:   String aKey
         value: Int aValue -> String
    func name:   String first, String last
         age:    Int aAge
         salary: aSalary Float -> Worker
\end{cyan}
have names \verb|"key:1 value:1"| and \verb|"name:2 age:1 salary:1"|.


A restricted form of \index{multi-methods} \index{method!multi} multi-methods is allowed in Cyan. In most languages, the receiver of a message determines the method to be called at runtime when the message is sent. In CLOS \cite{Seibel:2012:PCL:2339396},
all parameters of the message are taken into consideration (which includes what would be the ``receiver''\/). This is called multiple dispatch and the methods are called ``multi-methods''\/.

Cyan implements a restricted version of multi-methods: the method to be called is chosen based on the receiver and also on the runtime type of the parameters. This is called ``overloading''\/ in Cyan and the methods involved are called overloaded methods --- this is {\bf true} overloading. The compiler generates just one method for all
overloaded methods with the same name in a prototype. Then we sometimes say ``overloaded method''\/, in the singular.

An overloaded method is composed by one or more regular methods with the same name declared in the same prototype. The first method should be preceded by the Cyan keyword {\tt overload} (more on this soon).
\begin{cyan}
open
object MyBlackBoard
       // keyword 'overload' precedes an overloaded method declaration
    overload
    func draw: Square f   { ... }
    func draw: Triangle f { ... }
    func draw: Circle f   { ... }
    func draw: Shape f    { ... }
    private String name
end
\end{cyan}
Then in the example above there is one {\tt draw} overloaded method composed by four methods. In the example below there is another overloaded method {\tt draw} composed by two methods.
\begin{cyan}
object MyOtherBlackBoard extends MyBlackBoard
    func draw: Polygon f   { ... }
    func draw: Elipse e    { ... }
end
\end{cyan}

The first method of an overloaded method in a prototype hierarchy should be prefixed with the keyword ``\verb|overload|''\/. By ``first''\/ we mean the method that is higher in the hierarchy and that is textually before the others with the same name in its prototype. This first  overloaded method should not override another method and all methods that override it are also considered overloaded methods.  Overloaded methods can only be {\tt public}. \footnote{this may change later.} Keyword {\tt overload} should appear before the ``{\tt func}''\/ keyword.

The compiler checks  all methods of an overloaded  method of a prototype at once. The rules below are based on a single method but once no error is detected, all methods of the overloaded method of the prototype are considered correct (with relation to its signature). Before proceding, remember that the name of a method is composed by joining each of its keywords followed by the number of parameters followed by a white space. The last space is discarded. So the name of method\\
\verb|     func at: Int n with: String s, Long k add: Person p -> Boolean { ... }|\\
\noindent is\\
\verb|    "at:1 with:2 add:1"|\\
\noindent The signature of a method is composed by the keywords and the full name of the parameter types and return value type. Then the signature of the above method is\\
\verb|    at: Int with: String, Long add: main.Person -> Boolean|\\
\noindent Prototypes of package {\tt cyan.lang} are not preceded by the package name.

Consider a method {\tt m} not preceded by {\tt overload} whose signature is {\tt ms}. {\tt m} is declared in a prototype {\tt P}. During the semantic analysis, the Cyan compiler does some checkings on {\tt m}.
The compiler
search for all public and protected methods with the same name in {\tt P} and its superprototypes putting the list of methods found in {\tt msList}. The search includes {\tt m} itself. The compiler searches for all methods with the same name as {\tt m} in all interfaces implemented by {\tt P} putting the list of method signatures found in {\tt interMSList}. Then {\tt interMSList} contains zero or one signatures because interfaces cannot declare or inherit methods with different signatures and the same name. The list of all public and protected methods declared in {\tt P} with the same name as {\tt m} is put in list {\tt pmsList} (so {\tt pmsList} is a subset of {\tt msList} and it does not include inherited methods). The rules for the validity of the declaration of {\tt m} are given below.

\begin{enumerate}[(a)]
\item If {\tt msList} has just method {\tt m} and {\tt interMSList} is empty, the declaration of {\tt m} is correct. If {\tt interMSList} is not empty, it should have just one method since interfaces cannot declare overloaded methods. If the single method signature of {\tt interMSList} is different from the signature of {\tt m} then the declaration of {\tt m} is incorrect;

\item Suppose all methods of {\tt msList} are in {\tt P} ({\tt pmsList} is equal to {\tt msList}) and {\tt pmsList} has more than one element.  The declaration of the methods of {\tt pmsList} are correct if:
      \begin{enumerate}[(i)]
      \item each two methods of {\tt pmsList} have different signatures;
      \item all methods of {\tt pmsList} have the same return value type;
      \item the first textually declared method is preceded by keyword {\tt overload}. No other method is preceded by this keyword;
      \item if one method of {\tt pmsList} is final, all methods of this list should be final too. If it is not final, no method of the list should be final;

     \item no method of {\tt pmsList} is protected or abstract;

      \item {\tt interMSList} should be empty.
      \end{enumerate}

\item Suppose there is at least one method in {\tt msList} that is in a superprototype and:
\begin{enumerate}[(i)]
\item no method of {\tt msList} is preceded by {\tt overload};
\item there is just one method in {\tt pmsList}. It should be preceded by ``{\tt override}''\/;
\item  the return value type of {\tt m} is a subtype of the return value type of
{\tt m1}, which is the first method with name equal to {\tt m} found in a search starting in the superprototype of {\tt P} and continuing upwards;
\item all methods of {\tt msList} have the same signature except for the return value type. That would mean that each method of {\tt msList} is in a different prototype;
\item if the method of {\tt pmsList} is protected, so are all the methods of the list {\tt msList}.
\end{enumerate}
Then the declaration of the method of {\tt pmsList} is correct even if {\tt interMSList} contains an element.

\item Suppose there is at least one method in {\tt msList} that is in a superprototype and there are at least two methods of {\tt msList} that have different signatures. That includes the case in which {\tt pmsList} has two methods (since they have the same name,  they must have different signatures).
The declaration of the methods of {\tt pmsList} are correct if:
      \begin{enumerate}[(i)]
      \item each two methods of {\tt pmsList} have different signatures. Note that {\tt pmsList} may have just one element, {\tt m}, although {\tt msList} should have at least two elements;
      \item all methods of {\tt pmsList} are preceded by keyword ``{\tt override}''\/;
      \item all methods of {\tt pmsList} have the same return value type;
      \item let {\tt Q} be the first direct or indirect superprototype of {\tt P} that declares a method with the same name as {\tt m} and {\tt directSMList} be the list of methods of {\tt Q} that have the same name as {\tt m}. All methods of {\tt directSMList} have the same return value type {\tt R}. The return type of all methods of {\tt pmsList} should be subtype of {\tt R};
      \item let {\tt T} be the superprototype of {\tt P} that declares a method with the same name as {\tt m} and that is higher in the {\tt P} hierarchy. That is, no superprototype of {\tt T} declares a method with the same name as {\tt m}. Then the first textually declared method of {\tt T} should be preceded by keyword {\tt overload}. No other method in the {\tt P} hierarchy should be preceded by this keyword;
      \item either none or all methods of {\tt pmsList} are final;
      \item no method of {\tt pmsList} is protected or abstract;
      \item {\tt interMSList} should be empty.
      \end{enumerate}
\end{enumerate}

If a method {\tt m} of a prototype {\tt P} is declared with the {\tt overload} keyword, then:
\begin{enumerate}[(a)]

\item no method with the same name should have been declared textually before it in the prototype hierarchy. That includes {\tt P} and its superprototypes. That is, {\tt m} should not override a superprototype method;
\item no interface implemented by {\tt P} should declare a method with the same name as {\tt m};
\item no method with the same name in the prototype should be abstract;
\item if the method is final, all methods with the same name should be final too. If it is not final, no method with the same name can be final;
\item the return value type of all methods with the same name as {\tt m} in {\tt P} should be the same.
\end{enumerate}

Let us see some incorrect examples:
\begin{cyan}
object MyBlackBoard
    func draw: Square f   { ... }
    overload   // second method, compile-time error
    func draw: Triangle f { ... }
    func draw: Circle f   { ... }
    func draw: Shape f    { ... }
    private String name
end
\end{cyan}

\begin{cyan}
open
object A
    func draw: Square f   { ... }
end
object B extends A
    func draw: Shape p { ... }
end
\end{cyan}
Method {\tt draw} of {\tt A} should have been declared as {\tt overload}.

\begin{cyan}
interface I
    func draw: Shape
end
object A
    func draw: Square f   { ... }
end
object B extends A implements I
    func draw: Square p { ... }
    override
    func draw: Shape p { ... }
end
\end{cyan}
{\tt B} should implement ``\verb|draw: Shape|''\/ thus becoming {\tt draw} in an overloaded method. Then {\tt A} should have declared this method as an overloaded method.

In a message send, the method found at compile-time may be different from the method called at runtime. This is true regardless the method found at compile-time is an overloaded method or not. If the method found at compile-time is an overloaded method, the runtime search for a method is as usual starting from the runtime prototype of the receiver and proceding to the top of the hierarchy ({\tt Any}). However, there may be, in a prototype, two or more methods that have the name of the method found at compile-time. The compiler tests whether one of these methods, in the textual declaration order, can accepts parameters of the message send. To accept a parameter, each runtime argument should be subtype of the declared parameter of the method. If no method of the prototype can accept the runtime parameters, the search for a method continues at the superprototype.

To make the mechanism clearer, study the example below. Assume that {\tt Grass}, {\tt FishMeat}, and {\tt Plant} are prototypes that inherit from prototype {\tt Food}.

\label{cow}
\begin{cyan}
open
object Animal
    overload
    func eat: Food food { Out println: "eating food" }
end

object Cow extends Animal
     override
     func eat: Grass food { Out println: "eating grass" }
end

object Fish extends Animal
    override
    func eat: FishMeat food { Out println: "eating fish meat" }

    override
    func eat: Plant food  { Out println: "eating plants" }
end

object Program
    func run {
        var Animal animal;
        var Food food;
        animal = Cow;
        animal eat: Grass; // prints "eating grass"
        animal eat: Food;  // prints "eating food"
           // the next two message sends prints the same as above
           // the static type of the parameter does not matter
        food = Grass;
        animal eat: food; // prints "eating grass"
        food = Food;
        animal eat: food;  // prints "eating food"

        animal = Fish;
        animal eat: FishMeat; // prints "eating fish meat"
        animal eat: Plant;    // prints "eating plants"
        animal eat: Food;     // prints "eating food"
           // the next two message sends prints the same as above
           // the static type of the parameter does not matter
        food = FishMeat;
        animal eat: food; // prints "eating fish meat"
        food = Plant;
        animal eat: food;    // prints "eating plants"
        food = Food;
        animal eat: food;  // prints "eating food"

    }
end
\end{cyan}

\section{{\tt Nil} and {\tt Any}, the superprototype of Everybody}  \label{any}

{\tt Nil} is a prototype outside the type hierarchy. It is not supertype or subtype of any other prototype. Therefore a variable whose type is {\tt Nil} can only be assigned the value {\tt Nil}. And {\tt Nil} can only be assigned to a variable whose type is {\tt Nil}. But when using dynamic typing this rule should not be obeyed. As shown in Section~\ref{dynamictyping}, {\tt Nil} is also compatible with type {\tt Dyn}. Any expression can be assigned to a variable whose type is {\tt Dyn} and an expression whose type is {\tt Dyn} can be assigned to any variable. That is, {\tt Dyn} is supertype and subtype of anything, including {\tt Nil}. See the example.
\begin{cyan}
    var Nil myEmptyness;
    myEmptyness = Nil; // ok
    var String s;
    s = myEmptyness; // compile-time error
    s = Nil; // compile-time error
    myEmptyness = s; // compile-time error
    Dyn myDyn = Nil; // ok
\end{cyan}
The declaration of {\tt Nil} is given below. This prototype defines some basic methods that are not really necessary but were included to make {\tt Nil} and {\tt Any} have a common interface. Then there will never be an error if a message {\tt println} is sent to a {\tt Dyn} expression.
\begin{cyan}
package cyan.lang

object Nil

    func prototypeName -> String
    func asString -> String
    func asString: (Int ident) -> String
    func print
    func println
    func == (Dyn other)  -> Boolean
    func === (Dyn other)  -> Boolean
    func != (Dyn other)  -> Boolean
    func !== (Dyn other)  -> Boolean

end
\end{cyan}

Prototypes \index{Any} that are declared without explicitly extending a superprototype in fact extend an object called {\tt Any}. Therefore {\tt Any} is the superprototype of every other object but {\tt Nil}. It defines some methods common to all objects such as {\tt asString}, which converts the object data to a format adequate to printing. For example, \\
\verb@     Rectangle width: 100 height: 50@\\
\verb@     Rectangle set 0, 0;@\\
\verb@     Out println: (Rectangle asString);@\\
could print something like
\begin{cyan}
Rectangle {
   w: 100
   h: 50
   x: 0
   y: 0
}
\end{cyan}
Method {\tt asString: Int n} also converts its receiver to a {\tt String}. However, it does that with an identation of {\tt n} white spaces.  The indentation is made with {\tt defaultIdentNumber} white spaces. This is a shared variable declared in {\tt Any}.

The methods declared in {\tt Any} are given below. The method bodies are elided.
\begin{cyan}
package cyan.lang

public object Any

    func eq:  (Dyn other) -> Boolean
    func neq: (Dyn other) -> Boolean
    func prototype -> Any
    final func prototypeName -> String
    final func prototypeParent -> Any
    final func prototypePackageName -> String
    final func isInterface -> Boolean
    final func isA: (Any proto) -> Boolean
    final func notIsA: (Any proto) -> Boolean
    final func throw: (CyException e) -> Dyn
    func clone -> Any
    final
    func ++ (Any other) -> String
    func asString -> String = asString: 0;
    func asString: (Int ident) -> String
    func asStringThisOnly: Int ident -> String
    func asStringQuoteIfString -> String
    func == (Dyn other) -> Boolean
    func === (Dyn other) -> Boolean
    func != (Dyn other) -> Boolean
    func !== (Dyn other) -> Boolean
    func isCase: (Any other) -> Boolean
    func assertxx: (Boolean expr)
    func assertxx: Boolean expr, String message
    func print
    func println
    func toAny: Dyn elem -> Any
    final
    func featureList -> Array<Tuple<key, String, value, Any>>
    final
    func featureList: (String slotName) -> Array<Tuple<key, String, value, Any>>
    final
    func slotFeatureList -> Array<Tuple<slotName, String, key, String, value, Any>>
    final
    func annotList -> Array<Any>
    final
    func annotList: (String slotName) -> Array<Any>
    func doesNotUnderstand: (String methodName, Array<Array<Dyn>> args) -> Dyn
    final
    func functionForMethod: String signature -> Any
    final func functionForMethodWithSelf: String signature -> Any
    func hashCode -> Int

end
\end{cyan}

Method {\tt prototype} returns the prototype of an object. It is added to the compiler to every prototype --- the cannot be user defined. Its return value type is the prototype. Then in a prototype {\tt Student} method {\tt prototype} is
\begin{cyan}
    func prototype -> Student
\end{cyan}
Methods {\tt new} and {\tt new:} are added to the prototype if it defines the correspondent {\tt init} and {\tt init:} methods.

Note that some of the {\tt Any} methods are final and therefore they cannot be user-defined. As an example,
{\tt prototypeName} is final.

Method {\tt eq:} returns {\tt true} if {\tt self} and the parameter reference the same object, {\tt false} otherwise. This method is not final but it can only be overridden in the  basic subprototypes, including {\tt String} (a metaobject checks that). For the basic types, excluding {\tt Nil}, method {\tt eq:} compares the contents of the receiver object with the parameter. It is equivalent to {\tt ==}.

Method {\tt prototypeParent} returns the parent prototype of the receiver determined at runtime.
\begin{cyan}
var Person p = Person("fulano", 32);
assert p parent == Person prototypeParent;
\end{cyan}
Method {\tt prototypeParent} of an interface always return {\tt Any} even if the interface inherits from several other ones.  When the receiver of message {\tt prototypeParent} is {\tt Any}, the value returned is {\tt Any}. A future change is to make  {\tt prototypeParent} return \verb"Any|Nil". Then  {\tt prototypeParent} on {\tt Any} would return {\tt Nil}.

Method {\tt isInterface} returns {\tt true} if the receiver is an interface. It cannot be redefined.
Method {\tt isA:} returns true if the prototype of {\tt self} is the same as {\tt proto} or a descendent of it. Parameter {\tt proto} should be a prototype, which is checked by a metaobject.
Assuming that {\tt Circle} inherits from {\tt Elipse} that inherits from {\tt Any}, we have
\begin{cyan}
var Elipse e = Elipse(...);  // elided
var Circle c = Circle x: 100 y: 200 radius: 30;
assert c isA: Elipse && c isA: Circle;
assert c isA: Any && Circle isA: Any && Circle isA: Circle;
\end{cyan}
Method {\tt notIsA:} return the negation of {\tt isA:}.

Method {\tt throw:} throws the exception that is the parameter. See more on Chapter~\ref{ehs}.
{\tt hashCode} returns an integer that is the hash code of the receiver object (this needs to be better defined).

The compiler adds method  {\tt clone} to every prototype that does not define this method. If a prototype defines {\tt clone}, the compiler checks whether it has the correct method signature.
{\tt clone} returns a cloned copy of {\tt self}. It is used shallow copy.
Method {\tt asString} returns a string with the content of {\tt self}. It can and should be override to give a more faithful representation of the object. Method {\tt ==} returns the same as {\tt eq:} by default. But it can and should be user-defined. In the basic types, it returns {\tt true} if the values are equal.
Method \verb"!=" returns {\tt true} if {\tt ==} returns {\tt false} and vice-versa.
Method {\tt isCase:} is the same as {\tt ==} in {\tt Any}. This method will be used in a {\tt switch} statement that will be added to the language.

Method {\tt assertxx:} takes a boolean expression as parameter and throws exception {\tt ExceptionAssert} if {\tt expr} is false. Methods {\tt print} and {\tt println} print information on the receives using methods {\tt print:} and {\tt println:} of prototype {\tt Out}.

A feature is a metadata composed by a name and a value that can be attached to a prototype, field, shared variable, or method. {\tt Any} defines methods for retrieving features from the prototype and its methods and fields.
\begin{cyan}
package main
  // feature attached to a prototype
@feature(xmlRoot, "A_person")
@init(name, age, city)
object Person
    func getName -> String = name;
    func getAge -> Int = age;
    func getCity -> City = city;
      // feature attached to a field
    @feature(xmlElement, "PersonName")
    String name
    @feature(xmlElement, "PersonAge")
    Int age
    @feature(xmlElement, "PersonCity")
    City city


      // feature attached to a method
    @feature(aName, "a complex name")
    func at: Int n with: String s { }

    @feature(over1, "at(Int)")
    overload
    func at: Int n { }
    @feature(over2, "at(String)")
    func at: String s { }
end
\end{cyan}

Method {\tt featureList} returns an array with all features of the prototype.  Method
\begin{cyan}
    func featureList: (String slotName) -> Array<Tuple<key, String, value, Any>>
\end{cyan}
returns the feature list of slot {\tt slotName}, which may be the name of a field, method, or shared variable declared in the prototype. The method name includes the types of the parameters but not the return value type. This table gives several method names

\begin{tabular}{ll}
method          & method name for featureList: \\
\verb|+ -> Int| & \verb|"+"| \\
\verb|+ Int -> Int| & \verb|"+ Int"|\\
\verb|at:   Int -> Int| & \verb|"at: Int"|\\
\verb|at: cyan.lang.Int,| \\\verb| String with: Person p -> String| & \verb|"at: cyan.lang.Int, String with: Person p"|\\
\end{tabular}

Unfortunately, the parameter {\tt slotName} should be exactly as in the declaration. Which is not well specified. For example, a method \\
\verb|    add: Table t|\\
\nd has name \verb|"add: Table"| without any specification on the package of {\tt Table}. And if a method is declared as\\
\verb|    add: cyan.lang.Int elem|\\
\nd its name should be considered\\
\verb|"add: cyan.lang.Int"|\\
\nd and not the simpler\\
\nd \verb|"add: Int"|\\
\nd It is not necessary to say that this is a flaw in the language that will be corrected in a near future.

Method {\tt featureList} of {\tt Any} returns a list of features of the prototype of the receiver:
\begin{cyan}
for t in any featureList {
    Out println: "key = ", t key, " value = ", t value;
}
\end{cyan}

Method {\tt slotFeatureList} returns a list of all features of all slots of the prototype of the receiver.

Annotations are a special case of {\it features}. An attachment \\
\nd \verb!    @annot( #root )!\\
\nd is the same as\\
\nd \verb!    @feature("annot", #root)!\\
Method {\tt annotList} returns a list of annotation objects attached to the prototype. Method
\begin{cyan}
    func annotList: (String slotName) -> Array<Any>
\end{cyan}
returns the annotation list of slot {\tt slotName}. For methods, {\tt slotName} is the concatenation of the keywords

Method {\tt doesNotUnderstand:} is called whenever a message is sent to the object and it does not have an appropriate method for that message. The message name (as a symbol) and the arguments are passed as arguments to {\tt doesNotUnderstand:}. This method ends the program with an error message. The name of a message is the concatenation of its keywords. The name of message\\
\nd \verb"      ht at: i put: obj with: #first"\\
\nd is ``{\tt at:put:with:}''\/. The name is also called the ``selector'' of \index{message selector} the message.

Since Cyan is statically typed, regular message sends will never cause the runtime error ``method not found''\/. But that can occur with dynamic message sends such as
\begin{cyan}
var Int n = 0;
var Dyn d = 0;
n ?push: 10;       // runtime error
// runtime error if the previous line is commented
d at: 0 put: 10;
\end{cyan}

The second argument to method {\tt doesNotUnderstand:} is an array whose
elements are arrays, each one grouping the parameters of each keyword.
Then if method {\tt format:print:to:} does not exist in object {\tt x}, the call\\
\verb@    x format: "%d%s%i" print: n, name, age  to: output@\\
\nd will cause method {\tt doesNotUnderstand:} to be called with parameter\\
\begin{cyan}
    [
        [ first0 ],
        [ first1, name, age ],
        [ first2 ]
    ]
\end{cyan}
in which
\begin{cyan}
var Dyn first0 = "%d%s%i";
var Dyn first1 = n;
var Dyn first2 = output;
\end{cyan}
The last assignments assure that the array is of type \verb@Array<Dyn>@.

There are several missing methods in {\tt Any} related to reflective introspection.  These reflective introspection methods will be added to {\tt Any} in a near future.

\section{Abstract Prototypes}  \label{abstract}

Abstract prototypes in \index{object!abstract} \index{abstract} \index{keyword!abstract} Cyan are the counterpart of abstract classes of class-based object-oriented languages.
The syntax for declaring an abstract prototype is
\begin{cyan}
package myShapes

abstract object Shape
    func init: Int newColor {
       color: newColor;
       shapeColor = 0;
    }
    abstract func draw
    func color -> Int = shapeColor;
    func color: Int newColor { shapeColor = newColor }
    Int shapeColor
end
\end{cyan}
An abstract prototype is considered ``{\tt open}''. It is not necessary to put this word before the prototype declaration. However, to restrict the subprototypes to be package of the abstract prototype, one may use ``{\tt open(package)}''.

An abstract method is declared by putting keyword ``{\tt abstract}''\/ before ``{\tt func}''\/ and it can only be declared in an abstract prototype, which may also have non-abstract methods and fields. A subprototype of an abstract prototype may be declared abstract or not. However, if it does not define the inherited abstract methods, it must be declared as abstract.

Overloaded methods cannot be abstract. Future versions of Cyan may allow that.

It is a compile-time error to send a message {\tt new} or {\tt new:} to an abstract prototype. Since these methods can only be called through a prototype, no object will ever be created from an abstract prototype.
Methods {\tt new} and {\tt new:} cannot be called even using reflection:
\begin{cyan}
  var shape = Shape;
    // prints 'Any', the superprototype of Shape
  (shape ?new) prototypeName println;

     // compile-time error
  (Shape ?new) prototypeName println;
\end{cyan}

It is not a compile-time error to call other methods of an abstract prototype:
\begin{cyan}
  Shape prototypeName println; // ok, prints "Shape"
  Shape draw; // runtime error
  let Any any = Shape;
  any prototypeName println; // ok, prints "Shape"
  any draw; // runtime error if previous error is commented
  let Dyn dyn = Shape;
  dyn prototypeName println; // ok, prints "Shape"
  dyn draw; // runtime error if previous errors are commented
\end{cyan}
If the method called is abstract, as {\tt draw}, there will be a runtime error. An exception will be thrown because the compiler adds a body to every abstract method. This body
throws an exception {\tt ExceptionCannotCallAbstractMethod}:
\begin{cyan}
    func draw {
        throw: ExceptionCannotCallAbstractMethod("myShapes.Shape::draw")
    }
\end{cyan}

{\tt init} and {\tt init:} methods may be declared --- they may be called by subprototypes.

An abstract prototype can inherit from a non-abstract prototype. However, an abstract method cannot override an inherited method.

Objects are concrete things. It seems weird to call a concrete thing ``abstract''\/. However, this is not worse than to call an abstract thing ``abstract''\/. Classes are abstraction of objects and there are ``abstract classes''\/, an abstraction of an abstraction.

Suppose an abstract prototype {\tt Solid} inherits from abstract prototype {\tt Shape}. If {\tt Solid} declares an abstract method {\tt draw}, this should be preceded by keyword {\tt abstract}:
\begin{cyan}
abstract object Solid extends Shape
    abstract override
    func draw
end
\end{cyan}

\section{Types and Subtypes}  \label{typesandsubtypes}

A \index{type} type is a {\tt prototype} (when used as the type of a variable or return value) or an interface. Subtypes \index{subtype} are defined \index{object!type} inductively. {\tt S} is subtype of {\tt T} if:
\begin{enumerate}[(a)]
\item {\tt S} extends {\tt T} (in this case {\tt S} and {\tt T} are both prototypes or both interfaces);
\item {\tt S} implements {\tt T} (in this case {\tt S} is a prototype and {\tt T} is an interface);
\item {\tt S} is a subtype of a type {\tt U} and {\tt U} is a subtype of {\tt T}.
\end{enumerate}
Then, in the fake example below, {\tt I} is supertype of every other type, {\tt J} is supertype of {\tt I}, {\tt J} and {\tt D} are supertypes of {\tt E}, and {\tt B} is supertype of {\tt C}, {\tt D}, and {\tt E}.
\begin{cyan}
interface I   end
interface J extends I end
open
object A implements I end
open
object B extends A end
open
object C extends B end
open
object D extends C implements J end
open
object E extends D end
\end{cyan}

Considering that the static type or compile-time type of {\tt s} is {\tt S} and the static type of {\tt t} is {\tt T}, the assignment ``{\tt t = s}''\/ is legal if {\tt S} is a subtype of {\tt T}. Using the previous example, the following declarations and assignments are legal:
\begin{cyan}
var I i;
var J j;
var A a;
var B b;
var D d;
var E e;
i = j;     i = a;     a = e;    i = a;
j = d;     b = d;     j = e;
\end{cyan}

There is a predefined function {\tt typeof} evaluated at compile-time that return the type of an expression. In the example
\begin{cyan}
var Int x;
var typeof(x) y;
\end{cyan}
{\tt x} and {\tt y} have both the {\tt Int} type. {\tt typeof} works even with expressions, which are not evaluated at runtime:
\begin{cyan}
var typeof( 1 + 2 ) result;
assert result prototypeName == "Int";
\end{cyan}
{\tt result} will have the type \verb"Int".

\section{Union Types}  \label{uniontypes}

The language supports type unions that is the union of two or more types.
\begin{cyan}
var String|Int si;
si = 0;
assert si == 0;
si = "zero";
assert si == "zero";
\end{cyan}
An {\it union type} is a {\it virtual type}; that is, a type for which there is no prototype. There is a restriction in relation to an union {\tt T$_1$, T$_2$, ... T$_n$}: if $j \geqslant i$, then {\tt T$_i$} cannot be supertype of {\tt T$_j$}.
Hence,
\verb@Any|Int@ is illegal because {\tt Any} is supertype of {\tt Int}.
A special case of the second rule is that type cannot be repeated: \verb@Int|Int@ is illegal.

The type checking rules for unions are those expected:
\begin{enumerate}[(a)]
  \item {\tt A} is a supertype of {\tt T$_1$\verb@|@T$_2$\verb@|@ ...
  \verb@|@T$_n$} if {\tt A} is a supertype of every {\tt T$_i$} for $1 \leqslant i \leqslant n$;
  \item {\tt T$_1$\verb@|@T$_2$\verb@|@ ... \verb@|@T$_n$} is a supertype of {\tt A} if there is a {\tt T$_j$} that is supertype of {\tt A};
  \item {\tt T$_1$\verb@|@T$_2$\verb@|@ ... \verb@|@T$_n$} is a supertype of {\tt U$_1$\verb@|@U$_2$\verb@|@ ... \verb@|@U$_m$} if, for each {\tt U$_j$}, $1 \leqslant j \leqslant m$, there is a {\tt T$_i$}, $1 \leqslant i \leqslant n$, such that {\tt T$_i$} is a supertype of {\tt U$_j$}.
\end{enumerate}
Assume {\tt B} and {\tt C} inherits from {\tt A}, {\tt D} inherits from {\tt C}. All prototypes declare an {\tt init} method. The example shows some legal assignments.
\begin{cyan}
var Any any = Any;
var A a = A();
var B b = B();
var C c = C();
var D d = D();
var B|C bc = b;
a = bc;
// b = bc; // error
any = bc;
var D|C dc = c;
// var C|D cd; // D should come first
c = dc;
a = dc;
bc = d;
var B|D bd = b;
bc = bd;
var Any|Nil anil = bd;
anil = bc;
anil = Nil;
var Dyn dyn = anil;    // or anything
var B|Dyn bdyn = anil; // or anything
\end{cyan}

An expression whose type is an union can only receive messages \verb|==| and \verb|!=|. If other methods need to be called, use the {\tt type-case} statement. In this example, assume {\tt Shape} is a superprototype of {\tt Circle}.
\begin{cyan}
    var Circle|Shape join;
    ...
    type join
        case Circle aCircle {
            ("join is a Circle with radius " ++ aCircle radius) println
        }
        case Shape aShape {
            "join is a Shape" println
        }
\end{cyan}
The {\tt type-case} rules apply: the type of a {\tt case} clause cannot be a subtype of the type of a previous clause and each {\tt case} type should be a subtype of the expression type.

Unions are an alternative to method overloading. Instead of declaring several methods, one for each parameter type, there may be a single method:
\begin{cyan}
object Company
    ...

    func apply: Manager|Director futureEmployee { ... }

    Array<Manager | Director> employeeList;

end
\end{cyan}

\section{Tagged Unions} \label{taggedunions}

The generic prototype {\tt Union} of package {\tt cyan.lang} is used for {\it tagged unions}. An instantiation of this prototype should have the format\\
\verb|    Union<|{\tt id$_1$, T$_1$, ... id$_n$, T$_n$}\verb|>|\\
\noindent in which {\tt id$_i$} is a lower case identifier and {\tt T$_i$} is any Cyan type but {\tt Nil}. Each object keeps just one object at a time, tagged with the identifier. For each identifier {\tt id$_i$} associated with type {\tt T$_i$}, the union declares a method:\\
\verb|    |{\tt func id$_i$: T$_i$ elem -> Union\verb@<@ id$_1$, T$_1$, ... id$_n$, T$_n$}\verb|>|\\
{\tt id$_i$:} is used to store an element to the union object and associate it with {\tt id$_i$}. {\tt self} is the value returned by this method. The stored object can only be retrieved using statement {\tt type-case}.
\begin{cyan}
   // create the union object
var myUnion = Union<number, Int, numberStr, String> new;

// myUnion = "12"; // compile-time error if uncommented
  // associate 'number' to 12
myUnion number: 12; // ok
myUnion numberStr: "12"; // ok

type myUnion
    case Int number {
        (1 + 2*number) println
    }
    case String numberStr {
        if numberStr startsWith: "1" { numberStr println }
    }
\end{cyan}
The {\tt type-case} statement should have one {\tt case} clause for each tag of the union, in the order the tags appear in the {\tt Union} prototype. The variable name of each {\tt case} clause should be the tag.
Since the identifiers identify the value kept by the tagged union object, types can be repeated.
\begin{cyan}
var enery = Union<wattHour, Double, calorie, Double, joule, Double> new;

energy wattHour: 314.15;
type energy
    case Double wattHour {
       "watt-hours: $wattHour" println
    }
    case Double calorie {
        "I had $calorie calories" println
    }
    case Double joule {
        "I had $joule joules" println
    }
\end{cyan}

The types of union elements can be other tagged or non-tagged unions.
\begin{cyan}
var chother = Union<ch, Char, other, String>();
chother other: "aaa";
var  agename =
    Union<age, Int, name, Union<ch, Char, other, String>>();
agename name: chother;
type agename
    case Int age { printexpr age; }
    case Union<ch, Char, other, String> name {
        type name
            case Char ch { printexpr ch; }
            case String other { printexpr other; }
    }

var p = Union<age, Int, name, Char|String>();
p name: 'a';
type p
    case Int age { printexpr age; }
    case Char|String name {
        type name
            case Char ch { printexpr ch; }
            case String other { printexpr other; }
    }
\end{cyan}

\section{Interoperability with Java}

Not surprisingly, Cyan code call use Java classes. A Java class/interface can be the type of a field, local variable, parameter, or return value of a method. A Java package can be imported, a Java class can be imported.  Java objects can be created and messages can be sent to them using Cyan syntax. Objects of the wrapper classes of Java are automatically converted to the corresponding Cyan prototypes and vice-versa ({\tt Byte}, {\tt Integer}, {\tt Short}, {\tt Long}, {\tt Character}, {\tt Boolean}, {\tt Float}, {\tt Double}). Idem for values of the basic Java types ({\tt byte}, {\tt int}, {\tt short}, {\tt long}, {\tt char}, {\tt boolean}, {\tt float}, {\tt double}). Assignments from anything to {\tt java.lang.Object} are allowed.

\begin{cyan}
package javaInter.main

  // import Java package
import java.util
import java.lang
  // import Java class
import java.io.File

object JavaTest1
      // field whose type is a Java class
    var java.lang.StringBuffer sbWorker

       // return value is a Java class
    func retSB -> StringBuffer {
          // Cyan syntax for creating an object of
          // a Java class. There is automatic conversion of
          // "a" from Cyan to a Java string
        return StringBuffer("a");
    }
       // parameter type is a Java class
    func retSB: StringBuffer sb -> String {
          // automatic conversion from Java string to Cyan string
          // 'sb toString' is the sending of a message in Java.
        let String s = sb toString;
        // return s     OR just the line below
        return sb toString;
    }

    func test {
          // automatic conversion
        var Int n = Integer(0);
        assert n == 0;
        n = Integer new: 1;
        assert n == 1;
        var String s = self retSB toString;
        s println;
        (self retSB: StringBuffer("abc")) println;
        /*
           self retSB toString println
        does not work. The type of 'self retSB toString' is java.lang.String
        which does not have a 'println' method.
        */

           // conversion from 'int' of Java to 'Int' of Cyan
        n = java.lang.String("abc") length;
          // assignments from anything to java.lang.Object
          // are allowed
        var java.lang.Object obj = 0;
        obj = "ok";
        obj = String;

    }
end
\end{cyan}

There is automatic conversion from {\tt Integer} and {\tt int} to {\tt Int} in array indexing.
Java arrays can be declared as in Java. Only one-dimensional arrays are supported.
\begin{cyan}
package javaInter.main

  // import Java package
import java.util
  // import Java class
import java.io.File
import java.lang

object JavaTest2
    func test {
        var intArray = [ 7, 2, 3 ];

          // convertion from Java Integer(0) to Cyan 0
        assert intArray[ Integer(0) ] == 7;
        assert intArray[0] == 7;
          // convertion from Java Integer(9) to Cyan 9
        intArray[0] = Integer(9);
        assert intArray[0] == 9;
        assert intArray[Integer new: 2] == 3;

    }

    func messagePassingJavaTest {

          // File is a Java class
        var File file;
        file = File new: "C:\\Dropbox\\Cyan";
           // message send in Java
        if file isDirectory {
            Out println: (file getCanonicalPath), " is a directory"
        }
           // message send in Java
        file setReadable: true;
        let ok = true;
           // conversion of Boolean 'ok' in Cyan to boolean 'ok' in Java
        file setReadable: ok;
          // a Java array
        let File [] fileList = file listFiles;
           // for does not work with Java arrays yet
        for  n in 1..<fileList length {
               // indexing with a Java array
            let File ff2 = fileList[n];
            // Out println: ff2 getCanonicalPath;
        }


        var java.lang.StringBuffer sb = java.lang.StringBuffer();

        sb append: "Append in String Buffer!";
        let String ssss = sb toString;
        assert ssss == "Append in String Buffer!";

          // conversion of Cyan string to Java string
        var java.lang.String js = "abcdef";
        js = "a b c d";
           // automatic type for Java local variable
           // strArray has type java.lang.String[]
        var strArray = js split: " ";
           // conversion from Java string to Cyan string
        var String cyanStr = strArray[0];
        assert cyanStr == "a";
           // indexing of a Java array
        cyanStr = strArray[1];
        assert cyanStr == "b";
        cyanStr = strArray[2];
        assert cyanStr == "c";
        /* the code
               assert cyanStr[2] == "c";

           does not work. cyanStr[2] has type java.lang.String and therefore the '==' operator
           used is that of Java. It compares the pointer, not the contents as in Cyan.
        */

    }
end
\end{cyan}

Generic classes of Java can be used but there should be no restrictions on the type parameters. The real type parameters can be both Java classes and Cyan prototypes.
\begin{cyan}
package javaInter.main

  // import Java package
import java.util
  // import Java class
import java.lang

object JavaTest3

    func test {
           // parameter to ArrayList is Int, a Cyan prototype
        var ArrayList<Int> intArray = ArrayList<Int> new;
        intArray add: 88;
        intArray add: 99;
        var Int n = intArray get: 0;
        assert n == 88;
        n = intArray get: 1;
        assert n == 99;

           // parameter to ArrayList is StringBuffer, a Java class
        var ArrayList<StringBuffer> strArray = ArrayList<StringBuffer> new;
        strArray add: StringBuffer("aaa");
        strArray add: StringBuffer("bbb");
        var String s = strArray get: 0;
        assert s == "aaa";
        s = strArray get: 1;
        assert n == "bbb";

        var java.util.Set<Int> iset = java.util.HashSet<Int>();
        iset add: 0;
        iset add: 1;
        iset add: 2;
        var Boolean b = iset contains: 0;
        assert b;
        b = iset contains: 1;
        assert b;
        /*
            just
                 assert iset contains: 1;
            does not work. The return type of 'iset contains: 1' is java.lang.boolean
            which is not automatically cast to Boolean. Macro assert in fact sends
            unary message '!' to the expression.

        */

        b = iset contains: 3;
        assert !b;

    }

end
\end{cyan}

Generic prototypes of Cyan can take Java classes as real parameters. However, due to the fact that every Cyan prototype inherits from {\tt Any}, this will hardly work.

\begin{cyan}
package javaInter.main

open
object GP<T>

    func init: T elem { self.elem = elem }

    func get -> T = elem;
    func set: T elem { self.elem = elem }
    func getStr -> String {
        var java.lang.Object any = elem;
        var String s = any toString;
        return s
    }

    var T elem
end
\end{cyan}
Note the method {\tt getStr} of prototype {\tt GP}. The prototype expects a Java class as real parameter to the generic prototype. Therefore it uses special code to deal with Java objects. This prototype can be used as in the next example.

\begin{cyan}
package javaInter.main

import java.lang

object JavaTest4

    func test {

        let a56 = GP<Integer>(9);
        assert a56 getStr == "9";
        a56 set: Integer(5);
        assert a56 getStr == "5";

    }

end
\end{cyan}

Java {\tt Boolean} and {\tt boolean} values can be used in {\tt if}, {\tt while}, and {\tt repeat-until} statements.
\begin{cyan}
package javaInter.main

import java.lang

object JavaTest5

    func ifWhileTest {
        var java.lang.Boolean b = false;

        if b { assert false; }
        else {
            assert true;
        }
        b = true;
        if b { assert true; }
        else {
            assert false;
        }
        while b {
            b = false
        }
        var Boolean ok = b;
        assert !ok;
    }

end
\end{cyan}

There are some limitations on the use of Java inside Cyan code:
\begin{enumerate}[(a)]
\item whenever a prototype {\tt T} and a Java class or interface {\tt T} is imported, the compiler will use the prototype. The collision of names will not be considered an error. Then the word ``{\tt String}'' in a Cyan code will always mean the Cyan prototype even when package ``{\tt java.lang}'' is imported. To use the Java class, prefix it with the package: \\
    \verb|     java.lang.String|

\item the Java basic types, {\tt int}, {\tt char}, etc cannot be used in a Cyan code. They do not belong to package {\tt java.lang}, they are native to Java. Use the wrapper classes ({\tt Integer}, {\tt Character}, etc) instead. If a method returns an array of a basic type, use a Java class that casts this array to an array of a wrapper class.
    \begin{verbatim}
    class CastArray {
        public Integer[] intToInteger(int []v) {
            Integer []newV = new Integer[v.length];
            int i = 0;
            for ( int elem : v ) {
                newV[i] = elem;
                ++i;
            }
            return newV;
        }
    }
    \end{verbatim}
    This is awful;

\item {\tt null} of Java cannot appear in a Cyan code. Use prototype {\tt cyan.lang.Null} to get the {\tt null} value. It is returned by
    \begin{verbatim}
    Null getNull
    \end{verbatim}
    To compare an expression {\tt expr} to {\tt null}, use
    \begin{verbatim}
    Null equalNull: expr
    \end{verbatim}
    It returns a {\tt Boolean} object. Note that
    \begin{verbatim}
    expr == Null getNull
    \end{verbatim}
    will not work. The compiler will look for a method \verb|_equal_equal| (or something similar) in the class of {\tt expr}. It will not use the original {\tt ==} method of Java.

\item a Cyan prototype cannot inherit from a Java class because it must inherit, even indirectly, from {\tt Any};
\item {\tt for} statements do not accept Java types. They will do some day;

\item Java methods that take a variable number of parameters cannot be called in Cyan.
\end{enumerate}
The above limitations restric the usability of the mixing Cyan and Java. However, future versions of Cyan may improve the communication between the two languages.

Objects of basic Cyan types are used to do the communication between the Cyan code and the Java code inside Cyan code. Then the two codes are relatively separated from each other.

In order to use Java classes, it is necessary tell the compiler where to find the jar files with the Java packages --- the Java code must be in a jar files. This is made with the compiler option
``\verb|-cp|'' as in
\begin{cyan}
saci "C:\Dropbox\Cyan\cyanTests\master"
      -cyanlang "C:\Dropbox\Cyan\lib"
      -cp "C:\Dropbox\Cyan\external-Java-libs\javassist.jar"
\end{cyan}
Option ``\verb|-java|'' may be used if only the  packages of the basic libraries of Java, file ``rt.jar'', will be used.
\begin{cyan}
saci "C:\Dropbox\Cyan\cyanTests\master"
      -cyanlang "C:\Dropbox\Cyan\lib"
      -java
\end{cyan}

Caveat: interoperability with Java has not been   thoroughly  tested.

\section{Future Enhancements}

Operator \verb"|" may be used as a method name. It is in prototype {\tt Int}, for example. A future version of the language will use \verb"|" only for unions. It will not be possible to use it as a method name.

Type {\tt Dyn} is not a real prototype, it is {\it virtual} because there is not a source code associated to it. Then this type can appear as type where only types are expected, as variable type or return value type. But {\tt Dyn} cannot receive a message in an expression:
\begin{cyan}
var String s;
s = Dyn prototypeName;  // compile error
\end{cyan}
{\tt Dyn} is the only prototype of this kind in Cyan. Future versions of {\tt Cyan} will make every non-tagged {\tt Union} prototype virtual too. This saves the creation of an object for a union. The language will be faster.

The language may support the Elvis \index{Elvis operator} operator, {\tt Nil}-safe message sends, and {\tt Nil}-safe array access. They will be as follows.

The Elvis operator would be implemented as a  method {\tt ifNil:}, {\tt Nil}-safe message sends would have all keywords prefixed with {\tt ?.}, and {\tt Nil}-safe array access would be made with {\tt ?[} and {\tt ]?}. See the examples.  \index{ifNil}
\begin{cyan}
var String userName;
   // getUserName is a method name
var Nil|String gotUserName = UserDataBase getUserName;
userName =  ifNil gotUserName, "anonymous";
\end{cyan}
The last line is the same as
\begin{cyan}
    type gotUserName
        case String s99 {
            userName = s99
        }
        case Nil {
            userName = "anonymous"
        }
\end{cyan}

{\tt Nil}-safe message send: \index{?.}
\begin{cyan}
var Nil|IndexedList<String> v;
   // it may associate Nil to v
v = obj getPeopleList;
v ?.at: 0 ?.put: "Gauss";
\end{cyan}
The last line is the same as
\begin{cyan}
type v
    case IndexedList<String> v8 {
        v8 at: 0 put: "Gauss";
    }
\end{cyan}
Or the same as
\begin{cyan}
cast v8 = v {
        v8 at: 0 put: "Gauss";
}
\end{cyan}

There should not be any space between the {\tt ?.} and the keyword. And all keywords of a message should be preceded by {\tt ?.} in a {\tt Nil}-safe message send.

{\tt Nil}-safe array access: \index{?[}
\begin{cyan}
var Nil|Array<Person> clubMembers;
...
var firstMember = clubMembers?[0]?;
\end{cyan}
The last line is the same as
\begin{cyan}
var Person|Nil firstMember;
cast c2 = clubMembers {
    firstMember = c2[0];
}
else {
    firstMember = Nil
}
\end{cyan}

A code
\begin{cyan}
if clubMembers != Nil  {
    clubMembers[0] = "Newton"
}
\end{cyan}
is equivalent to
\begin{cyan}
clubMembers?[0]? = "Newton";
\end{cyan}

We can use all features at the same time:
\begin{cyan}
var Nil|Array<Nil|Person> clubMembers;
...
var String firstMemberName = ifNil (clubMembers?[0]? ?.name), "no member";
\end{cyan}

\chapter{Dynamic Typing}  \label{dynamictyping}

\index{dynamic typing}
A dynamically-typed language does not demand that the source code declares the type of variables, parameters, or methods (the return value type). This allows fast coding, sometimes up to ten times faster than the same code made in a statically-typed language. All type checking is made at runtime, which brings some problems: the program is slower to run and it may have hidden type errors. When a type error occur, an exception is thrown. Statically-typed languages produce faster programs and the type errors are caught at compile time. However, program development is slower.

The ideal situation is to combine both approaches: to develop the program using dynamic typing and, after the development ends, convert it to static typing. Cyan offers some mechanisms that help to achieve this objective, described next.

A message send whose keywords are preceded by  {\tt ?} is not checked at compile-time. That is, the compiler does not check whether the static type of the expression receiving that message declares a method with those keywords. For example, in the code below, the compiler does not check whether prototype {\tt Person} defines a method with keywords {\tt name:} and {\tt age:} that accepts as parameters a {\tt String} and an {\tt Int}.
\begin{cyan}
var Person p;
...
p ?name: "Peter"  ?age: 31;
\end{cyan}
The receiver of a message of this kind cannot be {\tt super}. That would not make sense because message sends to {\tt super} are static calls. We know which method will be called at compile-time.

This non-checked message send is useful when the exact type of the receiver is not known:
\begin{cyan}
    func openArray: (Array<Any> anArray) {
        anArray foreach: { (: Any elem :)
            elem ?open
        }
    }
\end{cyan}
The array could have objects of any type. At runtime, a message {\tt open} is sent to all of them. If all objects of the array implemented an {\tt IOpen} interface,\footnote{With a method {\tt open}.} then we could declare parameter {\tt anArray} with type \verb@Array<IOpen>@. However, this may not be the case and some kind of dynamic message send would be necessary to call method {\tt open} of all objects.

The expression that receives a {\tt ?}-message cannot be a prototype:
\begin{cyan}
    var a = A ?new;
\end{cyan}

If every message keyword (such as {\tt open} in the above examples) is preceded by a {\tt ?} we have transformed Cyan into a dynamically-typed language. If just some of the keywords are preceded by {\tt ?}, then the program will use a mixture of dynamic and static type checking.

Keyword {\tt Dyn} is used for a dynamic type in Cyan. {\tt Dyn} is not a prototype. It is a virtual type\footnote{Internally the compiler considers that {\tt Dyn} is a prototype declared in package ``{\tt cyan.lang}''\/.} that is supertype and subtype of every other prototype including {\tt Nil}. Therefore assignments to and from {\tt Dyn} are always legal at compile-time. At runtime there is a check in assignments from {\tt Dyn} to any other type (that includes, of course, parameter passing, which is a kind of assignment, and return value of methods). At runtime the {\tt Dyn} expression should refer to a prototype that is subtype of the type of the left-hand side
variable.\footnote{or indexing expression like ``{\tt a[0] = dynVar}''\/.}
\begin{cyan}
var Person p;
var Dyn dynVar;
...
p = dynVar;
\end{cyan}
In the assignment the compiler inserts a check to verify whether {\tt dynVar} refer to an object whose type is subtype of {\tt Person} (which includes {\tt Person}). {\tt Nil} can be assigned to a {\tt Dyn} variable and an expression whose type is {\tt Dyn} can be assigned to a variable whose type is {\tt Nil}.

Assignments whose left-hand side is {\tt Dyn} need not to be checked either at compile or runtime. Since {\tt Dyn} is not a prototype, it cannot be used as an expression:
\begin{cyan}
(Dyn prototypeName) println; // compile-time error
\end{cyan}

A message sent to a receiver whose type is {\tt Dyn} is not checked by the compiler. The return value type of the message send is considered to be {\tt Dyn} too. Then if the type of a variable is {\tt Dyn} we can send to it a regular message, without {\tt ?} preceding the keywords.
\begin{cyan}
var Dyn p = Person;
p name: "Peter"  age: 31;
\end{cyan}
The compiler will not do any checking. This is equivalent to declare {\tt p} with any other type and use {\tt ?} before the keywords. {\tt Dyn} is considered a supertype and a subtype of any prototype. Of course, it is a virtual type, there is no source file ``{\tt Dyn.cyan}''\/.

The return value type of a message send is considered to be {\tt Dyn} when the receiver expression has type {\tt Dyn}. Therefore the return value is not checked. In this example, the compiler consideres that {\tt get:} returns {\tt Dyn} and, since it is a subtype of {\tt Boolean}, there is no error.
\begin{cyan}
var Dyn t = MyHashtable<String, String>;
if t get: "one" == "1" {
   "found one" println
}
\end{cyan}

When the return value of a dynamic message is assigned to a variable declared without a type, the compiler considers that the type of the variable is {\tt Dyn}, as expected.
\begin{cyan}
    // n has type Dyn
var n = obj ?value;
\end{cyan}

Dynamic message sends, with keywords preceded by {\tt ?}, plus the reflective facilities of Cyan can be used to create objects with dynamic fields. Object {\tt DTuple} of the language library is a tuble initially without fields, which can be added dynamically:

\begin{cyan}
var t = DTuple new;
t ?name: "Carolina";
   // prints "Carolina"
Out println: (t ?name);
   // if uncommented the line below would produce a runtime error
//Out println: (t ?age);
t ?age: 1;
   // prints 1
Out println: (t ?age);
\end{cyan}
Here fields {\tt name} and {\tt age} are dynamically added to object {\tt t}. Whenever a message is sent to an object and it does not have the appropriate method, method {\tt doesNotUnderstand:} is called. The original message with the parameters are passed to this method. Every object has a {\tt doesNotUnderstand:} method inherited from {\tt Any}.

{\tt DTuple} keeps a list or hash table of pairs ``(fieldName, fieldValue)''\/. Each field has the name {\tt fieldName} and a value {\tt fieldValue}. When a {\tt DTuple} object receives a message {\tt id: aValue}, method {\tt doesNotUnderstand:} is called\footnote{Unless this is a method of {\tt Any}.} and a search is made in this list or hash table. If no field with name {\tt id} is found, one is created with value {\tt aValue}. If a field is found, its value is updated to {\tt aValue}.

When the {\tt DTuple} object receives a message {\tt id} and {\tt id} is not a method declared in {\tt DTuple} or {\tt Any}, method {\tt doesNotUnderstand:} is called. It searches in the list or hash table for {\tt id}. If it is not found, method {\tt doesNotUnderstand:} of {\tt Any} is called. If {\tt id} is found, its value is returned. For more information, see file {\tt DTuple.cyan} in package {\tt cyan.lang}.

The previous {\tt DTuple} example can be made more legible by declaring {\tt t} with type {\tt Dyn}.
\begin{cyan}
var Dyn t = DTuple new;
t name: "Carolina";
   // prints "Carolina"
Out println: (t name);
   // if uncommented the line below would produce a runtime error
//Out println: (t ?age);
t age: 1;
   // prints 1
Out println: (t age);
\end{cyan}

Cyan supports the \verb"`" operator (backquote, ASCII 96) for calling a method whose keywords are in string variables. Each keyword used at runtime is the contents of each variable.
\begin{cyan}
var String s = "print";
0 `s;
\end{cyan}
The last line sends message {\tt print} to {\tt 0}.

In a message send with parameters the variable names should be followed by {\tt :} as usual.
\begin{cyan}
var String s = "at";
var p = "put";

let IMap<String, Int> map = [ "two" -> 2 ];

map `s: "one"   `p: 1;

assert map get: "one" == 1
\end{cyan}
The method to be called has keywords \verb|s ++ ":"| and  \verb|p ++ ":"|, in which {\tt ++} is used for concatenating strings. That is, the method to be called is \verb|at:put:|. One may add {\tt :} at the end of the string too:
\begin{cyan}
var s = "at:";
var p = "put:";

let IMap<String, Int> map = [ "two" -> 2 ];

map `s: "one"   `p: 1;

assert map get: "one" == 1
\end{cyan}

Methods whose names are operators may also be called but the backquote variable should be followed by ``{\tt :}''\/.
\begin{cyan}
for op in [ "+", "*", "-" ] {
    Out println: (6 `op:  2);
}
// prints 8, 12, 4
\end{cyan}

Each keyword preceded by backquote should be a variable of type {\tt String}, {\tt CySymbol}, or {\tt Dyn}. It cannot be a field accessed through {\tt self} as {\tt self.name}. The receiver of a backquote message send cannot be {\tt super}.

The backquote operator cannot  be used in a chain of unary message sends. Then it is illegal to write either\\
\verb@    club `first `second@\\
\noindent or\\
\verb|    club members `second|\\
\noindent That is, a chain of message sends in which there is a backquote should have size one.

Language Groovy has this mechanism for message sends:
\begin{verbatim}
animal."$action"()
\end{verbatim}
The method of {\tt animal} called will be that of variable {\tt action}, which should refer to a {\tt String}.

During the design of Cyan, several decisions were taken to make the language support optional typing:
 \begin{enumerate}[(a)]

\item types are not used in order to decide how many parameters are needed in a message send. For example, even if method {\tt get:} of {\tt IMap} takes one parameter and {\tt put:} of {\tt MyTable} takes two parameters, we cannot write
    \begin{cyan}
    let IMap<String, Int> map = [ "one" -> 1 ];
    n = MyTable put: map get: "one", j
    \end{cyan}
    The compiler could easily check that the intended meaning is
    \begin{cyan}
    n = MyTable put: (map get: "one"), j
    \end{cyan}
    by checking the prototypes {\tt IMap} and {\tt MyTable}.

    However, if the type of {\tt map} is {\tt Dyn}, this checking would not be possible. The type information would not be availabe at compile time. Therefore Cyan consider that a message send includes all the keywords that follow the receiver and that are not in an expression within parentheses;

\item when a method is overloaded, the static or compile-time type of the real arguments are not taken into consideration to chose which method will be called at runtime. In the {\tt Animal}, {\tt Cow}, and {\tt Fish} example of page~\pageref{cow}, the same methods are called regardless of the static type of the parameter to {\tt eat:}. Therefore the use of static or dynamic typing does not change the semantics of message passing. That allows one to change from static to dynamic typing and vice-versa without fear of breaking the program.

    There is one more reason to employ the runtime search algorithm for methods that Cyan uses, which does not consider the static, compile-time type of the receiver and parameters:
    the exception system. Most exception handling systems of object-oriented languages are similar to the Java/C++ system. There are catch clauses after a try block that are searched for after an exception is thrown in the block. The catch clauses are searched in the declared textual order. In Cyan, these catch clauses are encapsulated in  {\tt eval} methods with are searched in the textual order too. The eval methods have parameters which correspond to the parameters of the catch clauses in Java/C++. The {\tt eval} methods are therefore overloaded. The search for an {\tt eval} method after an exception is made in the textually declared order of these methods, as would be made in any message send whose correspondent method is overload. This matches the search for a catch clause of a try block in Java/C++, which appear to be the best possible way of dealing with an thrown exception. And this search algorithm is exactly the algorithm employed in every message send in Cyan;

\item the Cyan syntax was designed in order to be clear and unambiguous even without types in the declaration of variables and parameters. For example, before a local variable declaration it is necessary to use ``{\tt var}''\/, which asserts that a list of variables follow, preceeded or not by a type. For example, the declaration of {\tt Int} variables in Cyan is \\
    \verb@     var Int a, b, c;@\\

    To declare these variables with type {\tt Dyn} one can write

    \verb@     var a, b, c;@\\

    If an expression is assigned to a variable in its declaration, its type will be the compile-time type of the expression if the type is not supplied:
    \begin{cyan}
    var count = 0;  // count has type Int
    var flavor = "vanilla";  // flavor has type String
    \end{cyan}

    If {\tt let} is used instead of {\tt var} the expression should always be supplied and therefore read-only variables will have the type of the expression.

    Remember that method parameters without types have type {\tt Dyn}:

    \begin{cyan}
    func add: key, value { ... }
    \end{cyan}
    Both {\tt key} and {\tt value} have type {\tt Dyn}. The return value type should always be supplied.

\end{enumerate}

\chapter{Generic Prototypes}
\label{generics}

\index{object!generic} \index{generics}
Generic prototypes in Cyan play the same role as generic classes and template classes of other object-oriented languages. Unlike other modern languages, Cyan takes a loose approach to generics. In many languages,
the compiler guarantees that a generic class  is type correct if the real parameter is subtype of a certain class specified in the generic class declaration. For example, a generic class {\tt Hashtable} takes a type argument {\tt T} which should be subtype of {\tt Hashable}, an interface with a single method
\verb@hashCode -> Int@ (using Cyan syntax). Then whenever one uses \verb"Hashtable<A>" and {\tt A} is subtype of {\tt Hashable}, it is guaranteed that \verb"Hashtable<A>" is type correct --- the compiler does not need to check the source code of \verb"Hashtable<A>" to assert that.

In Cyan, {\tt Hashtable} has to be compiled with real argument {\tt A} in order to assure the type correctness of the code. This has pros and cons. The pro part is that there is much more freedom in Cyan to create generic prototypes. The con part is that any changes in the code of a generic prototype can cause compile-time errors elsewhere. Cyan does not supports the conventional approach for two reasons: a) there would not be any novelty in it (no articles about it would be accepted for publishing) and b) the freedom given by the definition of Cyan generics makes them highly useful --- see the examples given here, in Section~\ref{synergy}, and in Chapter~\ref{ilo}.

There are several ways of declaring a generic prototype in Cyan. In the first and simplest way, a list of parameters is given between \verb|<| and \verb|>| after the prototype name:
\begin{cyan}
package ds

object P< T1,  T2, ...  Tn >
   ...
end
\end{cyan}
Parameters {\tt T1}, {\tt T2}, \ldots {\tt Tn} are called {\it \bf formal parameters} or {\it generic parameters} of the \index{formal parameters of generic prototype} generic prototype. Each of them starts with an upper-case letter.

There should be no space between the prototype name, {\tt P}, and the character ``\verb|<|''\/. Space may follow ``\verb|<|''\/ as in
\begin{cyan}
package ds

object Stack< T >

   func push: T elem { ... }
   func pop -> T { ... }
   func print {
       array foreach: { (: T elem :)
           elem print  // message print is sent to an object of type T
       }
   }
   ...
end
\end{cyan}
After importing package {\tt ds}, that declares {\tt Stack}, one may use {\tt Stack} if an argument is supplied:
\begin{cyan}
var Stack<Int>  intStack;
var Stack< Person >  personStack;
Stack< Stack<Int> > prototypeName print;

intStack push: 0;
personStack push: aPerson;
\end{cyan}
However, there should be no space between the generic prototype name and ``\verb|<|''\/. That would cause a compile-time error.
If there is a space between the object name and ``\verb@<@''\/, the compiler will consider ``\verb@<@''\/ as the operator ``less than''\/. Then in the code
\begin{cyan}
if  Stack < Int >  {
    "compile-time error in the line above" println
}
\end{cyan}
the compiler will consider that {\tt Stack} is receiving message ``\verb|<|''\/ with parameter {\tt Int} which is followed by ``\verb|>|''\/. The Cyan grammar does not allow multiple comparison operators in the same expression (that is, ``\verb|a < b < c > d|''\/ is illegal) and ``\verb|>|''\/ demands a parameter, which does not appear in the code above. Therefore there is a compile-time error even before the semantic analysis.

When the compiler finds ``\verb|Stack<Int>|''\/ in a source code that imports package {\tt ds}, it creates a brand new prototype whose name is ``\verb|Stack<Int>|''\/ by replacing the {\it generic parameter} {\tt T} in prototype \verb|Stack<T>| by {\tt Int}. This process is called {\it \bf instantiation} \index{instantiation} \index{instantiation of a generic prototype} of a generic prototype and {\tt Int} is called a \index{real parameter to generic prototype}
{\it \bf real parameter} or {\it \bf real argument} to the generic prototype. There are restrictions on where a formal parameter can appear in the source code of the prototype ({\tt Stack} in the example) and when it is replaced by a real parameter.

A formal parameter may appear as a keyword name (both in a method declaration and in a message send), type, identifier in an expression, parameter to an metaobject annotation, and after {\tt \#} (to define a symbol). No local variable or parameter will have the name of a {\bf formal parameter} because the former start with a lowercase letter and the latter with an uppercase letter. In any other case an identifier equal to a formal parameter is ignored in the process of instantiation of a generic prototype. That is, the formal parameter is not replaced by the real parameter in any other case.

More specifically, the compiler replaces a formal parameter by a real parameter if it is in a symbol literal or it is an Id or IdColon of the following grammar rules. Id in QualifId is only replaced if QualifId appears in the rules below. Only the part of the rules that matters are shown.

\p{QualifId} ::=  Id \{ ``.''\/ Id \}

\p{ExprPrimary} ::= QualifId  \{ ``\verb@<@''\/ TypeList ``\verb@>@''\/ \}+ [ ObjectCreation ] \verb@|@\\
\rr QualifId  \{ ``\verb@<@''\/ TypeList ``\verb@>@''\/ \}+  \verb@|@\\
\rr ``typeof''\/ ``(''\/ QualifId   [ ``$<$''\/ TypeList ``$>$''\/ ] ``)''\/

\p{MetSigUnary} ::= Id

\p{SelecWithParam} ::= IdColon \verb@|@ \\
\rr [ ``[]''\/ ] IdColon  ParamList

\p{InterMethSig2} ::= Id \verb@|@\\
\rr   \{ IdColon [ InterParamDecList ]  \}+

\p{SingleType} ::= QualifId  \{ ``\verb@<@''\/ TypeList ``\verb@>@''\/ \}  \verb@|@\\
\rr ``typeof''\/ ``(''\/ QualifId  [ ``$<$''\/ TypeList ``$>$''\/ ]  ``)''\/

\p{Annotation} ::= ``@''\/ Id\\
\rr [ ``(''\/ ExprLiteral [ ``,''\/ ExprLiteral ] ``)''\/ ] \\
\rr [ LeftCharString TEXT RightCharString ]

The rule ExprPrimary expands to three things:
\begin{enumerate}[(a)]
\item a generic prototype instantiation as \verb|Array<Int>|;
\item an object creation of a generic prototype instantiation such as\\
\verb|     Array<Int>(100)|

\item the {\tt typeof} compile-time function:
\begin{cyan}
typeof(x) prototypeName;
typeof(Array<Int>) prototypeName;
\end{cyan}
\end{enumerate}
ExprPrimary expands only inside expressions.

MetSigUnary and SelecWithParam are used to produce names of unary methods and keyword names of keyword methods. They expand to code like these:
\begin{cyan}
asString
read:
[] at: Int
between: Int start, Int theEnd
\end{cyan}
These rules are used to produce method signatures:
\begin{cyan}
func asString -> String { ... }
func read: -> Array<Byte> { ... }
func [] at: Int -> Int { ... }
func between: Int start, Int theEnd -> Array<Int> { ... }
\end{cyan}

InterMethSig2 produces signatures of methods in interfaces. The difference with SelecWithParam/MetSigUnary is that the type of parameters should appear (this rule is not shown). The examples are the same as those of SelecWithParam/MetSigUnary.

SingleType produces a type. It may expand to
\begin{cyan}
Person
Array<Int>
typeof(x)
typeof(cyan.lang.Int)
typeof( Array<Int> )
typeof( cyan.lang.Array<Int> )
\end{cyan}

Annotation expands to a metaobject annotation. It may expand to any of the following.
\begin{cyan}
@checkStyle
@annot("main")
@concept{* T implements I *}
\end{cyan}

The Id in these rules given above can be generic prototype formal parameters. Then if {\tt T} is a formal parameter, the following uses are legal. The lines do not compose a code, they are just a set of examples.
\begin{cyan}
myProto = Array<T>;
myArray = Array<T>(100);
var protoName = typeof(T) prototypeName;
var String protoName2 = typeof(Array<T>) prototypeName;
func T -> String { ... }
func T: -> Array<Byte> { ... }
func [] T: Int -> Int
func T: Int start, Int theEnd -> Array<Int> { ... }
T println;
[ 0 ] T println;
var T x;
var Array<T> tArray;
var typeof(T) aT;
var typeof(Array<T>) anotherTArray;
var typeof(cyan.lang.Array<T>) anotherTArray;
@annot(T)
@feature(T, T)
\end{cyan}
A formal parameter can also appear inside the text of an annotation:
\begin{cyan}
package main

@concept{*
    T has [ func next -> T ]
*}
object Element<T>
   ...
end
\end{cyan}

Let us see some examples. First a generic interface.

\begin{cyan}
package main

interface InterNice<T>

    func add: T -> Int

end
\end{cyan}

A superprototype used in the next example.
\begin{cyan}
package main

open
object SuperNice<T>

    func superMethod: Int n -> Int = n;

end
\end{cyan}

A nice example with all possible uses of generic prototypes
\begin{cyan}
package main

object Nice<T, R, S, AsString> extends SuperNice<T> implements InterNice<T>

    func init: T x {
        self.x = x;
        people = "people";
        tArray = Array<T> new;
    }

    @annot(T)
    @feature(AsString, T)
    func example {
        var typeof(T) aT;
        var typeof(Array<T>) aTArray;
        var typeof(cyan.lang.Array<T>) anotherTArray;

        aTArray = Array<T>();
        anotherTArray = Array<T>();

        var myProto = Array<T>;
        var myArray = Array<T>(100);
        var protoName = typeof(T) prototypeName;
        var String protoName2 = typeof(Array<T>) prototypeName;
        Out println: protoName, protoName2, myProto prototypeName,
           myArray prototypeName;

        var Boolean found = false;
        for elem in annotList {
            if elem == "person.Person" { found = true }
        }
        assert found;
    }

    override
    func AsString -> String {
            // after a # to define a symbol
        return #T ++ #R ++ #S;
    }

    // keyword name
    func R: Char p -> Int { return p asInt }

        // type
    override
    func add: (T p) -> Int {
        var S.Person  per = T("Carolina", 7);
        var per2 = S.Person("Livia", 11);
        var typeof(T) aPerson;
        var typeof(S.Person) otherPerson;

        return 0
    }

        // type
    T x
    var Array<T> tArray
    String people;

end
\end{cyan}

A prototype that uses {\tt Nice} is
\begin{cyan}
package main

import people

object Program

    func run {

        let Person livia = Person("Livia", 11);
        typeof(livia) prototypeName println;

        var nice = Nice<Person, charToInt, people, asString>(livia);

        assert nice asString == "people.PersoncharToIntpeople";
        assert (nice charToInt: 'a') == 97;
        nice add: Person("Carol", 7);
        assert (nice superMethod: 0) == 0;

    }
end
\end{cyan}

A formal parameter may be the name of a generic prototype:
\begin{cyan}
object NiceExample< T >
    public var T<Int> value
end
\end{cyan}
In the instantiation \verb|NiceExample<Empty>| the compiler checks whether there is an {\tt Empty} prototype. That is, a non-generic prototype called ``{\tt Empty}''\/. After the instantiation, when \verb|NiceExample<Empty>| is compiled, the compiler checks whether there is a generic prototype \verb|Empty| that takes one argument.

There is a compile-time error if the formal parameter is the name of a parameter because variables and parameter should start with a lowercase letter.
\begin{cyan}
object Wrong< T >
    func myError: (Int T) { } // compile-time error
end
\end{cyan}

A formal parameter appearing as a substring of a Cyan symbol is not replaced.
\begin{cyan}
object P<T>
    func print { #T1 print }
end
\end{cyan}
Prototype \verb|P<Int>| is
\begin{cyan}
object P<Int>
    func print { #T1 print }
end
\end{cyan}
because ``{\tt T}''\/ is just a substring of ``{\tt T1}''\/.

In the same way, package names and imported packages are not replaced.
\begin{cyan}
package T
import main.T
object P<T>
end
\end{cyan}

\verb|P<Person>| is

\begin{cyan}
package T
import main.T
object P<Person>
end
\end{cyan}

Currently there is no way of producing new symbols from formal parameters. There could be a \verb|+++| operator that is executed at compile-time to concatenate formal parameters and something else:
\begin{cyan}
object P<T>
    func print { #T +++ 1 print }
end
\end{cyan}
Prototype \verb|P<Int>| would be
\begin{cyan}
object P<Int>
    func print { #Int1 print }
end
\end{cyan}
Till now we have found no need for such operators or to compile-time commands such as ``static if''\/ of language D.


In a generic prototype instantiation, each real parameter should be a
type or an identifier starting with a lower-case letter (which is called \index{identifier parameter} {\it identifier parameter}). The first one one can be a generic prototype instantiation. Then the general format of a real parameter is given by rule {\tt RP} of the grammar below. IdLowerCase stands for ``identifier starting with a lower-case letter''.

\p{RP} ::= IdLowerCase \verb@|@ Type

\p{Type} ::= SingleType \{ \verb"|" SingleType \}

\p{SingleType} ::= QualifId  \{ ``\verb@<@''\/ TypeList ``\verb@>@''\/ \} \verb@|@ BasicType 

\p{TypeList} ::= Type \{ ``,''\/ Type \}

\p{QualifId} ::=  Id \{ ``.''\/ Id \}

Anyway, the real parameter starts with a Cyan identifier. If this identifier starts with an upper-case letter, the compiler considers that the real parameter is a type. Therefore this type should be visible in the place of the generic prototype instantiation or a compile-time error will be signalled. If the identifier starts with a lower-case letter, it is not considered a type,
the compiler does not do any checking in the place of the instantiation.

\begin{cyan}
var Stack<A> s; // compiler checks if "A" is a prototype declared or imported
var Stack< Set<Char> > s; // compiler checks if "Set<Char>" is legal
var Nice<myId> n; // compiler does not check if "myId" is a prototype
\end{cyan}

The instantiation  ``\verb@Wrong<Array>@''\/ causes a compile-time error because there is no prototype ``{\tt Array}''\/.
\begin{cyan}
object Wrong<T>
    T<String> myData
    ...
end
\end{cyan}
There is a generic prototype ``\verb@Array<T>@''\/ in package {\tt cyan.lang} which is not related to a non-existing non-generic {\tt Array} prototype.

A call to the compile-time function {\tt typeof} cannot be used as a parameter in a generic prototype instantiation.
\begin{cyan}
var Int count = 0;
var Stack<typeof(count)> intStack; // compile-time error
\end{cyan}
Because of this restriction, the grammar for {\tt RP} given above defines {\tt SingleType} differently from the grammar of Section~\ref{grammar}.

A generic prototype may declare more than one generic parameter:
\begin{cyan}
package cyan.lang

interface IMap<K, V> extends Iterable<Tuple<key, K, value, V>>

    func [] at: K key -> V|Nil
    func [] at: K key put: V value -> V|Nil
    ...
end
\end{cyan}
All formal parameters should have different names. Each of them should start with an upper-case letter and there should be no prototype in package {\tt cyan.lang} with the same name as the parameter. So a parameter cannot have name ``{\tt Tuple}''\/ or ``{\tt Interval}''\/. You are invited to use single letters to formal parameter names.

Currently there is no way of declaring a private generic prototype in Cyan (or a private regular prototype). The implementation of this feature would make the compiler more complex. We believe private generic prototypes would be rarely used and almost never necessary.

A real parameter to a generic prototype cannot be an integer number as in C++ \cite{Stroustrup:2013:CPL:2543987}. However, metaobject {\tt extract} can be used to simulate the passing of an {\tt Int} as parameter.

\begin{cyan}
package main

object Store<T>
    func set: Int elem {
        if elem > @extract(T) {
            throw: ExceptionStr("Number out of limits in Store prototype")
        }
        self.elem = elem
    }
    func get -> Int = elem;

    var Int elem = 0;

end
\end{cyan}
If {\tt T} has the form {\tt intN} or {\tt int\_N} in which {\tt N} is a literal {\tt Int}, then\\
\verb|    @extract(T)|\\
\nd results in {\tt N}.
\begin{cyan}
        var s100 = Store<int_100>();
        s100 set: 99; // ok
        s100 set: 200; // exception thrown
\end{cyan}

\section{Generic Prototypes with real arguments}

A prototype that is not generic can be declared using the generic prototype syntax:
\begin{cyan}
package ds

object Stack<Int>
   func push: Int elem { ... }
   func pop -> Int { ... }
   ...
end
\end{cyan}
There may be both the generic prototype \verb|Stack<T>| and this non-generic version in the same package. In this case, \verb|Stack<Int>| will refer to the non-generic version (the one above) and \verb|Stack<Char>| will be an instantiation of the generic prototype {\tt Stack}. The details of this combination will soon be explained.

We will refer to a non-generic prototype declared using the generic prototype syntax as ``{\it \bf generic prototype with real arguments}''\/. Each one of the parameters that appear inside \verb|<...>| will be called ``{\it \bf real argument}''\/.

A {\it real argument} of a {\it generic prototype with real arguments} can be:
\begin{enumerate}[(a)]
\item {\it identifier parameter}, which is a single identifier starting with a lower-case letter such as ``{\tt t}''\/ or ``{\tt add}''\/. For example,
    \begin{cyan}
    interface ISingle<write>
        func write: Char
    end
    \end{cyan}

\item a single identifier starting with an upper-case letter such that there is a prototype in package {\tt cyan.lang} with this same name. For example,
    \begin{cyan}
    object P<Int>
        func add: Int { ... }
    end
    \end{cyan}

\item a qualified identifier; that is, a sequence of identifiers separated by ``{\tt .}''\/ with at least one dot such as ``{\tt main.Person}''\/.  For example,
    \begin{cyan}
    package ds
    import main
    interface MyList<main.Person>
        func add: Person
    end
    \end{cyan}
    This qualifier identifier should be the full name of a prototype, which includes its  package name;

\item a generic prototype instantiation possibly preceded by a package name such as ``\verb|Tuple<Int, String>|''\/ or ``\verb|ds.Stack<main.Person>|''\/. For example,
    \begin{cyan}
    package ds

    object List< Tuple<key, String, value, Int>, ds.Map<String, main.Person> >
       ...
    end
    \end{cyan}
\end{enumerate}

By the above rules, a prototype can be used as a real argument if it is preceded by its package.
This demand is dropped in prototypes of package {\tt cyan.lang}. Therefore if {\tt Person} is in package {\tt main}, a prototype \verb|Stack<main.Person>| should be declared as
\begin{cyan}
package ds

object Stack<main.Person>
   func push: main.Person elem { ... }
   func pop -> main.Person { ... }
   ...
end
\end{cyan}
In this way the compiler knows whether an identifier that appears after \verb|<| is a formal parameter or a {\it real argument} of a {\it generic prototype with real arguments}.  If the parameter:
\begin{enumerate}[(a)]
\item is composed by a single identifier that starts with a lower-case letter it is a {\it real argument}. See a previous example of prototype {\tt ISingle} with parameter {\tt write};
\item is composed by a single identifier that starts with an upper-case letter and there is a prototype in {\tt cyan.lang} with this same name, then it is a {\it real argument};
\item is composed by a single identifier that starts with an upper-case letter and there is no prototype in {\tt cyan.lang} with this same name, then it is a {\it formal parameter};

\item  is qualified, with at least one ``{\tt .}''\/ in it, then it is a {\it real argument};
\item is a generic prototype instantiation, then it is a {\it real argument}.
\end{enumerate}

The non-generic version of a generic prototype is a completely independent prototype. It can have different methods, inheritance, and so on. This feature is used to define a prototype
\verb|Function<Boolean>| that represents a function that does not take parameters and return a {\tt Boolean}. This kind of function should support methods {\tt whileTrue:} and {\tt whileFalse:}
\begin{cyan}
var i = 0;
{ ^ i < 10 } whileTrue: {
    i println;
    ++i
};
\end{cyan}
No other function prototype should have these methods.

A {\it generic prototype with real arguments} may be useful for providing a more efficient
implementation for a given type. For example, a \verb|HashMap<Int, Int>| implementation could somehow be more efficient because {\tt Int}´s are used.

\section{Generic Prototype with a Varying Number of Parameters}

Generic prototypes with a variable number of parameters are supported. They are declared by putting a {\tt +} after the generic parameter name:
\begin{cyan}
object P<T+>
   ...
end
\end{cyan}
There should be just one formal parameter between ``\verb|<|''\/ and ``\verb|>|''\/ and there should be just one set of pairs ``\verb|<|''\/ and ``\verb|>|''\/. The generic prototype is used with any number of set of pairs ``\verb|<|''\/ and ``\verb|>|''\/. Then the prototype {\tt P} of the example above is used for all of the following instantiations:
\begin{cyan}
P<Int>   P<Int, String>  P<Int><Int>  P<Int><Char, Double>
P<Char, Int><Float, String, Char><Int><Int, Int><Nil>
\end{cyan}
Of course, the syntax is misleading for it induces one to think there is just one set of pairs ``\verb|<|''\/ and ``\verb|>|''\/.

There is no way to use formal parameter like {\tt T} using regular Cyan syntax. The only way of doing that is through metaprogramming, using metaobjects (See ``The Cyan Metaobject Protocol'' in the Cyan site). For example, prototype {\tt Tuple} is declared as
\begin{cyan}
package cyan.lang

@createTuple
object Tuple<T+>
end
\end{cyan}
In an instantiation of {\tt Tuple}, as \verb|Tuple<Int, String>|,
metaobject {\tt createTuple} has access to the list of real argument, {\tt Int} and {\tt String}. Basead on these parameters, {\tt createTuple}
generates Cyan code that replaces the metaobject annotation. That is, ``{\tt @createTuple}''\/ is replaced by declarations of methods and fields produced by {\tt createTuple}.

It is tempting to add language constructions to handle a variable number of real arguments. However, that would be a mistake. The number of constructions needed to do something useful would be large. Since this kind of feature will be rarely used, they are best left for metaobjects. It is important to note that several library prototypes of Cyan are implemented using generic prototypes with a varying number of arguments: {\tt Tuple}, {\tt Union}, and {\tt Function}.

\section{Multiple Parameter Lists}

A generic prototype may have more than one \verb|<...>| list. Inside each list, there may appear more than one parameter as before.
\begin{cyan}
package example

object Test<T1, T2><U1, U2>
end
\end{cyan}

It is illegal to mix different kinds of parameters. All parameters should be one of the following:
\begin{enumerate}[(a)]
\item real arguments;
\item formal parameters without a {\tt +} operator;
\item a formal parameter (just one) with a {\tt +} operator.
\end{enumerate}
Then there are three possible ways of declaring a generic prototype:

\begin{cyan}
package example

object Test<Int, Char><main.Person>
end
\end{cyan}

\begin{cyan}
package example

object Test<T1, T2><U1, U2><R>
end
\end{cyan}

\begin{cyan}
package example

object Test<T+>
end
\end{cyan}

There will be a compile-time error if the different kinds of parameters are mixed as in
\begin{cyan}
package example

object Test<T+><Int><U>  // error
end
\end{cyan}

\section{Source File Names}

Cyan has rules for associating file names to prototypes. As seen,  a public prototype {\tt P} should be in a file called ``{\tt P.cyan}''\/. A generic prototype with {\bf real arguments}
\begin{cyan}
package pack

object P<T1, T2, ... Tn>...<U1, U2, ... Um>
   ...
end
\end{cyan}
should be in a file \\
\verb@    P(T1,T2,...Tn)...(U1,U2,...Um).cyan@\\
There should be no space in the file name. For example, consider the source file below.

\begin{cyan}
package example

object Test<Int, Char><main.Person>
end
\end{cyan}
It should be in file\\
\noindent \verb|    Test(Int,Char)(main.Person)|\\
\noindent
The generic prototype
\begin{cyan}
package pack

object P<T1, T2, ... Tn>...<U1, U2, ... Um>
   ...
end
\end{cyan}
should be in file ``\verb|P(n)...(m).cyan|''\/. All parameters are formal ones.

The generic prototype
\begin{cyan}
package pack

object P<T+>
   ...
end
\end{cyan}
should be in file ``\verb|P(1+).cyan|''\/.

As examples of declarations and file names, see the table.

\vspace*{2ex}
\begin{tabular}{ll}
\verb|P<Int, Char, main.Person>|  & \verb|P(Int,Char,main.Person).cyan| \\
\verb|P<R, S, T><U, V><W>|  & \verb|P(3)(2)(1).cyan| \\
\verb|P<T+>|  & \verb|P(1+).cyan| \\
\verb|P< Tuple<key, String, value, main.Person> >|  & \verb|P(Tuple(key,String,value,main.Person)).cyan| \\
\verb|ISingle<write>|  & \verb|ISingle(write).cyan| \\
\end{tabular}
\vspace*{2ex}

\section{Combining Generic Prototypes}

A package may declare a non-generic prototype and several generic prototypes with the same name. A generic prototype may have formal parameters, real arguments, or a varying number of parameters. All source files should be in the same package directory which means the source file names are different.

Suppose an imported package declares several prototypes with name  {\tt P} --- at most one is non-generic and the others are generic ones. When the compiler finds an instantiation \\
\verb|    P<T1, ... Tn>...<U1, ... Um>|\\
\noindent it tries to find a generic prototype {\tt P} with real arguments that match exatly the real arguments {\tt T1}, ... {\tt Tn}, ... {\tt U1}, ... {\tt Um}.
If none is found, the compiler searches for a generic prototype whose number of parameters in each \verb|<>| list is equal to the instantiation. If none is found, it searches for a generic prototype {\tt P} with a variable number of parameters (file \verb|P(1+).cyan|). If no adequate generic prototype is found, the compiler signals an error.

For example, suppose that the instantiation is \\
\verb|    Tuple<key, Int, value, String>|\\
\noindent First the compiler searches for a prototype \verb|Tuple<key, Int, value, String>| which should be in a file\\
\verb|    Tuple(key,Int,value,String).cyan|\\
\noindent If this prototype does not exist, it searches for a generic prototype\\
\verb|    Tuple<T1, T2, T3, T4>|\\
\noindent with four formal parameters. This should be in a file\\
\verb|    Tuple(4).cyan|\\
\noindent If there is no such prototype, the compiler searches for\\
\verb|    Tuple<T+>|\\
\noindent which should be in a file {\tt Tuple(1+).cyan}.

If the instantiation uses other instantiations the process is recursive. In
\\
\verb|    Tuple<key, Tuple<Array<Int>, Union<Int, Char>> >|\\
the compiler searches for a prototype with this name which should be in file\\
\verb|    Tuple(key,Tuple(Array(Int),Union(Int,Char))).cyan|\\

\section{Concepts}

A generic prototype may assume that one of its parameters, say {\tt T}, is a prototype that defines some methods, is subprototype of some other prototype, implements a certain interface, and so on.
\begin{cyan}
package main

object Test<T>
    func run {
        var T x = T();
        x open;
        x write: "not all is ok";
        var IHas<String> h = x;
        if h has: "ok" {
            "has ok" println;
        }
    }
end
\end{cyan}
In this example, the code assumes that {\tt T}:
\begin{enumerate}[(a)]
\item  is a prototype;
\item has a method {\tt init} without parameters;
\item has an unary method {\tt open};
\item has a method {\tt write:} that can accept a string as parameter;
\item implements interface \verb|IHas<String>|.
\end{enumerate}

If {\tt Test} is instantiate with a prototype that does not define {\tt init}, {\tt open}, {\tt has: String}, {\tt write: Any}, or {\tt write: String}, a compilation error will occur. The compiler will show a stack of generic prototype instantiations:
\begin{verbatim}
In file C:\Dropbox\Cyan\cyanTests\negTestsJose\t10\main\--tmp\Test(Int).cyan
       (line 9 column 9)
object/interface main.Test<Int>
Method open was not found in prototype Int or its superprototypes
Stack of generic prototype instantiations:
    main.Test<Int> line 9 column 9
    main.OtherTest<Int> line 8 column 30
    main.Program line 5 column 21

        x open;
\end{verbatim}
This message says that in line {\tt 5}, column {\tt 21}, of {\tt main.Program} prototype \verb|main.OtherTest<Int>| was instantiated. Then in line {\tt 8}, column {\tt 30}, of \verb|main.OtherTest<Int>| prototype \verb|main.Test<Int>| was instantiated. Then in line {\tt 9}, column {\tt 9}, of \verb|main.Test<Int>| there was the error ``{\it Method open was not found in prototype Int or its superprototypes}''\/. The last line shows the offending line:\\
\verb|        x open;|\\

Although the error message helps to trace the error, it shows code internal to the generic prototype.
The error message will not be easily understood by a user of the generic prototype. Instead of allowing this kind of error to happen one can use Concepts \cite{Gregor:2006:CLS:1167515.1167499}. They can be used to specify restrictions that the real arguments to a generic prototype should have. If the real argument does not obey the restrictions, an error message is issued in the instantiation of the generic prototype. Code internal to the generic prototype is not shown.

Concepts are predicate on types and values. In Cyan, concepts are implemented using a metaobject and are predicates on types and
{\it identifier parameter}.\footnote{Identifiers starting with lowercase letters passed as parameters to generic prototypes, like {\tt speed} in {\tt BinaryTree<Int, speed>}.} The concepts of a generic prototype are made based on the semantic needed in the prototype. For example, suppose a generic prototype {\tt GroupWork} has a generic parameter which should be a Group.\footnote{A group is a set G with an operation $*$ in such a way that, if $a, b, c \in G$, $a*b \in G$,  $a*(b*c) = (a*b)*c$, there exist $e \in G$ (the unit of G) such that $a*e = e*a = a$ for all $a \in G$, and for all $a \in G$ there exists an element $b$ called the inverse of $a$ such that $a*b = b*a = e$.}

In the methods of {\tt GroupWork} it is assumed that {\tt T} has methods {\tt inverse}, {\tt unit}, and binary {\tt *}. And that these methods obey the semantics expected for a Group.
\begin{cyan}
package main

object GroupWork<T>

    func work: T a {
        printexpr a;
        printexpr a inverse;
        printexpr a unit;
        printexpr a * a unit;

        assert a * a inverse == T unit;
        assert T unit == a inverse * a;
        assert a unit == T unit;
        assert a * T unit == a;

    }

    func workout: T a, T b, T c {
        printexpr (b inverse * a inverse) * a * b;
        printexpr (c inverse * b inverse * a inverse) * a * b * c;

        printexpr b inverse * b;
        printexpr c * b * a * ( a inverse * b inverse * c inverse );

        assert c * b * a * ( a inverse * b inverse * c inverse ) == T unit;
        assert a*(b*c) == (a*b)*c;
        assert c*(b*a) == (c*b)*a;

    }
end
\end{cyan}
As said before, prototype {\tt GroupWork} may be instantiated with any type, which will result in a compile-time error if the real argument does not support the methods expected in the prototype body.
And there will be runtime errors if the methods do not have the expected semantics.

The error messages can be made clearer with the use of metaobject {\tt concept}. An annotation should be attached to the generic prototype:

\begin{cyan}
package main

@concept{*
    // both kinds of comments are allowed

    T has [ func * (T other) -> T
            func unit -> T
            func inverse -> T ],
        "T should have methods *, unit, and inverse in order to be considered element of a Group",

    axiom opTest: T a, T b, T c {%

        if (a * (b * c) != (a * b) * c) ||
           (c * (b * a) != (c * b) * a {
            return "T is not associative"
        }
        return Nil
    %},

    axiom unitTest: T a, T b, T c {%

        if (a * a unit != a unit * a) ||
           (b * a unit != b unit * b) ||
           (a unit * b unit != c unit * c unit) {
            return "The unit element of T is not an identity"
        }
        return Nil
    %},

    axiom inverseTest: T a, T b, T c {%

        if (a * a inverse != b unit) ||
           (a unit != b inverse * b) ||
           (c inverse * c != T unit) {
            return "The inverse operation is not working properly"
        }
        return Nil
    %}


*}
object GroupWork<T>
    func work: T a, T b, T c {
        printexpr a asInt;
        printexpr a inverse asInt;
        printexpr a unit asInt;
        printexpr (a * a inverse) asInt;
        printexpr (a * a unit) asInt;
        printexpr ((b inverse * a inverse) * a * b) asInt;
        printexpr ((c inverse * b inverse * a inverse) * a * b * c) asInt;

        printexpr (b inverse * b) asInt;
        printexpr (c * b * a * ( a inverse * b inverse * c inverse )) asInt;

    }


    func workout: T a, T b, T c {
        printexpr ((b inverse * a inverse) * a * b) asInt;
        printexpr ((c inverse * b inverse * a inverse) * a * b * c) asInt;

        printexpr (b inverse * b) asInt;
        printexpr (c * b * a * ( a inverse * b inverse * c inverse )) asInt;

        let tunit = T unit;
        assert c * b * a * ( a inverse * b inverse * c inverse ) == tunit;
        assert a*(b*c) == (a*b)*c;
        assert c*(b*a) == (c*b)*a;

    }

end

\end{cyan}
The Domain Specific Language (DSL) of the {\tt concept} metaobject annotation of this example specifies the restrictions the generic parameters should have. The first line,\\
\verb|    T has [ func * ... |\\
\nd means that parameter {\tt T} should have the methods between {\tt [} and {\tt ]}. The string that follows, \\
\verb|    "T should have methods * ..."|\\
\nd is the error message issued by the metaobject if {\tt T} does not define the methods. Each method can have its own error message:
\begin{cyan}
package main

@concept{*
    T has [ func * (T other) -> T  "T should support operator * in order to be a Group",
            func unit -> T
            func inverse -> T ],
        """T should have methods *, unit, and inverse in order to be considered element of a Group""",
    ...
*}
object GroupWork<T>
    ...
end
\end{cyan}
If the message after the method signature is not given, the message after the predicate is used. If there is no message after the predicate, a standard message is used. In this last example, if a type {\tt MyGroup} does not define method {\tt *} in \verb|GroupWork<MyGroup>|, message\\
\verb|    "MyGroup should support ..."|\\
 \nd is issued. If {\tt MyGroup} does not define method {\tt unit}, message\\
\verb|    "MyGroup should have methods *, unit, and ..."|\\
\nd is issued. If there was no message \\
\verb|    "T should have methods *, unit, and  ..."|\\
\nd then the standard message would be used in this last case.

{\tt axiom} is a keyword of the {\tt concept} DSL. It defines a method whose body appears between \verb|{%|
and \verb|%}| in this example (any left char sequence can be used. See the definition of left char sequence for metaobjects). If the metaobject annotation has parameter {\tt test} as in
\begin{cyan}
...
@concept(test){*
   ...
*}
object GroupWork<T>
    ...
end
\end{cyan}
the metaobject will create test packages, prototypes, and methods. In particular, for each concept annotation there will be a test prototype. This prototype will have a method for each axiom and it will be in a test directory created in a directory \verb|--test| of the program. For example, suppose the program is in directory\\
\verb|    C:\Dropbox\Cyan\cyanTests\simple|\\
\nd Package {\tt main} of this program contains the {\tt GroupWork} prototype. The metaobject {\tt concept} will create the path \\
\verb|    C:\Dropbox\Cyan\cyanTests\simple\main_ut\|\\
\verb|       groupwork_lt_main_d_intgroupplus_gt__axiom_test|\\
\nd if {\tt GroupWork} is instantiated with parameter {\tt IntGroupPlus} (a prototype). The path above is shown in two lines to fit in the page.
Inside this directory there will be a prototype\\
\verb|    GroupWork_lt_main_d_IntGroupPlus_gt__Axiom_Test|\\
\nd with the axioms. It is shown next.
\begin{cyan}
package main_ut.groupwork_lt_main_d_intgroupplus_gt__axiom_test

object GroupWork_lt_main_d_IntGroupPlus_gt__Axiom_Test

    func opTest_0: main.IntGroupPlus a, main.IntGroupPlus b, main.IntGroupPlus c -> String|Nil {
        if (a * (b * c) != (a * b) * c) ||
           (c * (b * a) != (c * b) * a {
            return "main.IntGroupPlus is not associative"
        }
        return Nil
    }

    func unitTest_1: main.IntGroupPlus a, main.IntGroupPlus b, main.IntGroupPlus c -> String|Nil {


        if (a * a unit != a unit * a) ||
           (b * a unit != b unit * b) ||
           (a unit * b unit != c unit * c unit) {
            return "The unit element of main.IntGroupPlus is not an identity"
        }
        return Nil
    }


    func inverseTest_2: main.IntGroupPlus a, main.IntGroupPlus b, main.IntGroupPlus c -> String|Nil {
        if (a * a inverse != b unit) ||
           (a unit != b inverse * b) ||
           (c inverse * c != main.IntGroupPlus unit) {
            return "The inverse operation is not working properly"
        }
        return Nil
    }
end
\end{cyan}
Each axiom gives origin to a method with the same name with a suffix number.

Another path will be created:\\
\nd \verb|    --test\main_ut\groupwork_test|\\
\nd It will contain usually one prototype for each formal parameter of the generic prototype and a test prototype. In this example, the prototypes will be \verb|GroupWork_Test| and \verb|T|.
\begin{cyan}
package main_ut.groupwork_test

object GroupWork_Test
    func run {
        var main.GroupWork<T> testVar;

    }

end
\end{cyan}

Usually but not always a prototype is created for each parameter with the restrictions it should have. In the example, the real argument for {\tt T} should have some methods such as \verb|*|, {\tt unit}, and {\tt inverse}. Then prototype {\tt T} is declared with these methods:
\begin{cyan}
package main_ut.groupwork_test

object T
    func * T other -> T = T;
    func unit -> T = T;
    func inverse -> T = T;

end
\end{cyan}
{\tt T} could have other restrictions as to implement a certain interface {\tt IMyInter}. If that was the case, prototype {\tt T} would be declared as\\
\verb|    object T implements IMyInter|\\

To test whether the DSL used in the metaobject annotation {\tt concept} of {\tt GroupWork} is enough, one should compile \verb|GroupWork_Test| as a program. It instantiates {\tt GroupWork} using prototype {\tt T} above. If {\tt GroupWork} assumes that its formal parameter {\tt T} has a method not described in the concept DSL, a compilation error will occurs when compiling \verb|GroupWork_Test|. This is the goal of creating prototypes in package \verb|main_ut.groupwork_test|.

The test cases are not inserted in the program that uses the {\tt concept} metaobject. They have to be separated compiled. The code of the axioms are not checked by metaobject {\tt concept}. They may contain invalid Cyan code. Each axiom should return {\tt Nil} if the there is no error or an error message as a {\tt String}.

The valid predicates of metaobject {\tt concept} are given below. We use {\tt T}, {\tt U}, and {\tt S} for types and {\tt I} for identifier parameters. These types may be anyone, including generic parameter.

\vspace*{3ex}
\begin{tabular}{l|l}
predicate & meaning \\ \hline
{\tt T is U}   & {\tt T} should be equal to {\tt U} \\ \hline
{\tt  T implements U}  & {\tt T} should implement interface {\tt U}\\ \hline
{\tt S subprototype T}    & {\tt S} should be subprototype of {\tt T}\\ \hline
{\tt S superprototype T}   & {\tt S} should be superprototype of {\tt T}\\ \hline
{\tt T interface}  & {\tt T} should be an interface\\ \hline
{\tt T noninterface} & {\tt T} should be a prototype that is not an interface\\ \hline
{\tt T has [ list of methods ]}   & {\tt T} should have the methods in the list    \\ \hline
{\tt T in [ list of prototypes ]}& {\tt T} should be one of the prototypes in the list  \\ \hline
{\tt I in [ list of identifiers]}& {\tt T} should be one of the prototypes in the list  \\ \hline
{\tt T identifier} & {\tt T} should be an identifier parameter \\ \hline
{\tt ! any of the predicates} & the opposite of the predicate should be true \\ \hline
{\tt axiom axiomMethod} & generates the test case {\tt axiomMethod}, which is similar\\ & to a method declaration\\ \hline
\end{tabular}
\vspace*{3ex}

The compile-time function {\tt typeof} can be used in the DSL of a concept. It may even be recursive:
\begin{cyan}
package main

@concept{*
     T has [
         func search: typeof(R get) -> typeof(T get: Int)
            // recursion
         func at: Int -> typeof(R at: Int)
         func get: Int -> Double
         ],
     R has [
        func get -> Program
            // recursion
        func at: Int n -> typeof(T at: 0)
     ],
     typeof(R at: Int) in [ Int, Long, typeof(Int asString) ]
*}
object Strange<T, R>
end
\end{cyan}

However, there will be an error if parameter {\tt test} is passed to this {\tt concept} metaobject annotation. The metaobject will not be able to generate the test prototype because of the {\tt typeof} compile-time function.

There may be errors in the predicates. For example, the DSL of a {\tt concept} annotation may:
\begin{enumerate}[(a)]
\item demand that {\tt T}, the parameter, is both a prototype (not interface) and an interface;
\item demand that {\tt T} inherits from {\tt A} and that {\tt A} inherits from {\tt T};
\item have inconsistences such as demand that {\tt T} should be in a list of prototypes and that implement a certain interface but no prototype in the list implements the interface;
\item requires that {\tt T} has incompatible methods such as \\
\nd \verb|    func get: Int -> Int|\\
\nd \verb|    func get: Double -> Double|\\
\end{enumerate}

Most of these errors are not caught by the metaobject {\tt concept}.

A list of predicates can be put in a file and reused. For example, the whole DSL that appears between \verb|{*| and \verb|*}| of metaobject annotation {\tt concept} of {\tt GroupWork} can be put in a file\\
\verb|    group(T).concept|\\
\nd of the directory \verb|--data| of the directory of package {\tt main}. Now the example can be written as

\begin{cyan}
package main

@concept{*
    main.group(T)
*}
object GroupWork<T>
    // as before
    ...
end
\end{cyan}
The concept file ``{\tt group(T).concept}''\/ can be used by other packages of the programa. The file name should be preceded by the package name:
\begin{cyan}
package other

@concept{*
    main.group(T)
*}
object MyGroupWork<T>
    ...  // elided
end
\end{cyan}
This prototype is in package {\tt other}. Note that it is not necessary to import the package {\tt main} in order to use {\tt group(T).concept}.

Package {\tt cyan.lang} has several concept files:

\nd {\tt addable(T).concept},\\
{\tt arithmetic(T).concept},\\
{\tt comparable(R,S).concept},\\
{\tt container(T,R).concept},\\
{\tt equatable(T).concept},\\
{\tt init(T).concept},\\
{\tt init(T,R).concept},\\
{\tt init(T,R, S).concept},\\
{\tt iterator(T).concept},\\
{\tt iteratorSize(T).concept},\\
{\tt lessThan(T).concept},\\
{\tt predicate(T).concept}\\

 To discover an up-to-date description of each of them, open the files in a directory \\
\verb|    cyan\lang\--data|\\

Whenever one uses a concept file its axioms are incorporated in the test prototype of the concept that used it. Then if  a concept attached to a prototype uses concept  {\tt arithmetic(T)} of cyan.lang, its prototype test will have the axioms of the concept.

Metaobject {\tt concept} may be used with non-generic prototypes. This is useful to enforce that a prototype should obey some restrictions and that some test cases should be generated for it. See the example

\begin{cyan}
package algebra

@concept{*
    cyan.lang.arithmetic(Matrix),
    cyan.lang.init(Matrix, Int, Int)
*}
object Matrix
    ...  // elided
end
\end{cyan}
Test cases would be generated for {\tt Matrix}.

Currently it is not possible to pass a generic prototype as a parameter to a concept file:
\begin{cyan}
package structures

@concept{*
    cyan.lang.init(Vector<T>)  // error
*}
object Vector<T>
    ...  // elided
end
\end{cyan}

\section{Message Sends To Generic Prototype Instantiations}

Compile-time messages can be send to a generic prototype instantiation through the syntax\\
\verb|    Function<String, Int, Char> .# writeCode|\\
\noindent Currently only message ``{\tt writeCode}''\/ can be send. This message calls a virtual method\footnote{It does not exist really.} {\tt writeCode} at {\it compile-time}. The generic prototype created for this instantiation is in file\\
\verb|    Function(String,Int,Char).cyan|\\
\nd Virtual method {\tt writeCode}
writes to file \\
\noindent \verb|     full-Function(String,Int,Char).cyan|\\
\noindent of the directory of the project. This code has all the parts added by the compiler and metaobjects. Method {\tt writeCode} is very useful to discover what is inside the real generic prototype. When there is a metaobject annotation in the generic prototype, as in {\tt Function}, errors may be difficult to discover without the full code of the prototype.

The syntax \verb|.#| only works if the generic prototype instantiation is where a type is expected as in a variable declaration. One can check the final version of prototype {\tt Program} and \verb|Function<Int, Int>| using the example that follows.
\begin{cyan}
package main

object Program
    func run {
        var Program .# writeCode p;
        var Function<Int, Int> .# writeCode f;
    }
end
\end{cyan}

\section{Future Enhancements}

The compile-time message send using \verb|.#| will be replaced by metaobject annotations attached to types. Then \\
\verb|    Function<String, Int, Char> .# writeCode|\\
\noindent will be replaced by \\
\verb|    Function<String, Int, Char>@writeCode|\\
\noindent Now {\tt writeCode} can be used even inside an expression:
\begin{cyan}
package main

object Program
    func run {
          // Program is an expression here
        Out println: Program@writeCode prototypeName;
          // Function<Int, Int> is a type here
        var Function<Int, Int>@writeCode f;
    }
end
\end{cyan}

Cyan does not support {\it generic methods}. However, it is very probably it will do in the future. We then give a first definition of this construct and show the characteristics it should have in the language.

A generic method would be declared by putting the generic parameters after keyword {\tt func} as in
\begin{cyan}
object MySet
    final
    func<T> T add: (T elem) { ... }
    ...
end
\end{cyan}
When the compiler finds a message send using {\tt add:} of {\tt MySet}, as in\\
\verb@    p = MySet add: p@\\
\nd it creates a specific method for that type using the compile-time type of {\tt p}. This method could not override any superprototype method and it could not be redefined in subprototypes. It should be a {\tt final} method.

The difference between using a generic method {\tt add:} and declaring a method \\
\verb@    func add: (Any elem) -> Any@\\
\nd is that the compiler checks the relationships between the parameter and the return value. As another example, a generic method \\
\verb@    public func<T>  relate: (T first, T second)@\\
\nd demands that the arguments to the method be of the same compile-time type.

\chapter{Important Library Objects}  \label{ilo}

This Chapter describes some important library objects of the Cyan basic library. All the objects described here are automatically imported by any Cyan program. They are in a package called cyan.lang.

\section{System}

\index{System} \index{exit}
Prototype {\tt System} has methods related to the runtime execution of the program. It is equivalent to the {\tt System} class of {\tt Java}. Its methods are given below. Others will be added in due time.
\begin{cyan}
       // ends the program
    func exit
       // ends the program with a return value
    func exit: (Int errorCode)
       // runs the garbage collector
    func gc
       // current time in milliseconds
    func currentTime -> Long
       // prints the stack of called methods in the
       // standard output
    func printMethodStack

        // execute a command
    func exec: String command
    func exec: Array<String> commandList
        // see the Java method System.exec for help
    func exec: Array<String> commandList, Array<String> envpList, String dir
    @doc{*
        this method can be used as a dynamic storage for global variables
    *}
    func globalTable -> IMap<String, Dyn> = mapGlobalVariables;


\end{cyan}

\section{Input and Output}

Prototype {\tt In} and {\tt Out} \index{In} \index{Out} are used for doing input and output in the standard devices, usually the keyboard and the monitor.
\begin{cyan}
public object In
    func readInt -> Int
    func readFloat -> Float
    func readDouble -> Double
    func readChar -> Char
    func readLine -> String
    ...
end

public object Out
    func (println: (Any)*)
end
\end{cyan}

\section{Tuples}

A tuple is an object with methods for getting and setting a  set of values of possibly different types. A literal tuple is defined in Cyan between ``\verb@[.@''\/  and ``\verb@.]@''\/ as in: \index{tuple} \index{literal!tuple}

\begin{cyan}
var t = [. name = "Lívia", age = 4 .];
Out println: "name: " ++ t name ++ " age: " ++ t age;
\end{cyan}
This literal object has type \verb@Tuple<name, String, age, Int>@. This is a tuple in which the fields have user-defined names.

A literal tuple may also have unnamed fields which are further referred as {\tt f1}, {\tt f2}, etc:
\begin{cyan}
var t = [. "Lívia", 4 .];
Out println: "name: ", t f1, " age: ", t f2;
\end{cyan}
The type of this literal tuple is \verb@Tuple<String, Int>@ which is exactly the prototype \\
 \verb@    Tuple<f1, String, f2, Int>@

Prototype {\tt Tuple} can take any number of parameters. A metaobject is responsible for creating its methods and fields. The real prototype created from \verb@Tuple<name, String, age, Int>@ is below.
\begin{cyan}
package cyan.lang

public final object Tuple<name, String, age, Int>


    func init: (String g1, Int g2) {
        _name = g1;
        _age = g2;
    }
    func name: String g1 age: Int g2 -> Tuple<name, String, age, Int> {
        return Tuple<name, String, age, Int> new:  g1,  g2;
    }
    @annot( #name ) var String _name
    func name -> String = _name;
    func name: String other { _name = other }

    @annot( #age ) var Int _age
    func age -> Int = _age;
    func age: Int other { _age = other }

    override
    func == (Dyn other) -> Boolean {
        if other isA: Tuple<name, String, age, Int> {
            var Tuple<name, String, age, Int> another;
            @javacode{*             _another = (_Tuple_LT_GP__name_GP_CyString_GP__age_GP_CyInt_GT ) _other;
            *}
            if name != (another name) {  return false }
            if age != (another age) {  return false }
            return true
        }
        else {
            return false
        }
    }

    override    func asString -> String {
         return "[. name = " ++ name asStringQuoteIfString ++ ", age = " ++
             age asStringQuoteIfString ++ " .]"
    }

    func copyTo: (Any other) { }

end

\end{cyan}
Metaobject \verb"annot" \index{annot} \label{annotfield} attaches to a field, shared variable, method, prototype, or interface a feature given by its parameter. This feature can be retrieved at runtime by method {\tt featureList:} of the object.

The \verb|Tuple<name, String, age, Int>| prototype has methods for getting and setting each tuple field:
\begin{cyan}
var Tuple<name, String, age, Int> t;
t name: "Carolina" age: 1;
Out println: (t name);
t name: "Lívia";
t age: 4;
Out println: "name: ", t name, " age: ", t age;
\end{cyan}

An empty tuple is illegal:
\begin{cyan}
var t = [. .]; // compile-time error: empty tuple
var anotherError = [..]; // unidentified symbol '[..]'
\end{cyan}

\subsection{Future Enhancements}

Object {\tt Tuple} will have a method {\tt copyTo:} that copies the information of the tuple into a more meaningful object. We will shown how it works using an example. We want to copy a tuple of type\\
\verb@    Tuple<String, Array<String>, String, Int>@\\
\nd into an object of {\tt Book}.
\begin{cyan}
package main

@init(name, authorList, publisher, year)
object Book

    @annot( #f1 )
    @property String name

    @annot( #f2 )
    @property Array<String> authorList

    @annot( #f3 )
    @property String publisher

    @annot( #f4 )
    @property String year

    override
    func asString -> String {
        return authorList[0] ++ " et al. " ++ name ++ ". Published by " ++ publisher ++ ". " ++ year ++ ".";
    }

end
\end{cyan}

However, {\tt copyTo:} has to know to which field of {\tt Book} it should copy field {\tt f1} of the tuple. This method cannot choose one field based on the types --- there are two of them whose type is {\tt String}. We should use annotations for that:
Now the following code will work as expected.
\begin{cyan}
var Tuple<String, Array<String>, String, Int> t;
var b = Book("", [ "" ], "", 0);
t = [. "Philosophiae Naturalis Principia Mathematica",
        [ "Isaac Newton" ],
        "Royal Society",
        1687
     .];
t copyTo: b;
b println;
\end{cyan}

Tuples inside tuples are copied recursively. The {\tt Manager} example is
\begin{cyan}
@init(person, company)
object Manager

    @annot( #f1 )
    @property Person person

    @annot( #f2 )
    @property String company

end

@init(name, age)
object Person

    @annot( #f1 )
    @property String name
    @annot( #f2 )
    @property Int age
end
...

var manager = Manager new: Person, 0;
var john = [. [. "John", 28 .], "Cycorp" .];
john copyTo: manager;
assert john person name == "John" &&
       john person age == 28 &&
       john company == "Cycorp";
\end{cyan}

Method {\tt copyTo:} can be used in grammar methods to store the single method argument into a meaningful object:
\begin{cyan}
object BuildBook
    @grammarMethod{*
        (bookname: String (author: String)* publisher: String year: Int)
    *}
    func build: Tuple< String, Array<String>, String, Int>  t -> Book {
        var book = Book new("", [ "" ], "", 0);
        t copyTo: book;
        return book
    }
\end{cyan}
This method accepts as arguments all the important book information: name, authors, publisher, and publication year:
\begin{cyan}
var prin = BuildBook bookname = "Philosophiae Naturalis Principia Mathematica"
                  author = "Isaac Newton"
                  publisher = "Royal Society"
                  year = 1687;
\end{cyan}

\section{Dynamic Tuples} \label{DTuple}

Object {\tt DTuple} is a dynamic \index{tuple!dynamic} \index{DTuple} tuple. When an object of {\tt DTuple} is created, it has no fields. When a dynamic message ``{\tt ?attr: value}''\/ is sent to the object, a field {\tt attr} whose type is the same as {\tt value} is created. The value of this field can be retrieved by sending the message ``{\tt ?attr}''\/ to the object. See the example:
\begin{cyan}
var t = DTuple new;
t ?name: "Carolina";
   // prints "Carolina"
Out println: (t ?name);
   // if uncommented the line below would produce a runtime error
//Out println: (t ?age);
t ?age: 1;
   // prints 1
Out println: (t ?age);
\end{cyan}

Object {\tt DTuple} is the object
\begin{cyan}
package cyan.lang

object DTuple

    func init {  ... }

    override
    func doesNotUnderstand: (String methodName, Array<Array<Dyn>> args) -> Dyn ...
    func contains: String fieldName -> Boolean  ...
    func size -> Int = fieldList size;

    func getFieldList -> Array<String> = fieldList;

    // elided
end
\end{cyan}  \label{AddFieldDynamicallyMixin} \index{AddFieldDynamicallyMixin}

{\tt DTuple} redefines method {\tt doesNotUnderstand:} in such a way that a field is added dynamically if it does not exist. When a non-existing method ``{tt f:}''\/ in a message send ``{\tt f: value}''\/ is called on the object, {\tt doesNotUnderstand:}  simulates the addition to the receiver of
 a field \verb@_f@ and methods {\tt f: T} and \verb@T f@. The methods set and get the field. {\tt T} is the type of {\tt value}.

\section{Intervals}
\index{intervals}

A interval is the return value of methods  \verb".." and \verb"..<" of the types  {\tt Byte}, {\tt Short}, {\tt Int}, {\tt Char}, and {\tt Boolean}. Then if {\tt first} and {\tt last} are integers,
\verb"first..last" returns an interval with all integers numbers between {\tt first} and {\tt last}, including this last one. And \verb"first..< last" returns an interval with all integers between {\tt first} and {\tt last - 1} --- it is equivalent to \verb"first..(last - 1)". If \verb"last < first" the return is a valid interval but without elements.

\begin{cyan}
        var Interval<Int> inter;
        inter = 3..5;
            // this code prints numbers 0 1 2
        0..2 foreach: { (: Int i :)
            Out println: i
        };
            // this code prints numbers 3 4 5
        inter repeat: { (: Int i :)
            Out println: i
        };
            // prints the alphabet
        'A'..'Z' foreach: {
            (: Char ch :)
            Out println: ch
        };
        var anArray = [ 0, 1, 2, 3 ];
        0..<anArray size foreach: { (: Int n :) n
            println
        };
\end{cyan}
Operator ``{\tt ..}''\/ has smaller precedence than the arithmetical operators and greater precedence than the logical and comparison operators. So, the lines
\begin{cyan}
i+1 .. size - 1 repeat: { ... }
if 1..n == anInterval { ... }
\end{cyan}
are equivalent to
\begin{cyan}
((i+1) .. (size - 1)) repeat: { ... }
if (1..n) == anInterval  { ... }
\end{cyan}

Prototype {\tt Interval} is defined as follows. Generic parameter {\tt T} can only be instantiated with types {\tt Byte}, {\tt Short}, {\tt Int}, {\tt Long}, and {\tt Char}.
\begin{cyan}
package cyan.lang

   // T should be one of these types. Otherwise a compiler error is issued
@concept{*
    T in [ Byte, Short, Int, Long, Char ],
        "The parameter 'T' to this generic prototype instantiation should be Byte, Short, Int, Long, or Char"
*}
object Interval<T> implements Iterable<T>
    ...
       // method bodies elided
    override
    func == (Dyn other) -> Boolean
    func asArray -> Array<T>
    func times: Function<Nil> b
    func repeat: Function<T, Nil> b

    override
    func foreach: Function<T, Nil> b
    func filter: Function<T, Boolean> f -> Array<T>
    func filter: Function<T, Boolean> f foreach: Function<T, Nil> b
    func map: Function<T, T> f -> Array<T>
    func |> Function<Interval<T>, Interval<T>> f -> Interval<T>
    func + Iterable<T> other -> Iterable<T>
       // Smalltalk-like injection
    func inject: (T initialValue)
        into: Function<T, T, T> b
        -> T
       // injection method to be used with context object.
       // the initial value is private to injectTo
    func to: (T max)
        do: (InjectObject<T> injectTo)
        -> T
    func size -> Int
    func first -> T
    func last  -> T

    func apply: (String message) -> Dyn
    func .* (String message)
    func .+ (String message) -> Any
    override
    func iterator -> Iterator<T>
    // elided
end



package cyan.lang

interface Iterable<T>
    func iterator -> Iterator<T>
    func foreach: Function<T, Nil>
end

package cyan.lang

abstract object InjectObject<T> extends Function<T, Nil>
    override
    abstract func eval: T
    abstract func result -> T
end
\end{cyan}

Intervals can be used with method {\tt in:} of the basic types:
\begin{cyan}
        var String s = "";
        var Int age = In readInt;
        if age in: 0..2 {
            s = "baby"
        }
        else if age in: 3..12 {
            s = "child"
        }
        else if age in: 13..19 {
            s = "teenager"
        }
        else {
            s = "adult"
        }
\end{cyan}

\chapter{Grammar Methods}  \label{grammarmethods}

\index{grammar method} \index{method!grammar}

Many languages support methods with a varying number of parameters. These parameters are usually accessed as an array:
\begin{verbatim}
      // Java
    public void print(String format, Object... args) {
        ...
    }
\end{verbatim}
This method could be used as
\begin{verbatim}
    out.print("Color %s %f", "red", 33.0);
\end{verbatim}
Cyan goes beyond by allowing a varying number of parameter, a varying number of keywords, optional parameters, optional keywords, and much more. The pattern of a message passing can be given by a regular expression. That all is made using metaobject {\tt grammarMethod} whose annotation should be attached to a method. The DSL of the metaobject annotation describes the pattern of possible message passings through a regular expression containing message keywords, types, and regular expression operators. The valid operators are \verb"|" (``or''\/), \verb|+| (one or more repetitions), \verb|*| (zero or more repetitions), and \verb|?| (optional).

\begin{cyan}
package grammar

object Car
    @grammarMethod{*
        (do:
           (on: | off: | left: | right: | move: Int)+
        )
    *}
    func carPlay: Tuple< Any,
                         Array< Union<f1, Any, f2, Any, f3, Any, f4, Any, f5, Int> >
                   >   t -> String {

        var s = "";
        for elem in t f2 {
            type elem
                case Any f1 { s = s ++ "car on " }
                case Any f2 { s = s ++ "car off " }
                case Any f3 { s = s ++ "car left " }
                case Any f4 { s = s ++ "car right " }
                case Int f5 { s = s ++ "car move($f5) " }
        }
        return s
   }

end
\end{cyan}
The annotation for metaobject {\tt grammarMethod} in this example is attached to method {\tt carPlay:}. The attached method should always take one single parameter whose type is based on the DSL of the annotation. Latter we will describe how to calculate this type. Anyway, It is not necessary to know how to build the parameter type from the DSL. Simply declare the method, {\tt carPlay:} in the example, without the type of the parameter. The metaobject will sign an error and tell you which should be the type of the parameter. Copy and paste this type to your code. There is no restriction on the method return type.

A {\tt Car} object may receive messages that match the regular expression of the DSL of the metaobject annotation {\tt grammarMethod}. The regular expression should be between parentheses. The \verb"|" between the keywords mean ``or''\/. The \verb|+| mean ``one or more''\/. The the regular expression mean ``{\tt do:} followed by zero or more of the following keywords: {\tt on:}, {\tt off:}, {\tt left:}, {\tt right:}, or {\tt move:} (with an {\tt Int} parameter).
Then the message passings below are legal:
\begin{cyan}
    let car = Car();
    car do: on:;
    car do: on: off:;
    car do: on: move: 50 left: move: 20 right: off:;
\end{cyan}
Prototype {\tt Car} does not define methods {\tt on:}, {\tt on: off:}, and {\tt on:move:left:move:right:off:}. Each of the message passings above should cause a compile-time error. They almost do. Before signalling the error the compiler searches for a metaobject whose annotation, attached to the prototype or a method, implements interface\\
\verb|    IActionMethodMissing_dsa|\\
\nd This interface has a method that returns the code that should replace the message passing. It will be a message {\tt carPlay:} with the appropriate parameter. The three messages of the last example will be replaced by the code that follows. How the parameter is generated will soon be explained.
\begin{cyan}
    car carPlay:  [. Any,
         [  ( Union<f1, Any, f2, Any, f3, Any, f4, Any, f5, Int>() f1: Any )  ]
       .];
    car carPlay:  [. Any,
         [  ( Union<f1, Any, f2, Any, f3, Any, f4, Any, f5, Int>() f1: Any ) ,
            ( Union<f1, Any, f2, Any, f3, Any, f4, Any, f5, Int>() f2: Any )  ]
       .];

    car carPlay:  [. Any,
         [  ( Union<f1, Any, f2, Any, f3, Any, f4, Any, f5, Int>() f1: Any ),
            ( Union<f1, Any, f2, Any, f3, Any, f4, Any, f5, Int>() f5: 50 ),
            ( Union<f1, Any, f2, Any, f3, Any, f4, Any, f5, Int>() f3: Any ),
            ( Union<f1, Any, f2, Any, f3, Any, f4, Any, f5, Int>() f5: 20 ),
            ( Union<f1, Any, f2, Any, f3, Any, f4, Any, f5, Int>() f4: Any ),
            ( Union<f1, Any, f2, Any, f3, Any, f4, Any, f5, Int>() f2: Any )  ]
       .];
\end{cyan}

A method {\tt add:} that accepts any number of real arguments that are subtypes of type {\tt T} can be declared as
\begin{cyan}
package grammar

object GMTest<T>
    @grammarMethod{*
        (add: (T)+ )
    *}
    func addAll: Array<T> args {
        all addAll: args
    }

    func init { all = Array<T>(); }

    override
    func asString -> String = all asString;

    func getAll -> Array<T> = all;

    let Array<T> all

end
\end{cyan}
The compiler will replace {\tt T} inside the DSL of the metaobject annotation {\tt grammarMethod} by the real argument.
\begin{cyan}
    let GMTest<Int> ti = GMTest<Int>();
    ti add: 0, 1, 2, 3;
    assert ti getAll == [ 0, 1, 2, 3 ];
\end{cyan}

The \verb@+@ means one or more real arguments of type {\tt T} or its subtypes. We could have used \verb@*@ instead to mean ``zero or more real arguments''\/. In this case, the following code would be legal.
\begin{cyan}
    let GMTest<Int> ti = GMTest<Int>();
    ti add: ;
    assert ti getAll == Array<Int>();
\end{cyan}

Instead of using keyword {\tt add:} just one time, we may want to use a keyword {\tt each:} before each element added in the array.
\begin{cyan}
package grammar

object GMTest<T>
    @grammarMethod{*
        (add: (T)* )
    *}
    @grammarMethod{*
        (each: T)+
    *}
    func addAll: Array<T> args {
        all addAll: args
    }

    func init { all = Array<T>(); }

    override
    func asString -> String = all asString;

    func getAll -> Array<T> = all;

    let Array<T> all

end
\end{cyan}
Both metaobject annotations are attached to method {\tt addAll:}. This is possible because both demand exactly the same parameter, \verb|Array<T>|. This is not usually the case. Now elements may be added with one or more {\tt each:} keyword:
\begin{cyan}
    let GMTest<Int> ti = GMTest<Int>();

    ti add: 0, 1, 2, 3;
    assert ti getAll == [ 0, 1, 2, 3 ];

    ti each: 4 each: 5 each: 6;
    assert ti getAll == [ 0, 1, 2, 3, 4, 5, 6 ];
\end{cyan}

Here we should use \verb@+@ because we cannot have zero ``{\tt each: value}''\/ elements. If we used \verb|*| the metaobject would issue the error\\
\verb|    This regular expression matches an empty input, which is illegal|\\

More than one keyword may be repeated as in
\begin{cyan}
package grammar

object StringHashTable
    func init { map = HashMap<String, String>(); }

    @grammarMethod{*
        (key: String value: String)+
    *}
    func multKeyValue: Array<Tuple<String, String>> list  {
        for t in list {
            map[ t f1 ] = t f2
        }
    }

    func getMap -> IMap<String, String> = map;

    let IMap<String, String> map
end
\end{cyan}
Part ``\verb@key: String, value: String@''\/ is represented by \verb!Tuple<key, String, value, String>!. Since there is a plus sign after this part, the whole method takes a parameter of type\\
\verb@    Array<Tuple<String, String>>@\\
An example of use is
\begin{cyan}
    let ht = StringHashTable();
    ht key: "John" value: "Professor"
       key: "Mary" value: "manager"
       key: "Peter" value: "designer";
\end{cyan}

\section{Matching Message Sends with Methods}

A prototype may have one or more methods with an attached {\tt grammarMethod} prototype. It may have other metaobject annotations that implement interface\\
\verb|    IActionMethodMissing_dsa|\\
\nd These annotations may be attached to the prototype, to fields, local variables, etc or not attached to anything. The prototype may inherit from a prototype that has annotations of metaobjects that implement this interface.

The semantic analysis of a message passing starts with the semantic analysis of the receiver and real arguments. Then the compiler searches for an adequate method in the prototype that is the type of message receiver, say {\tt T}, and its superprototypes. How this search is done is discussed elsewhere (page~\pageref{messagepassing}).

If no adequate method is done, the compiler puts in a list all metaobject annotations of {\tt T} of metaobjects that implement \\
\verb|    IActionMethodMissing_dsa|\\
\nd This list is ordered according to the textual order in which the annotations appear in the prototype {\tt T}. The first element appears in the smaller line number of the source code of {\tt T}.

Then the compiler calls method \\
\verb|    dsa_analyzeReplaceKeywordMessage|\\
\nd of each metaobject. If two or more of them return a non-{\tt null} value, an error is issued: the call is ambiguous. If one of them returns a non-{\tt null} value, this value is a tuple with the code that should replace the original message passing. For example, in the example
\begin{cyan}
    ti add: 0, 1, 2, 3;
\end{cyan}
of prototype \verb|GMTest<T>|, the compiler will replace this message passing by
\begin{cyan}
    ti addAll: [ 0, 1, 2, 3 ];
\end{cyan}
The code returned by the metaobject {\tt grammarMethod}, in this case, is \\
\verb|    "ti addAll: [ 0, 1, 2, 3 ]"|

The call to method \\
\verb|    dsa_analyzeReplaceKeywordMessage|\\
of all metaobjects may return null. That is, metaobjects corresponding to metaobject annotations in {\tt T}.
In this case, the compiler searches for metaobject annotations of the superprototypes such that the metaobject implement interface\\
\verb|    IActionMethodMissing_dsa|\\

If there is no superprototype an error message is issued.

The {\tt grammarMethod} annotation cannot be attached to {\tt init:} methods because {\tt init:} methods cannot be called by sending messages to  expressions.

A grammar method annotation will try to match as much as possible the message with its regular expression.
Then the second {\tt add:}  keyword of prototype {\tt MyOddArray} will never be used. The second tuple element, an array, will always have size zero.
\begin{cyan}
package grammar

object MyOddArray
    @grammarMethod{*
        ( (add: Int)+ (add: Int)* )
    *}
    func addAll: Tuple<Array<Int>, Array<Int>> t {
        all addAll: t f1;
        assert t f2 size == 0;
        all addAll: t f2;
    }

    func init { all = Array<Int>(); }

    override
    func asString -> String = all asString;

    func getAll -> Array<Int> = all;

    let Array<Int> all

end
\end{cyan}

Note that:
\begin{enumerate}[(a)]
\item  there could be other metaobject annotations that may change message passing. A future restriction would be to restrict to public places an annotation of a metaobject that implements interface\\
\verb|    IActionMethodMissing_dsa|\\
\nd Currently an annnotation internal to a method, for example, is taken into consideration;

\item interfaces are not searched for. That is, it is legal to attach, to an interface or method signature of an interface, an annotation of a metaobject whose class implements \\
\verb|    IActionMethodMissing_dsa|\\
\nd But it will not be taken in consideration;

\item the grammar method metaobject does not do any further anaylysis in the annotation DSL. The in the prototype {\tt MyOddArray}, there will not be any warning that the second {\tt add:} keyword is never used. The only checking is whether the regular expression accepts the empty string;

\item a method that has a metaobject annotation {\tt grammarMethod} may be overridden in a subprototype. Then the method called may be that of the subprototype. See this example:
\begin{cyan}
package grammar

object SubMyOddArray extends MyOddArray
    @grammarMethod{*
        ( (addThis: Int)+ (addOther: Int)* )
    *}
    override
    func addAll: Tuple<Array<Int>, Array<Int>> t {
        let Array<Int> array = getAll;
        for elem in t f1 {
            array add: elem + 1
        }
    }
end
\end{cyan}
The compiler replaces the message send based on the compile-time type.
\begin{cyan}
        var MyOddArray array = SubMyOddArray();
          // this is replaced by a call to addAll
          // the method called is that of SubMyOddArray
        array add: 0 add: 1;
        array println;
        assert array asString == "[ 1, 2 ]";
        // if uncommented, there would be a compile-time error
        // array addThis: 0 addThis: 1;
\end{cyan}

\item {\tt init:} and {\tt new:} are not allowed as keywords in the regular expression of a grammar method annotation.
\end{enumerate}

\section{Unions and Optional Keywords}

Unions are used to compose the type of the parameter of grammar methods that use the regular operator ``\verb!|!''\/. The signature ``\verb!A | B!''\/ means {\tt A} or {\tt B} (one of them but not both).
\begin{cyan}
package grammar

object EnergyStore
    @grammarMethod{*
        (add: (wattHour: Double | calorie: Double | joule: Double))
    *}
    func addEnergy: Tuple<Any, Union<f1, Double, f2, Double, f3, Double>> t {
        addAmount: t f2
    }

    func addAmount: Union<f1, Double, f2, Double, f3, Double> value {
        type value
            case Double f1 {
                amount = amount + f1*3600.0
            }
            case Double f2 {
                amount = amount + f2*4.1868
            }
            case Double f3 {
                amount = amount + f3;
            }
    }

        // keeps the amount of energy in joules
    @property var Double amount = 0.0;
end
\end{cyan}

{\tt Any} is the type associated to keywords without parameters such as {\tt add:} of this example. We can use this prototype as
\begin{cyan}
    var EnergyStore store = EnergyStore new;
    store add: wattHour: 5.0;
    store add: joule: 10.0;
    store add: calorie: 3.0;
    store getAmount println;
\end{cyan}

The optional keywords may be repeated using  ``\verb@+@''\/ (one or more) or ``\verb@*@''\/. We used ``\verb@+@''\/ in the annotation of {\tt addEnergyList:}.
\begin{cyan}
package grammar

object EnergyStore

    @grammarMethod{*
        (add: (wattHour: Double | calorie: Double | joule: Double)+)
    *}
    func addEnergyList: Tuple<Any, Array<Union<f1, Double, f2, Double, f3, Double>>> t {
        for elem in t f2 {
            addAmount: elem
        }
    }

    @grammarMethod{*
        (add: (wattHour: Double | calorie: Double | joule: Double))
    *}
    func addEnergy: Tuple<Any, Union<f1, Double, f2, Double, f3, Double>> t {
        addAmount: t f2
    }

    func addAmount: Union<f1, Double, f2, Double, f3, Double> value {
        type value
            case Double f1 {
                amount = amount + f1*3600.0
            }
            case Double f2 {
                amount = amount + f2*4.1868
            }
            case Double f3 {
                amount = amount + f3;
            }
    }

        // keeps the amount of energy in joules
    @property var Double amount = 0.0;
end
\end{cyan}
Now we can write things like
\begin{cyan}
EnergyStore add:
    wattHour: 100.0
    calorie: 12000.0
    wattHour: 355.0
    joule: 3200.67
    calorie: 8777.0;
\end{cyan}
This is transformed in a method call to {\tt addEnergyList:}.

As another example, a stub of a prototype {\tt MyFile} could be
\begin{cyan}
package grammar

object MyFile
    @grammarMethod{*
        ( open: String (read: | write: ) )
    *}
    func openReadWrite: Tuple<String, Union<f1, Any, f2, Any>> t {
        let String name = t f1;

        type t f2
            case Any f1 {
                "open '$name' for reading" println;
            }
            case Any f2 {
                "open '$name' for writing" println;
            }

    }

end
\end{cyan}
Method {\tt openReadWrite:} is called twice at runtime in this example:
\begin{cyan}
        var MyFile myfile = MyFile();
        myfile open: "AAAA" read:;
        myfile open: "BBBB" write:;
\end{cyan}

Optional parts should be enclosed by parentheses and followed by ``\verb!?!''\/, as in
\begin{cyan}
package grammar

@init(name, age)
object Person

    @grammarMethod{*
        ( name: String
                 (age: Int)? )
    *}
    func set: Tuple<String, Union<some, Int, none, Any>> t {
        self.name = t f1;
        type t f2
            case Int some {
                self.age = some
            }
            case Any none {
            }
    }

    override
    func asString -> String = "Person($name, $age)";

    @property var String name = "";
    @property var Int age = 0;
end
\end{cyan}
The type associated to \verb|(age: Int)?| is \\
\verb|    Union<some, Int, none, Any>|\\
\nd The type associated to \verb|R?| will be \\
\verb|    Union<some, type of R, none, Any>|\\
\nd Method {\tt set:} of {\tt Person} is called by the message passings
\begin{cyan}
    let Person p = Person("Carolina", 7);
    assert p getName == "Carolina";

      // call set:
    p name: "Carol";
    assert p getName == "Carol";

      // call set:
    p name: "Carolina" age: 7;
    assert p getName == "Carolina";
\end{cyan}

In a grammar annotation, it is possible to use more than one type between parentheses separated by ``\verb@|@''\/:
\begin{cyan}
package grammar

object ArrayIS

    @grammarMethod{*
        (add: (Int | String)*)
    *}
    func addMany: Array<Union<Int, String>> unArray {
        for intStr in unArray {
            type intStr
                case Int elem    { array add: elem }
                case String elem { array add: elem }
        }
    }


    override
    func asString -> String {
        let Array<Any> anyArray = Array<Any>();
        for intStr in array {
            type intStr
                case Int elem    { anyArray add: elem }
                case String elem { anyArray add: elem }
        }
        return anyArray asString
    }


    @property let Array<Int|String> array = Array<Int|String>();
end
\end{cyan}
Message {\tt add:} with a variable number of {\tt Int} and {\tt String} parameters can be sent to an {\tt ArrayIS} object:
\begin{cyan}
    let ArrayIS isArray = ArrayIS();
    isArray add: 0, "zero", 1, "one", 2, "three";
\end{cyan}

Care must be taken with alternative keywords in the DSL code of a grammar method.
\begin{cyan}
package grammar


object ArrayIS

    @grammarMethod{*
        (addElem: Any | addElem: Int | addElem: String)+
    *}
    func addManyElem: Array<Union<f1, Any, f2, Int, f3, String>> unArray {
        for anyIntStr in unArray {
            type anyIntStr
                case Any f1 { "found Any" println }
                case Int f2 { array add: f2 }
                case String f3 { array add: f3 }
        }
    }


    @grammarMethod{*
        (add: (Int | String)*)
    *}
    func addMany: Array<Union<Int, String>> unArray {
        for intStr in unArray {
            type intStr
                case Int elem    { array add: elem }
                case String elem { array add: elem }
        }
    }


    override
    func asString -> String {
        let Array<Any> anyArray = Array<Any>();
        for intStr in array {
            type intStr
                case Int elem    { anyArray add: elem }
                case String elem { anyArray add: elem }
        }
        return anyArray asString
    }


    @property let Array<Int|String> array = Array<Int|String>();
end
\end{cyan}
Using this prototype, one can write
\begin{cyan}
    isArray addElem: 0
            addElem: "zero"
            addElem: Any
            addElem: 0.0
            addElem: 'a'
            addElem: 1;
    assert isArray getArray size == 0;
\end{cyan}
No element is inserted in the field {\tt array} of {\tt ArrayIS} because the first keyword of the grammar method, \\
\verb|    addElem: Any|\\
\nd is always chosen.

\section{Refining the Definition of Grammar Methods}

It is time to describe precisely the type of the parameter of a method that has a metaobject annotation {\tt grammarMethod}. The association of regular expressions with types is given by the following table.  {\tt T1}, {\tt T2}, ..., {\tt Tn} are types and {\tt R} is part of the signature of the code of the DSL of the metaobject annotation. For example, {\tt R} can be
\begin{cyan}
    add:
    add: Int
    at: Int put: String
    add: Int | sub: Int
    (add: Int)*
\end{cyan}
Whenever there is a list of {\tt R}´s, assume that the types associated to them are {\tt T1}, {\tt T2}, and so on. For example, in a list {\tt R R R}, assume that the types associated to the three {\tt R}´s are {\tt T1}, {\tt T2}, and {\tt T3}, respectively.
We used {\tt typeof(S)} for the type associated, by this same table, to the grammar element {\tt S}.

\vspace*{3ex}
\begin{tabular}{l|l}  \hline
rule & type\\ \hline
T1   & T1 \\ \hline
R R  ... R  & \verb@Tuple<T1, T2, ..., Tn>@ \\ \hline
Id ``:''\/ R R  ... R  & \verb@Tuple<T1, T2, ..., Tn>@ \\ \hline
Id ``:''\/ &  Any \\ \hline
Id ``:''\/ T  & T, which must be a type \\  \hline
Id ``:''\/  ``(''\/ T ``)''\/  ``$*$''\/ &  \verb@Array<T>@\\  \hline
Id ``:''\/  ``(''\/ T ``)''\/  ``$+$''\/ &  \verb@Array<T>@\\  \hline

``(''\/ R ``)''\/  &  \verb@typeof(R)@  \\ \hline

``(''\/ R ``)''\/  ``$*$''\/  &  \verb@Array<typeof(R)>@  \\ \hline

``(''\/ R ``)''\/  ``$+$''\/  &  \verb@Array<typeof(R)>@  \\ \hline

``(''\/ R ``)''\/  ``$?$''\/  &  \verb@Union<some, typeof(KeywordUnitSeq), none, Any>@  \\ \hline

T1 ``\verb@|@''\/ T2 ``\verb@|@''\/ ... ``\verb@|@''\/ Tn & \verb@Union<f1, T1, f2, T2, ..., fn, Tn>@ \\ \hline

R ``\verb@|@''\/ R ``\verb@|@''\/ ... ``\verb@|@''\/ R & \verb@Union<f1, T1, f2, T2, ..., fn, Tn>@ \\ \hline

\end{tabular}
\vspace*{3ex}

We will give now the precise definition of the type of the parameter of the method based on the grammar of the DSL of the metaobject annotation. It will be used ``{\tt typeof(P)}''\/ for the type associated to the grammar production {\tt P}.

The productions will be divided in cases.

\p{KeywordGrammar} ::= ``(''\/ KeywordUnitSeq  ``)''\/   ``$*$''\/

\p{KeywordGrammar} ::= ``(''\/ KeywordUnitSeq  ``)''\/    ``$+$''\/

\verb@typeof(KeywordGrammar) = Array<typeof(KeywordUnitSeq)>@

\p{KeywordGrammar} ::= ``(''\/ KeywordUnitSeq  ``)''\/

\verb@typeof(KeywordGrammar) = typeof(KeywordUnitSeq)@

\p{KeywordGrammar} ::= ``(''\/ KeywordUnitSeq  ``)''\/  ``?''\/

Now \verb@typeof(KeywordGrammar) = Union<some, typeof(KeywordUnitSeq), none, Any>@

\p{KeywordUnitSeq} ::= KeywordUnit

\verb@typeof(KeywordUnitSeq) = typeof(KeywordUnit)@\\

When there are at least two KeywordUnit´s:

\p{KeywordUnitSeq} ::= KeywordUnit KeywordUnit \{ KeywordUnit \}

\verb@typeof(KeywordUnitSeq) = Tuple<typeof(KeywordUnit1), ..., typeof(KeywordUnitn)>@\\

\noindent in which \verb@typeof(KeywordUniti)@ is the {\tt i}$^th$ production.

\vspace*{3ex}

When there are at least two KeywordUnit´s separated by ``\verb@|@''\/

\p{KeywordUnitSeq} ::=  KeywordUnit ``\verb@|@''\/ KeywordUnit \{ ``\verb@|@''\/ KeywordUnit \}

\begin{verbatim}
typeof(KeywordUnitSeq) = Union<f1, typeof(KeywordUnit1), ...,
                                fn, typeof(KeywordUnitn)>
\end{verbatim}
\noindent in which \verb@typeof(KeywordUnit)@ is the {\tt i}$^{th}$ production.

\p{KeywordUnit} ::=  SelecGrammarElem

\verb@typeof(KeywordUnit) = typeof(SelecGrammarElem)@

\p{KeywordUnit} ::=  KeywordGrammar

\verb@typeof(KeywordUnit) = typeof(KeywordGrammar)@

\p{SelecGrammarElem} ::= IdColon

\verb@typeof(SelecGrammarElem) = Any@

\p{SelecGrammarElem} ::= IdColon Type1, Type2, ... Typen

\verb@typeof(SelecGrammarElem) = Tuple<Type1, Type2, ..., Typen>@

if IdColon is followed by two or more types or

\verb@typeof(SelecGrammarElem) = Type1@

if

\p{SelecGrammarElem} ::= IdColon Type1

\vspace*{3ex}

\p{SelecGrammarElem} ::= IdColon  ``(''\/ Type ``)''\/  ( ``$*$''\/ \verb@|@ ``$+$''\/ )

\verb@typeof(SelecGrammarElem) = Array<typeof(Type)>@

Note that a Type may be an union type. Then the type of\\
\verb@    Int | String@\\
\noindent is \verb@Union<Int, String>@.

Let us see some examples of associations of signatures of grammar methods with types:

\vspace*{3ex}
\begin{tabular}{l|l}  \hline
{\tt Int}  & Int \\ \hline
{\tt add: Int} & Int \\ \hline
{\tt add: Int, String} & \verb"Tuple<Int, String>" \\ \hline
{\tt add: (Int)*} & \verb"Array<Int>" \\ \hline
{\tt add: (Int)+} & \verb"Array<Int>" \\ \hline
{\tt (add: Int)*} &  \verb"Array<Int>" \\ \hline
{\tt (add: Int)+} &  \verb"Array<Int>" \\ \hline
\verb"(add: Int | String)"  &  \verb"Union<Int, String>" \\ \hline
\verb"(add: (Int | String)+)"  &  \verb"Array<Union<Int, String>>" \\ \hline
\verb"(add: Int | add: String)"  &  \verb"Union<f1, Int, f2, String>" \\ \hline
{\tt key: Int value: Float} & \verb"Tuple<Int, Float>" \\ \hline
\verb"nameList: (String)* (size: Int)?" & \verb"Tuple<Array<String>, Union<some, Int, none, Any>>" \\ \hline
\verb"coke:" & \verb"Any"  \\ \hline
\verb"coke: | guarana:" & \verb"Union<f1, Any, f2, Any>"  \\ \hline
\verb"(coke: | guarana:)*" & \verb"Array<Union<f1, Any, f2, Any>>"   \\ \hline
\verb"(coke: | guarana:)+" & \verb"Array<Union<f1, Any, f2, Any>>"   \\ \hline
\verb"((coke: | guarana:)+)?" & \verb"Union<some, Array<Union<f1, Any, f2, Any>>, none, Any>"   \\ \hline
\verb"((coke: | guarana:)?)+" & \verb"   Array<Union<some, Union<f1, Any, f2, Any>, none, Any>>" \\ \hline
\verb"amount: (gas: Float | alcohol: Float)" & \verb"Tuple<Any, Union<f1, Float, f2, Float>>" \\ \hline
\end{tabular}
\vspace*{3ex}

It is possible to have an annotation that does not use any regular operator. That is legal
\begin{cyan}
    @grammarMethod{*
        (format: (String form) print: (String s))
    *}
    func formatPrint: Tuple<String, String> t {
        ...
    }
\end{cyan}

\section{Domain Specific Languages}   \label{dsl}

\index{Domain Specific Language} \index{DSL}
Grammar methods make it easy to implement  domain specific languages (DSL). A small DSL can be implemented in Cyan in a fraction of the time it would take in other languages. The reasons for this efficiency are:
\begin{enumerate}[(a)]
\item the lexical analysis of the DSL is implemented using grammar methods is the same as that of Cyan;
\item the syntactical analysis of the DSL is given by a regular expression, the signature of the grammar method, and that is easy to create;
\item the program of the DSL is a grammar message send. The Abstract Syntax Tree (AST) of such a program is automatically built by the compiler. The tree is composed by tuples, unions, arrays, and prototypes that appear in the definition of the grammar method. The single method parameter refer to the top-level object of the tree;
\item code generation for the DSL is made by interpreting the AST referenced by the single grammar method parameter. Code generation using AST´s is usually nicely organized with code for different structures or commands being generated by clearly separated parts of the compiler;

\end{enumerate}

To further exemplify grammar methods, we will give more examples of them.
\begin{cyan}
@init(from, to)
object Edge
    @annot( #f1 ) @property Int from
    @annot( #f2 ) @property Int to
end

@init(numVertices, edgeArray)
object Graph

    @annot( #f1 ) @property Int numVertices Int
    @annot( #f2 ) @property Int edgeArray Array<Edge>
end

object MakeGraph
    @grammarMethod{*
        (numVertices: Int (edge: Int, Int)* )
    *}
    func  make: Tuple<Int, Array<Tuple<Int, Int>> t -> Graph {
        let edgeArray = Array<Edge>();
        for elem in t f2 {
            edgeArray add: Edge(elem f1, elem f2);
        }
        return Graph new: t f1, edgeArray;
    }
end
\end{cyan}
A call
\begin{cyan}
var g = MakeGraph numVertices: 5
               edge: 1, 4
               edge: 3, 1
               edge: 1, 2
               edge: 2, 4;
\end{cyan}
would produce and return an object of type {\tt Graph} properly initialized.

Flower \cite{flower:2014:Online} gives an example of a DSL used to control a camera which is in fact a window of visibility over a larger image. As an example, we can have a 1600x900 image but only 200x100 pixels can be seen at a time (this is the camera size). Initially the ``camera''\/ shows part of the image and a program in the DSL moves the camera around the larger image, showing other parts of it. The DSL grammar is
\begin{verbatim}
<Program> ::= <CameraSize> <CameraPosition> <CommandList>
<CameraSize> ::= "set" "camera" "size" ":" <number> "by" <number> "pixels" "."
<CameraPosition> ::= "set" "camera" "position" ":" <number> "," <number> "."
<CommandList> ::= <Command>+
<Command> ::= "move" <number> "pixels" <Direction> "."
<Direction> ::= "up" | "down" | "left" | "right"
\end{verbatim}
{\tt CameraSize} is the size of the window visibility of the camera. {\tt CameraPosition} is the initial position of the camera in the larger image (lower left point of the window). {\tt CommandList} is a sequence of commands that moves the camera around the larger image. The site \cite{flower:2014:Online} shows an annimation of this.

A grammar method implementing the above grammar is very easy to do:
\begin{cyan}
package grammar

object Camera
    @grammarMethod{*
        (sizeHoriz: Int sizeVert: Int
                positionX: Int positionY: Int
                (move: Int (up: | down: | left: | right:) )+ )
    *}
    func camera:
       Tuple<Int, Int, Int, Int,
            Array<
                Tuple<Int,
                    Union<f1, Any, f2, Any, f3, Any, f4, Any>>>> t {
       // here comes the commands to actually change the camera position
    }
end
\end{cyan}
This method could be used as
\begin{cyan}
Camera sizeHoriz: 1600 sizeVert: 900
       positionX: 0    positionY: 0
       move: 100 up:
       move: 200 right:
       move: 500 up:
       move: 150 left:
       move: 200 down;
\end{cyan}
It takes seconds, not minutes, to codify the signature of this grammar method given the grammar of the DSL. Other easy-to-do examples are a Turing machine and a Finite State Machine.

A future work is to design a library of grammar methods for paralel programming that would implement some commom paralel patterns. We could have calls like:
\begin{cyan}
Process par: { Out println: 0 }, { Out println: 1 }
        seq: { Out println: 2 }, { Out println: 3 }
        par: (Graphics getMethod: "convert"), (Printer getMethod: "print");
\end{cyan}
Functions after {\tt par:} would be executed in any paralel. Functions after {\tt seq:} would be executed in the order they appear in the message send. Then {\tt 1} may appear before {\tt 0} in the output. But {\tt 2} will always come before {\tt 3}. Remember methods are u-functions.

\chapter{Functions} \label{functions}

Functions of Cyan are similar to blocks of \index{block} \index{anonymous functions} \index{functions}
Smalltalk or anonymous functions of other languages. A function is a literal object --- an object declared explicitly, without being cloned of another object.
A function may take arguments and can declares local variables. The syntax of a literal function is:\\
\verb@    { (: ParamRV  :) code }@\\
{\tt ParamRV} represents the declaration of parameters and the return value type (optional items).
A function is very similar to a method definition --- it can take parameters and return a value. For example,\\
\verb@    b = {  (: Int x -> Int :) ^ x*x };@\\
declares a function that takes an {\tt Int} parameter and returns the square of it. Symbol \verb@^@ is used for returning a value. However, to {\tt b} is associated a function, not a return value, which depends of the parameter. Functions are objects and therefore they support methods. The function body is executed by sending to the message {\tt eval:} with the parameters the function demands or {\tt eval} if it does not take parameters. For example, \\
\verb@    y = b eval: 5;@\\
assigns {\tt 25} to variable {\tt y}. The {\tt eval:} methods are similar to Smalltalk´s {\tt value} methods. We have chosen a method name different from that of Smalltalk because in Cyan a function may not return a value when evaluated. In Smalltalk, it always does.

The function \verb@{ (: Int x :) ^ x*x }@ is similar to the object
\begin{cyan}
object LiteralFunction001
    func eval: (Int x) -> Int {
       return x*x;
    }
end
\end{cyan}
For every function the compiler creates a prototype like the above, although one that inherits from yet-to-be-seen prototypes \verb|Function<...>|. Then two identical functions give origin to two different prototypes. There are important differences between the function and this prototype which will be explained in due time.

The return value type of a function can be omitted. In this case, it will be the same as the type of the return value of the expression returned --- all returned values should be of the same type. For example,
\begin{cyan}
    { (: Int x, Int y :)
        var Int r;
        r = sqrt: ((x-x0)*(x-x0) + (y-y0)*(y-y0));
    ^ r }
\end{cyan}
declares a function which takes two parameters, {\tt x} and {\tt y}, declares a local or temporary variable {\tt r},\footnote{Which of course can easily be removed as the function can return the expression itself.} and returns the value of {\tt r} (therefore the return value type is {\tt Int}). Assume that this function is inside an object which has a method called {\tt sqrt:}. Variables {\tt x0} and {\tt y0} are used inside the function but they are neither parameters nor declared in the function. They may be fields of the object or local variables of functions in which this literal function is nested. These variables can be changed in the function.

The language does not demand that the return value type of a function be declared. In some situations, the compiler may not be able to deduce the return type:
\begin{cyan}
var b = { ^b };
\end{cyan}
To prevent this kind of error, when a function is assigned to a variable {\tt b} in its declaration, as in this example, {\tt b} is only considered declared after the compiler reaches the beginning of the next statement. Then in this code the compiler would sign the error ``{\tt b} was not declared''\/. In the general case, in an assignment ``{\tt var v = e}''\/ variable {\tt v} cannot be used in {\tt e}.

Generic arrays of Cyan have a method {\tt foreach} that can be used to iterate over the array elements. The argument to this method is a function that takes a parameter of the array element type. This function is called once for each array element:
\begin{cyan}
var Array<Int> firstPrimes = [ 2, 3, 5, 7, 11 ];
    // prints all array elements
firstPrimes foreach: { (: Int e :)
    Out println: e
};
var sum = 0;
    // sum the values of the array elements
firstPrimes foreach: { (: Int e :)
    sum = sum + e
};
Out println: sum;
\end{cyan}

An statement \verb@^ expr@ is equivalent to {\tt return expr} when it appears in the level of method declaration; that is, outside any function inside a method body. See the example:   \label{returnfunction} \index{function return}
\begin{cyan}
   func aMethod: Int x, Int y -> Int {
      var b = { ^ x < 0 || y < 0 };
          // method does not return in the next statement
      (b eval) ifTrue: { Error signal: "wrong coordinates" };
         // method returns in the next statement
      return sqrt: ((Math sqr: x) * (Math sqr: y));
   }
\end{cyan}

A Cyan function at runtime is a \index{closure} closure, a literal object that can close over the variables visible where it was defined. More rigorously, the syntax \verb@{ (: params :) stats }@ creates a closure at runtime for the linking with the instance and local variables is only made dynamically. An object is created each time a function appears at runtime. Therefore the code
\begin{cyan}
var Int x;
var Function<Int> a, b, c;
a = {^ i*i + x };
b = {^ i*i + x };
c = {^ i*i + x };
\end{cyan}
creates three functions, each of which captures variable {\tt x}.

Functions can be curried; that is, we can supply some of the parameters and get a new functions with the remaining parameters: \index{curry}
\begin{cyan}
    var Function<Int, Int, Int> mult = { (: Int a, Int b :) ^a*b };
    var Function<Int, Int>  doubleNum = mult curry: 2;
    var Function<Int> six = mult curry: 2, 3;
        // print 6
    (doubleNum eval: 3) println;
    six eval println; // print 6
\end{cyan}
{\tt doubleNum} is the function
\begin{cyan}
{ (: Int b :) ^2*b }
\end{cyan}
A function that takes $n$ parameters has {\tt curry:} methods
that take from $1$ to $n$ parameters.

\section{Problems with Anonymous Functions}

Anonymous functions \index{anonymous function}  are extremely useful features. They are supported by many functional and object-oriented languages such as Scheme, Haskell, Smalltalk, D, and Ruby. However, this feature causes a runtime error when
\begin{enumerate}[(a)]
\item an anonymous function accesses a local variable that is destroyed before the function becomes inaccessible or is garbage collected. Then the body of the function may be executed and the non-existing local variable may be accessed, causing a runtime error;
\item a function with a return statement live past the method in which it was declared. When the anonymous function body is executed, there will be a return statement that refers to a method that is no longer in the call stack. For the time being, return statements inside an anonymous function is prohibited in Cyan. But the examples of this section will show what would happen if they are allowed.
\end{enumerate}
We will give examples of these errors. Assume that ``\verb"Function<Nil>"''\/ is the type functions that does not take parameters and returns nothing.
\begin{cyan}
object Test
    func init {
        function = { }
    }
    func run {
        prepareError;
        makeError;
    }
    func prepareError {
        function = { return  };
        return;
    }
    func makeError {
        function eval;
    }
    Function<Nil> function
end
\end{cyan}
Suppose the execution starts at method {\tt run} that calls {\tt prepareError} that stores an anonymous function in field {\tt function}.
In {\tt makeError}, the function stored in the field receives message {\tt eval} and statement {\tt return} of this function is executed. This is a return from method {\tt prepareError} that is no longer in the stack. There is a runtime error.

\begin{cyan}
object Test
    func run {
        returnFunction eval
    }
    func returnFunction -> Function<Nil> {
        return { return  };
    }
end
\end{cyan}
Here {\tt returnFunction} returns a function which receives message {\tt eval} in {\tt run}. Again, statement {\tt return} of the function is executed in method {\tt run} and refers to {\tt returnFunction}, which is not in the call stack anymore.

\begin{cyan}
object Test
    func init {
        function = { ^0 }
    }
    func run {
        prepareError;
        makeError;
    }
    func prepareError {
        var x = 0;
        function = { ^x };
    }
    func makeError {
        Out println: (function eval);
    }
    Function<Int> function
end
\end{cyan}
Suppose method run is called on an object of {\tt Test}.
In statement ``{\tt function eval}''\/ in method {\tt makeError}, the function body is executed which accesses variable {\tt x}. However, this variable is no longer in the stack. It was when the function was created in {\tt prepareError} because {\tt x} is a local variable of this method. There is again a runtime error.

\begin{cyan}
object Test
    func run {
        var a1 = 1;
        var Function<Nil> b1;
        if a1 == 1 {
            var a2 = 2;
            b1 = { Out println: a2 };
        }
        b1 eval
    }
end
\end{cyan}
Here a function that uses local variable {\tt a2} is assigned to variable {\tt b1} that outlives {\tt a2}. After the if statement, {\tt a2} is removed from the stack and message {\tt eval} is sent to {\tt b1}, causing an access to variable {\tt a2} that no longer exists.

\verb"Function< Function<Int> >" is the type of functions that return objects of type \verb"Function<Int>".
\begin{cyan}
object Test
    func run {
        var a1 = 1;
        var Function< Function<Int> > b1;
        if a1 == 1 {
            var a2 = 2;
            b1 = { ^{ ^a2 } }
        }
        (b1 eval) eval;
    }
end
\end{cyan}
After the execution of ``\verb#var b1 = { ^{ ^a2 } }#''\/, {\tt b1} refers to a function that refers to local variable {\tt a2}. In statement {\tt (b1 eval) eval}, variable {\tt a2}, which is no longer in the stack, is accessed causing a runtime error.

There are some unusual use of functions that would not cause runtime errors:
\begin{cyan}
object Test
    func run {
        var a1 = 1;
        var Function<Int> b1;
        if a1 == 1 {
            var b2 = {
                b1 = { ^a1 }
            };
            b2 eval;
        }
        b1 eval
    }
end
\end{cyan}
No error occurs here because {\tt b1} and {\tt a1} are create and removed from the stack at the same time.

\begin{cyan}
object Test
    func run {
        var a1 = 1;
        var Function<Function<Int>> b1;
        if a1 == 1 {
            b1 = { ^{ ^a1 } }
        }
        Out println: b1 eval eval
    }
end
\end{cyan}
Here {\tt b1 eval eval} will return the value of {\tt a1} which is in the stack. No error will occur.

\begin{cyan}
object Test
    func run {
        Out println: test
    }
    func test -> Int {
        var Function<Nil> b1;
        {
          var b2 = {
              b1 = { return 0 };
          };
          b2 eval;
        } eval;
        b1 eval;
        Out println: 1
    }
end
\end{cyan}
After message send  ``{\tt b2 eval}''\/ a function is assigned to {\tt b1}. After ``{\tt b1 eval}''\/ statement ``{\tt return 0}''\/ is executed and method test returns. The last statement is never reached. Note that function\\
\verb!    { return 0 }!\\
\nd is a function that does not return a value. Therefore its type is {\tt Function<Nil>}.

Currently the Cyan compiler allows a anonymous function to access any visible variable. There is never a runtime error because the local variables accessed inside a anonymous function are allocated in the heap. Note that fields belong to the {\tt self} object and are always in the heap. The {\tt return} statatement cannot be used inside an anonymous function.

\section{Functions with Multiple Keywords}  \label{functionswithmultiple}

\index{function!multiple keywords}
Regular functions only have one keyword, which is  {\tt eval:} or {\tt eval} (when there is no parameter). It is possible to declare a function with more than one {\tt eval:} keyword. One can declare

\noindent \verb@var b = { (: eval: (T@$_{11}$ {\tt p$_{11}$, T$_{12}$ p$_{12}$, ..., T$_{1_{k_1}}$ p$_{1_{k_1}}$)}\\
\verb@             eval: (T@$_{21}$ {\tt p$_{21}$, T$_{22}$ p$_{22}$, ..., T$_{2{k_2}}$ p$_{2{k_2}}$)}\\
\verb@             ...@\\
\verb@             eval: (T@$_{n1}$ {\tt p$_{n1}$, T$_{n2}$ p$_{n2}$, ..., T$_{n{k_n}}$ p$_{n{k_n}}$)}\\
\verb@            -> R :) {@\\
\verb@     // function body@\\
\verb@};@\\
in which $k_i \geqslant 0$ for each $i$.

Consider a function with a method composed by {\tt n} {\tt eval:} keywords, each of them with at least one parameter.  The {\tt i}$^{th}$ {\tt eval:} keyword has {\tt k$_i$} parameters. This function inherits from prototype\\
\nd  \verb"     Function<"{\tt T$_{11}$, T$_{12}$, ..., T$_{1k_{1}}$\verb"><"T$_{21}$, T$_{22}$, ... T$_{2k_{2}}$>...<T$_{n1}$, T$_{n2}$, ... T$_{nk_n}$\verb", R>"}\\
This prototype declares just one abstract method, which is \\
\verb@abstract@\\
\verb@func eval: ( T@$_{11}$ {\tt p$_{11}$, T$_{12}$ p$_{12}$, ..., T$_{1{k_1}}$ p$_{1{k_1}}$)}\\
\verb@    eval: ( T@$_{21}$ {\tt p$_{21}$, T$_{22}$ p$_{22}$, ..., T$_{2{k_2}}$ p$_{2{k_2}}$)}\\
\verb@           ...@\\
\verb@    eval: ( T@$_{n1}$ {\tt p$_{n1}$, T$_{n2}$ p$_{n2}$, ..., T$_{n{k_n}}$ p$_{n{k_n}}$)} \verb| -> R |\\

As an example, one can declare a function
\begin{cyan}
var Function<String><Int, Nil> b;
b = { (: eval: String key eval: Int value :)
        Out println: "key $key is $value"
};
    // prints "key One is 1"
b eval: "One" eval: 1;
\end{cyan}

An {\tt eval:} method can have zero parameters. In this case, {\tt none} is used in the place of the type in the generic prototype. That is, function
\begin{cyan}
     { (: eval: Int i, Int j
          eval:
          eval: Char ch :) ^ ch ++ i ++ j }
\end{cyan}
has type
\begin{cyan}
     Function<Int, Int><none><Char, String>
\end{cyan}

\section{Methods as Functions} \label{typeofmethods}

Method {\tt functionForMethod} of {\tt Any} takes a literal string with the name of a method and
returns a function that call that method. The name of a unary method is the unary method. The name of a non-unary method is the joining of each keyword followed by its number of parameters, separated by a single white space. Then the names of the methods
\begin{cyan}
object Test
    func run2 { ... }
    func run: Array<String> { ... }
    func aa: String s0, Int s1, Char s2
         bb: Int s3, Char s4
         cc: Char s5, String s6 -> String  {
         return asString ++ s0 ++ s1 ++ s2 ++ s3 ++ s4 ++ s5 ++ s6
    }

    func * Int n -> Char = n asChar;
    func - -> String = "000";
    ...
end
\end{cyan}
are
\begin{cyan}
    run2
    run:1
    aa:3 bb:2 cc:2
    *1
    -
\end{cyan}
As an example, the code calls the methods of prototype {\tt Test} given above
\begin{cyan}
        let Function<Nil> mRun = Test functionForMethod: "run2";
        mRun eval;

        let Function<Array<String>, Nil> mRun1 = Test functionForMethod: "run:1";
        mRun1 eval: [ "0", "1", "2" ];

        let Function<String, Int, Char><Int, Char><Char, String, String> mabc =
            Test functionForMethod: "aa:3 bb:2 cc:2";
        (mabc eval: "0", 1, '2' eval: 3, '4' eval: '5', "6") println;

        let Function<Int, Char> mMult = Test functionForMethod: "*1";
        (mMult eval: 0) println;

        let Function<String> mMinus = Test functionForMethod: "-";
        mMinus eval println;

\end{cyan}
The function returned by {\tt functionForMethod:}, when receives an {\tt eval} or {\tt eval:} message, send the original message to the receiver of {\tt functionForMethod:}. For example, \\
\verb|            mRun1 eval: [ "0", "1", "2" ]|\\
\noindent sends message \verb| [ run: "0", "1", "2" ]| to object {\tt Test} because {\tt Test} is the receiver of message \\
\verb|    functionForMethod: "run:1"|\\

The metaobject {\tt changeFunctionForMethod} whose annotation is attached to {\tt functionForMethod} changes \\
\verb|    Test functionForMethod: "run:1"|\\
\noindent to\\
\verb|    { (: Array<String> p0 :) Test run: p0 }|\\
\noindent This metaobject can only be applied to methods {\tt functionForMethod:} and {\tt functionForMethodWithSelf:}.

Method {\tt functionForMethodWithSelf:} return a function as {\tt functionForMethod:} does but with one difference: there is an additional first parameter whose type is the type of the receiver of the message. In the function returned, the message is sent to this first parameter.

Both methods return a function that works like a {\bf method} of the receiver, but with an important difference. {\tt functionForMethod:} returns what could be called a ``{\it object method}''\/, a method specific to the object that is the receiver. And {\tt functionForMethodWithSelf:} returns a more generic method, one that demands that the receiver be passed as parameter. Let us see how {\tt functionForMethodWithSelf:} works.

\begin{cyan}
        let t0 = Test();
        let t1 = Test();
        let Function<Test, Nil> mRunSelf = Test functionForMethodWithSelf: "run2";
        mRunSelf eval: t0;
        "aaa" println;
        let Function<Test><Array<String>, Nil> mRun1Self = Test functionForMethodWithSelf: "run:1";
        mRun1Self eval: t1 eval: [ "0", "1", "2" ];
        "bbb" println;
        let Function<Test><String, Int, Char><Int, Char><Char, String, String> mabcSelf =
            Test functionForMethodWithSelf: "aa:3 bb:2 cc:2";
        (mabcSelf eval: t0 eval: "0", 1, '2' eval: 3, '4' eval: '5', "6") println;
        "ccc" println;
        let Function<Test><Int, Char> mMultSelf = Test functionForMethodWithSelf: "*1";
        (mMultSelf eval: t1 eval: 0) println;
        "ddd" println;
        let Function<Test, String> mMinusSelf = Test functionForMethodWithSelf: "-";
        (mMinusSelf eval: t1) println;
        "eee" println;
\end{cyan}
{\tt mRunSelf} is a function that takes a {\tt Test} as parameter because {\tt run2} is an unary method. This function is returned in a call to {\tt Test} (line 3) but it is called in line 4 on object {\tt t0}.
{\tt mRun1Self} takes two parameters, one for each {\tt eval:} keyword. If the literal string argument of
{\tt functionForMethodWithSelf:} specifies a  non-unary method, the returned function will have an {\tt eval:} keyword taking a single parameter whose type is the type of the receiver of this message. Then\\
\verb|    Test functionForMethodWithSelf: "run:1"|\\
\noindent returns
\begin{cyan}
    { (: eval: Test myself
         eval: Array<String> p0 :) ^myself run: p0
      }
\end{cyan}

If {\tt Test} had a method\\
\verb|     func open: String s read: close: -> String|\\
\noindent the function returned by \\
\verb|    Test functionForMethod: "open:1 read:0 close:0"|\\
\noindent would be
\begin{cyan}
    { (: eval: String p0
         eval:
         eval:  :) ^Test open: p0 read: close:
      }
\end{cyan}
whose type is \verb|Function<String><none><none, String>|.

The function returned by {\tt functionForMethodWithSelf:} would be
\begin{cyan}
    { (: eval: Test myself
         eval: String p0
         eval:
         eval:  :) ^myself open: p0 read: close:
      }
\end{cyan}
whose type is \verb|Function<Test><String><none><none, String>|.

Using methods as objects is very convenient in creating \index{graphical user interfaces} \index{GUI} graphical user interfaces. Listeners can be regular methods. See the example.
\begin{cyan}
object MenuItem
    func onMouseClick: (Function<Nil> b) {
        ...
    }
end

object Help
    func show { ... }
    ...
end

object FileMenu
   func open { ... }
end

...
var helpItem = MenuItem new;
helpItem onMouseClick: (Help functionForMethod: "show");
var openItem = MenuItem new;
openItem onMouseClick: (FileMenu functionForMethod: "open");
...
\end{cyan}

\section{Methods of Functions for Decision and Repetition}

Object {\tt Function<Boolean>} defines some methods used for decision and iteration statements. The code of these methods is shown below.

\index{whileTrue:} \index{whileFalse:}
\begin{cyan}
package cyan.lang

abstract object Function<Boolean>
    abstract func eval -> Boolean
    func whileTrue: (Function<Nil> aFunction)  {
       (self eval) ifTrue: {
           aFunction eval;
           self whileTrue: aFunction
       }
    }
    func whileFalse: (Function<Nil> aFunction)  {
       (self eval) ifFalse: {
           aFunction eval;
           self whileFalse: aFunction
       }
    }
end
\end{cyan}

An function that does not take any parameters and does not return a value inherits from \verb|Function<Nil>|.
\begin{cyan}
package cyan.lang

abstract object Function<Nil>
    abstract func eval
    func loop {
        while true {
            self eval
        }
    }

    func repeatUntil: (Function<Boolean> test) {
        self eval;
        while ! (test eval) {
            self eval
        }
    }

    @createCatchMethodsForFunctionNil

    @checkCatchParameter
    @grammarMethod{*
        ( (catch: Any)+ (finally: Function<Nil>)? )
    *}
    func catchFinally: Tuple<Array<Any>, Union<some, Function<Nil>, none, Any>> t {
    }

    func hideException {
        {
            self eval
        } catch: { (: CyException e :)
        };
    }
    // other methods

end
\end{cyan}

Method {\tt loop} implements an infinite loop and {\tt repeatUntil:} implements a loop that ends when the function parameter evaluates to {\tt true}. There are many {\tt catch:} methods that are not shown. They are produced by metaobject {\tt createCatchMethodsForFunctionNil}.

\section{Future Enhancements}  \label{fefunctions}

This section proposes a future change to anonymous functions that would allow the compiler to allocate local variables in the stack even if they are used inside a function.

Metaobject annotations can to be attached directly to types as in
\begin{cyan}
var Char@letter ch;
ch = 'A'; // ok
ch = '0'/ // compile-time error
\end{cyan}
Here a metaobject {\tt letter} is attached to {\tt Char} and controls the type checking of {\tt ch}. This feature will be used with {\tt Functions}. An annotation of metaobject {\tt rf} will be attached to prototype {\tt Function} to give the precise type of a function:
\begin{cyan}
func Int test {
    var Int a1 = 0;
    var f1 = {
        ++a1;
    };
    var f2 = { ^0 };
    (f2 eval) println;
    f1 eval;
    f0 eval;
}
\end{cyan}
The types of the function variables will be:


\vspace*{2ex}
\begin{tabular}{ll}
{\tt f1} & \verb|Function<Nil>@rf(1)| \\ \hline
{\tt f2} & \verb|Function<Int>@rf(-1)| \\ \hline
{\tt f2} & \verb|Function<Int>| \\ \hline
\end{tabular}
\vspace*{2ex}

Note that {\tt Function} without the metaobject will be equal to \verb|Function<Int>@rf(-1)|.

Before studying functions in depth, it is necessary to define what is ``scope''\/, ``variable of level k''\/, and ``function of level k''\/.  Each identifier is associated to a scope, the  region of the source code in which the identifier is visible (and therefore it can be used).  A scope can be the region of a method or of a function, both delimited by  {\tt \{} and {\tt \}}. The scope of a local variable starts just after its declaration and goes to the enclosing ``\}''\/ of the function in which it was declared.
A scope will be called ``level 1''\/ if the delimiters {\tt \{} and {\tt \}} are that of a method. ``level 2''\/ is the scope of a function inside level 1. In general, scope level $n+1$ is a function inside scope $n$:
\begin{cyan}
func test: (Int n) {
       // scope level 1
    var Int a1 = n;
    (n < 0) ifFalse: {
          // scope level 2
        var Int a2 = -a1;
        (n > 0) ifTrue: {
              // scope level 3
            var a3 = a2 + 1;
            Out println: "> 0", a3
        }
        ifFalse: { Out println: "= 0" }
    }
} // a1 and n are removed from the stack here
\end{cyan}
We will call ``variable of level k''\/  a variable defined in scope level ``k''\/. Therefore variable {\tt ai} of this  example is a variable of level {\tt i}. The level of parameters is considered -1. There is no variable of level {\tt 0}.

The variables {\it external} to a function are those declared outside the code between {\tt {} and {\tt }} that delimits the function. For example, {\tt a1} is external to the function passed as parameter to keyword {\tt ifFalse:} in the previous example (any of the {\tt ifFalse:} keywords). And {\tt a1} and {\tt a2} are external to the function that is argument to the keyword {\tt ifTrue:}.

In the discussion that follows, it is important to remember that Cyan currently does not allow return statements inside anonymous functions.
A function is called ``function of level {\tt -1}''\/ if it does not access non-constant local variables. This kind of function can access only parameters, variables that are constants, and fields.

By ``{\it access}''\/ a  variable we mean that a local variable appear anywhere between the function delimiters, which includes nested functions. In the example that follows, the function that starts at line {\tt 3} and ends at line {\tt 7} {\it accesses} local variable {\tt a1} which is external to the function. This access is made in the function of line {\tt 5} which is inside the function of lines {\tt 4-6} which is inside the {\tt 3-7} function. Therefore {\tt 3-7} is not a function of level {\tt -1}. And neither is the function of lines {\tt 4-6} or the function of line {\tt 5}. However, the function that is the body of method {\tt test} (lines {\tt 1-8}) is a function of level {\tt -1}.
\begin{cyannum}
func test {
    var a1 = 1;
    { var a2 = 2;            // start
       { var a3 = 3;
         { ++a1 } eval;
       } eval;
    } eval                // end
}
\end{cyannum}

The following function, between lines {\tt 3-5}, is a function of level -1.
\begin{cyannum}
func make: Int n {
    let Int p = In readInt;
    var f = { (: Int k :)
        ^k + p + n;
    };
    (f eval: 0) println;
}
\end{cyannum}

Let {\tt v1}, {\tt v2}, ..., {\tt vn} be the external non-constant {\it local} variables accessed in a function B --- B is a function, not a variable that refers to a function. Fields and parameters are not considered. If m is the level in which B is defined, then B can only access external local variables defined in levels $\leq$ m. But not all variables of levels $\leq$ m are visible in B for some of them may belong to sister functions or they may be defined after the definition of B. Variables defined in levels $>$ m are either inaccessible  or  internal to the function. The following example explains these points.
\begin{cyan}
func test {
    // level 1
    var a1 = 1;
    {  // level 2
       var a2 = 2;           // start of function B1
       { // level 3
         var a31 = 31;       // start of function B2
         { ++a1 } eval;   // function B3
       } eval;            // end B2
       var a22 = 2;
       { // level 3
         var a32 = 32;       // start of function B4
         {                // start of function B5
            // level 4
            var a5 = 5;
            a2 = a1 + a2 + a5
         } eval           // end B5
       } eval;            // end B4
    } eval                // end B1
}
\end{cyan}
Function B2 is defined at level 2 but it cannot access variable {\tt a22} of level 2 --- it is defined after B2. Variable {\tt a5} defined at level 4 is not visible at function B2.

\begin{figure}
\center
\includegraphics[scale=1]{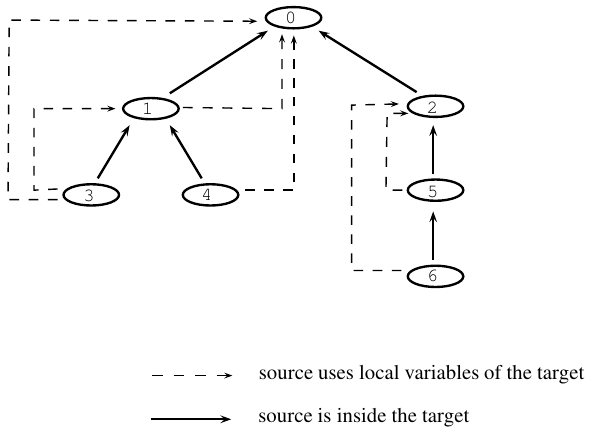}
\caption{Nesting of functions}
\label{functionnesting}
\end{figure}

The important thing to remember is ``B defined at level m can only access external local variables defined in levels $\leq$ m''\/, although not all variables of levels $\leq$ m are accessible at B.
The example of Figure~\ref{functionnesting} should clarify this point. Ellipses represent functions. A solid arrow from function C to function B means that C is inside B. A dashed arrow from C to B means that C uses local variables declared in B.

This Figure represents the functions of the example that follows. The root is the function of the method itself which is represented by the top-level ellipse in the Figure. The numbers that appear in the ellipses are the return values of the functions. This number is used to identify the functions (we will say function {\tt 0} for the function that returns {\tt 0}). The values returned by all the functions are not used (the return value of a method may be ignored. Statements like ``{\tt 1 + 2}''\/ are legal).
\begin{cyan}
func test {
    var v0 = 0;
    {
        var v1 = 1;
        {
            ++v1;
            ++v0;
            ^3
        } eval;
        var v11 = 2;
        {
            ++v0;
            ^4
        } eval;
        ^1
    } eval;
    {
        var v2 = 2;
        { ^5
            {
                ++v0;
                ++v2;
                ^6
            } eval;
        } eval;
        ^2
    } eval;
    return 0
}
\end{cyan}
By the scope rules of Cyan, a function B can only access its own local variables or variables from functions that are ancestors of B.\footnote{X is an ancestor of Y if Y is textually inside X.}
Only variables declared before B are accessible.
In this example, the function that returns {\tt 3} cannot access {\tt v11} even though this variable is declared in an outer function (because the declaration appears after the declaration of function 3). In the Figure, a function B may access local variables of function A if there is a path in solid arrows from A to B (we will write just path from A to B).

When method {\tt eval} or {\tt eval:} of a function A is called, the runtime system pushes to the stack the local variables of A. Till the method returns, these local variables are there and they can be accessed by functions declared inside A. Using the Cyan example above and the Figure, when method {\tt eval} of function 0 is called, it pushes its local variables to the stack. Then function 1 is called and this function calls function 4 that accesses variable {\tt v0} declared at function 0. No error occurs because {\tt v0} is in the stack. To call function 4 it was first necessary to call function 1 and, before this, function 0 which declares variable {\tt v0}.

However, this example could be modified in such a way that function 3 is assigned in function 1 to a variable {\tt b1} declared at function 0 (suppose this is legal --- it is not as we will see).
\begin{cyan}
   // unimportant functions were removed
func test {
    var v0 = 0;
    var Function<Int> b1;
    {
        var v1 = 1;
        b1 = {
            ++v1;
            ++v0;
            ^3
        };
        var v11 = 2;
        ^1
    } eval;
       // compile-time correct, runtime error
    b1 eval;
    return 0
}
\end{cyan}
{\tt  b1} is visible in function 1 by the scope rules of Cyan. After functions 3 and 1 are removed from the stack and control returns to method {\tt eval} of 0, {\tt b1} receives an {\tt eval} message. Since {\tt b1} refers to function 3, the method called will try to access variable {\tt v1} declared in function 1. This variable is no  longer in the stack. There would be a runtime error. However, the rules of Cyan will not allow function 3 be assigned to variable {\tt b1} of function 0. A function variable {\tt b} will never refer to a function that uses external variables that live less than {\tt b}.

Inner functions may be assigned to variables of outer functions without causing runtime errors:
\begin{cyan}
func test {
    var a1 = 1;
    var Function<Nil> b;
    {
        var a2 = 2;
        {
            {
                b = { ++a1; }
            } eval
        } eval;
        c eval;
    } eval;
    b eval;
}
\end{cyan}
Here a function  \verb"{ ++a1 }" is assigned to variable {\tt b} declared at level 1. This does not cause errors because the function only refer to variables of level 1. Variable {\tt b} and {\tt a1} will be removed from the stack at the same time. There is no problem in this assignment. In this example, if the function used {\tt a2} instead of {\tt a1}, there would be a runtime error at line ``{\tt b eval}''\/. Variable {\tt a2} that is no longer in the stack would be accessed.
To prevent runtime errors of the kind ``reference to a variable that is no longer in the stack''\/ Cyan only allows an assignment ``{\tt b = B}''\/, in which {\tt B} is a function, if the variables accessed in {\tt B} will live as much as {\tt b}. This is guaranteed by the rules given in the next section.

Functions are classified according to the external local variables they access.
To a function B is associated a number bl(B) called ``the level of function B''\/ found according to the following rules.
\begin{enumerate}
\item A function that do not access non-constant local variables
are called ``functions of level -1''\/. This kind of function may access parameters and variables declared as constants.

\item Functions that access at least one external non-constant local variable
 have their number bl(B) calculated as\\
    \nd \verb@    bl(B) = max{ lev(v1), lev(v2), ..., lev(vn) }@\\

    {\tt lev(v)} is the level of variable {\tt v}.

    \noindent {\tt v1}, {\tt v2}, ..., {\tt vn} are the external {\it local} non-constant variables accessed in function B.  The ``{\it function level}''\/ of B is bl(B).


\end{enumerate}
A function that has a reference to a external non-constant local  variable of level {\tt k} is at least a function of level {\tt k}. However, it may be a function of level $\geq$ {\tt k} (if it accesses an external non-constnat variable of a superior level). The use of fields, constants, or parameters is irrelevant to the calculus of the level of a function. Fields are not created with the method or when the function receives message {\tt eval} or {\tt eval:}. And parameters and constants are read only.

The definition of bl(B), the level of a function, is different from the definition ``function defined or declared at level {\tt k}''\/ used previously. A function defined at level {\tt k} is a function that is textually at level {\tt k}. The  {\it function level} of a function depends on the external local variables that appear in its body (including the nested functions inside it).

The higher the level of a local variable a function accesses, the more restrictive is the use of the function. For example, function B3 in the next example can be assigned to any of the local function variables {\tt bi} of this example. But {\tt B4} cannot. If it is assigned to {\tt b2}, for example, the message send ``{\tt b2 eval}''\/ would access a local variable {\tt a31} that is no longer in the stack.
\begin{cyan}
func test {
        // level 1
    var a1 = 1;
    var Function<Nil> b1;
    {                        // start of function B1
           // level 2
       var a2 = 2;
       var Function<Nil> b2;
       {                     // start of function B2
           // level 3
         var a31 = 31;
         var Function<Nil> b31;
         var b31 = { ++a1 };    // function B3
         var Function<Nil> b32;
         b32 = { ++a31 };    // function B4
         b2 = { ++a2 };
       } eval;
       b2 eval;
    } eval;
    b1 eval;
}
\end{cyan}

The example below should clarify the definition of ``function level {\tt k}''\/.
\begin{cyan}
func test: (Int p) -> Int {
       // level 1
    var a1  = 1;
    var b1_1 = { ^a1 };  // function of level 1, defined at level 1
    var b1_2 = { ^0 };   // function of level -1, defined at level 1
    var b1_3 = { // function of level 1 because it uses a1
             // level 2
        var a2  = 2;
        var c2 = 0;
        var b2_1 = { ^a1 };                  // function of level 1, defined at level 2
        var b2_2 = { a2 = 1 };               // function of level 2, defined at level 2
        var b2_4 = { Out println: c2 };      // function of level -1, defined at level 2
        var b2_5 = { // function of level 2 because it uses a2
                // level 3
            var a3   = 3;
            var b3_1 = { b1_1 eval };          // function of level 1, defined at level 3
            var b3_3 = { ^a3 };              // function of level 3, defined at level 3
            var b3_5 = { ^p };               // function of level -1, defined at level 3
        }
    };
    b1_3 eval;
    return 0
}
\end{cyan}

A function of level -1 may access constants, parameters, and fields. Functions of level -1 are called  {\it u-functions} \index{u-function} \index{function!u-function} \index{function!unrestricted-use} or {\it unrestricted-use functions}. There is no restriction on the use of u-functions: they may be passed as parameters, returned from methods, returned from functions, assigned to fields, or assigned to any variable. They only have the type restrictions of regular objects.

Functions of levels 0 and up are called {\it r-functions} \index{r-functions} \index{function!r-function} \index{function!restricted-use} or {\it restricted-use functions}. There are limitations in their use: they cannot be stored in fields, returned from methods and functions, and there are limitations on the assignment of them to local variables. This will soon be explained.

An function of level that takes parameters of types {\tt T1}, {\tt T2}, ..., {\tt Tn} and returns a value of type {\tt R} inherits from  prototype  \index{Function}
\begin{cyan}
abstract object Function<T1, T2, ..., Tn, R>
    abstract func eval: (T1, T2, ..., Tn) -> R
    // other methods --- explained later
end
\end{cyan}
If the function is restricted, a metaobject annotation {\tt rf} should be attached to its type:
\begin{cyan}
var Function<Int, String>@rf(2) f;
\end{cyan}
The parameter is the level number.

Every function has its own prototype that inherits from  \verb"Function<...>" objects. When the compiler finds a function\\
\nd \verb@    { ^n }@\\
\nd it creates a prototype {\tt Function001} that inherits from \verb"Function<Int>" (assume that {\tt n} is a local {\tt Int} variable). The name {\tt Function001} was chosen by the compiler and it can be any valid identifier. If this function is assigned to a  variable in an assignment,\\
\nd \verb"    var b = { ^n }"\\
\nd the type of {\tt b} will be \verb|Function<Int>@rf(k)| in which {\tt k} is a literal integer representing the level of the function.

\nd  As another example, the type of variable {\tt add} in\\
\verb@   var add = { (: Int n :) ^n + 1 };@\\
\nd could be {\tt UFunction017}. Since this function inherits from \verb"Function<Int, Int>"  we can declare {\tt add} before assigning it a value as
\begin{cyan}
var Function<Int, Int> add;
add = { (: Int n :) ^n + 1 };
\end{cyan}
Or as
\begin{cyan}
var Function<Int, Int>@rf(-1) add;
add = { (: Int n :) ^n + 1 };
\end{cyan}
Since this function does not access any local variable, its level is {\tt -1}.

Methods of functions are called \index{primitive methods} \index{method!primitive} {\it primitive} methods. A primitive method is not an object. The only allowed operation on a primitive method is to call it.

Functions are them a special kind of object, \index{method!primitive} \index{primitive method} one that has only primitive methods. However, all prototypes that extend prototype {\tt Function} can be passed as an argument to a method that expect a {\tt Function} as a real argument. For example, an {\tt Int} array defines a {\tt foreach:} method that expects an r-function as parameter that accepts an {\tt Int}  parameter and returns {\tt Nil}. One can pass as parameter a regular object:
\begin{cyan}
object Sum extends Function<Int, Nil>
    func init { sum = 0 }
    @property var Int sum

    func eval: (Int elem) {
        sum = sum + elem
    }
end
...
var Array<Int> v = [ 2, 3, 5, 7, 11, 13 ];
v foreach: Sum;
Out println: "array sum = " ++ Sum getSum;
\end{cyan}

\subsection{Type Checking Functions}  \label{typecheckingfunctions}

Now it is time to unveil the rules that make functions statically typed in Cyan.
The rules are:
\begin{enumerate}[(a)]
\item there is no restriction on the use of u-functions and variables whose type is
 \verb!Function<..., R>! or \verb!Function<..., R>@rf(-1)!. A field can have type \verb!Function<..., R>!;

\item fields cannot have type \verb!Function<T1, ... Tn, R>@rf(k)! with {\tt k} greater than {\tt -1};
\item methods and functions cannot have \verb!Function<T1, ... Tn, R>@rf(k)! as the return type if
{\tt k} is greater than {\tt -1};
\item a variable {\tt r} declared at level {\tt k} whose type is \verb!Function<T1, ... Tn, R>@rf(kp)! may receive in assignments:
 \begin{itemize}
    \item a variable {\tt s} of level {\tt m} if {\tt m} $\leq$ {\tt k} and the type of {\tt s} is
    \verb!Function<T1, ... Tn, R>@rf(mp)!  or one of its subtypes, including
    \verb!Function<T1, ... Tn, R>@rf(-1)!;

    \item an r-function of level {\tt m} if {\tt m} $\leq$ {\tt k} and this r-function extends prototype \verb!Function<T1, ... Tn, R>@rf(m)!;

    \item an u-function that extends prototype \verb!Function<T1, ... Tn, R>@rf(-1)!;

 \end{itemize}

\item a parameter whose type is  \verb!Function<T1, ... Tn, R>!  is considered a variable of level 0. The real argument corresponding to this parameter may be a variable or function of any level. Of course, the type of the variable or function should be \verb!Function<T1, ... Tn, R>!  or one of its subtypes;

\item a variable or parameter whose type is \verb!Any!  {\it \bf cannot} receive as real argument any r-function. Unfortunately this introduces an exception in the subtype hierarchy: a subprototype may not be a sub-type. For example, \verb|Function<Int>| is not subtype of {\tt Any}. Although a function like \verb"{ ^0 }" inherits from {\tt Any} (indirectly), its type is not considered subtype from {\tt Any}. The only way of correcting this is allocating the local variables in the stack. But that is inefficient to say the least.

\end{enumerate}

Based on the rules for type checking functions, one can conclude that:
\begin{enumerate}[(a)]
\item fields can be referenced by both u-functions and r-functions;


\item the restriction ``methods and functions can have \verb!Function<T1, ... Tn, R>!  as the return type''\/ (but not \verb!Function<T1, ... Tn, R>@rf(k)! with {\tt k} $\geqslant 0$  could be changed to ``a method can only return u-functions and a function defined at level {\tt k} can only return a function if it is of level {\tt m} with {\tt m} $\leq$ {\tt k}''\/. In the same way, a function defined at level {\tt k} can have a variable as the return value if this variable is of level {\tt m} with {\tt m} $\leq$ {\tt k}. However, we said ``{\it could}''\/, these more liberal rules are not used in Cyan;

\item since parameters are read-only, it is not possible to assign a variable or function to any of them;

\item both r-functions and u-functions can access fields since their use do not cause any problems --- fields belong to objects allocated in the heap, a memory space separated from the stack. Then it is legal to return a function that accesses a field or to assign such a function to any {\tt UFunction} variable:
\begin{cyan}
@init(name, age)
object Person
    func init { }
    private func functionCompare -> UFunction<Person, Boolean> {
        return { (: Person p :) ^age > (p getAge) }
    }
    @property String name = "noname";
    @property Int age = 99;
end
...
var myself = Person new;
   // methods setName: String  and  setAge: Int are automatically created
myself setName: "José";
myself setAge: 14;
if (Person functionCompare) eval: myself {
    Out println: "Person is older than José";
}
\end{cyan}

\item the type of a field or return method value cannot be an r-function. But it can be an u-function. Therefore there will never be a field referring to a function that has a reference to a local variable. And a function returned by a method will never refer to a local method variable;

\item a parameter that has type \verb!Function<T1, ... Tn, R>! cannot be assigned to any variable of the same type because this variable is of level at least 1 and the parameter is of level -1;

\item the generic prototype \verb"Array<T>" declares a field of type {\tt T}. Therefore the generic array instantiation \verb"Array<Function<T1, ... Tn, R>>" causes a compile-time error --- r-functions cannot be types of fields. In the same way, \verb"Function<T1, ... Tn, R>" cannot be the parameter to most generic containers (yet to be made) such as {\tt Hashtable}, {\tt Set}, {\tt List}, and so on.

    This is regrettable. We cannot, for example, create an array of r-functions:
    \begin{cyan}
    var sum    Float = 0;
    var prod   Float = 0;
    var sumSqr Float = 0;
    mySet applyAll: [ { (: Float it :) sum = sum + it   },
                      { (: Float it :) prod = prod*it   },
                      { (: Float it :) sumSqr = sumSqr + it*it }  ];
    \end{cyan}  \label{functionproposal}
    Of course, this restriction applies to a version of Cyan that uses the functions as defined in this section. The current version of Cyan allows this array.
\end{enumerate}

The rules for checking the use of r-functions are embodied in metaobject {\tt rf}. The compiler passes the control to this metaobject when type checking r-functions. It then implements the above rules.

\subsection{Examples}

In this example, an r-function is passed as a parameter. There is no runtime error.
\begin{cyan}
object A
   func aMethod {
       var Int x;
       x = In readInt;
       Out println: (anotherMethod: { (: Int y :) ^y + x });
   }
   func anotherMethod: (Function<Int, Int> b) ) -> Int {
       return yetAnotherMethod: b;
   }
   func yetAnotherMethod: (Function<Int, Int> b) -> Int {
       return b eval: 0;
   }
   ...
end
\end{cyan}
Method {\tt aMethod} calls {\tt anotherMethod} which calls {\tt yetAnotherMethod}. No reference to function
\verb@{ (: Int y :) ^y + x }@ last longer than local variable {\tt x}.

A parameter of type {\tt Any} cannot receive an r-function as real argument. If it could, a runtime error would occur. \begin{cyan}
object Test
   func test {
       { var n = 0;
            // function passed as parameter. The
            // real argument has type Any
         do: { ++n }
       } eval;
       makeError
   }
   func do: (Any any) {
       self.any = any
   }
   func makeError {
          // access to local variable n
          // that no longer exists
       any ?eval
   }
   Any any
end
\end{cyan}

\subsection{Why Functions are Statically-Typed in Cyan}

This section does not present a proof that functions in Cyan are statically typed. It just gives evidences of that.

To introduce our case we will use functions B$_0$, B$_1$, ..., B$_n$ in which  B$_{i}$ is defined at level $i$ and B$_{i+1}$ is defined inside B$_i$. So there is a nesting\\
\nd \verb"     "{\tt B$_n \subset $ B$_{n-1} \subset \ldots \subset $ B$_1 \subset $ B$_0$  }\\
\nd It was used $\subset$ to mean ``{\it nested in}''\/. Function B$_j$ declares a local variable v$_j$. Note that B$_0$ is the body of a method (functions of level 0 are always methods).

Suppose B$_n$ uses external local variables v$_{i_1}$, v$_{i_2}$, ..., v$_{i_k}$ of functions B$_{i_1}$, B$_{i_2}$, ..., B$_{i_k}$ with $i_1 < i_2 < \ldots i_{k-1} < i_k$. It is not important whether B$_n$ uses or not more than one variable of each function.

Let us concentrate on B$_{i_k}$ which defines variable v$_{i_k}$ accessed by B$_n$. Since there is a nesting structure, functions B$_{i_{k}+1}$, B$_{i_{k}+2}$, ..., B$_{n-1}$ also have references to v$_{i_k}$ (because B$_n$ is nested inside these functions). This fact is used in the following paragraph.

B$_n$ can be assigned to a function variable of B$_j$ with $i_k \leq j < n$. This does not cause a runtime error because a function B$_j$ with $i_k \leq j < n$ is only called when B$_{i_k}$ is in the stack. B$_j$ cannot be assigned to a variable b$_t$ of level $t$ with $t < i_k$ because B$_j$ also has a reference to v$_{i_k}$ and, by the rules, it can only be assigned to variables that appear in function B$_t$ with $i_k \leq t < j$.

B$_n$ also has a reference to variable v$_{i_{k-1}}$ of B$_{i_{k-1}}$. Therefore B$_n$ could not be assigned to function variables of functions B$_j$ with $j < i_{k-1}$. Considering all cases, B$_n$ cannot be assigned to function variables of functions B$_j$ with\\
\nd \verb@    @$j < i_1$\\
\nd \verb@    @$j < i_2$\\
\nd \verb@    @$\ldots$\\
\nd \verb@    @$j < i_{k-1}$\\
\nd \verb@    @$j < i_k$\\

Since $i_1 < i_2 < \ldots i_{k-1} < i_k$, we conclude that B$_n$ cannot be assigned to a function variable of function B$_{i_k}$.
Then B$_n$ can only be assigned to a function variable of function B$_j$ with $j \geq i_k$. This is what one of the rules of Section~\ref{typecheckingfunctions} says. Therefore these rules prevent any runtime errors of the kind ``access to a function variable that does not exist anymore''\/ related to the assignment of r-functions to local variables. It is not difficult to see that the other rules prevent all of the other kinds of errors related to r-functions such as the passing of parameters, assignment of functions to {\tt Any} variables, assignment of r-functions to fields (not allowed), and so on.

\subsection{Adding Methods to Objects}  \label{contextfunctions}

This sections describes a possible future feature of Cyan, the possibility of adding methods to an object.

We add a grammar method to prototype {\tt Any} (Section~\ref{any}) \index{function!context} \index{context functions} for dynamically adding methods to prototypes. It is necessary to specify each keyword, the types of all parameters, the return value type, and the method body. This grammar method has the signature
\begin{cyan}
    func (addMethod:
                 (keyword: String ( param: (Any)+ )?
                   )+
                  (returnType: Any)?
                  body: Any)
\end{cyan}

Suppose we want to add a {\tt print} method dynamically to prototype {\tt Box}:
\begin{cyan}
object Box
     func get -> Int { return value }
     func set: (Int other) { value = other }
     var Int value = 0
end
\end{cyan}
We want to add a {\tt print} method to every object created from {\tt Box} or that has already been created using this prototype using {\tt new} or {\tt clone} (with the exception to those objects that have already added a {\tt print} method to themselves). This method, if textually added to {\tt Box}, would be
\begin{cyan}
     func print { Out println: get }
\end{cyan}

Note that {\tt Any} already defines a {\tt print} method. However, the method {\tt print} we define has a behavior different from that of the inherited method.

A first attempt would to add {\tt print} dynamically would be    \label{addMethod}
\begin{cyan}
Box addMethod:
    keyword: #print
    body: { Out println: get };
\end{cyan}
However, there is a problem here: it is used {\tt get} in the function that is parameter to keyword {\tt body:}. The compiler will search for a {\tt get} identifier in the method in which this statement is, then in the prototype, and then in the list of imported prototypes, constants, and interfaces. Anyway, {\tt get} will not be considered as a method of {\tt Box}, which is what we want. A second attempt would be
\begin{cyan}
Box addMethod:
    keyword: #print
    body: { Out println: (Box get) };
\end{cyan}
Here it was used {\tt Box get} instead of just ``{\tt get}''\/. But then the {\tt print} method of every object created from {\tt Box} will use the {\tt get} method of {\tt Box}:
\begin{cyan}
var myBox = Box new;
myBox set: 5;
Box   set: 0;
   // prints 0
Box ?print;
   // prints 0 too !
myBox ?print;
\end{cyan}
Since the {\tt print} method was dynamically added, it has to be called using {\tt ?}. In this example, both calls to {\tt print} used the {\tt get} method of {\tt Box}, which returns the value {\tt 0}.

This problem cannot be solved with regular functions. It is necessary to define a new kind of function, {\it context function} to solve it.  A {\it context function} is declared as\\
\nd \verb@    { (: T self, parameters and return type :) body }@\\
\nd Part ``\verb!T self!''\/ is new. It means that inside the method body {\tt self} has type {\tt T}. The identifiers visible inside the function body are those declared in the function itself, those accessible through {\tt T} (but using ``{\tt self}''\/), external parameters, and external constant local variables. For each parameter or constant the function declares a variable with the same type and name. At the function creation, the values of the external parameters and constants are copied to these function variables. ``{\tt super}''\/ cannot be used inside a context function. All restrictions given above apply to regular function nested inside a context function.
\begin{cyan}
var cf = { (: Person self, Int n :)
    var f = { get print  // error: should use "self get"
        super println;   // error: 'super' cannot be used in a context function
    };
    n println;
};
\end{cyan}


Methods of the current object can be accessed by means of a local variable:
\begin{cyan}
let mySelf = self;
var b = {
      (: Any self :)
   Out println: (myself age)
   };
\end{cyan}
Fields of the current object can be indirectly accessed by means of get and set methods of local variables such as {\tt mySelf}.

With context functions, the {\tt print} method of one of the previous example can now be adequately added to {\tt Box}.
\begin{cyan}
Box addMethod:
    keyword: #print
    body: { (: Box self :) Out println: (self get) };
\end{cyan}
The ``{\tt self}''\/ before ``{\tt get}''\/ is mandatory.
Now the {\tt print} method will send message {\tt get} to the object that receives  message {\tt print}:
\begin{cyan}
var myBox = Box new;
myBox set: 5;
Box   set: 0;
   // prints 0
Box ?print;
   // prints 5
myBox ?print;
\end{cyan}

Method {\tt addMethod: ...} checks \index{addMethod} whether the context object passed in keyword {\tt body:} matches the keywords, parameters, and return value.
\begin{cyan}
   // error: function with parameter, keyword without one
Box addMethod:
    keyword: #print
    body: { (: Box self, Int n :) Out println: n };
   // error: function has no Int parameter
   // and return value should be Int
Box addMethod:
    keyword: #add
    param: Int
    returnType: Int
    body: { (: Box self -> String :) ^(self get) asString };
\end{cyan}

The type of the {\it context function}
\begin{cyan}
    { (: S self, T1 t1, T2 t2, ..., Tn tn -> R :) ... }
\end{cyan}
is
\begin{cyan}
     ContextFunction<S, T1, T2, ..., Tn, R>
\end{cyan}
Interface {\tt ContextFunction} is defined as
\begin{cyan}
interface ContextFunction<S, T1, T2, ..., Tn, R>
    func bindToFunction: S -> UFunction<T1, T2, ..., Tn, R>
end
\end{cyan}

Therefore the type of
\begin{cyan}
    { (: S self, T1 t1, T2 t2, ..., Tn tn  :) ... }
\end{cyan}
is
\begin{cyan}
     ContextFunction<S, T1, T2, ..., Tn, Nil>
\end{cyan}
Assuming that there is no statement ``\verb|^ expr|''\/ in the body of the context function or there is such a statement but the type of {\tt expr} is {\tt Nil}.

The compiler creates a {\it context object}\footnote{Chapter~\ref{contextobjects} define context objects, which are a generalization of functions.} from a context function. From
\begin{cyan}
    { (: S self, T1 t1, T2 t2, ..., Tn tn -> R :) ... }
\end{cyan}
that uses local variables and parameters {\tt v1}, {\tt v2}, ... {\tt vk} the compiler creates
\begin{cyan}
object ContextFunction001(V1 v1, ..., Vk vk)
           implements ContextFunction<S, T1, T2, ..., Tn, R>

    func bindToFunction: (S newSelf) -> UFunction<T1, T2, ..., Tn, R> {
        return { (: T1 t1, T2 t2, ..., Tn tn -> R :)
              // body of the context function with
              // self replaced by newSelf
            ...
            }
    }
end
\end{cyan}

For example, from the context function of the code
\begin{cyan}
    let Int i = 0;
    var b = { (: Box self :) Out println: (i + (self get)) };
    (b bindToFunction: Box) eval;
\end{cyan}
\nd the compiler creates a regular object
\begin{cyan}
object ContextObject001(Int i)
           implements ContextFunction<Box, Nil>

    func bindToFunction: Box -> UFunction<Nil>
        return {
                Out println: (i + (newSelf get));
            }
    }
end
\end{cyan}
And
\begin{cyan}
    var b = { (: Box self :) Out println: (i + (self get)) };
\end{cyan}
becomes
\begin{cyan}
    var b = ContextObject001(i);
\end{cyan}

As another example, consider
\begin{cyan}
var Person p = Person("Carol", 5);
let Int otherAge = 8;

var cf = { (: Person self, Int age2 -> Boolean :)
      self name println;
      self name prototypeName println;
      ^ (self age) == age2 && (self age) == otherAge;
  };
\end{cyan}
The compiler generates, for this context object, the following prototype:
\begin{cyan}
object CFun_1__(Int otherAge)
   implements ContextFunction<Person, Int, Boolean>
   func bindToFunction: (Person newSelf__) -> UFunction<Int, Boolean> {
      return { (: Int age2 -> Boolean :)
          newSelf__ name println;
          newSelf__ name prototypeName println;
          ^ (newSelf__ age) == age2 && (newSelf__ age) == otherAge;
      }
   }
end
\end{cyan}

A context function with multiple keywords is a context function with multiple {\tt eval:} keywords:
\begin{cyan}
    { (: S self, eval: T11 t11, ... T1n t1n  eval: T21 t21, ... T2m t2m,
          ... eval: ... Tkp tkp -> R :)
          ...
    }
\end{cyan}
The type of this context function is
\begin{cyan}
interface ContextFunction<S, T11, ..., T1n><T21, ... T2m>...<Tk1, ... Tkp, R>
    func bindToFunction: S -> UFunction<T11, ..., T1n><T21, ... T2m>...<Tk1, ... Tkp, R>
end
\end{cyan}
The {\tt UFunction} returned by {\tt bindToFunction:} is defined in Section~\ref{functionswithmultiple}.
The type of a context function with multiple keywords that does not return a value is defined similarly.

In what follows, we will specify the checks made when calling {\tt addMethod:} to add a method with a single keyword. In a call
\begin{cyan}
obj  addMethod:
     keyword: sel
     param: T1  param: T2 ... param: Tn
     returnType: R
     body: expr
\end{cyan}
metaobject  {\tt checkAddMethod}  checks whether:
\begin{enumerate}[(a)]
\item the parameters to all keywords of {\tt addMethod: ...} but {\tt body:} are literals;
\item the keyword {\tt sel} is a valid method name;
\item the keyword {\tt sel} ends with ``{\tt :}''\/ if \verb|n > 0|;
\item the keyword {\tt sel} does not end with  ``{\tt :}''\/ if \verb|n == 0|;
\item the type of {\tt expr} is subtype of \verb|ContextFunction<S, T1, T2, ..., Tn, R>| in which {\tt S} is  supertype of {\tt typeof(obj)} (the compile-time type of {\tt obj}).

\end{enumerate}
Even with these checkings there may be an error when the method {\tt addMethod: ...} is called. For example, {\tt obj} may refer to a {\tt B} object although {\tt typeof(obj)} is {\tt A}. There is a final method {\tt sel} in {\tt B} that is not defined in {\tt A}. The metaobject cannot detect that a final method is being changed. In case of error, method {\tt addMethod: ...} throws exception {\tt ExceptionAddMethod}.

It is possible that in future versions of Cyan all checking be postponed to runtime. At least if some of the parameters are not literals.

Let us see more examples of use of {\it context functions}.
\begin{cyan}
var myContextFunction = { (: Box self, Int p -> Int :)  ^(self get) + p };
Box set: 5;
var Function<Int, Int> b = myContextFunction bindToFunction: Box
assert: (b eval: 3) == 8;

var anotherBox = Box new;
anotherBox set: 1;
b = myContextFunction bindToFunction: anotherBox;
assert: (b eval: 3) == 4;
\end{cyan}

In one of the examples given above, a {\tt print} method is added to prototype {\tt Box} through ``{\tt addMethod: ...}''\/. When this grammar method is called at runtime, method {\tt print} will be added to all instances of {\tt Box} that have been created and that will created afterwards. However, if an instance of {\tt Box} has added another {\tt print} method, it is not affected:
\begin{cyan}
var myBox = Box new;
myBox set: 10;
myBox addMethod:
    keyword: #print
    body: { (: Box self :)
        Out println: "value = ", self get;
    };
Box addMethod:
    keyword: #print
    body: { (: Box self :) Out println: (self get) };
    // will print "value = 10" and not just "10"
myBox print;
\end{cyan}

Another method that takes a parameter and returns a value can be added to {\tt Box}:
\begin{cyan}
Box addMethod:
    keyword: #returnSum
    param: Int
    returnType: Int
    body: { (: Box self, Int p -> Int :) ^(self get) + p };
\end{cyan}
The metaobject whose annotation is attached to this grammar method checks whether the number of keywords (one), the parameter type, and the return value type matches the context function. It does in this case.
\begin{cyan}
var myBox = Box new;
myBox set: 5;
assert (myBox ?returnSum: 3) == 8;
\end{cyan}

As another example,  one can add methods to change the color of a shape:
\begin{cyan}
object Shape
    @property Int color
    abstract func draw
    ...
end
...

var colors =       [ "blue", "red",      "yellow",    "white",    "black" ];
    // assume that hexadecimal integer numbers can
    // be given in this way
var colorNumbers = [ ff_Hex, ff0000_Hex, ffff00_Hex,     ffffff_Hex, 0 ];
var i = 0;
colors foreach: {
    (: String elem :)
    Shape addMethod:
          keyword: elem
          body: { (: Shape self :) self color: colorNumbers[i] };
    ++i;
};
\end{cyan}
Methods {\tt blue}, {\tt red}, {\tt yellow}, {\tt white}, and {\tt black} are added to {\tt Shape}. So we can write
\begin{cyan}
var Shape myShape;
...
myShape ?blue;
   // draws in blue
myShape draw;
myShape ?red;
   // draws in red
myShape draw;
  // Square is a subprototype of Shape
var Square sqr = Square new;
...
sqr ?black;
  // draws in black
sqr draw;
\end{cyan}
Assume that {\tt draw} of subprototypes use the {\tt color} defined in {\tt Shape}.

We could have got the same result as above by adding all of these methods to {\tt Shape} textually. For example, method {\tt blue} would be
\begin{cyan}
    func blue { color: ff_Hex }
\end{cyan}

Regular objects may be used as parameters to keyword ``{\tt body:}''\/.
\begin{cyan}
Box addMethod:
    keyword: #print
    body: PrintBox;
\end{cyan}
{\tt PrintBox} is a regular prototype.
\begin{cyan}
object PrintBox
       implements ContextFunction<Box, Nil>

    func bindToFunction: (Box newSelf) -> UFunction<Nil> {
        return { Out println: (newSelf get) }
    }
end
\end{cyan}

There could be libraries of context objects that implement methods that could be added to several different prototypes. For example, there could be a {\tt Sort} context object to sort any object that implements an interface
\begin{cyan}
interface Indexable<T>
     func at: Int -> Int
     func at: Int put: T
     func size -> Int
end
\end{cyan}

A context object used to add a method to an object could have more methods than just {\tt bindToFunction:}.
\begin{cyan}
object PrintFormatedBox
          implements ContextFunction<Box, Nil>

    func bindToFunction: (Box newSelf) -> UFunction<Nil> {
        return { Out println: (format: (newSelf get))  }
    }
        /* one could declare a context function
           with one more method like format:
           this method fills the first positions
           with 0. Then
              format: 123
           should produce "0000000123"
        */
    private func format: (Int n) -> String {
         var strn = (n asString);
         return ("0000000000" trim: (10 - strn size)) ++ strn
    }
end
\end{cyan}
After {\tt PrintFormatedBox} is added to {\tt Box} as in
\begin{cyan}
Box addMethod:
    keyword: #print
    body: PrintFormatedBox;
\end{cyan}
the {\tt print} method puts zeros before the printed number, if necessary.
{\tt format:} is a method that can only be used by the method {\tt print} added to {\tt Box}. It is like a private method of {\tt print}.

Suppose you want to replace a method by a context function that calls the original method after printing a message.
 Using the {\tt Box} prototype, we would like something like this:
\begin{cyan}
object Box
     func get -> Int { return value }
     func set: (Int other) { value = other }
     var Int value = 0;
end

...
Box set: 0;
Box addMethod:
    keyword: #get
    returnType: Int
    body: { (: Box self :)
        Out println: "getting 'value'";
        self get
    };
\end{cyan}
It is a pity this does not work. In a call ``{\tt Box get}''\/ made after the call to {\tt addMethod: ...}, the context function will be called. It prints \\
\nd \verb@    getting 'value'@\\
\nd as expected but them it calls {\tt get}, which is a recursive call. There is an infinity loop. What we would like is to call the original {\tt get} method. That cannot be currently achieved in Cyan. However, it will be possible if context functions are transformed into ``{\it literal dynamic mixins}''\/ (LDM) or ``{\it literal runtime metaobjects}''\/ (LRM). This feature is not yet supported by Cyan. But the description of it would be as follows.

The syntax of LRM's would be the same as that of context functions except that ``{\tt super}''\/ could be used as receiver of messages. Calls to {\tt super} are calls to the original object. Then the code above can be written as
\begin{cyan}
Box set: 0;
Box addMethod:
    keyword: #get
    returnType: Int
    body: { (: Box self :)
        Out println: "getting 'value'";
        super get
    };
\end{cyan}
In this way a call ``{\tt Box get}''\/ would print ``\verb|getting 'value'|''\/ and the original {\tt get} method would be called. Exactly what we wanted.

We are unaware of any language that allows literal runtime metaobjects.

This feature has not been introduced into Cyan because:
\begin{enumerate}[(a)]
\item it seems to be difficult to implement (which may not be a good reason). The compiler being built generates Java code and literal runtime metaobjects probably demand code generation at runtime, which would be difficult with Java (although not impossible);
\item there are some questions on what is the type of a LDM/LRM. This is the same question of ``what is the type of a mixin prototype?''\/.
\end{enumerate}

\chapter{Context Objects} \label{contextobjects}

A Cyan function becomes a closure at runtime for it can access variables from its context as in the example: \index{object!context} \index{context objects}
\begin{cyan}
   // sum the vector elements
var sum = 0;
v foreach: { (: Int x :) sum = sum + x };
\end{cyan}
Here the sum of the elements of vector {\tt v} is put in variable {\tt sum}. But {\tt sum} is not a local variable or parameter of the  function. It was taken from the environment. Then to use a function it is necessary to bind (close over) the free variables to some variables that are visible at the function declaration. {\tt self} is visible in the function and messages can be sent to it:
\begin{cyan}
v foreach: { (: Int x :) sum = sum + (self calc: x) };
\end{cyan}

Although functions are tremendously useful, they cannot be reused because they are {\it literal} objects. A function that accesses local and fields  is specific to a location in the source code in which those variables are visible. Even if the programmer copy-and-past the function source code it may need to be modified because the variable names in the target environment may be different. A generalization of functions would make the free variables and the message sends to {\tt self} explicit. That is what context objects do.

In Cyan it is possible to define a {\it context object} with free variables that can be bounded to produce a workable object. For example, the context object
\begin{cyan}
object Sum( Int &sum ) extends Function<Int, Nil>
    override
    func eval: (Int x) {
       sum = sum + x
    }
end
\end{cyan}
defines method {\tt eval:} and uses a free  {\tt Int} variable {\tt sum}
which is binded in the object creation.
\begin{cyan}
var v = [ 1, 2, 3 ];
var Int s = 0;
v foreach: Sum(s);
assert: (s == 6);
\end{cyan}
The syntax {\tt Sum(s)} means the same as\\
\nd \verb@     (Sum new: s)@\\
\nd which is the creation of an object from {\tt Sum} passing {\tt s} as a parameter. However, this is not a regular parameter passing --- it is passing by reference as we will soon discover.

A context object cannot define {\tt init}, {\tt init:}, or {\tt clone} methods. The only way of creating a context object is by using a {\tt new:} method created by the compiler.

When the type of a context object parameter is preceded by {\tt \&}, the real argument should be a local variable or field. It cannot be a parameter of the current method. Context objects can be inherited. In the code that follows, {\tt Manager} extends {\tt Employee}. In {\tt Employee}, the context parameters {\tt name} and {\tt age} are not transformed into fields because they are passed as parameters to the constructor of {\tt Person}. The same happens with all context parameters of {\tt Manager}.
\begin{cyan}
package people

open
object Person
    func init: String name, Int age {
        self._name = name;
        self._age = age
    }

    func name: String name age: Int age {
        self._name = name;
        self._age = age;
    }

    @property var String _name
    @property var Int _age
end

open
object Employee(String name, Int age, Int salary, String companyName,
         String &outp, Int &sum)
         extends Person(name, age)

    func getSalary -> Int = salary;
    func getCompanyName -> String = companyName;

    func doSum {
        outp = outp ++ name;
        sum = sum + salary
    }
end

object Manager(String name, Int age, Int salary, String companyName,
              String &outp, Int &sum)
       extends Employee(name, age, salary, companyName, outp, sum)

    func getSectionName -> String = sectionName;
    func setSectionName: String sn { sectionName = sn }

    var String sectionName = "";
end

\end{cyan}

\section{Passing Parameters by Copy}

When the {\tt \&} do not precede the parameter of a {\it context object}, a copy of the real argument is made when creating the object. Just like in the creation of a regular object.
\begin{cyan}
object DoNotSum(Int sum)
   func eval: (Int x) {
      sum = sum + x
   }
end
...

var Int s = 0;
v foreach: DoNotSum(s);
assert: (s == 0);
\end{cyan}
Here a copy of the value of {\tt s}, {\tt 0}, is passed as a parameter to the context object. This ``parameter''\/ is then changed. But the value of the original variable {\tt s} remains unchanged. Parameters that are not preceded by {\tt \&} will be called \index{context objects!copy parameters} \index{copy parameters} ``{\it copy parameters}''\/. Parameters preceded by {\tt \&} will be called \index{context objects!reference parameters} \index{reference parameters} ``{\it reference parameters}''\/ or {\tt \&} parameters.

A context object with a copy parameter may have any expression as real argument:
\begin{cyan}
v foreach: DoNotSum(0);
[ 0, 1, 2 ] foreach: DoNotSum(Math fatorial: 5);
\end{cyan}
Therefore, method parameters can be real arguments to {\tt DoNotSum}.

\section{Passing Parameters by Reference}

Some languages such as C++ support passing of parameters by reference. In this case, changes in the parameter are reflected in the real argument, which should be a variable (it cannot be an expression).
Cyan does not support directly this construct. However, it can be implemented using the generic context object {\tt Ref}:
\begin{cyan}
object Ref<T>(T &v)
    func value  -> T { return v }
    func value: (T newValue) { v = newValue }
end
\end{cyan}
    Now if you want to pass a parameter by reference, use {\tt Ref}:
\begin{cyan}
private object CalcArea
       // it is as if parameter to keyword area: were by reference
    func squareSide: (Float side) area: (Ref<Float> refSqrArea) {
           // by calling method value: we are changing the parameter
           // of the context object
        refSqrArea value: side*side
    }
end

public object Program
    func run {
        var side = In readFloat;
        var Float sqrArea;

           /* encapsulate the reference parameter inside a
              context object. That is, use "Ref<Float>(sqrArea)"
              instead of just "sqrArea".
              Local variable "sqrArea" is changed inside
              method squareSide:area: of prototype CalcArea when message
              value: is sent to refSqrArea
           */
        CalcArea squareSide: side area: Ref<Float>(sqrArea);

        Out println: "Square side = $side";
        Out println: "area = $sqrArea"
    }
end
\end{cyan}
Of course, the ``passing by reference''\/ syntax in Cyan is not straightforward. However, it has two advantages:
\begin{enumerate}[(a)]
\item it does not need a special syntax;
\item and, most importantly, it is type-safe. Context objects use the same rules as the static functions of Cyan. That means, for example, that a field of prototype {\tt Calc} cannot refer to a parameter of type \verb"Ref<Float>". That guarantees there will never be a reference to local variable of {\tt run} of {\tt Program} after this method is removed from the stack.
\end{enumerate}
    There will never be an error in Cyan equivalent to the following error in a C program, in which pointer {\tt mistake} refers to a local variable that has been removed from the stack.
\begin{verbatim}
#include <stdio.h>

const float pi = 3.141592;

float *mistake;
void calc(float radius, float *area) {
    mistake = area;
    *area = pi*radius*radius;
}
void run() {
    float area;
    calc(1, &amp;area);
}
float useStack() { float ten = 10; return area; }
int main() {
    run();
    useStack();
        // mistake refers to a variable that has been
        // removed from the stack
        // 10 is printed in some compilers
    printf("%f\n", *mistake);
    return 0;
}
\end{verbatim}

\section{Should Context Objects be User-Defined?}

An alternative definition of Cyan could get rid of context objects. They could not be defined as shown in this text. Instead, one could use {\it reference types} like \verb"&Int" to declare a restricted prototype directly. So the programmer could define a prototype like
\begin{cyan}
object Sum extends Function<Nil, Int>

    func init { }
    func new: (Int &sum) -> Sum {
        var newSum = Sum new;
        newSum bind: sum;
        return newSum
    }
    public bind: (Int &sum) {
        self.sum = sum
    }
    Int &sum
    override
    func eval: (Int x) {
       sum = sum + x
    }
end
\end{cyan}
This new version of Cyan would have a concept called ``{\it restricted type}''\/ defined inductively as:
\begin{enumerate}[(a)]
\item a {\it reference type} is a {\it restricted type};
\item any prototype that declares a field of a restricted type is a {\it reference type}.
\end{enumerate}
All the restriction on the use and type checking defined nowadays for context objects would apply to {\it reference types}.

With this feature, the programmer herself would explicitly create her own context objects.

\section{More Examples}

The example of trees of page~\pageref{easy} can be made even more compact with context objects:
\begin{cyan}
open
object Tree
end

object BinTree(@property Tree left, @property Int value, @property Tree right) extends Tree
end

object No(Int value) extends Tree
end
...

var tree = BinTree( No(-1), 0, BinTree(No(1), 2, No(3)) );
Out println: ((tree left) value);
\end{cyan}
When the compiler finds a class like {\tt BinTree}, it creates a regular class with fields {\tt left}, {\tt value}, and {\tt right}:
\begin{cyan}
object BinTree extends Tree
    func init { }
    func new: (Tree left, Int value, Tree right) -> BinTree {
        var newObj = BinTree new;
        newObj bind: left, value, right;
        return newObj
    }
    func bind: (Tree left, Int value, Tree right) {
        self.left  = left;
        self.value = value;
        self.right = right;
    }
    @property Tree left
    @property Tree value
    @property Tree right
end
\end{cyan}

Suppose there is a sport {\tt Car} prototype that has two doors, left and right. The colors of these doors should always be the same as the main color of the car. One way of assuring that is declaring in the {\tt CarDoor} prototype a field that is a reference (a C-language pointer) to the field of the {\tt Car} that keeps the color. Since Cyan does not have C-like pointers, we can use context objects.
\begin{cyan}
object CarDoor(Int &color)
    func getColor -> { return color }
    func setColor: Int newColor { color = newColor }
    ...
end

object Car
    func init: Int aColor {
        _color = aColor;
        _leftDoor  = CarDoor(_color);
        _rightDoor = CarDoor(_color);
    }
    func color: (Int newColor) { _color = newColor }
    func color -> Int = _color;
    var Int _color
    @property CarDoor _leftDoor
    @property CarDoor _rightDoor
end

...
Car color: 255;
   // prints "color = 255"
Out println: "color = ", Car leftDoor getColor;

(Car rightDoor) setColor: 0;
   // prints "color = 0"
Out println: "color = ", Car color;
\end{cyan}

\verb@inject:into:@ methods in Smalltalk are used to accumulate a result over a loop. For example,\\
\nd \verb@    var sum = (1 to: 10) inject: 0 into: { (: Int total, Int elem :) total + elem }@\\
\nd accumulates the sum from {\tt 1} to {\tt 10}. Initially {\tt total} receives {\tt 0}, the argument to the keyword {\tt inject:}. Then the function is called passing {\tt total} and the current index (from {\tt 1} to {\tt 10}). In each step, the value returned from the function, {\tt total + elem}, is assigned to {\tt total} (Smalltalk returns the last block expression).

The basic types of Cyan support a Smalltalk-like {\tt inject} method and another form made to be used with context objects.
\begin{cyan}
object InjectInto<T>(T total) extends InjectObject<T>
    override
    func eval: (T elem) {
        total = total + elem
    }
    override
    func result -> T = total; end
\end{cyan}
Now the total is kept in the context object and we can write
\begin{cyan}
var inj = InjectInto<Int>(0);
1 to: 10 do: inj;
Out println: "Sum = ", inj result;
\end{cyan}
print the sum of the numbers from {\tt 1} to {\tt 10}.

\section{Future Enhancements}

This Section describes future enhancements to context objects.

\subsection{Type Checking Context Objects}

Context objects will be type-checked as functions will be. See Section~\ref{fefunctions}.

There are two kinds of context objects:
\begin{enumerate}[(a)]
\item the ones with at least one reference parameter such as {\tt Sum}. These are called {\it restricted} context objects, r-co for short;
\item the ones with no reference parameter.  These are called {\it unrestricted} context objects, u-co for short.
\end{enumerate}

There is no restriction on the use of unrestricted context objects (as expected!). They can be types of variables, fields, return values, and parameters. u-co are a generalization of u-functions.

Restricted context objects are a generalization of r-functions. Both suffer from the same problem: a context object could refer to a dead local variable:
\begin{cyan}
var Sum mySum;
var b = {
   var Int sum1 = 0;
   mySum = Sum(sum1);
};
b eval;
mySum eval: 1;
\end{cyan}
The message send ``{\tt b eval}''\/ makes {\tt mySum} refer to a context object that has a reference to {\tt sum1}. In the last message send, ``{\tt mySum eval: 1}''\/, there is an access to {\tt sum1}, which no longer exists.

Another error would be to return a r-co from a method:
\begin{cyan}
object Program
    func run {
        [ 1, 2, 3 ] foreach: makeError
    }
    func makeError -> Sum {
        var sum = 0;
        return Sum(sum);
    }
\end{cyan}
Here {\tt Sum(sum)} has a reference to a local variable {\tt sum}. When {\tt foreach:} calls method {\tt eval:} of the object {\tt Sum(sum)}, variable {\tt sum} is accessed causing a runtime error.

To prevent this kind of error, r-co have exactly the same set of restrictions as r-functions. In particular, the compiler would point an error in the assignment ``{\tt mySum = Sum(sum1)}''\/ of the example above.

A context object that does not inherit from anyone inherits from {\tt Any}, as usual.
Both r-co´s and u-co´s can inherit from any prototype and implement any interface. However, there are restrictions on assignments mixining restricted and unrestricted types. A r-co {\tt RCO} that inherits from an unrestricted prototype {\tt P} or implements an unrestricted interface I is not considered a subtype of {\tt P} or {\tt I}. That is,  if {\tt p} is a variable of type {\tt P} or {\tt I}, an assignment\\
\nd \verb@     p = RCO;@\\
\nd is illegal.

Apart from the rules for type checking, context objects are regular objects. For example, they may be abstract, have shared variables, and inherit from other prototypes. Inheritance demands some explanations. When a context object with a field or reference parameter {\tt x} is inherited by another context object, this last one should declare {\tt x} in its list of parameters with the same symbol preceding the parameter (none or {\tt \&}) as the superprototype. {\tt x} should precede the parameters defined only in the subprototype. After the keyword ``{\tt extends}''\/ there should appear the superprototype with its parameters.
\begin{cyan}
open
object A(Int &x)
   ...
end

object B(Int &x, Int y, String &z) extends A(x)
   ...
end
\end{cyan}
Since {\tt A} is a r-co, {\tt B} is a r-co too. A context object cannot be inherited by a regular prototype.

Note that context objects that use only copy parameters are regular prototypes. Therefore subprototypes need not to obey the rules given above. The subprototype does not even need to be a context prototype.

A context object can also be a generic object. {\tt Sum} can be generalized:
\begin{cyan}
object Sum<T>(T &sum) extends Function<T, Nil>
    override
    func eval: (T x) {
       sum = sum + x
    }
end

...
var intSum = 0;
var Float floatSum = 0;
var String abc = "";
[ 1, 2, 3 ] foreach: Sum<Int>(intSum);
[ 1.5, 2.5, 1 ] foreach: Sum<Float>(floatSum);
assert: (floatSum == 5);
assert: (intSum == 6);
\end{cyan}

\subsection{Adding Context Objects to Prototypes}

Section~\ref{contextfunctions} (Future Enhancements) explain how to use the {\tt addMethod: ...} grammar method of {\tt Any} to add methods to a prototype.
\begin{cyan}
    func (addMethod:
                 (keyword: String ( param: (Any)+ )?
                   )+
                  (returnType: Any)?
                  body: Any)
\end{cyan}
A context object can be used instead of a context function. One has just to extends the appropriate {\tt ContextObject} prototype.

\begin{cyan}
object Car
    func addDoorColor {
        leftDoor addMethod:
            keyword: #getColor
            returnType: Int
            body: GetColor(color);
        leftDoor addMethod:
            keyword: #setColor
            param: Int
            body: SetColor(color);
    }
    ...
    public Door leftDoor, rightDoor
    Int color
end

object GetColor(Int &color)
       implements ContextFunction<Door, Int>

    func bindToFunction: (Door newSelf) -> UFunction<Int> {
        return { ^color }
    }
end

object SetColor(Int &color)
       implements ContextFunction<Door, Int, Nil>

    func bindToFunction: (Door newSelf) -> UFunction<Int, Nil> {
        return { (: Int newColor :) color = newColor }
    }
end
\end{cyan}
After \\
\noindent \verb|    Car addDoorColor|\\
\noindent the left door will share a color with the car. Changes in one will reflect in the other.

\chapter{The Exception Handling System}   \label{ehs}

The exception handling system of Cyan has underwent big changes. Now it is a regular language statement instead of a message passing. Soon this manual will be updated.

Exception handling systems (EHS) \index{EHS} \index{Exception Handling System} allow the signalling and handling of errors or abnormal situations. There is a separation from the detection of the error and its treatment which can be in different methods or modules. The exception handling systems of almost all object-oriented languages are very similar. An exception is thrown \index{throw} by a statement such as ``{\tt throw e}''\/ or ``{\tt raise e}''\/ and caught by one or more \index{catch clause} catch clauses. We will show an example in Java. Assume there is a {\tt MyFile} class with methods for opening, reading and closing a file and that methods {\tt open} and {\tt readCharArray} of this class may throw exceptions {\tt ExceptionOpen} and {\tt ExceptionRead}.
\begin{cyannum}
char []charArray;
MyFile f = new MyFile("input.txt");
try {
   f.open();
   charArray = f.readCharArray();
   if ( charArray.length == 0 )
       throw new ExceptionZero();
}  catch ( ExceptionOpen e ) {
      System.out.println("Error opening file");
   }
   catch ( ExceptionRead e ) {
      System.out.println("Error reading file");
   }
   finally {
      f.close();
   }
\end{cyannum}
An exception is thrown by statement {\tt throw} (see line 7). We can also say that an error is signalled by a {\tt throw} statement. The class of the object following {\tt throw} should be a direct or indirect subclass of class {\tt Throwable}.
In this example, all statements that can throw exceptions are put in \index{try block} a {\tt try} block (which is between lines 4 and 7). The exceptions thrown inside the try block at runtime will be treated by the {\tt catch} clauses that follow the try block. There are two catch clauses and one {\tt finally} clause. Each catch clause accepts a parameter and treats the error associated to that parameter. Therefore\\
\nd \verb"     catch ( ExceptionOpen e ) { ... }"\\
\nd will treat the error associated to the operation of opening a file.

If file {\tt f} cannot be read, method {\tt readCharArray} throws  exception {\tt ExceptionRead} with a statement\\
\verb@     throw new ExceptionRead(filename);@\\
\nd After that, the runtime system starts a search for an appropriate handler \index{handler} for this exception. A handler is a piece of code, given in a {\tt catch} clause, that can treat the exception. This search starts  in method {\tt readCharArray} which does not have any catch clauses. It continues in the stack of called methods. Therefore an appropriate handler (or catch clause) is looked for in the code above. The runtime system checks whether the first catch clause can accept an object of {\tt ExceptionRead}, the one thrown by the {\tt throw} statement. It cannot. Then it checks whether the second catch clause can accept this object as parameter. Tt can. Then method {\tt readCharArray} is terminated and control is transferred to the catch clause
\begin{cyan}
   catch ( ExceptionRead e ) {
      System.out.println("Error reading file");
   }
\end{cyan}
Parameter {\tt e} receives the object ``\verb@new ExceptionRead(filename)@''\/ which was the parameter to statement {\tt throw} and the body of the clause is executed. After that the execution continues in the {\tt finally} clause, which is always executed --- it does not matter whether an exception is thrown or not in the try block.
When an exception is thrown, the stack of called methods is unwound till an appropriated catch clause is found and the control is transferred to this catch clause.

The exception handling system (EHS) \index{EHS} of Cyan is similar in several aspects of the model just described. However, it was based on the object-oriented exception handling system of \index{Green} Green \cite{journals/cj/Guimar04} \cite{Guimaraes:2006:GL:1646590.1646595} and it is object-oriented in nature. The throwing of an exception is a message send, exception treatment(catch clauses) can be put in prototypes and inherited, and polymorphism applies to exception treatment. All the arsenal of object-oriented programming can be used with exception signalling and treatment, which is not possible possible, to our knowledge, in other languages but Green. The exception handling system (EHS) of Cyan goes well beyond that of Green which is awkward to use if local variables should be accessed to treat the error. In Cyan the EHS is both easy to use and powerful. However, it is not a checked exception system like that of Java or Green. An exception may be thrown and not caught as in C++ or C\#.

The Java example in Cyan would be
\begin{cyannum}
var Array<Char> charArray;
var f = MyFile new: "input.txt";
try
   f open;
   charArray = f readCharArray;
   if charArray size == 0  {
       throw ExceptionZero
   }
catch { (: ExceptionOpen e :) Out println: "Error opening file" }
catch { (: ExceptionRead e :) Out println: "Error reading file" }
finally {
      f close
}
\end{cyannum}
An exception is thrown by statement  {\tt throw} as shown in line 7:
\begin{cyan}
       throw ExceptionZero
\end{cyan}
{\tt ExceptionZero} is a prototype that inherits from {\tt CyException}, \index{CyException} \index{object!CyException} the superprototype of all exception objects. Since this exception does not demand any useful additional information, the prototype does not have any fields:
\begin{cyan}
object ExceptionZero extends CyException
end
\end{cyan}
Every exception prototype should inherit from {\tt CyException}, which  inherits from {\tt Any} and  does not define any methods.

In the above Cyan example, the {\tt try-catch-finally} statement catches the exceptions thrown during the execution of the statements between {\tt try} and the first {\tt catch} {\it clause} (or {\tt finally}, if there is no {\tt catch}).
That is almost the same as in the Java code. When an exception is thrown in the function body, as {\tt ExceptionRead}, the runtime system searches for an adequate handler in the expressions of the {\tt catch} clauses. First it checks whether method {\tt eval:} of the first function,\\
\nd \verb@     { (: ExceptionOpen e :) Out println: "Error opening file" }@\\
\nd can accept an object of {\tt ExceptionRead} as real argument. It cannot. Then the search continues in the second {\tt catch} clause. Since \\
\nd \verb@     { (: ExceptionRead e :) Out println: "Error reading file" }@\\
\nd can accept an {\tt ExceptionRead} object, message {\tt eval} is sent to this function with the thrown exception as argument. After that, the {\tt finally} statements are executed and the execution continues in the statement after {\tt try-catch-finally}.

This works exactly the same as the exception system of Java/C++ and many other object-oriented languages. In Cyan, there may be one or more {\tt catch} clauses and an optional {\tt finally} clause. Every {\tt catch} accepts as argument an object that has at least one method\\
\verb"    eval: (E e)"\\
\nd in which {\tt E} is a prototype that inherits from {\tt CyException} (directly or indirectly). Functions
\begin{cyan}
    { (: ExceptionOpen e :) Out println: "Error opening file" }
    { (: ExceptionRead e :) Out println: "Error reading file" }
\end{cyan}
satisfy these requirements. For example, the first function has a method\\
\nd \verb@     eval: (ExceptionOpen e) { Out println: "Error opening file" }@\\

It is not necessary that the expression following a {\tt catch} be a function or be a subprototype of any function.

\section{Using Regular Objects to Treat Exceptions}

Each {\tt catch:} keyword may receive as argument an object that has more than one {\tt eval:} method. \index{catch object} \index{object!catch}
\begin{cyan}
object ExceptionCatchFile
    overload
    func eval: (ExceptionOpen e)   { Out println "Error opening file" }
    func eval: (ExceptionRead e)   { Out println "Error reading file" }
    func eval: (ExceptionWrite e)  { Out println "Error writing to file" }
end
\end{cyan}
Prototype {\tt ExceptionCatchFile} treats all errors associated to opening, reading, and writing to files (but not to closing a file). This kind of object, to treat exceptions, will be \index{catch objects} called {\it catch objects}. It can be used as in the next example.
\begin{cyan}
var Array<Char> charArray;
var f = MyFile new: "input.txt";
try
   f open;
   charArray = f readCharArray;
   if charArray size == 0 {
       throw ExceptionZero
   }
catch ExceptionCatchFile
finally {
   f close
}
\end{cyan}
When an exception is thrown, the runtime system starts a search for an {\tt eval:} method (a handler) in the nearest {\tt catch} clause, which is {\tt ExceptionCatchFile}. Supposing that there was a read error, the correct {\tt eval:} method should accept a {\tt ExceptionRead} object as argument. The runtime system searches for the {\tt eval:} method in {\tt ExceptionCatchFile} using the same algorithm used for searching for a method after a message is send to an object. That is, the runtime system tries to send message {\tt eval:} with a {\tt ExceptionRead} as argument to object {\tt ExceptionCatchFile}. By the regular algorithm, the second textually declared method of {\tt ExceptionCatchFile}, \\
\nd \verb@         func eval: (ExceptionRead e)   { Out println "Error reading file" }@\\
\nd is found and called. After that,  the statements of the {\tt finally} clause are executed and computation continues after the {\tt try-catch-finally} statement.

\section{Selecting an eval Method for Exception Treatment}

A Cyan program starts its execution in a method called {\tt run} of a prototype designed at compile-time. For this example, suppose this prototype is {\tt Program}. To start the execution, method {\tt run} is called inside a function that receives a {\tt catch:} message:
\begin{cyan}
{
   Program run: args
} catch: RuntimeCatch;
\end{cyan}
Method {\tt eval:} of prototype {\tt RuntimeCatch} just prints the stack of called methods:
\begin{cyan}
object RuntimeCatch
    func eval: (CyException e) {
        /* prints the stack of called methods and ends the program
        */
    }
    ...
end
\end{cyan}
Maybe we may will add a {\tt finally:} keyword to the {\tt catch:} message allowing some code to be executed before the program ends.

Let us now explain what happens conceptually when an exception is thrown and caught. The implementation need not to be as described next.

When a message with at least one {\tt catch:} keyword is sent to a function, a grammar method is called. We will call this grammar method {\tt catch-finally} (this is just a name for explaining this text).
Method {\tt catch-finally} pushes the parameters to {\tt catch:} in a stack {\tt CatchStack} in the reverse order in which they appear in the call. So
\begin{cyan}
{
   ...
} catch: c1
  catch: c2
  catch: c3;
\end{cyan}
pushes {\tt c3}, {\tt c2}, and {\tt c1} into the stack, in this order. Therefore {\tt c1} is in the top. When an exception is thrown by the message send {\tt throw: obj}, method {\tt throw:} of {\tt Any} searches the stack {\tt CatchStack} from top to bottom until it finds an {\tt eval:} method that accepts {\tt obj} as parameter. Inside each stack object the search is made from the first declared {\tt eval:} method (in textual order) to the last one. {\tt CatchStack} is a prototype that just implements a stack.

Consider the catch objects\footnote{Objects with {\tt eval:} methods that treat exceptions.} and the example that follow. The prototypes are show as if they were in a single file.
\begin{cyan}
   // number < 0, == 0, > 1000, or even
open
object ExceptionNum extends CyException
end

   // when the number is == 0
object ExceptionZero extends ExceptionNum
end

   // when the number is < 0
object ExceptionNeg extends ExceptionNum
end

   // when the number is > 1000
object ExceptionBig extends ExceptionNum
end

   // when the number is even
object ExceptionEven extends ExceptionNum
end

   // when the number is 5
object ExceptionFive extends ExceptionNum
end


object CatchZeroBig
    overload
    func eval: (ExceptionZero e) {
        Out println: "zero number";
    }
    func eval: (ExceptionBig e) {
        Out println: "big number";
    }
end

object CatchNeg
    func eval: (ExceptionNeg e) {
        Out println: "negative number";
    }
end

object CatchEven
    func eval: (ExceptionEven e) {
        Out println: "even number";
    }
end

object CatchNum
    func eval: (ExceptionNum e) {
        Out println: "number < 0, == 0, > 1000, or even";
    }
end

object Program
    let Int MaxN = 1000;

    func run: Array<String> args {
           // 1
        var n = In readInt;
        {  // 2
           process: n
        } catch: CatchZeroBig
          catch: CatchEven
          catch: CatchNum;
        // 5
        Out println: "this is the end"
    }
    private func process: (Int n) {
        {  // 3
           check: n;
           if n > MaxN {
               throw: ExceptionBig
           }
        } catch: CatchNeg
        // 6
    }
    private func check: (Int n) {
        // 4
        if n == 0 {
            throw: ExceptionZero
        }
        if n < 0 {
            throw: ExceptionNeg
        }
        if n%2 == 0 {
            throw: ExceptionEven
        }
    }
end
\end{cyan}
There are four exceptions, {\tt ExceptionZero}, {\tt ExceptionNeg}, {\tt ExceptionBig}, and {\tt ExceptionEven} that inherit from {\tt ExceptionNum} and four catch objects, {\tt CatchZeroBig}, {\tt CatchEven}, {\tt CatchNeg}, and {\tt CatchNum}. The program execution starts at point ``\verb"// 1"''\/. At line \verb"// 2", message {\tt catch:catch:catch:} has been send and the function that has just ``{\tt process: n}''\/ has been called. At point \verb"// 2", {\tt CatchStack} has objects {\tt CatchNum}, {\tt CatchEven}, and {\tt CatchZeroBig} (last on top).

Inside the function that starts at \verb"// 2", if message ``\verb@throw: exc@''\/ is sent to {\tt self}, the search for a method would start at {\tt CatchZeroBig} and proceeds towards {\tt CatchNum} at the bottom of the stack. First method {\tt throw:} would check whether object {\tt exc} is subprototype of {\tt ExceptionZero}. If it is not, it would test whether object {\tt exc} is a subprototype of {\tt ExceptionBig}. If it is not, the search would continue in {\tt CatchEven}.

At line marked as \verb"// 3", object {\tt CatchNeg} has already been pushed into the stack {\tt CatchStack}. At point \verb"// 4" in the code, if statement\\
\nd \verb"    throw: ExceptionEven"\\
\nd is executed, there is a search for an {\tt eval:} method that can accept {\tt ExceptionEven} as parameter, starting at the {\tt CatchNeg} object. This method is found in object {\tt CatchEven} pushed in the {\tt run:} method. Therefore control is transfered to the first statement after the message send
\begin{cyan}
        {  // 2
           process: n
        } catch: CatchZeroBig
          catch: CatchEven
          catch: CatchNum;
\end{cyan}
which is ``\verb@Out println: "this is the end"@''\/.  This is exactly like the exception handling system of almost all object-oriented languages.

Before returning, the {\tt throw:} method of {\tt Any} removes the  objects pushed into {\tt CatchStack} together and after {\tt CatchEven}.

Every function of type \verb|Function<Nil>| has a method
\begin{cyan}
    @checkCatchParameter
    func ((catch: Any)+  finally: Function<Nil>) t {
        ...
    }
\end{cyan}
responsible for catching exceptions. The metaobject {\tt checkCatchParameter}, whose annotation is attached to this
method, checks whether each parameter to a {\tt catch:} keyword has at least one
{\tt eval:} method,  each of them accepting
one parameter whose type is subprototype of {\tt CyException}.

\section{Other Methods and Keywords for Exception Treatment}

Functions of type \verb|Function<Nil>|
have a method {\tt hideException} that just eats every exception thrown in them: \index{hideException} \index{method!hideException}
\begin{cyan}
n = 0;
{
   n = (In readLine) asInt
} hideException;
\end{cyan}
Of course, this method should be rarely used.

Keywords {\tt retry} or {\tt retry:} \index{retry} \index{method!retry} may be used after all {\tt catch:} keywords in order to call the function again if an exception was caught by any object that is argument to any of the {\tt catch:} keywords. If keyword {\tt retry:} is used, it should have a function as parameter that is called before the main function is called again.
\begin{cyan}
   // radius of a circle
Float radius;
{
  radius = In readFloat;
  if radius < 0 {
      throw: ExceptionRadius(radius)
  }
} catch: CatchAll
  retry: {
      Out println: "Negative radius. Type it again"
  };
\end{cyan}
{\tt CatchAll} has a method\\
\nd \verb@    func eval: (CyException e) { }@\\
\nd that catches all exceptions. This prototype is automatically included in every file. It belongs to package cyan.lang.

One can just write  {\tt retry:} without any {\tt catch:} keywords. If any exception is thrown in the function, the {\tt eval} method of the argument to {\tt retry:} is called and the function is called again.
\begin{cyan}
   // radius of a circle
var Float radius;
{
  radius = In readFloat;
  if radius < 0 {
      throw: ExceptionRadius(radius)
  }
  else if radius == 0 {
      // end of input
      return 0
  }
} retry: {
      Out println: "Negative radius. Type it again"
  };
\end{cyan}

Keyword {\tt tryWhileTrue:} \index{tryWhileTrue} \index{method!tryWhileTrue} may be put after the {\tt catch:} keywords in order to control how many times the function is retrieved. The argument to {\tt tryWhileTrue:} should be a \verb@Function<Boolean>@ function. If an exception was thrown in the function and the argument to {\tt tryWhileTrue:} evaluates to {\tt true}, the function is called again.
\begin{cyan}
numTries = 0;
{
      // may throw an exception ExceptionConnectFail
   channel connect;
   ++numTries;
} catch: CatchAll
  tryWhileTrue: {^ numTries < 5 };
\end{cyan}
The above code tries to connect to a channel five times. Each time the connection fails an exception is thrown by method {\tt connect}. Each time the function after {\tt tryWhileTrue:} is evaluated. In the first five times it returns {\tt true} and the main function is called again. If no exception is thrown by {\tt connect}, the argument to {\tt tryWhileTrue:} is not called. Again, the {\tt catch:} keywords are optional.  Keyword {\tt tryWhileFalse:} \index{tryWhileFalse} \index{method!tryWhileFalse} is similar to {\tt tryWhileTrue}.

Prototype {\tt CatchIgnore} could be used instead of {\tt CatchAll}. This generic prototype ignores the exceptions that are parameters to it. Any number of exceptions can be used. An instantiation of this prototype with parameter
 \index{CatchIgnore} \index{object!CatchIgnore} {\tt ExceptionConnectFail} would be something like
\begin{cyan}
object CatchIgnore<ExceptionConnectFail>
    func eval: ExceptionConnectFail e { }
end

...
numTries = 0;
{
      // may throw an exception ExceptionConnectFail
   channel connect;
   ++numTries;
} catch: CatchIgnore<ExceptionConnectFail>
  tryWhileTrue: {^ numTries < 5 };
\end{cyan}

This example can be made more compact with the use of a context object to count the number of attempts:
\begin{cyan}
object Times(Int numTries) extends Function<Boolean>
    func eval -> Boolean {
        --numTries;
        return numTries > 0;
    }
end

...
{
      // may throw an exception ExceptionConnectFail
   channel connect;
} tryWhileTrue: Times(5);
\end{cyan}

Using union types, we can catch several exceptions with a single function:
\begin{cyan}
{
   ...
} catch: { (: ExceptionEmptyLine | ExceptionLineTooBig e :)
           Out println: "Limit error in line " ++ line
  }
  catch: { (: ExceptionWhiteSpace | ExceptionRead e :)
           Out println: "Other error happened"
  };
\end{cyan}

A future improvement to the EHS of Cyan would be to make it support features of the EHS of Common Lisp \index{Lisp} \index{Common Lisp} (conditions and restarts). That would be made by allowing communication between the error signaling and the error handling. This could be made using a variable ``{\tt exception}''\/.
A catch object could have other meaningful methods besides ``{\tt eval: T}''\/. For example, a catch object could have an ``{\tt getInfo}''\/ method describing the error recovery to be chosen afterwards:
\begin{cyan}
object CatchStrategy
    func getInfo -> CySymbol = #retry;
end

object Test
    func test {
        {
            connectToServer;
            buildSomething
        } catch: CatchStragegy
    }
    func connectToServer {
        {
           var Boolean fail = true;
           ...
              // if connection to server failed, signal
              // an exception
           if fail {
               throw: ExceptionConnection
           }
        } catch: { (: ExceptionConnection e :)
               // if connection to server failed,
               // consult getInfo for advice.
            if exception getInfo == #retry {
                connectToServer
            }
        }
    }
    ...
end
\end{cyan}
Maybe there should be another method that obeys automatically instructions given by objects like {\tt CatchStrategy}. Maybe {\tt catch} itself should automatically retry when ``{\tt exception getInfo}''\/ demands it:
\begin{cyan}
    func connectToServer {
        {
           var Boolean fail = true;
           ...
              // if connection to server failed, signal
              // an exception
           if fail {
               exception eval: ExceptionConnection
           }
        } catch: CatchIgnore<ExceptionConnection>
    }
\end{cyan}

\section{Why Cyan Does Not Support Checked Exceptions?}

\index{exception!checked} \index{exception!unchecked}
Cyan does not support checked exceptions as Java in which the exceptions a method may throw are described in its declaration:
\begin{verbatim}
       // this is how method "check" of Program
       // would be declared in Java
    private void check(int n)
             throws ExceptionZero, ExceptionNeg,
                    ExceptionEven  {
        // 4
        if ( n == 0 )
            throw new ExceptionZero();
        if ( n < 0 )
            throw new ExceptionNeg();
        if ( n%2 == 0 )
            throw new ExceptionEven();
    }
\end{verbatim}
Here method {\tt check} may throw exceptions {\tt ExceptionZero}, {\tt ExceptionNeg}, and {\tt ExceptionEven}. We could add a syntax for that in Cyan following language Green \cite{journals/cj/Guimar04}:
\begin{cyan}
    private func check: (Int n)
                EvalZeroNegEven exception {
        // 4
        if  n == 0 {
            exception eval: ExceptionZero
        }
        if n < 0 {
            exception eval: ExceptionNeg
        }
        if n%2 == 0 {
            exception  eval: ExceptionEven
        }
    }
\end{cyan}
Pseudo-variable {\tt exception} would be declared after all regular method parameters. Inside the method this variable is type-checked as a regular variable. Then there would be an error if there was a statement\\
\nd \verb"    exception eval: ExceptionRead"\\
\nd in method {\tt check} because there is no {\tt eval:} method in {\tt EvalZeroNegEven} that can accept a {\tt ExceptionRead} object as parameter.
Interface {\tt EvalZeroNegEven}\footnote{Note that currently Cyan does not support the declaration of overloaded methods in interfaces.} is
\begin{cyan}
interface EvalZeroNegEven
    func eval: ExceptionZero
    func eval: ExceptionNeg
    func eval: ExceptionEven
end
\end{cyan}
Green employs a mechanism like this, which works perfectly in a language without functions.

But think of method {\tt ifTrue:} of functions of types \verb"Function<Boolean, Nil>":
\begin{cyan}
    func ifTrue: (Function<Nil> b)
               T exception {
        if self == true {
            b eval
        }
    }
\end{cyan}
What is the type {\tt T} of {\tt exception}? In
\begin{cyan}
(i < 0) ifTrue: {
    throw: ExceptionRead;
}
\end{cyan}
{\tt T} should be
\begin{cyan}
interface InterfaceExceptionRead
    func eval: ExceptionRead
    // possibly more methods
end
\end{cyan}
But in another call of this method {\tt T} should be different:
\begin{cyan}
(i <= 0) ifTrue: {
    if openError {
        throw: ExceptionOpen
    }
    else if i == 0 {
        throw: ExceptionZero
    }
}
\end{cyan}
In this case {\tt T} should be
\begin{cyan}
interface InterfaceOpenExceptionZero
    func eval: ExceptionOpen
    func eval: ExceptionZero
    // possibly other methods
end
\end{cyan}
Then the type of {\tt T} depends on the exceptions the function may throw. We have a solution for that but it is too complex to be added to a already big language. Without explaining too much, method {\tt ifTrue:} would be declared as
\begin{cyan}
    func ifTrue: (Function<Nil> b)
               (b getMethod: "eval") .exception exception {
        if self == true {
            b eval
        }
    }
\end{cyan}
The declaration means that the type of {\tt exception} in {\tt ifTrue:} is the type of variable {\tt exception} of the method {\tt eval} of function {\tt b} at the call site. If {\tt ifTrue:} could throw exceptions by itself, these could be added to the type ``\verb@(b getMethod: "eval") .exception@''\/ using the type concatenator operator ``{\tt ++}''\/ (introduced just for this use here).

For short, we could have checked exceptions in Cyan but it seems they are not worthwhile the trouble.

\section{Synergy between the EHS and Generic Prototypes} \label{synergy}

Package {\tt cyan.lang} defines two generic prototypes that accept any number of prototype parameters: {\tt CatchExit} and {\tt CatchWarning}. The first one is used to catch exceptions and end the program. The later just issues a warning message. If they had just one parameter, they would be as shown below.
\begin{cyan}
object CatchExit<T>
    func eval: (T e) {
       Out println: "Fatal error: exception " ++ T prototypeName ++
          " was thrown";
       System exit
    }
end

object CatchWarning<T>
    func eval: (T e) {
       Out println: "Exception " ++ T prototypeName ++ " was thrown"
    }
end
\end{cyan}
These prototypes can be used to exit the program for some exceptions or just issue a message for others.
\begin{cyan}
...

{
  line = In readLine;
  if line size == 0 {
      throw: ExceptionEmptyLine
  } else if line size > MaxLine {
      throw: ExceptionLineTooBig(line)
  }
  Out println "line = " ++ line
} catch: CatchExit<ExceptionLineTooBig>
  catch: CatchWarning<ExceptionEmptyLine>;
\end{cyan}
Object \verb@CatchExit<ExceptionLineTooBig>@ treats exception {\tt ExceptionLineTooBig} because it has an {\tt eval:} method that accepts this exception as parameter. This method prints an error message and ends the program execution.

Object \verb@CatchWarning<ExceptionEmptyLine>@ treats exception {\tt ExceptionEmptyLine}. Method {\tt eval} of this object just prints a warning message.

Generic object \index{CatchIgnore} \index{object!CatchIgnore} {\tt CatchIgnore} accepts any number of parameters. The {\tt eval:} methods of this object do nothing. The definition of {\tt CatchIgnore} with two parameters {\it would be}
\begin{cyan}
object CatchIgnore<T1, T2>
    func eval: T1 e1 { }
    func eval: T2 e2 { }
end
\end{cyan}
If we want to ignore two exceptions and treat a third one, we can write something like
\begin{cyan}
{
  line = In readLine;
  if line size == 0 {
      throw: ExceptionEmptyLine
  } else if line size > MaxLine {
      throw: ExceptionLineTooBig(line)
  } else if line[0] == ' ' {
      throw: ExceptionWhiteSpace
  }
  Out println "line = " ++ line
} catch: CatchIgnore<ExceptionLineTooBig, ExceptionEmptyLine>
  catch: { (: ExceptionWhiteSpace e :)
      Out println: "line cannot start with white space";
      System exit
      };
\end{cyan}

With generic prototypes, it is easy to implement the common pattern of encapsulating some exceptions in others. That is what prototype {\tt ExceptionConverter} does. This prototype is defined in package {\tt cyan.lang} and accepts any number of even parameters. With two parameters it would be equivalent to:
\begin{cyan}
object ExceptionConverter<Source, Target>
    func eval: (Source e) {
        throw: Target()
    }
end
\end{cyan}
In the example that follows, when an exception {\tt ExceptionNegNum} is thrown, a {\tt catch:} method captures it and throws a new exception from prototype {\tt ExceptionOutOfLimits}.

\begin{cyan}
...
{
  if i < 0 { throw: ExceptionNegNum };
  s = v[i];
  s println;
} catch: ExceptionConverter<ExceptionNegNum, ExceptionOutOfLimits>;
\end{cyan}

Another common pattern of exception treatment is to encapsulate exceptions in an exception container. This is what does generic prototype {\tt ExceptionEncapsulator}. It takes any number of parameters (but at least two). The last one should be the container.
 {\tt ExceptionEncapsulator} with two parameters would be:
\begin{cyan}
object ExceptionEncapsulator<Item, Container>
    func eval: (Item e) {
         throw: Container(e)
    }
end
\end{cyan}
This prototype could be used as in this example.
\begin{cyan}

...
{
  if i < 0 { throw: ExceptionNegNum(i) };
  s = v[i];
  s println;
} catch: ExceptionEncapsulator<ExceptionNegNum, ExceptionArithmetic>;
\end{cyan}
Whenever {\tt ExceptionNegNum} is thrown in the function, it is packed into an exception of {\tt ExceptionArithmetic} and thrown again.

\section{More Examples of Exception Handling}

One can design a {\tt MyFile} prototype in which the error treatment would be passed as parameter:
\begin{cyan}
object MyFile
    func new: (String filename) { ... }
    func catch: (ExceptionCatchFile catchObject) do: (Function<String, Nil> b) {
        {
            open;
                // readAsString read the whole file and put it in a String,
                // which is returned
            b eval: readAsString;
            close;
        } catch: catchObject
    }
end
\end{cyan}

Context object {\tt Throw} of package {\tt cyan.lang} has an {\tt init:} method that throws the exception that is its parameter.
\begin{cyan}
object Throw
    func init: CyException e {
        throw: e
    }
end
\end{cyan}
It makes it easy to throw some exceptions:
\begin{cyan}
{
  line = In readLine;

  if line size == 0           { Throw(ExceptionEmptyLine) }
  else if line size > MaxLine { Throw(ExceptionLineTooBig(line)) }
  else if line[0] == ' '      { Throw(ExceptionWhiteSpace) }

  Out println "line = " ++ line
} catch: CatchIgnore<ExceptionLineTooBig, ExceptionEmptyLine>
  catch: { (: ExceptionWhiteSpace e :)
      Out println: "line cannot start with white space";
      System exit
      };
\end{cyan}

Prototype {\tt CatchWithMessage} catchs all exceptions. It prints a message specific to the exception thrown and prints the stack of called methods:
\begin{cyan}
object CatchWithMessage
    func eval: (CyException e) {
        Out println: "Exception ", e prototypeName, " was thrown";
        System printMethodStack;
        System exit
    }
end
\end{cyan}

An exception prototype may define an {\tt eval:} method in such a way that it may be used as a catch parameter:
\begin{cyan}
object ExceptionZero extends CyException
    func eval: (ExceptionZero e) {
        Out println: "Zero exception was thrown";
        System exit
    }
end
...

    // inside some method
    {
       n = In readInt;
       if n == 0 { throw: ExceptionZero }
       ...
    } catch: ExceptionZero;
\end{cyan}
This is confusing. But somehow it makes sense: the exception, which represents an error, provides its own treatment (which is just a message). Guimarães \cite{Jose:GreenSite:Online} suggests that a library that may throw exceptions should also supply catch objects to handle these exceptions. It could even supply an hierarchy of exceptions for each set of related exceptions. For example, if the library has a prototype for file handling, it should also has a catch prototype with a default behavior for the exceptions that may be thrown. And subprototypes with alternative treatments and messages.

Since exceptions and theirs treatment are objects, they can be put in a hash table used for choosing the right treatment when an exception is thrown.
\begin{cyan}
object CatchTable
    func init {
        table = [
            Any -> Any, // just to set the type of the table
            ExceptionZero -> CatchWarning<ExceptionZero>,
            ExceptionNeg  -> CatchAll,
            ExceptionBig -> { (: ExceptionBig e :)
                Out println: "Number " ++ e number ++ " is too big"
                },
            ExceptionNum -> CatchNum

            ];
    }
    func eval: (CyException e) {
        type table[e prototype]
            case Any any {
                any ?eval: e
            }
            case Nil nil {
                throw: ExceptionStr("Exception " ++
                   (e prototypeName) ++ " is not supported " ++
                   "by table")
            }
    }
    IMap<Any, Any> table
end
\end{cyan}
{\tt CatchTable} can be used as the catch object:
\begin{cyan}
    // inside some method
    {
       ...
    } catch: CatchTable;
\end{cyan}
If an exception is thrown in the code ``{\tt ...}''\/, method {\tt eval:} of {\tt CatchTable} is called (its parameter has type {\tt CyException}, the most generic one). In this method, the hash table referenced by variable ``{\tt table}''\/ is accessed using as key ``{\tt e prototype}''\/, the prototype of the exception. As an example, if the exception is an object of {\tt ExceptionTriangle}, ``{\tt e prototype}''\/ will return {\tt ExceptionTriangle}. By indexing {\tt table} with this value we get {\tt CatchTriangle}. That is, \\
\verb@     assert: table[e prototype] == CatchTriangle@\\
\nd in this case. Here {\tt table[elem]} returns the value associated to {\tt elem} in the table.

Message \verb"?eval: e" is then sent to  object {\tt CatchTriangle}. That is, method {\tt eval:} of {\tt CatchTriangle} is called. The result is the same as if {\tt CatchTriangle} were put in a {\tt catch:} keyword as in the example that follows.
\begin{cyan}
object ExceptionTriangle(public Double a, public Double b, public Double c)
end

object CatchTriangle
    func eval: (ExceptionTriangle e) {
            // "e a" is the sending of message "a" to object "e"
            // that returns the side "a" of the triangle
        Out println: "There cannot exist a triangle with sides ", e a, ", ", e b, ", and ", e c
    }
end


    // inside some method
    {
       ...
       if a >= b + c || b >= a + c || c >= a + c {
           throw: ExceptionTriangle(a, b, c)
       }
       ...
    } catch: CatchTriangle;
\end{cyan}
Then we can replace {\tt catch: CatchTriangle} in this code by ``{\tt catch: CatchTable}''\/. However, if an exception that is not in the table is thrown, exception {\tt ExceptionTable} is thrown. Assume that {\tt Nil} is returned by indexing the hash table when the key is not found. That is, ``\verb@table[e prototype]@''\/ returns {\tt Nil} if the prototype is not found in the table.

Exception {\tt ExceptionStr} of package {\tt cyan.lang} is used as a generic exception which holds a string message. \label{ExceptionStr}
\begin{cyan}
package cyan.lang

object ExceptionStr(String _message) extends CyException

    public func eval: ExceptionStr e {
        Out println: (e message);
    }

    func message -> String = _message;

end
\end{cyan}
It can be used as
\begin{cyan}
{
   var s = In readLine;
   if s size < 2 {
       throw: ExceptionStr("size should be >= 2")
   } else if s size >= 10 {
       throw: ExceptionStr("size should be < 10")
   }
} catch: ExceptionStr;
\end{cyan}

\chapter{The Cyan Language Grammar}
\label{grammar}

This Chapter describes the language grammar. The reserved words  \index{grammar, Cyan}
and symbols of the language are shown between `` and  ''.
Anything between
\begin{itemize}
\item \{ and \} can  be repeated zero or more times;
\item \{ and \}$+$ can be repeated one or more times;
\item \verb"[" and \verb"]"  is optional.
\end{itemize}
The program
must be analyzed by unfolding the rule
``CompilationUnit''. ScriptCyan programs are produced by rule ``ScriptCompilationUnit''\/.

There are two kinds of comments:
\begin{itemize}
 \item anything between \verb@/*@ and \verb@*/@. Nested comments
are allowed.
\item anything after \verb@//@ till the end of the line.
\end{itemize}
Of course, comments are not shown in the grammar.

The rule CharConst is any  character between a single
quote \verb"'". Escape characters are allowed.
 The rule Str is a string of zero or more characters
surrounded by
double quotes \verb!"!. The double quote itself can be put in a string
preceded by the backslash character \verb@\@. Rule AtStr is {\tt @"} followed by a string ended by double quotes. The backslash character cannot be used to introduce escape characters in this kind of string.

A literal number starts with a number which can be followed by numbers and underscore (\verb@_@). There may be a trailing letter defining its type:\\
\verb@     35b   // Byte number@\\
\verb@      2i   // integer number@\\
There should be no space between the last digit and the letter. User-defined literal numbers start with a digit and may contain digits, letters, and underscore:\\
\nd \verb"    100Reais   2_3_5_7_prime_0_2_4_even"\\

All words that appear between quotes in the grammar are reserved Cyan
keywords. Besides these words, there are other keywords cited in Section~\ref{keywords} that are not currently used by the language.

{\tt Id} is an identifier composed by a sequence of letters, digits, and underscore, beginning with a letter or underscore. But a single underscore is not a valid identifier. IdColon is an Id followed by a ``{\tt :}''\/, without space between them, such as ``{\tt ifTrue:}''\/ and ``{\tt ifFalse:}''\/. InterIdColon is an Id  followed by a ``:''\/ and preceded by  ``?''\/ as in ``{\tt ?at:}''\/  (dynamic unchecked message send). InterId is an Id preceded by ``?''\/ such as ``{\tt ?name}''\/.
TEXT is a terminal composed by any number of characters. Symbol {\tt `} is terminal BACKQUOTE, ASCII 96.
InterDotIdColon is an Id followed by a ``:''\/ and preceded by  ``?.''\/ as in ``{\tt ?.at:}''\/. ({\tt nil}-safe message send).  InterDotId is an Id preceded by  ``?.''\/ as in ``{\tt ?.name}''\/.

LeftCharString is any sequence of the symbols \\
\begin{verbatim}
= !  $ % & * - + ^ ~ ? / : . \  | (  [  {  <
\end{verbatim}
Note that \verb">", \verb")", \verb"]", and \verb"}" are missing from this list.
RightCharString is any sequence of the same symbols of LeftCharString but with
\verb">", \verb")", \verb"]", and \verb"}" replacing
\verb"<", \verb"(", \verb"[", and \verb"{", respectively.
The compiler will check if the closing RightCharString of a LeftCharString is the inverse of it.
That is, if LeftCharString is \\
\nd \verb"    (*=<["\\
\nd then its corresponding RightCharString should be \\
\nd \verb"    [>=*)"\\

SymbolLiteral is a literal symbol (see page~\pageref{symboldef} for definition).
There are limitations in the sequences of symbols that are considered valid for literal objects.
They cannot start with {\tt ((}, {\tt ))}, {\tt ([}, {\tt ])}, {\tt [[}, {\tt ]]},  {\tt (:},  {\tt \{(:}, \verb@>(@, \verb@{^@, \verb@:[@, \verb@:(@, \verb@{.@, and {\tt ::}. For short, they cannot start with any sequence of symbols which can appear in a valid Cyan program. For example, {\tt [(:} is illegal because we can have a function declared as\\
\nd \verb@    { (: Int n :) ^ n*n }@\\

\vspace{4ex}

\p{CompilationUnit} ::=  PackageDec ImportDec   \{ AnnotList ProgramUnit \}

\p{ScriptCompilationUnit} ::= [ ImportDec ] ( StatementList \verb@|@ \{ SlotDec \} )

\p{PackageDec} :: ``package''\/ QualifId  [ ``;''\/ ]

\p{ImportDec} ::= \{ ``import''\/ IdList [ ``;''\/ ] \}

\p{ProgramUnit} ::=  [ QualifProgUnit ] ( ObjectDec \verb@|@ InterfaceDec )

\p{QualifProtec} ::=  ``private''\/  \verb@|@ ``public''\/  \verb@|@ ``protected''\/

\p{QualifProgUnit} ::=  ``public''\/  \verb@|@ ``package''\/

\p{AnnotList} ::= \{ Annotation \}

\p{Annotation} ::= ``@''\/ Id\\
\rr [ ``(''\/ ExprLiteral [ ``,''\/ ExprLiteral ] ``)''\/ ] \\
\rr [ LeftCharString TEXT RightCharString ]


\p{ObjectDec} ::=  [  ``abstract''\/ \verb@|@ ``final''\/ ] \\
\rr ``object''\/ Id \{ TemplateDec \}\\
\rr [ ContextDec ] \\
\rr [ ``extends''\/ Type [ ``(''\/ IdList ``)''\/ ] ]\\
\rr [ ``implements''\/ TypeList ]\\
\rr \{ SlotDec \}\\
\rr ``end''\/


\p{TemplateDec} ::= ``\verb@<@''\/ TemplateVarDecList ``\verb@>@''\/

\p{TemplateVarDecList} ::= TemplateVarDec \{ ``,''\/ TemplateVarDec \}

\p{TemplateVarDec} ::= Id [ ``$+$''\/ ]

\p{ContextDec} ::=  ``(''\/ CtxtObjParamDec  \{ ``,''\/ CtxtObjParamDec \}``)''\/

\p{CtxtObjParamDec} ::=  AnnotList [ ``public''\/ \verb@|@ ``protected''\/ \verb@|@ ``private''\/ ] Type  \\
\rr [ ``\&''\/  ] Id

\p{Type} ::= SingleType \{ ``\verb@|@''\/  SingleType \}

\p{SingleType} ::= QualifId  \{ ``\verb@<@''\/ TypeList ``\verb@>@''\/ \} \verb@|@ BasicType \verb@|@\\
\rr ``typeof''\/ ``(''\/ QualifId  [ ``\verb@<@''\/ TypeList ``\verb@>@''\/ ]  ``)''\/

\p{BasicType} ::=   [ ``cyan.lang'' ] BasicTypeNoPackage

\p{BasicTypeNoPackage} ::=   ``Byte'' \verb@|@ ``Short''   \verb@|@ ``Int'' \verb@|@ ``Long'' \verb@|@\\
\rr  ``Float'' \verb@|@ ``Double'' \verb@|@ ``Char'' \verb@|@ ``Boolean''

\p{SlotDec} ::= AnnotList QualifProtec ( ObjectVariableDec\\
\rr  \verb@|@ MethodDec  )

\p{ConstDec} ::=   [ ``shared''\/ ] ``let''\/ Type Id ``=''\/ Expr  [ ``;''\/ ]

\p{MethodDec} ::=  [ ``final''\/ ] [ ``override''\/ ]  [ ``abstract''\/ ]  \\
\rr [ ``overload''\/ ]  ``func''\/ MethodSigDec \\
\rr ( MethodBody \verb@|@ ``=''\/ Expr  ``;''\/  )

\p{MethodSigDec} ::=  ( MetSigNonGrammar \verb@|@  MetSigUnary \verb@|@\\
\rr MetSigOperator  ) [ ``\verb|->|''\/ Type ]

\p{MetSigNonGrammar} ::=  \{ SelecWithParam \}+

\p{MetSigUnary} ::= Id

\p{MetSigOperator} ::= UnaryOp \verb@|@ BinaryOp ( ParamDec \verb@|@ ``(''\/ ParamDec ``)''\/ )

\p{SelecWithParam} ::= IdColon \verb@|@ \\
\rr [ ``[]''\/ ] IdColon  ParamList

\p{TypeOrParamList} ::= TypeList \verb@|@ ParamList

\p{TypeList} ::= Type \{ ``,''\/ Type \}

\p{ParamList} ::= ParamDec \{ ``,''\/ ParamDec \}  \verb@|@\\
\rr  ``(''\/ ParamDec \{ ``,''\/ ParamDec \} ``)''\/

\p{ParamDec} ::=  [ Type ] Id


\p{MethodBody} ::= ``\{''\/ StatementList ``\}''\/

\p{ObjectVariableDec} ::=  [ ``shared''\/ ] ``var''\/  Type Id \{ ``,''\/ Id \} [ ``;''\/ ] \verb@|@\\
\rr [ ``shared''\/ ] ``var''\/  Type Id [ ``=''\/ Expr ] [ ``;''\/ ] \verb@|@ \\
\rr [ ``shared''\/ ] [ ``let''\/ ]  Type Id [ ``=''\/ Expr ] [ ``;''\/ ] \verb@|@\\
\rr [ ``shared''\/ ] [ ``let''\/ ]  Type Id [ ``,''\/ Id ]  [ ``;''\/ ]

\p{FunctionDec} ::= ``{''\/ [ ``\verb@(:@''\/  FuncSignature ``\verb@:)@''\/ ] StatementList ``}''\/

\p{FuncSignatureRet} ::= FuncSignature [ ``\verb@->@''\/ Type ]

\p{FuncSignature} ::=    ( Type Id \verb@|@ Type ``self''\/ ) \{ ``,''\/ Type Id \}   [ ``\verb@->@''\/ Type ] \verb"|" \\
\rr                      [ Type ``self''\/ ``,''\/ ]  IdColon  \{  Type Id \} \{ ``,''\/ Type Id \}

\p{QualifId} ::=  Id \{ ``.''\/ Id \}

\p{IdList} ::= Id \{ ``,'' Id \}

\p{InterfaceDec} ::= ``interface''\/ Id [ TemplateDec ] [ ``extends''\/ TypeList ] \\
\rr \{ ``func''\/ InterMethSig \}\\
\rr ``end''\/

\p{InterMethSig} ::=  InterMethSig2  [ ``\verb|->|''\/ Type ] \/

\p{InterMethSig2} ::= Id \verb@|@\\
\rr   \{ IdColon [ InterParamDecList ]  \}+ \verb@|@\\
\rr UnaryOp \verb"|" \\
\rr BinaryOp ( SingleInterParamDec \verb@|@ ``(''\/ SingleInterParamDec ``)''\/ )

\p{InterParamDecList} ::= WithoutParentDecList \verb"|" WithParentDecList

\p{WithoutParentDecList} ::= ParamTypeDecList  \{ ``,''\/ ParamTypeDecList  \}

\p{ParamTypeDecList} ::= Type [ Id ]

\p{WithParentDecList} ::= ``(''\/ WithoutParentDecList ``)''\/

\p{SingleInterParamDec} ::= Type Id

\p{StatementList} ::=  Statement \{ ``;''\/ Statement \}  \verb@|@ $\epsilon$

\p{Statement} ::= ExprAssign  \verb@|@  ReturnStat  \verb@|@  VariableDec \verb@|@ Annotation \verb@|@ \\
\rr IfStat \verb@|@ WhileStat \verb@|@ ForStat \verb@|@ NullStat\\
\rr PlusPlusStat \verb@|@ MinusMinusStat \verb@|@ CastStat \verb@|@ \\
\rr TypeStat \verb@|@ TryStat  \verb@|@ ThrowStat

\p{VariableDec} ::=  ``var''\/ [ Type ] Id [ ``=''\/ Expr ] \{ ``,''\/ Type Id  [ ``=''\/ Expr ] \}  \verb@|@ \\
\rr ``let''\/ [ Type ] Id  ``=''\/ Expr \{ ``,''\/ Type Id  ``=''\/ Expr \}

\p{ReturnStat} ::= ``return''\/ Expr \verb@|@ ``\verb@^@''\/ Expr

\p{ForStat} ::= ``for''\/ Id ``in''\/ Expr StatListBracket

\p{IfStat} ::= ``if''\/ Expr  StatListBracket   \\
\rr \{ ``else''\/ ``if''\/  Expr  StatListBracket  \}\\
\rr [ ``else''\/ StatListBracket ]

\p{CastStat} ::=  ``cast''\/ [ Type ] Id  ``=''\/ Expr  \{ [ Type ] Id  ``=''\/ Expr \} StatListBracket\\
\rr [ ``else''\/ StatListBracket ]

\p{WhileStat} ::= ``while''\/  Expr StatListBracket

\p{TryStat} ::= ``try''\/ StatementList \{ ``catch''\/ Expr \} [ ``finally'' StatListBracket ]

\p{ThrowStat} ::= ``throw'' Expr ``;''

\p{TypeStat} ::= ``type''\/ Expr \{ CaseClause \}  [ ``else''\/ StatListBracket ]

\p{CaseClause} ::= ``case''\/ Type [ Id ] StatListBracket

\p{StatListBracket} ::=  ``\{''\/ StatementList ``\}''\/

\p{NullStat} ::= ``;''\/

\p{PlusPlusStat} ::= ``++''\/ Id

\p{MinusMinusStat} ::= ``--''\/ Id

\p{ExprAssign} ::= Expr [ Assign ]

\p{Assign}  ::= \{ ``,''\/ Expr \}  ``=''\/ Expr

\p{Expr} ::= ExprOr [ MSendNonUn ]  \verb"|" \\
\rr [ Annotation ] MSendNonUn  \verb"|"

\p{MSendNonUn} ::=  \{ [ BACKQUOTE ] IdColon [ RealParameters ] \}+ \verb"|"\\
\rr \{ InterIdColon [ RealParameters ] \}+ \verb"|"\\
\rr \{ InterDotIdColon [ RealParameters ] \}+

\p{BinaryOp} ::= ShiftOp \verb@|@ BitOp \verb@|@ MultOp \verb@|@ AddOp \verb@|@ RelationOp \verb@|@\\
\rr ``\verb@~||@''\/ \verb@|@   ``\verb@..@''\/ \verb@|@ ``\verb@..<@''\/

\p{RealParameters} ::= ExprOr \{ ``,''\/ ExprOr \}

\p{ExprOr} ::= [ Annotation ] ExprXor \{ ``\verb@||@''\/ ExprXor \}

\p{ExprXor} ::= ExprAnd \{ ``\verb@~||@''\/ ExprAnd \}

\p{ExprAnd} ::= ExprEqGt \{ ``\verb@&&@''\/ ExprEqGt \}

\p{ExprEqGt} ::= ExprExc \{ (``\verb@=>@'' \verb@|@ ``\verb@=>@'') \/ ExprExc \}

\p{ExprExc} ::= ExprRel \{ ``\verb@&&@''\/ ExprRel \}


\p{ExprRel} ::= ExprMSendNonUn [ RelationOp ExprMSendNonUn ]

\p{ExprMSendNonUn} :: = MSendNonUn \\
\rr ExprOrGt [ MSendNonUn ] \\
\rr ``super''\/  MSendNonUn

\p{ExprOrGt} ::= ExprBPP \{ ``|>'' ExprBPP \}

\p{ExprBPP} ::= ExprInter \{ (``++'' \verb@|@ ``--'') ExprInter \}

\p{ExprInter} ::= ExprAdd [ ( ``\verb|..|''\/ \verb@|@ ``\verb|..<|''\/ ) ExprAdd ]

\p{ExprAdd} ::= ExprMult \{ AddOp ExprMult \}

\p{ExprMult} ::= ExprBit \{ MultOp  ExprBit \}

\p{ExprBit} ::= ExprShift \{ BitOp  ExprShift \}

\p{ExprShift} ::= ExprColonColon [ ShiftOp ExprColonColon ]

\p{ExprColonColon} ::= ExprDotOp \{ ``::''\/ ExprDotOp \}

\p{ExprDotOp} ::= ExprUnaryUnMS \{ DotOp ExprUnaryUnMS \}

\p{DotOp} ::= ``\verb|.*|''\/ \verb"|"  ``\verb|.+|''\/

\p{ExprUnaryUnMS} ::= ExprUnary \{ UnaryId \}

\p{UnaryId} := [ BACKQUOTE ] Id \verb"|" InterId \verb"|" InterDotId

\p{ExprUnary} ::= [ UnaryOp ]  ExprPrimaryIndexed

\p{ExprPrimaryIndexed} ::= ExprPrimary \{ Indexing \}

\p{Indexing} ::= ``[''\/ Expr ``]''\/  \verb"|"  ``?[''\/ Expr ``]?''\/

\p{UnaryOp} ::=  ``$+$'' \verb@|@ ``$-$''   \verb@|@ ``\verb"!"''\/ \verb@|@ ``\verb"~"''\/

\p{ExprPrimary} ::= ``self''\/ [ ``.''\/ Id ] \verb@|@\\
\rr ``self''\/  \verb@|@\\
\rr ``super''\/ UnaryId \verb@|@\\
\rr QualifId  \{ ``\verb@<@''\/ TypeList ``\verb@>@''\/ \}+ [ ObjectCreation ] \verb@|@\\
\rr ``typeof''\/ ``(''\/ QualifId   [ ``$<$''\/ TypeList ``$>$''\/ ] ``)''\/ \\
\rr ExprLiteral \verb@|@  ``(''\/ Expr ``)''\/

\p{ObjectCreation} ::= ``(''\/ [ Expr \{ ``,''\/ Expr \} ] ``)''\/

\p{ExprLiteral} ::= ByteLiteral \verb@|@ ShortLiteral \verb@|@ IntLiteral  \verb@|@\\
\rr LongLiteral \verb@|@ FloatLiteral \verb@|@  DoubleLiteral \verb@|@  CharLiteral \verb@|@\\
\rr BooleanLiteral \verb@|@ Str \verb@|@ AtStr \verb@|@ SymbolLiteral \verb@|@ ``Nil''  \verb@|@ \\
\rr LiteralArray \verb@|@ FunctionDec \\
\rr LeftCharString TEXT RightCharString \verb@|@ LiteralTuple

\p{BooleanLiteral} ::= ``true''\/ \verb@|@ ``false''\/

\p{LiteralArray} ::= ``[''\/ [ Expr ``,''\/  \{ Expr \} ] ``]''\/

\p{LiteralTuple} ::= ``[.''\/  TupleBody \verb@|@ UTupleBody ``.]''\/

\p{TupleBody} ::= (IdColon \verb@|@ Id ``:''\/ )  Expr \{ ``,''\/ IdColon Expr \}

\p{UTupleBody} ::= Expr \{ ``,''\/ Expr \}

\p{ShiftOp} ::= ``\verb@<.<@'' \verb@|@ ``\verb@>.>@'' \verb@|@ ``\verb@>.>>@''

\p{BitOp} ::= ``\verb@&@'' \verb@|@  ``\verb@|@''  \verb@|@ ``\verb@~|@''

\p{MultOp} ::= ``$/$'' \verb@|@ ``$*$'' \verb@|@ ``\%''

\p{AddOp} ::= ``$+$'' \verb@|@ ``$-$''

\p{RelationOp} ::= ``$==$'' \verb@|@ ``$<$'' \verb@|@ ``$>$'' \verb@|@ ``$<=$''
\verb@|@ ``$>=$'' \verb@|@  ``$!=$''  

\chapter{Opportunities for Collaboration}  \label{collaboration}

There are many research projects that could be made with Cyan and on Cyan:

\begin{enumerate}[(a)]
\item to implement metaobjects \verb"dynOnce" and \verb"dynAlways" and
to design algorithms that help the transition of dynamically-typed Cyan to statically-typed Cyan. There are a great deal of work here, at least several master thesis. This work can involve the discovery of types statically (at least most of them), the use of a profiler to discover some types at runtime, the combination of static and dynamic type information, refactorings directed by the user (he/she chooses the type of each troublesome variable/parameter/return type, for example), help by the IDE, etc.

     It would be very important to have a language in which the programmer could develop a program without worrying about types in variables/parameters/return values and then convert this program to statically-typed Cyan. I would say that this is one of the central points of the language;


\item implement some Design Patterns using compile-time metaobjects;

\item implement some literal objects which are the code of some small languages such as AWK and SQL. It would be nice if Cyan code could be used inside the code of the language;

\item to use Cyan to implement a lot of small Domain Specific Languages;

\item to use Cyan to investigate language-oriented programming \cite{Ward95languageoriented};

\item to add parallelism \index{parallelism} to the language and to design a library for distributed programming. That includes the implementation of patterns for parallel programming;

\item to design code optimization algorithms for Cyan;

\item to program the Cyan basic libraries for handling files, data structures, and so on;

\end{enumerate}

\chapter{Future Enhancements}

Some Cyan features may be changed and others may be added. This is a partial list of them:

\begin{enumerate}

\item more methods to intervals:
\begin{cyan}
    (2..500 but: 100..200, 37 select: Prime) print;
\end{cyan}
{\tt Prime} here is a context object:
\begin{cyan}
object Prime extends Function<Int, Boolean>
    func eval: Int elem -> Boolean {
        return elem prime
    }
end
\end{cyan}

\item intervals with {\tt Floats} and {\tt Doubles}:
\begin{cyan}
    1.0..5.0 step: 0.01  repeat: { (: Float elem :)
        graphFun plot: elem, (sin: elem);
        };

    -1.0..1.0 but: 0.0 step: 0.1 repeat: { (: Float elem :)
        graphFun plot: elem, (sin: elem);
        };
\end{cyan}
\verb|"-1.0..10 but: 0.0"|  is an object of \verb|Interval<Float>|. So is \verb|"-1.0..10 but: 0.0"| and
\verb|"-1.0..10 but: -0.1..0.1"|. This prototype has methods {\tt step:repeat:},
{\tt but:step:repeat:}, {\tt contains:}, etc (see prototype {\tt Int}). Maybe {\tt but:} should be replaced by binary minus (\verb|"-"|).

With {\tt GraphFun} there could be supplied a context object:
\begin{cyan}
object Plot<FunToPlot>(GraphFun graphFun) extends Function<Float, Nil>
    func eval: Float elem {
        graphFun plot: elem, (FunToPlot: elem);
    }
end
\end{cyan}

A literal object could be aware of the context:
\begin{cyan}
    var GraphFun graphFun = GraphFun new;
    graphFun  x: -10, 10 y: -40, 40;
    1.0..5.0 step: 0.01  repeat: @plot(sin);
\end{cyan}
{\tt @plot} would produce
\begin{cyan}
    Plot<sin>(graphFun)
\end{cyan}
It would discover that there is a local variable {\tt graphFun} of the correct type.
If there are two, {\tt plot} would sign a compilation error;

\item private generic prototypes, which are currently illegal;

\item {\tt package} qualifier for prototypes and methods. These could only be used in the current package;

\item {\tt typeof} may be legal as a real argument in a generic prototype instantiation:
\begin{cyan}
var Int count = 0;
var Stack<typeof(count)> intStack; // ok
\end{cyan}

\item A {\tt finally:} keyword may be added to the initial function that starts the program execution. That would allow finalizers, code that is called when the program ends. There could be a list of methods to be called when the program ends. This is odd, but someone will certainly like it.

    \begin{cyan}
    {
    } catch: RuntimeCatch
      finally: {
          DoomsdayWishList foreach: { (: Function<Nil> elem :) elem eval };
      };
    \end{cyan}

    In some other place:
    \begin{cyan}
     DoomsdayWishList add: { "Good bye!"  print };
    \end{cyan}

\end{enumerate}

The features given below were removed from Cyan. They may be added later maybe in a different form.

\section{Runtime Metaobjects or Dynamic Mixins}  \label{runtimemetaobjects}

Mixin prototypes can also be dynamically attached to objects. Returning to the {\tt Window}-{\tt Border} example, assume {\tt Window} does not inherit from {\tt Border}. This mixin can by attached to {\tt Window} at runtime by the statement:  \label{attachMixin}
\begin{cyan}
Window attachMixin: Border;
\end{cyan}
Effectively, this makes {\tt Border} a metaobject with almost the same semantics as shells of the Green language \cite{conf/ecoop/Guimaraes98}. Any messages sent to {\tt Window} will now be searched first in {\tt Border} and then in {\tt Window}. When {\tt Window} is cloned or a new object is created from it using {\tt new}, a new {\tt Border} object is created too.

As another example, suppose you want to redirect the {\tt print} method of object {\tt Person} so it would call the original method and also prints the data to a printer. This can be made with the following mixin:
\begin{cyan}
mixin(Any) object PrintToPrinter
    override func print {
        super print;
            // print to a printer
        Printer print: (self asString)
    }
end
\end{cyan}
``{\tt self asString}''\/ returns the attached object as a string, which is printed in the printer by method {\tt print:}. This mixin can be added to any object adding a {\tt print} method to it:  \index{attachMixin}
\begin{cyan}
object Person
    @property String name
    @property Int age
    override func asString -> String {
        return "name: $name age: $age"
    }
end
...
var p = Person new;
p name: "Carol";
p age: 1;
p attachMixin: PrintToPrinter;
    // prints both in the standart output and in the printer
p print;
Person name: "fulano";
Person age:  127;
    // print only in the standard output
Person print;
\end{cyan}

Note that {\tt attachMixin} is a special method of prototype {\tt Any}: it is added by the compiler and it can only be called by sending messages to the prototype. These dynamic mixins are runtime metaobjects. Probably they can only be efficiently implemented by changing the Java Virtual Machine (but I am not so sure). Maybe efficient implementation is possible if the metaobjects (dynamic mixins) that can be attached to an object are clearly identified:
\begin{cyan}
object(PrintToPrinter) Person
    @property String name
    @property Int age
    override func asString -> String {
        return "name: $name age: $age"
    }
end
\end{cyan}
Then only {\tt PrintToPrinter} metaobjects can be dynamically attached to {\tt Person} objects.

The last dynamic mixin attached to an object is removed by method {\tt popMixin} defined in prototype {\tt Any}.  It returns {\tt true} if there was a mixin attached to the object and {\tt false} otherwise. Therefore we can remove all dynamic mixin of an object {\tt obj} using the code below.
\begin{cyan}
while obj popMixin {
}
\end{cyan}

The above definition of runtime mixin objects is similar to the definition of runtime metaobjects of Green \cite{conf/ecoop/Guimaraes98}. The semantics of both are almost equal, except that Green metaobjects may declare a {\tt interceptAll} method that is not supported by mixin objects (yet).

\section{Multiple Assignments}

Cyan supports a restricted form of \index{assignment!multiple} multiple assignments. There may be any number of comma separated assignable expressions in the left-hand side of ``{\tt =}''\/ if the right-hand side is a tuple (named or unnamed) with the compatible types. That is, it is legal to write\\
\nd \verb@    v1, v2, ..., vn = tuple@\\
\nd if {\tt tuple} is a tuple with at least {\tt n} fields and the type of field number {\tt i} (starting with {\tt 1}) is a subtype of the type of {\tt vi}.
\begin{cyan}
var Float x, y;
x, y = [. 1280, 720 .];
var tuple = [. 1920, 1080 .];
x, y = tuple;
\end{cyan}
However, a variable cannot be declared in a multiple assignment:\\
\nd \verb@    var x, y = [. 1280, 720 .]@\\
\nd The compiler would sign an error in this code.

The assignment ``\verb@v1, v2, ..., vn = tuple@''\/ is equivalent to
\begin{cyan}
   // tuple may be an expression
var tmp = tuple;
vn = tmp fn;
...
v2 = tmp f2;
v1 = tmp f1;
\end{cyan}
A multiple assignment is an expression that returns the value of the first left-hand side variable, which is {\tt v1} in this example.

A method may simulate the return of several values using \index{assignment!tuple} \index{tuple} tuples.
\begin{cyan}
object Circle
    func getCenter -> Tuple<Float, Float> {
        return [. x, y .]
    }
    ...
    private Float x, y  // center of the circle
    private Float radius
end
...

var Float x, y;
x, y = Circle getCenter;
\end{cyan}

\bibliography{C:/Dropbox/research/myArticles/bibtex/allbib}

\vspace*{4ex}
\clearpage
\appendix

\chapter{The Compiler}

\label{thecompiler}

The Cyan site, \url{www.cyan-lang.org}, has a download tab from which file {\tt lib.zip} can be downloaded. Put this file in a directory, say, \verb|C:\Dropbox\Cyan|. Uncompress it resulting in directory \\
\verb|    C:\Dropbox\Cyan\lib|\\
\noindent Directory {\tt lib} contains several jar files, the Cyan runtime libraries, the compiler (saci.exe), and the source code of the basic libraries.

Set the system environment variable {\tt CYAN\_HOME} to directory {\tt lib}.
Set the system environment variable {\tt JAVA\_HOME\_FOR\_CYAN} to the JDK directory of Java 8. Then the value of this variable can be\\
\verb|    C:\Program Files\Java\jdk1.8.0_241|\\
\noindent

The compiler should be called as follows.
\begin{verbatim}
    saci projectDirectoryOrName compilerOptions
\end{verbatim}
The compiler name is {\tt saci.exe}.
{\tt projectDirectoryOrName} is the file name of the project or the directory name in which the project is.
If {\tt projectDirectoryOrName} is a file name, it should have extension ``{\tt .pyan}''\/. Its contents should be as described in Section~\ref{pyan}. If {\tt projectDirectoryOrName} is a directory, the compiler will create a project file ``\verb@projectDirectoryOrName\project.pyan@''\/. It will consider that the program consists of all directories inside {\tt projectDirectoryOrName}. Each directory given origin to a package. The {\tt .cyan} files inside each directory contain the prototypes of the package.

The easy way of compiling a Cyan program is to put all packages in the same super-directory --- see the examples in the Cyan site. Then this directory should be {\tt projectDirectoryOrName}.

{\tt compilerOptions} is a list of options. The valid options are:
\begin{itemize}
\item \verb|-noexec|, the Java code produced by saci is compiled by the Java compiler but it is not executed;
\item \verb|-nojavac|, the Java compiler is not called for compiling the Java code produced by the Cyan compiler;
\item \verb|-args argList|, arguments to the Cyan program. The arguments that follow '-args', argList, will be passed to the Cyan program if it is to be executed. It is not an error to have both options
    \verb|-noexec| and  \verb|-args argList|. Of course, this should be the last option in the command line;
\item \verb|-sourcePath aPath| for supplying 'aPath' for the Java compiler. This option can appear any number of times. Each time can be composed by multiple paths, separated by ';'.
\item \verb|-cp aPath| for supplying 'aPath' for the Java interpreter. This option can appear any number of times. Each time can be composed by multiple paths, separated by ';';
\item \verb|-es filename| for interpreting the remaining arguments as Cyan code. After the interpretation, the compiler exits;
\item \verb|-ef filename| for interpreting the statements of file 'filename'. After the interpretation, the compiler exits. Example:\\
    \verb|    saci -ef "C:\Dropbox\tests\first.syan"|\\
    \noindent The recommended file extension is ``tyan'' which stands for ``inTerpreted cYAN''.
\end{itemize}
As an example of calling the compiler, we can have
Example:
\begin{verbatim}
saci "C:\Dropbox\Cyan\cyanTests\simple"
   -cp "C:\Dropbox\Cyan\externalLibs"
   -args 0 "C"
\end{verbatim}

The compiler can be also be called programmatically by calling methods {\tt parseSingleSource} and {\tt compileProject} of a {\tt saci.Saci} object --- see the source files of the Cyan compiler. These methods are usually called by the IDE.  An IDE plugin should create a single object of class {\tt Saci} for each Cyan project. It is important that a single {\tt Saci} object is used because this object will keep information from the last compilation that will speed up the next one.

The IDE can call method {\tt compileProject} to compile the whole project. The parameters to this method correspond to the parameters passed in the command line when calling the compiler. The source files will be read from the file system.
\begin{verbatim}
    public void compileProject( String projectDirectoryOrName,
            String cyanLangDir,
            String javaLibDir,
            boolean exec, boolean callJavac, boolean parseOnly )
\end{verbatim}
Before calling this method, the IDE plugin should save all source files associated to this project that are being edited.

To simulate incremental compilation, the compiler should be called whenever the user changes the source code being edited. However, the Cyan compiler goes through a long and complex compilation process and it is not feasible to
compile a project or even a single source file after each editing command. However, an IDE  plugin can call the Cyan compiler to parse the source being edited several times a second. That will take just a few miliseconds.

To parse a single source file the IDE plugin should call method {\tt parseSingleSource}. Only parsing will be done.
\begin{verbatim}
    public boolean parseSingleSource( String cyanLangDir, String javaLibDir,
            String packageName, String prototypeName, char []sourceCodeToParse,
            String projectDirectoryOrName,
            char []sourceCodeProject,
            boolean loadProjectFromFile )
\end{verbatim}
Most of the parameters speak for themselves. {\tt sourceCodeToParse} is the source code of prototype {\tt prototypeName} of package {\tt packageName}. The IDE plugin should not read it from disk. It should be retrieved from the IDE editor to make parsing faster.
{\tt projectDirectoryOrName} is the directory of the project or the project file name (ending in ``{\tt .pyan}''\/). The source code of the project is given by {\tt sourceCodeProject}. It is the contents of the project file which is either {\tt projectDirectoryOrName} or \verb@projectDirectoryOrName\project.pyan@.

{\tt packageName}, {\tt prototypeName}, and {\tt sourceCodeToParse} cannot be {\tt null}. If {\tt sourceCodeProject} is {\tt null},  {\tt loadProjectFromFile} is true, or the {\tt Saci} object has not kept the ``project object''\/ from the last compilation, then the project file is reload from disk to memory and compiled, producing a {\tt saci.Project} object (see classes of the Cyan compiler). This object can be retrieved by calling method {\tt getProject()} of {\tt Saci}.  Through this object one can get the source code of the project that should be passed to the call to {\tt parseSingleSource} in the next time it is called. The flow of control would be:
\begin{verbatim}
      // flow, not source code --- you got the idea
   Saci aSaci = new Saci();
   char []sourceCodeProject = null;
   ...
   aSaci.parseSingleSource(..., sourceCodeProject, true);
      // now the project does exist, if there was not any errors
   Project p = aSaci.getProject();
   if ( p != null )
       sourceCodeProject = p.getText();
   aSaci.parseSingleSource(..., sourceCodeProject, false);
   ...
\end{verbatim}

The IDE plugin should call {\tt parseSingleSource} of {\tt Saci} when the text is changed by adding or removing
characters (and after the user pauses typing). Method {\tt eventChangeSourceCodeBeingEdited} of {\tt Saci} should be called by the IDE plugin whenever the source code being edited is changed to a new one. For example, the user was editing a file {\tt Program.cyan} and she changes the focus to file {\tt Test.cyan}. The new file is parsed and old data is discarded.

After the IDE plugin calls {\tt parseSingleSource}, it can retrieve information on the symbols produced by the lexical analyzer. This is made by calling method {\tt getSymbolList()} of {\tt Saci}. This list has size {\tt getSizeSymbolList()} and it can be used to highlight keywords, change the color of strings, etc.
{\tt getSymbolList()} returns an object of {\tt Symbol[]}. See package {\tt lexer} of the Cyan compiler for
subclasses of {\tt Symbol}. In general, {\tt symbol.getSymbolString()} returns the symbol as a
String.

When the mouse is over the source code being edited, the IDE should call method\\
\verb|    |{\tt searchCodegAnnotation(int line, int column)} \\
\noindent or \\
\verb|    |{\tt searchCodegAnnotation(int offset)}\\
\noindent
of {\tt Saci}. The parameters of the first method are the line and column of the text over which the mouse pointer is. The parameters of the second method is the offset in the text over which the mouse pointer is.
If the mouse is over a Codeg, any of these methods will return an object of {\tt CyanMetaobjectWithAtAnnotation} of package {\tt meta} of the Cyan compiler (or {\tt null} if the mouse is not over a Codeg). This is a class of the AST and the object represents an annotation, the Codeg annotation.
This
object should be passed to method {\tt eventCodegMenu(Saci, CyanMetaobjectWithAtAnnotation)} of {\tt Saci} that is responsible
to call method {\tt getUserInput} of the Codeg. This method should open a menu and accept user input
through the mouse and keyboard. After the input is done, button \verb|"Ok"| should be pressed (there should be such a button). If {\tt moAnnotation} is a reference to the Codeg passed as parameter to {\tt eventCodegMenu},
{\tt moAnnotation.getUserInput(...)} returns a byte array that is the information of the Codeg (after showing a menu and getting user input). That is,
user input through mouse and keyboard is translated into a byte array that is returned by {\tt getUserInput}. This byte array is written to a special file in disk that is managed by the compiler. This array may be a DSL when converted to text (a String) --- or it may be not. The IDE plugin should not call {\tt getUserInput()}. This method is called by the compiler inside {\tt eventCodegMenu()}. The byte array returned by {\tt getUserInput} is passed as the second parameter in the next time this method is called on the same Codeg. In this way the Codeg has the previous information --- the previous chosen color of Codeg {\tt color}, for example. In the first time {\tt getUserInput} is called, the second parameter is {\tt null}.

After calling {\tt eventCodegMenu}, the byte array returned by {\tt getUserInput} can be retrieved by the IDE plugin by calling method {\tt getCodegInfo()} of the Codeg annotation object. The flow of control would be like
\begin{verbatim}
CyanMetaobjectWithAtAnnotation moAnnotation = searchCodegAnnotation(line, column);
if ( moAnnotation != null ) {
    eventCodegMenu(moAnnotation);
    byte []codegInfo = moAnnotation.getCodegInfo();
    ...
}
\end{verbatim}

The IDE plugin may show a list of Codegs of the current source file in a window. By clicking in one Codeg of the list, {\tt eventCodegMenu} could be called.

To avoid access to the hard disk, information on the Codegs of the current source file is kept in memory as well as in files. When the information is updated by calling {\tt getUserInput} it is written to disk. But when the information in only read it is read from memory. Then if some external source changes the Codeg files the compiler will not know that. It will continue to use the values that are in memory. A complete recompilation is necessary. When a source code is parsed the compiler reads Codeg information that is in files. The file associated to a Codeg annotation has the name composed by the annotation name and the first parameter to the annotation. It is created in a directory \verb|--prototypeName| of the directory of the project. Only the compiler knows where it keeps these files.

Method {\tt parseSingleSource} read Codeg information from the files in the first time it is called. From the second time on it reads the information from a hash table the compiler keeps with Codeg information of the current source code. Whenever the graphical interface of a Codeg changes the information a disk file is updated. When the Codeg information is just read it is read from the hash table in memory.

It is time to give some low-level details of handling Codegs and symbols.\\
\verb|     Symbol sym = aSaci.getCodegList().get(0).getFirstSymbol();|\\
\noindent {\tt sym} is the first symbol of the first codeg of the last source code parsed (if there is one Codeg in the last source code parsed). By ``first''\/ we mean the Codeg with smaller (line, column) in the text of the last source code parsed. ``{\tt sym.getLineNumber()}''\/ is the line of the Codeg call and ``{\tt sym.getColumnNumber()}''\/ is the column of the ``\verb|@|''\/ symbol. ``{\tt sym.getSymbolString()}''\/ will return ``{\tt color}''\/ if the Codeg call is that of the line\\
\verb|    var Int c = (@color(red));|\\
\noindent Although the type of {\tt sym} is {\tt Symbol}, the runtime class of it is {\tt lexer.SymbolCyanMetaobjectAnnotation}.

The call\\
\verb|     | {\tt aSaci.getCodegList().get(0).getLastSymbol()}\\
\noindent returns a object describing the symbol {\tt )} in the Codeg call \verb|@color(red)|. Then the IDE plugin have sufficient information to highlight the Codeg call.

Assume method {\tt parseSingleSource} of object referenced by {\tt aSaci} has been called. Then the list of symbols of the current source code is\\
\verb|    aSaci.getSymbolList()|\\
The IDE should color symbol referenced by variable {\tt sym} with color {\tt sym.getColor()}, that returns an {\tt int}. The  symbol is in line {\tt sym.getLineNumber()} and column {\tt sym.getColumnNumber()}. Its number of characters is usually\\
\verb|    sym.getSymbolString().length()|\\
\noindent In some cases, method {\tt getSymbolString()} does not return the string of characters of the symbol. For example, this happens in a metaobject annotation. Some metaobject annotations may be represented as  symbols but {\tt getSymbolString} does not return the whole text of the call.

Metaobjects are used to implement Domain Specific Languages (DSL) that may need specific color schemes.
\begin{cyan}
    @demands{*
        T in [ "Int", "Short", "Byte", "Char", "Long" ],
        U implements Savable,
        R subtype Person
    *}
    object Proto<T, U, R>
        ...
    end
\end{cyan}
To allow that, a metaobject class may redefine the inherited method
\begin{verbatim}
    ArrayList<Tuple4<Integer, Integer, Integer, Integer>>
    getColorList()
\end{verbatim}
This method returns a list of tuples. Each tuple has the format \\
\verb|    (colorNumber, lineNumber, columnNumber, size)|\\
\noindent It means that the characters starting at {\tt lineNumber} and {\tt columnNumber} till {\tt columnNumber + size} should be highlighted with color {\tt colorNumber}.

The line number is relative to the first character of the metaobject annotation. Any metaobject class may redefine method {\tt getColorList}, including number literals (example: {\tt 0101bin}) or string literals (example: \verb|r"[a-z]+0*"|).

The compiler class {\tt saci.IDEPluginHelper} defines a method {\tt getColorList} that returns a list of tuples that can be used for syntax highlighting of the last source file parsed (method {\tt parseSingleSource} should have been called first). {\tt getColorList} takes a {\tt Saci} object as parameter.

After compiling a Cyan program or parsing a Cyan source file, the errors can be retrieved by calling two {\tt Saci} methods: {\tt getProjectErrorList()} and {\tt getCyanErrorList()}. The first method returns a list of errors of class {\tt ProjectError}. A {\tt ProjectError} is an error that occurs outside a Cyan source file. For example, an error in the project file (``.pyan''\/), in the arguments of {\tt parseSingleSource}, and so on.

Method {\tt getCyanErrorList()} return a list of errors of class {\tt UnitError}. Every object of this class describes an error that occurred inside a Cyan source file.

\addcontentsline{toc}{chapter}{Index}
\printindex

\section{Separate Compilation}

A Cyan package can be packed into a jar file except its generic prototypes, which should continue to be in the package directory. To force the compilation of package {\tt cyan.lang}, attach the following annotation to {\tt program} in the project file:
\begin{cyan}
    @feature("compilePackageCyanLang", true)
    program
        ...
\end{cyan}
Unless this annotation is in the project file, the compiler will look for a {\tt cyan.lang.jar} file in directory given by the environment varible {\tt CYAN\_HOME}.

To compile a package named {\tt aaa.bbb} that is not {\tt cyan.lang} for separate compilation, create a project with the following contents:
\begin{cyan}
    program
        @annot("compilePackage")
        package aaa.bbb
        // possibly other packages
\end{cyan}
Now call the Java compiler to compile {\bf all} of the Java files for the package. Saci will call the Java compiler but only the files reachable from the main class will be compiled. Therefore it is necessary to call the Java compiler again. Finally, call the {\tt jar} program to create a jar file. More instructions are given in the Cyan site.

The jar file should be in the {\it base directory} of the package. That is, for package {\tt aaa.bbb}, the directory structure should be as shown below.
\begin{cyan}
    aaa.bbb.jar
    aaa\
        bbb\
            // package prototypes
\end{cyan}
Whenever a package is referenced, the Cyan compiler will look for a jar file in the base directory of the package. That is, if the the project file is
\begin{cyan}
    program
        package ccc.ddd at "C:\Dropbox\tests\00\ccc\ddd"
        // possibly other packages
\end{cyan}
then the compiler will look for \\
\verb|    C:\Dropbox\tests\00\ccc.ddd.jar"|\\
\noindent
If there is no such file, Saci will compile the files of the package directory. Note that even if there is a jar file, generic prototypes of package {\tt ccc.ddd} should be kept in the package directory. They will be searched for in that directory.

\section{Known Compiler Errors}  \label{compilererrors}

This is a list of known compiler errors.

\begin{enumerate}

\item a Java class that is not imported can be used if it is in file ``rt.jar'';

\item The code
\begin{cyan}
    var Int n;
    repeat
        if true {
            break
        }
        else {
            n = 0
        }
        n println
    until true;
\end{cyan}
should not cause the error ``local variable n may not have been initialized'' in line\\
\verb|    n println|

However, it does.

\item literal hash tables (maps, dictionaries) with union types cause a Java compilation error.
\begin{cyan}
        let Int|String is95 = 0;
        let myMap = [ is95 -> is95, "zero" -> 0, 0 -> "zero" ];
\end{cyan}

\item the line number of the error below is incorrect because of annotation \texttt{insertCode}

\begin{cyan}
package main

object Program

    @insertCode{*
        for num in [ 2, 3, 4, 5 ] {
            insert: "    func multBy$num: Int n -> Int = n*$num;"
        }
    *}

    func test {
        zeroTen = 11;  // no variable zeroTen
    }
end
\end{cyan}

\end{enumerate}

\end{document}